# THE NATIONAL UNDERGROUND SCIENCE AND ENGINEERING LABORATORY AT HOMESTAKE: PROJECT BOOK, REFERENCE DESIGN STAGE

A Facility for Physics, Astrophysics, EarthLab,
Applied Science and Engineering, and Outreach/Education

July, 2003

submitted by

The Homestake Collaboration

## The Homestake Collaboration

**Executive Committee**
Baha Balantekin, University of Wisconsin
Thomas Bowles, Los Alamos National Laboratory
Janet Conrad, Columbia University
Sherry Farwell, South Dakota School of Mines and Technology
Wick Haxton, University of Washington
Ken Lande, University of Pennsylvania
Kevin Lesko, Lawrence Berkeley Laboratory
Bill Marciano, Brookhaven National Laboratory
Marvin Marshak, University of Minnesota
Tullis Onstott, Princeton University
Michael Shaevitz, Columbia University
John Wilkerson, University of Washington

**Scientific Advisor to the Executive Committee**

John Bahcall, Institute for Advanced Study

**Earth Science Steering Committee**

Brian McPherson, New Mexico Tech
Tullis Onstott, Princeton
Tommy Phelps, ORNL
Bill Roggenthen, SDSM&T
Herb Wang, Wisconsin
Joe Wang, LBNL

**Collaboration Members**

Craig Aalseth, Pacific Northwest National Laboratory
Daniel S. Akerib, Case Western Reserve University
Steven Anderson, Black Hills State University
Elena Aprile, Columbia University
Frank T. Avignone III, University of South Carolina
Tom Barket, South Dakota Science Teachers Association
Laura Baudis, Stanford University
John F. Beacom, Fermilab
Mark Boulay, Los Alamos National Laboratory
George Brimhall, UC Berkeley
Len Bugel, Fermilab
Thomas Campbell, SDSM&T
Art Champagne, University of North Carolina
Juan I. Collar, University of Chicago
F.S. Colwell, Idaho National Engineering and Environmental Laboratory
Peter Doe, University of Washington
Michael Dragowsky, Case Western Reserve University
Ed Duke, SDSM&T
Dan Durben, Black Hills State University
Tom Durkin, SDSM&T
Hiro Ejiri, RCNP, Osaka University
Steve Elliott, University of Washington
Royce Engstrom, University of South Dakota
Joseph Formaggio, University of Washington

Jim Fredrickson, Pacific Northwest National Laboratory
George M. Fuller, University of California, San Diego
Richard Gaitskell, Brown University
Maury Goodman, Argonne National Laboratory
Uwe Greife, Colorado School of Mines
Alec Habig, University of Minnesota Duluth
Andre Hamer (deceased), Los Alamos National Laboratory
Frank Hartmann, Max Planck Institute, Heidelberg
Karsten M. Heeger, Lawrence Berkeley National Laboratory
Andrew Hime, Los Alamos National Laboratory
Zbignew (Ziggy) J. Hladysz, Mining Engineering Program, SDSM&T
Chang Kee Jung, The State University of New York at Stony Brook
Jon Kellar, SDSM&T
Thomas L. Kieft, New Mexico Institute of Mining and Technology
Sally Koutsoliotas, Bucknell University
Robert Lanou, Brown University
Barbara Sherwood Lollar, University of Toronto
Clark McGrew, SUNY at Stony Brook
Harry Miley, Pacific Northwest National Laboratory
Jeffrey S. Nico, National Institute of Standards and Technology
Bob Noiva, University of South Dakota
Peter Parker, Yale University
Tommy J. Phelps, Oak Ridge National Laboratory
Andreas Piepke, University of Alabama
Alan Poon, Lawrence Berkeley National Laboratory
Lisa M. Pratt, Indiana University
Jan Puszynski, SDSM&T
Bill Roggenthen, SDSM&T
Bernard Sadoulet, University of California, Berkeley
Ben Sayler, Black Hills State University
Richard Schnee, Case Western Reserve University
Kate Scholberg, MIT
Tom Shutt, Princeton University
Panagiotis Spentzouris, Fermilab
Robert Svoboda, LSU
Joseph S. Y. Wang, Lawrence Berkeley National Laboratory
Peter J. Wierenga, University of Arizona
Raymond Wildung, Pacific Northwest National Laboratory
Paul Wildenhain, University of Pennsylvania
Patrick R Zimmerman, Institute of Atmospheric Sciences, SDSM&T

**Collaboration Engineers**

Jerry Aberle, Lead
John Marks, Lead
Gary Kuhl, Skyline Engineering
Jamie Stampe, Skyline Engineering

## A. PROJECT SUMMARY

This proposal is a condensed version of the **Project Book for the National Underground Science Laboratory, Homestake, at the Reference Design stage**. This submission represents a major step beyond the conceptual proposal our group submitted in June 2001. Much of the past two years was invested in strengthening and broadening the science case – an effort that involved not only our collaboration, but many of our colleagues in the broader science community who are interested in underground science. The science case itself has changed, with major new discoveries in neutrino physics occurring since June 2001, and with the development of a compelling program of NUSEL earth science. The community's arguments for NUSEL are summarized in the **Science Book**. In parallel with the science effort, our group has learned a great deal about the Homestake site and how it can be best adapted to meet the science requirements. This has led to the **Science Timeline and Reference Design**, and a facilities development plan much improved over that of the conceptual proposal. We describe this design – the access to underground, the underground and surface campuses, and the options remaining to be explored – and the engineering studies that allow us to assign costs and contingencies. We also describe the work remaining to be done and the program plan for producing a **Baseline Definition** of the Homestake project.

This project began in September 2000, when the INT hosted a group of 200 neutrino physicists in Seattle to discuss, in connection with the NSAC Long-Range Plan (LRP) for nuclear physics, possible priorities for this subfield. The deliberations of one of the meeting's working groups, on Underground Science Laboratories, was dramatically influenced by a proposal Lande made at this meeting, conversion of the Homestake Gold Mine into a National Underground Science and Engineering Laboratory (NUSEL). The availability of this very deep site, with massive shafts and lifts, sophisticated utilities, ventilation, and communications systems, and established operations costs, prompted the NSAC Town Meeting group to make the creation of NUSEL its highest priority. The National Science Foundation and Department of Energy responded by supporting an *ad hoc* study group, the Bahcall Committee, to consider the scientific case for NUSEL and the suitability of possible sites. The Committee's membership included leading underground scientists from particle, nuclear, and astrophysics. Its consultants included experts in earth science and large project management. The Bahcall Report, submitted to NSAC as a LRP White Paper, made a compelling scientific case for NUSEL and identified Homestake as the recommended site. In its final deliberations the NSAC LRP group made creation of a deep underground science laboratory its highest midscale construction priority for the next decade. Because Homestake closure plans limited the window in time when this site would be available, NSAC also wrote the NSF, urging the agency to proceed immediately with NUSEL-Homestake.

Following this decision a national group of underground scientists, several of whom had taken part in the Bahcall Committee and the Seattle, Oakland, and Santa Fe NSAC LRP meetings, collaborated on a NUSEL-Homestake proposal. This MRE proposal was submitted to the NSF in early June 2001, and was reviewed by the Physics Division later that summer. The results of the reviews – the Physics Division conducted two panel reviews in addition to soliciting written reviews – were shared with the proposers in October 2001.

Since submission of the original proposal, the creation of NUSEL and the associated science have generated extensive discussions in the scientific community. In HEPAP's long-range plan, which was debated over the last six months of 2001, neutrino physics, dark matter searches, and other underground science received strong support. Two high-level NRC committees reviewed the science arguments for NUSEL, both concluding that a deep US laboratory is needed. Major changes in the science have resulted from discoveries subsequent to June, 2001, including the Sudbury Neutrino Observatory demonstration that the solar neutrino flux is dominated by heavy-flavor neutrinos, the KamLAND verification of solar neutrino oscillations, the K2K results, and the identification of thermophilic methanogens at the 8000 ft level of Homestake. New communities, most notably those advocating NUSEL-Homestake as an "EarthLab" for earth science, geomicromiology, and rock mechanics/engineering and those concerned with applications of new detector technologies to a variety of post-9/11 issues, have joined the collaboration. A series of conferences and workshops – the Lead meetings on Underground Science and on Geomicrobiology, the Aspen Workshop on Underground Science, and the NSF-sponsored NESS02 conference – not only contributed to broadening the science, but also clarified the technical requirements (depth, space, utility needs) and readiness of proposed experiments. (These meetings were either organized by the NUSEL-Homestake Collaboration or strongly supported by our members. The materials from these meetings – talks presented, working group white papers – have been preserved on the NUSEL-Homestake web page, http://int.phys.washington.edu/NUSEL/.) Major improvements in the NUSEL design have occurred, allowing us to avoid the costly Yates shaft extension while providing a more versatile laboratory that meets the needs of our broader collaboration. Finally, we have continued to develop partnerships nationally and in the region with the goal of enhancing the public outreach potential of NUSEL-Homestake. This includes exploiting Homestake's unique location and history – the US mine most identified with the opening of the American west, located in a major tourist area – and developing links to K-12 and regional and national college and university students and educators. Of particular importance are the opportunities to work with NativeAmerican educators, through both established NSF programs at the tribal colleges and new ones, and to provide a major-science "anchor" for the EPSCoR states of the Northern Great Plains, which currently lack such a focus.

This "version #2" proposal summarizes the progress the Homestake Collaboration has made in developing and broadening the science case for NUSEL; in defining the technical requirements that NUSEL must meet; in integrating this project into national efforts in nuclear and particle physics, astrophysics, earth science and geomicrobiology, outreach and education, and applications to materials science and national security; in formulating a baseline design for NUSEL construction that takes maximal advantage of the extraordinary existing infrastructure of the Homestake Mine; in developing a project management plan; and in producing a Work Breakdown Structure sufficiently detailed to justify project funding in FY06. The full Project Book contains the following elements:

- An **Overview (Section B)** of the project, with special emphasis on our work and that of the community since June 2001. The overview summarizes major changes in the science scope of NUSEL and how these changes impact the technical aspects of the proposal. The overview reviews the conclusions of the various review committees that have dealt with NUSEL. The status of the Homestake site is also summarized.
- The **Science Book (Section C)**, in which the science case for NUSEL is updated. Important contributions to the Science Book came from community meetings organized by or strongly supported by our collaboration: the Lead physics, outreach, and geomicrobiology meetings; the Aspen underground science program; and NESS02. The Science Book contains chapters on NUSEL earth science, applied science, and outreach.
- The **Science Timeline (Section D)** section summarizes the technical needs and readiness of proposed experiments, and thus the schedule and parameters that NUSEL-Homestake should meet. We provide a strawman ``Timeline'' of such experiments, correlating this with our proposed NUSEL schedule.
- A **Program Plan and Reference Design (Section E)** of the NUSEL-Homestake project, in which we describe the proposed conversion of the Homestake Mine into the world's deepest and most flexible underground science laboratory. We propose a specific plan for meeting the needs of envisioned and far-future underground science, including optimizing prospects for a megadetector important to very-long-baseline neutrino oscillation experiments and next-generation proton decay searches. We describe the work remaining to be done to arrive at a **Baseline Definition** for the NUSEL-Homestake project, and some of our plans for reaching this milestone. This section concludes with a suggested **Management Plan**.
- A **Work Breakdown Structure (Section F)**, including rather detailed description of the underground campus developments at the 8000, 7400, and 4850 ft levels and of the plans for optimizing access to the underground and the capacity for future hall expansions.
- A last short section, **Mine Status: Dewatering and Site Transfer Issues (Section G)**, in which we provide a qualitative assessment of some of the nonscientific issues that must be solved before NUSEL-Homestake can proceed.
- **Appendices** containing detailed cost/schedule engineering spreadsheets for NUSEL construction and operations

The current submission to the NSF includes somewhat shortened versions of the Overview, Science Timeline, Program Plan and Reference Design, and Work Breakdown Structure sections. The Science Book and Appendices are not included. The full Reference Design Project Book can be found at http://int.phys.washington.edu/NUSEL/, and on the LANL archive. Hard copies have been sent to selected NSF and community officials.

## B. PROJECT OVERVIEW

### I. NUSEL-Homestake Science and Outreach

**What is the physics justification for a National Underground Science and Engineering Laboratory?**
*(Summary, Chapter I of the Science Book)* There is growing recognition that some of the most profound problems in subatomic physics, astrophysics, and cosmology must be attacked with new types of experiments, ones involving massive, ultraclean detectors mounted deep underground to escape backgrounds from cosmic rays. In this overview we outline the prospects for paradigm-shifting discoveries in four areas, neutrino physics, dark matter, nucleon stability, and neutrino astronomy.

*Neutrino Physics:* Over the past decade experiments done underground have provided some of the most significant new results in physics. The discovery that atmospheric and solar neutrinos oscillate proves that neutrinos have mass, the first demonstration of physics beyond the standard electroweak model. The most naive interpretation of that mass – the so-called seesaw mechanism – suggests we are probing physics characterized by energies ~ $10^{15}$ GeV, near the grand unified scale and many orders of magnitude beyond the direct reach of accelerators. The discovery that neutrino mixing angles are nearly maximal, unlike those of the quarks, suggests that the neutrino mass generation mechanism differs fundamentally from that of the other fermions. There is every expectation that, as we identify the masses, mixing angles, and CP phases of neutrinos, the emerging pattern will point theorists in the direction of a new standard model, one that explains many of the apparently arbitrary parameters of our low-energy world.

The pace of discovery in neutrino physics is extraordinary. Since the submission of our original proposal in June 2001, the Sudbury Neutrino Observatory has demonstrated that heavy-flavor neutrinos make up two-thirds of the solar neutrino flux. This result eliminates all but one (the large-mixing-angle) proposed oscillation solution to the solar neutrino problem, and vindicates the standard solar model by showing that the total flux (independent of flavor) was correctly predicted. KamLAND showed that the oscillation phenomena responsible for the solar neutrino problem could be tested in a controlled terrestrial experiment, and significantly narrowed the range of neutrino mass differences that could account for the oscillations. These results have had an immediate impact on cosmology. Because of the large mixing angles, Big Bang nucleosynthesis now severely limits neutrino chemical potentials in the early universe, regardless of flavor. We have identified the first component of particle dark matter: neutrinos are at least as important as the visible stars in the universe's matter budget. Yet the allowed range for the absolute scale of neutrino mass – from about 0.05 eV to a few eV – is still too broad. WMAP's very recent measurement and analysis of temperature anisotropies in the cosmic microwave background radiation, probing the structure of our universe 400,000 years after the big bang, proved sensitive to neutrino masses slightly below 1 eV. Thus precision cosmology now is linked to the progress being made in neutrino physics.

One measure of the community's regard for this science comes from last year's Nobel Prize to Raymond Davis, Jr., who founded the field of neutrino astrophysics when he constructed, with the help of the Homestake Corporation, the first solar neutrino detector.

It is important to recognize that these discoveries are the first step in an effort that promises to resolve some of the deepest questions in science. The field's excitement comes from the knowledge that many more steps can be taken. As mentioned in connection with WMAP, we do not know the absolute scale of neutrino mass: the oscillation results provide only a lower bound. We do not know whether the neutrino has a distinct antiparticle. We do not know the mass hierarchy, that is, whether the nearly degenerate pair of mass eigenstates responsible for solar neutrino oscillations is lighter or heavier than the third mass eigenstate. We do not know $\theta_{13}$, the crucial mixing angle that will determine whether neutrino oscillations affect the explosion of and nucleosynthesis within a core-collapse supernova. Perhaps most exciting, the pattern of neutrino mass now being revealed appears consistent with a novel mechanism that accounts for the existence of matter itself, leptogenesis: CP violation long hidden in neutrinos may account for dominance of matter over antimatter in our universe.

All of these unresolved questions appear answerable. The absolute scale of neutrino mass and the particle-antiparticle nature of the neutrino are connected in a fascinating way. The seesaw mass-generation mechanism exploits the fact that neutrinos, unlike other standard-model fermions, lack charges or other additive quantum numbers that distinguish particles from antiparticles. This allows neutrinos to have, in addition to ordinary Dirac masses, lepton-number-violating Majorana masses. The breaking of this basic standard model symmetry leads to neutrinoless double beta decay, an exotic nuclear decay mode that explicitly violates lepton number. The rates for neutrinoless double beta decay are proportional to the squares of the masses of the neutrinos that couple to the electron. One of the most exciting consequences of recent neutrino physics discoveries is the likelihood, in two of the three most popular neutrino mass scenarios, that neutrinoless double beta decay will soon be discovered if experiments 100 times larger than those currently operating can be mounted. In most scenarios this would then determine the overall scale of neutrino mass. While such experiments are challenging – they generally require ton quantities of an enriched isotope, a very deep underground location, and excellent energy resolution – several of the proposed detectors appear practical. It is likely that a next-generation double beta decay experiment will be one of the early NUSEL-Homestake experiments.

Progress on neutrino mixing angles is expected to come from a combination of solar, reactor, and accelerator experiments. Despite the success of SNO and Super-Kamiokande, 99.99% of the solar neutrino flux has not been seen in direct detection experiments. While there are many reasons for measuring the low-energy solar neutrino flux, one important goal is to better constrain the solar neutrino mixing angle $\theta_{12}$. This requires a well-calibrated neutrino source. Arguably the pp solar neutrino flux is the best candidate, astrophysical or terrestrial, because the normalization is known to 1%. Reactor neutrino experiments using multiple detectors and accelerator experiments using either broad-band or off-axis neutrino beams have been proposed to measure the unknown mixing angle $\theta_{13}$. This crucial measurement will determine in part the size of CP-violating effects in long-baseline neutrino oscillation experiments.

A further step in the above program addresses one of the deepest questions in physics, why does our universe contain matter? The puzzle exists because, in the Big Bang, one would naively expect matter and antimatter to have been produced equally, only to annihilate each other later when the universe cooled sufficiently. Yet our universe (at least locally) contains matter, but virtually no antimatter. Theory tells us that a necessary condition for baryogenesis (an excess of baryons over antibaryons) is the breaking of CP, a symmetry combining parity reversal with charge conjugation. While there is CP violation in the standard model outside the neutrino sector, almost all theorists believe it is too weak to account for the matter asymmetry. However the patterns emerging from recent neutrino physics – the masses and the large mixing angles – suggest that the necessary asymmetry could be produced by CP violation among the neutrinos, which would then be transferred to the baryons, a process call leptogenesis. Very long baseline neutrino oscillation experiments (~ 2000 km) using neutrino superbeams (or ultimately the beam from a neutrino factory) and megaton detectors could measure aspects of the CP violation, providing our first clues to the origin of matter in the Big Bang.

*Dark Matter:* The results of WMAP and other cosmic microwave background probes, of large-scale structure surveys, and of supernova probes of the Hubble expansion all indicate that the vast majority of matter and energy in our universe is unknown to us, dark and undiscovered. The best global fits to cosmological parameters lead to remarkable conclusions. The total energy density of the universe is very close to the critical value, the minimum energy density guaranteeing that our universe will not recollapse, as inflationary cosmologies predict. And most of that energy density is in hidden components: 73% is dark energy --usually identified either with Einstein's cosmological constant or with a time-evolving pervasive scalar field -- and 27% is matter. While a small part (~ 4%) of this matter is ordinary baryons (and only part of the baryonic matter is visible in stars and gas clouds), the vast majority is dark, nonbaryonic, and cold, that is, something new, outside the standard model of particle physics. As our cosmological colleagues struggle to understand the nature and equation of state of the dark energy, underground scientists will mount a new generation of detectors to find the particles that comprise the cold dark matter.

The dark matter/dark energy problem is indeed remarkable. We are aware of only a small fraction of the matter/energy in our universe (and therefore in our local region of the Milky Way galaxy). The unseen components nevertheless shaped, through their gravitational and other interactions, the structure of our visible universe. The goal of this field is to identify and understand the properties of the unseen matter and energy, thereby enabling cosmologists to understand how the universe evolved from the Big Bang into its present form.

We have noted that one component of particle dark matter, neutrinos, was recently identified. But we know neutrinos are a bit of a spice in the dark matter mix, not the main ingredient. Perhaps the strongest candidate for the dark matter comes from the predictions of supersymmetry, which is invoked in particle physics to explain the stability of the electroweak scale with respect to radiative corrections. Supersymmetry predicts new particles around the TeV scale, the lightest of which could well be stable. This particle would have been produced in the Big Bang, and would now be affecting the expansion of our universe.

Cross sections for supersymmetric particles interacting with ordinary matter can be estimated, and generally lie between $10^{-6}$ and $10^{-10}$ pb. (A $10^{-10}$ pb cross section corresponds to about one event/100 kg/y.) This is an encouraging result for this young field, as existing detectors can be scaled up to one-ton masses to search all of this allowed range. Experimenters envision detector improvements of about four orders of magnitude. Accompanying these improvements will be the need for deep underground sites, up to 4500 mwe, to escape neutron and other cosmic-ray-induced backgrounds.

The particle dark matter quest illustrates a wonderful connection between physics at the largest scales – cosmological structure – and that at the smallest scales, ultra-high-energy accelerator physics. A major goal of the Large Hadron Collider, which will probe particle interactions at distances of 1/100,000$^{th}$ of a fermi, is to discover supersymmetry directly by producing supersymmetric particles, just as in the Big Bang. The LHC timeline hopefully predicts such a discovery by 2012, about the time particle dark matter experimenters will reach their event rate goal of 1 event/100kg/y.

*The Stability of Matter*: Almost all of the known elementary particles have finite lifetimes, decaying into other lighter particles. When this does not occur – when a particle is absolutely stable – there is an explanation based on conservation laws. For example, the electron is stable because, as the lightest charged particle, there is no possible decay channel that conserves electric charge. The other component of ordinary matter, nucleons, also appears to be stable, at least on time scales so far measured. The standard model accounts for this stability by assigning to nucleons a conserved charge called baryon number, first introduced by Stuckelberg in 1938.

However, efforts to go beyond the standard model, to unify the strong, electromagnetic, and weak interactions, predict that baryon number is not exactly conserved. That is, ordinary matter will eventually decay at some unimaginably late epoch in the life of our universe. The original predictions of such Grand Unified theories were based on the gauge group SU(5) and predicted lifetimes for the proton between $10^{28}$ and $10^{32}$ years. The largest detectors ever built underground were designed to detect favored decay modes, such as $p \rightarrow e^+\pi^0$: Super-Kamiokande experimenters have spent a decade watching a dark cylinder containing 50 kilotons of ultrapure water, waiting for the light from the decay of a single nucleon. Their negative results excluded these simplest Grand Unified models.

Modern versions of Grand Unified theories are very attractive, accounting for the masses and mixings of all the quarks and leptons (including the neutrinos) and for the matter-antimatter asymmetry through leptogenesis. They also predict nucleon decay, though at rates suppressed relative to early SU(5) models. In particular, currently popular supersymmetric SO(10) models predict a favored $\nu K^+$ proton decay mode within an order of magnitude of present limits, as well as a branch for the $e^+\pi^0$ mode of about $10^{35}$ years. We have the technology to construct a detector ten times larger than Super-Kamiokande to reach such lifetimes. NUSEL-Homestake would allow us to place this detector at unprecedented depths, where it would be 100 times quieter than Super-Kamiokande. (Low cosmic-ray backgrounds are important in the identification of certain proton decay modes as well as in the ancillary uses of the detector, e.g., supernova, atmospheric, and solar neutrino physics.) Such a megadetector nucleon decay experiment is expected to be a keystone of the NUSEL-Homestake program.

Proton decay is another example of an ultrasensitive low-energy experiment that, while indirect, reaches far beyond current accelerators in its quest for new physics. The processes responsible for proton decay in SO(10) Grand Unified models probe interactions at $10^{16}$ GeV, corresponding to length scales of $10^{-17}$ fermis.

*Supernova Neutrino Observatories*: The nucleon-decay megadetector just described, operating for decades in a multidisciplinary underground laboratory, could serve many communities. We noted that the quest for CP violation in neutrino oscillations has led to several proposals for neutrino superbeams, and requirements for targets placed ~ 2000 km away. (The distances from Homestake to FNAL and BNL are 1290 and 2530 km, respectively.) The requisite target mass for such long-baseline experiments is comparable to that needed for next-generation nucleon decay. Within two decades the international high-energy community will likely produce a neutrino factory and contemplate experiments with even longer baselines, comparable to the earth's radius.

A Type-II (or core-collapse) supernova is one of the most spectacular events in nature. This occurs when a massive star has finished its burning cycles, producing in the final Si-burning stage an iron core. When that core reaches the Chandresekar mass, it collapses under its own gravity to form a hot neutron star, initially on the order of 50 kilometers in radius. Via a mechanism not well understood, but believed to depend on both the hydrodynamic shock wave generated by the collapse to nuclear density and by neutrino heating of the dissociated nuclear matter left in the shock wave's wake, the outer mantle of the star is ejected. This ejection enriches the interstellar medium in the metals produced during hydrostatic burning as well as in rarer elements produced in the explosion itself. Type II supernovae are major engines driving the long-term chemical evolution of the galaxy. In particular, approximately half of the elements heavier than iron are produced in the rapid-neutron-capture process, which requires the explosive conditions and high neutron fluences found in neutrino-driven winds of supernovae.

Just as solar neutrinos allow us to probe the details of the sun's core, complete measurements of the spectrum, flavor, and time evolution of the supernova "neutrino light curve" will provide detailed information on the explosion mechanism, on the behavior of neutron star matter at several times nuclear density, on supernova nucleosynthesis, and on new aspects of neutrino physics that might not be testable elsewhere (such as the effect of an intense neutrino background on neutrino oscillations).

A megadetector of the type discussed for proton decay and long-baseline measurements, located at sufficient depth, would be an incomparable observatory for galactic and extragalactic supernovae, complementing optical and gravitational wave instruments. A massive detector could follow the "neutrino light curve" of a galactic supernova for tens of seconds, the period in which the puffy protoneutron star radiates its lepton number, becoming a compact neutron star 10 kilometers in radius. At such late times nuclear matter phase transitions might be revealed through sudden changes in the light curve. Such a detector could also see the constant supernova "neutrino background," the integrated effect of all past supernova. There is extraordinary information in the flux and redshifts of this spectrum, to which the earliest stars, now believed to have formed just one billion years after the big bang, contributed. Plausible models of star formation predict first detection is just a factor of three beyond current Super-Kamiokande limits. Such a detector would also reach beyond our galaxy: A supernova in Andromeda would produce a few tens of events.

Supernova neutrino detectors are also crucial to the "supernova watch" program – the plan to correlate optical, gravitational wave, neutrino, and other signals from the next galactic supernova. Supernova neutrino detectors have a 100% detection rate – intervening matter cannot obscure the neutrino signal. This signal is prompt, coming typically several hours to a day before the optical display, which begins when the shock wave reaches the outer envelope. .

**How will NUSEL contribute to earth science and geomicrobiology?** *(Summary, Chapter II of Science Book)* One of the most important developments in the justification for NUSEL-Homestake since the June 2001 proposal has been the recognition of the laboratory's importance to earth science and geomicrobiology. The earth-science component of the NUSEL-Homestake collaboration, led by a steering committee [Brian McPherson (New Mexico Tech), Tullis Onstott (Princeton), Tommy Phelps (ORNL), Bill Roggenthen (South Dakota School of Mines and Technology), Herb Wang (Wisconsin), and Joe Wang (LBNL)], has dubbed this science **EarthLab**.

EarthLab is a proposal to utilize Homestake as a subterranean laboratory and observatory in the study of geomechanical, hydrological, geochemical, and biological processes that modify earth from its surface to the limit of habitable depths. Opportunities for earth scientists to observe directly deep subsurface changes – microbes that precipitate minerals and generate gas, migrating fluids that transport drinking water and weaken earthquake-generating faults, and stresses and strains that cause rock to deform slowly or break catastrophically – are currently very limited. The complex coupling of biogeochemical processes, fluid flow, rock-water interaction, and rock deformation is largely unexplored. A more complete understanding of these processes and their coupling is crucial in advancing disciplinary research ranging from earthquake engineering to bioremediation. The EarthLab proposers also envision partnerships with industry, including practical applications of its biosphere research to bioremediation, biotechnology, and pharmaceuticals, as well the development of new geophysical and geochemical tools for subsurface characterization and new geological mapping, rock drilling, and other engineering technologies for exploration and construction. By partnering with NASA, these new technologies could be adapted for subsurface life exploration on other planets.

EarthLab's goal of studying complex geologic processes *in situ* with 3D access for continuous observations and controlled experiments requires a very large, instrumented rock volume and access to great depths. As discussed in more detail below, we believe there is only one proposed NUSEL site that satisfies these requirements. The EarthLab science program is focused on five major themes, the nature of life at depth, fluid flow and transport at depth, rock deformation and failure at depth, mineral resources and environmental geochemistry, and scientific and engineering innovation underground. The EarthLab program will explore the interrelation of complex physical, environmental, and microbial processes. For example, tectonic forces bend and fracture rock, in turn altering the permeability and porosity of the rock, and therefore the pressures, directions, and rates of fluid movement. Changes in fluid pressures alter the elastic response of rock to deforming forces, which govern movement along faults and thus the frequency and magnitude of earthquakes. Fluid flow is also important to the distribution of environmentally and economically important minerals and compounds in the crust, many of which are dissolved in and precipitated from hot rocks.

*Life at Depth:* The relatively recent discovery of a subsurface biosphere – deep subsurface microbial communities – may provide crucial insights into how life on this and other planets may have originated and evolved. The diversity of geologically isolated extremophiles is thought to provide our best guide to possibilities for life beneath the surface of Mars, for example. Subsurface microorganisms also play crucial roles in the dissolution and formation of minerals. Yet the coupling between geomicrobiological and biogeochemical processes and the earth science, chemistry, and physics governing rock masses and fluids is poorly understood. Certain minerals may provide nutrients for microorganisms, and fluids may transport microorganisms to those nutrients. The microorganisms in turn may precipitate minerals or generate gas, altering the permeability and hence the flow of fluids. Fluid flow may dictate the temperature at depth, thereby determining whether thermophiles and hyperthermophiles can survive. If the temperature change is rapid compared to migration times for microorganisms, then the microorganism must adapt or expire. Fracture propagation may also impact fluid flow, exposing fresh mineral surfaces to redox reactions and providing aqueous and gaseous energy sources to the microorganisms.

EarthLab would be the only facility in the world where the coupling of microbial, chemical, and physical processes could be explored from surface to the great depths defining the limits of life.

*Fluid Flow and Transport at Depth:* Very little is currently known about fluid flow and transport at depth because drill-hole experiments are so limiting, providing small rock samples altered by the drilling process. EarthLab could revolutionize the field by allowing temporal and spatial data far beyond that currently available. The goal would be to achieve a quantitative understanding of recharge and infiltration, fracture permeability, the physics of multiphase flow, flow in fracture networks, verification of well and tracer test models, the coupling of flow, stress, and heat, and the storativity and transmissivity of tight rocks. This science is central to important societal issues, including the stability of water supplies, hazardous waste disposal, and the remediation of contaminated groundwater.

Borehole measurements are fundamentally limited because they provide no direct information about the large volume of rock residing between tested points. Results often erroneously reflect the spatial scale of the separate point measurements. EarthLab would, for the first time, allow hydraulic, tracer, and geophysical imaging over a 3D array encompassing approximately 9 cubic kilometers of rock, thereby characterizing the structure, fracture connectivity, and transport properties, as well as their variability with scale, depth, and distance across the excavated zone.

*Rock Deformation at Depth:* Rock deformation deep in the subsurface traditionally has been characterized through proxy methods, such as seismic tomography. Apart from a few deep mines outfitted with extensometers, active rock strain must be deduced from surface methods such as InSAR (satellite-generated Interferometic Synthetic Aperture Radar) and GPS (Global Positioning System) data. EarthLab will allow continuous direct measurements of rock strain and the variables that govern that strain. As with permeability, past studies of strain and stress have generally been compromised because the deep rock being characterized was too small in volume.

EarthLab experiments will be able to test the hypothesis that the Earth's crust is "critically stressed," close to failure by fracture. Repeated shearing of such fractures can keep flow paths open that would otherwise be closed by mineral cementation, enhancing rock permeability. By mapping fractures, stress, and fluid flow within the subsurface, the theory of critical stress can be tested quantitatively.

This research will benefit from the interdisciplinary nature of NUSEL-Homestake: the large detector cavities excavated for physics detectors are ideal laboratories for monitoring 3D+time deformation on unprecedented scales. Hydraulic, tracer, and geophysical measurements of transport parameters will be made during cavern enlargement, and instrumentation arrays will be installed in the finished cavities to monitor the long-term passive response in the stable zone surrounding the detectors.

*Mineral Resources and Environmental Geochemistry:* Many economically important mineral resources are formed or concentrated by fluid flow in the subsurface. For example, oil, natural gas, and some brines are localized in the crust largely by their physical response to fluid flow. Most metals, such as iron, copper, and gold, are localized chemically, by mineral dissolution and subsequent deposition. Although considerable progress has been made in understanding mineral deposition by observation of fossil systems, opportunities for studying these processes in an active environment are exceedingly rare. For example, most geothermal areas are too hot and deep for direct study. EarthLab would fill this void, allowing direct testing of many aspects of mineral deposition.

Fluid flow through rocks governs the release and concentration of metals and organic compounds of environmental concern. Most such releases are triggered when rock from deep in the crust is exposed to water and oxygen, causing minerals formed at depth to decompose. The most widely known of these processes, acid mine drainage, results when pyrite ($FeS_2$) is oxidized by near-surface waters. Most studies of acid mine drainage are confined to points where water has reached the surface. EarthLab would permit these processes to be observed at early stages and at depth, perhaps leading to better strategies for control.

*Science Technology and Engineering Innovation:* The technological and engineering impacts expected from EarthLab include:
- New generic materials, novel microorganisms, and biotechnology applications
- Analytic techniques for geomicrobiology and exobiology
- Natural resource recovery
- Drilling and excavation technology
- Novel uses of underground space
- Mine safety, including large cavern stability analyses
- Subsurface imaging
- Environmental remediation

EarthLab instrumentation itself will break new ground: in the early days of NUSEL-Homestake, when access to most of the 600 km of available drifts is still possible, instrumentation will be installed in coreholes adjacent to excavated blocks, and continuously monitored thereafter for excavation-induced displacements, microseismic activity, temperature, and fluid pressures. Aqueous and particulate samples will be taken from accessible areas. New sensor technologies will be designed for many of these applications.

A specific set of mine engineering research and development issues will be addressed by the **Hard Rock Mining Training and Research Center** of NUSEL. The research component of HRMTRC will focus on experiments in stoping and rock breaking, with the goal of accident reduction and productivity and mine safety optimization; on the application of practical digital mine mapping methods; and on mine engineering, especially advances in stoping methods, long-hole mining, bench mining robotics, and production optimization.

**<u>What is the relevance of NUSEL to post-9/11 challenges and other applied-science issues?</u>** *(Summary, Chapter III of Science Book)* In the past year various advisory committees have stressed the need for the scientific community to help with urgent national security issues. For example, the 2002 National Academy Study "Making the Nation Safer: The Role of Science and Technology in Countering Terrorism," states *"Indeed, America's historical strength is science and engineering is perhaps its most critical asset in countering terrorism without degrading our quality of life."* An important workshop was recently hosted by the Department of Energy on the "Role of the Nuclear Physics Research Community in Combating Terrorism." Research community members from DOE and NSF laboratories and universities met with government counterterrorism experiments to discuss technological solutions to national security challenges. While the scope of the resulting report was quite broad (**http://www.sc.doe.gov/henp/np/homeland/descript.html)**, strong emphasis was placed on the past and future role of low-level-counting techniques – techniques developed for basic underground science.

The active interrogation of luggage, vehicles, and cargo containers for conventional explosives and for special nuclear materials such as highly enriched uranium requires sensitive, large-volume detectors for recording the responses of materials to x-ray, gamma-ray, and neutron probes. Two technologies highlighted in the DOE report are high-resolution intrinsic germanium crystals and very-large-volume scintillation and water Cerenkov detectors. Much of the recent progress in increasing the efficiency and resolution of Ge detectors has come from nuclear spectroscopists and from double beta decay experimentalists – groups that are now collaborating on plans for next-generation double beta decay experiments. The need for rapid imaging of large cargo containers – ships can unload cargo at the rate of a container a minute – has raised the need for very large neutron and photon detectors. The DOE report stresses the kiloton-volume detectors developed by nuclear and particle physicists for neutrino detection as a possible solution. These include detectors modeled after LSND, large scintillation tanks viewed by photomultipliers, as well as water Cerenkov detectors. The report notes that the neutrino physics need for versatile, relatively inexpensive detector technologies has prepared the way for the adaptation and wide application of these techniques to national security problems.

The report also notes that underground experiments to search for extremely rare processes has led to an impressive confluence of techniques that can now be used to monitor our environment for potential hazards:

- few-atom, high-purity chemical separations
- ultra-low-level gas counting systems constructed of materials free of natural radioactive contaminants
- near-zero background germanium and scintillation detectors, and
- deep underground laboratory facilities where cosmic-ray backgrounds are reduced by many orders of magnitude (thereby allowing samples to be counted with high sensitivity).

The applications – sampling high-risk areas, followed by expedited counting in specialized underground facilities – are numerous. The report notes, because of radiochemical solar neutrino experiments, this technology is largely in hand: the applications to national security should be further developed.

An important change from our original proposal is a much expanded **low-level counting facility** for national security and other applications, as well as for many basic science needs (such as determining the activities of materials before they are used in detector construction). Examples of other applications include the development and fabrication of high-purity materials; materials analysis; and the extraction of activities from materials (such as the reduction of Th and U to levels of $10^{-16}$ g/g in SNO, Borexino, and KamLAND). Such technical steps result in shorter collection times, smaller samples sizes, less restrictive transport times, and much improved sensitivity to short-lived isotopes, impacting both basic science experiments (such as the radiochemical neutrino experiments) and applications, such as atmospheric Xe monitoring for national security and trace-element analysis for materials dating. The microelectronics industry, which currently uses underground laboratory space, is interested in measuring cosmic-ray-induced error rates in chips as a function of depth. Several additional applications – novel microorganisms and their biotechnology applications, resource recovery and environmental monitoring, drilling and excavation technology – connected with earth science are addressed in detail in Chapter III.

**<u>What is NUSEL's potential for enhancing science outreach and education?</u>** (*Summary, Chapter IV of the*

*Science Book*) The proposers believe that NUSEL-Homestake has remarkable potential for regional (and national) outreach and education. The reasons include its location (near the geographic middle of five contiguous EPSCoR states); its potential for drawing visitors from among the three million tourists who come to the Black Hills yearly; the links between Black Hill's geology and the region's history (both Native American and mining); access to scientifically underserved K-12 and tribal college students; and Homestake's significance as the birthplace of neutrino astronomy:

- *The opportunity to create a major-science focus for the Northern Great Plains states.* Homestake is located near South Dakota's Wyoming border. The five states within 120 miles of Homestake – South Dakota, North Dakota, Montana, Wyoming, and Nebraska – are **EPSCoR** states. Several others – Idaho, Nevada, Kansas – are within ~ 500 miles. The presence, for the first time, of a major basic research center affords many opportunities to "grow" new research programs in the regional universities. The collaboration's initial success in introducing South Dakota scientists to NUSEL indicates some of the potential. In the conceptual stages of this project (2000 and 2001) several of the senior underground scientists visited state universities to give seminars and to hold discussions with interested faculty. Following submission of the June 2001 proposal, the protocollaboration's first activities were the NUSEL physics and earth science workshops held in Lead. One result was the recruitment of 16 South Dakota scientists as working members of the collaboration, several assuming leadership roles; approximately 30 others have participated in specific collaboration activities, such as formulating NUSEL outreach plans. The collaboration has also contacted key university scientists in neighboring states, expressing our interest in partnering with them. We are encouraging the regional universities to consider an association, perhaps similar to Oak Ridge Associated Universities, in which these institutions can advance more effectively science and education by partnering with NUSEL.

- *Undergraduate and graduate education.* These efforts to integrate South Dakota and regional scientists into the collaboration and to encourage a regional science/education consortium are important to the undergraduate and graduate students in the region. In Chapter IV we describe why we believe a strong NUSEL-regional universities partnership is important to NUSEL's **Research Experience for Undergraduates** and **summer school** efforts, and in follow-up activities that could keep regional undergraduates involved in NUSEL science during the academic year. This partnership is essential in realizing NUSEL's potential for graduate education in the region, as active involvement of graduate faculty advisors is clearly the most efficient way to bring these students to NUSEL. Of course, NUSEL scientists will come from throughout the US and overseas. While NUSEL will endeavor to be of particular help to Northern Great Plains institutions, it will also serve the customary national laboratory role in supporting visiting graduate and undergraduate students from the broader community.

  One program unique to NUSEL will be the training component of the Hard Rock Mining Training and Research Center. Fewer schools are training mining geologists and engineers even though, as US open mines are reaching economic limits with depth and increased stripping ratio, more deep underground mines are being planned to extend mine lifetimes. The HRMTRC would serve as a centralized national training and research center, working to support the educational mission of the nation's mining schools. Many of these schools no longer have the staffing levels or student enrollments to justify the operation of local training mines.

- *Public outreach.* Three million visitors come to the Black Hills annually, some attracted by the outdoor recreation and some by the region's colorful gold mining and Native American history. This opens wonderful opportunities for interesting these visitors in science, particularly if the science is presented well and connected to the region's history and geology. In our original proposal we estimated that a well-designed NUSEL visitor center could draw perhaps 100,000 visitors per year. This was based in part on the success of the Soudan Mine visitor program that, though operating in a smaller and more remote site, attracts 40,000 visitors per year. Regional tourism experts, including the superintendent of Mt. Rushmore, have advised us that our initial estimate may have been far too conservative, with 400,000/yr being a better guess. If the visitor total exceeds 100,000/year, the NUSEL outreach program would be the largest among pure-science research sites. (We distinguish NUSEL from technology centers, such as Cape Kennedy.)

  The opportunities for constructing an exciting visitor center seem almost limitless: the geology, cultural history,

and science are connected in ways that should fascinate the public. The Black Hills are sacred to the Sioux because of their unique physical features, features that reflect the region's remarkable geology. The Black Hills are literally a peep hole through the Great Plains into the ancient geologic past. The domal uplifts that exposed this island of Precambrian rock, 1.9-2.5 billion years old, occurred 530 and 65 million years ago, the last after the great inland sea retreated. The accompanying erosion scoured off in excess of 5000 feet of the sedimentary rock that covers other parts of the Great Plains, exposing the dome. The extraordinary mineralization that formed the Homestake gold deposit occurred beneath ancient seas as a result of submarine hot springs activity, forming the largest gold deposit of its kind in the world.

Our vision is to explain how this geology influenced the lives of the people who lived in this region, from Native Americans, to the gold rush days and the opening of the American west, to modern times where science now needs the great overburdens provided by Homestake. The NUSEL geomicrobiology program is directly coupled to the geologic history, as isolated pockets of ancient water may hold the most surprising extremophiles. The basic earth science studies – understanding the coupling of subsurface and surface phenomena, the relation between hydrology and mineralization, and environmental science – tie nicely to the history. Part of the history is also the creation of a new field, neutrino astronomy, at Homestake with the start of the chlorine experiment in 1964. This "mining for neutrinos" is then the door to questions such as the origin of the Cosmos, the nature of the dark matter and energy, the stability of matter, supernovae and the origins of the elements, and other compelling questions that interest both the scientist and the public.

Chapter IV of the Science Book describes our plans for the **Visitor Experience Center** program and the near-surface underground tour, as well as an option for deep underground tours to see NUSEL detectors. There is also an important issue of historical preservation. We have had several discussions with Barrick and the State of South Dakota about establishing a **museum/archive** within the existing foundry building at Homestake – a building interesting historically and architecturally. Homestake is arguably the most influential mine in US history, with many important mining inventions originating there. The company wants to preserve this legacy, as well as the library of charts, maps, and employment records dating back to the 1880s. (As many families migrated to the US to work at Homestake, the employment records are an important genealogical resource.) The Barrick Corporation and we have discussed how the company could help support the operations of the museum/archive, and how these operations could be coordinated with those of the visitor center.

- *K-12 education and distance education.* There are many examples of successful **K-12 outreach programs** built around visitor centers, e.g., the Lawrence Hall of Science in Berkeley. NUSEL will host classroom groups and help classroom teachers integrate visits into a broader science experience by making available various preparatory and follow-up science experiences. NUSEL intends to partner with state and regional institutions expert in K-12 education, such as the Black Hills State University's Center for the Advancement of Mathematics and Science Education, to prepare materials appropriate to various age groups. (CAMSE, a South Dakota Board of Regents center of excellence, currently distributes science curriculum kits to about 300 science teachers in western South Dakota.) These materials would be a form of distance education by NUSEL, which could prepare students for their visits to NUSEL. The Laboratory will encourage its scientists and visitors to reinforce these efforts by visiting school rooms, by participating in science fairs as mentors and judges, etc.

  The changing demographics of the Great Plains – outward migration has reduced large portions of the Northern Great Plains to "frontier" population densities of less than six per square mile -- was the subject of the book "Buffalo Commons" by Frank and Deborah Potter, published some 15 years ago. This phenomenon has challenged educators in this region as many school districts fall below a critical mass, and as school closures impose longer student commutes. South Dakota has responded to this challenge by creating one of the most "wired" K-12 educational systems in the nation. NUSEL hopes to build on this opportunity by creating interactive web sites for science students and by encouraging advanced students to join distance education programs utilizing data from ongoing NUSEL experiments.

- *Undergraduates in tribal colleges.* We expect South Dakota and regional tribal colleges to be important NUSEL partners in science outreach to the Native American community. Two of the South Dakota tribal colleges, Sinte Gleska (with about 1000 students) and Oglala Lakota, are well known for their computer science, pre-engineering, and environmental science curriculum. Sinte Gleska University is a partner with the National Science Foundation and the South Dakota School of Mining and Technology in NAMSEL, the Native American Mathematics and Science Educational Leadership, a program that encourages teacher leadership skills, addresses the needs of Native American students, and promotes school cultures which support systemic change. Sinte Gleska and Oglala Dakota are partners in an NSF Model Institutes for Excellence (MIE) grant, which supports the science curriculum developments mentioned above, as well as distance learning efforts involving three other South Dakota tribal colleges (Si Tanka College on the Cheyenne River Reservation, Sitting Bull College on the Standing Rock Reservation, and Sisseton Wahpeton Community College in northeastern South Dakota). Together these five institutions comprise the Oyate Consortium. Currently 94% of OLC and 85% of SGU graduates are employed or seeking graduate degrees; this compares to reservation unemployment rates that frequently reach 80-85%. The goal of the MIE initiative is to integrate traditional tribal values into new programs of study for Native American students. Among these are environmental science, information technology, computer science, pre-engineering, and life science. As several of these programs aim to prepare students for more advanced studies at other institutions, NUSEL-based relationships will help facilitate some of the MIE goals. In short, NUSEL will be located in a center for progressive Native American science and mathematics education and has an opportunity, by partnering with and supporting the tribal colleges, to significantly enhance diversity in science.

Several tribal college educators discussed possible partnerships with the proposers at the time the initial NUSEL proposal was submitted, then took part in the Education and Outreach Lead workshop. The tribal colleges will be a particular focus of the Research Experiences for Undergraduates Program discussed above.

## II. Why is a next-generation National Underground Science and Engineering Laboratory needed?

In this section we describe why a deep, multipurpose underground science and engineering laboratory is needed by the US and international scientific community. This laboratory must differ from those now in existence. As background requirements in many neutrino and dark matter experiments are improving by more than an order of magnitude per decade, the site must be very deep to accommodate the experiments we foresee in the next twenty years. As detectors an order of magnitude larger than any yet constructed are likely to be housed there, the site must be practical for large cavity construction. It must be suitable for experiments involving large quantities of cryogens or flammables: safety considerations place stringent constraints on ventilation, and make advisable multiple access routes and possibilities for isolating such experiments. Multiple depths can be important, allowing experiments to balance background needs against the additional time and cost to go deeper. The earth science needs are very special. Ideally the laboratory will provide very large volumes of well-characterized rock, with 3D access. The geomicrobiology becomes more interesting with depth, with the age of rock, and with interesting hydrology.

In this section we discuss the existing sites for underground science, the need for a next generation facility, and the many reasons the Homestake Mine presents an opportunity to create a laboratory unequaled in quality and in flexibility.

**What underground sites are currently available?** The program of science outlined in Section I has a rich heritage. Arguably the field of neutrino astronomy began with the Homestake chlorine experiment, which began in 1964. The interest in this field intensified in the early 1980s, stimulated by the SU(5) predictions of observable proton decay rates. The realization that the solar neutrino problem was a neutrino physics issue, the growing certainty of particle dark matter, and the demonstration in 1987 that supernova explosions could be probed with neutrino detectors all contributed to current excitement in this field.

The three multipurpose underground physics laboratories now operating all were conceived around 1980. The first of these, the Baksan Laboratory in Russia, was excavated in the late 1970s, under the direction of Chudakov and Zatsepin. It was the first deep laboratory built for physics, providing 4700 mwe of cover. The laboratory was built to study the penetrating components of the cosmic rays – muons and neutrinos – and was the site of the SAGE solar neutrino experiment, the first experiment sensitive to the low-energy pp neutrinos. This research continues, though under difficult conditions.

Zichichi proposed the world's largest multipurpose facility, the Gran Sasso National Laboratory in Italy, in 1981. Built off a highway tunnel during the early 1980s, the laboratory provides about 3800 mwe in overburden. Gran Sasso hosts the GALLEX/GNO and Borexino solar neutrino experiments, two long-baseline experiments to detect neutrinos from CERN, the LVD kiloton supernova detector, the double beta decay experiments Heidelberg-Moscow and Cuoricino, the dark matter detector DAMA, and two low-energy accelerators for nuclear astrophysics cross section measurements.

At about the same time Japan began an underground science program that has produced exceptional results. The two sites are an abandoned rail tunnel near Oto (1400 mwe) and the operating Kamioka mine (2700 mwe). The Oto site houses kilogram-scale double beta decay and dark matter experiments. The shallowness of the site will limit future efforts to improve sensitivity. Kamioka housed the Kamiokande water Cerenkov detector, a proton decay experiment that also made historic measurements of the $^{8}$B solar neutrino flux and of the neutrinos from Supernova 1987A (along with IMB), and its currently operating successor, the 50-kiloton experiment Super-Kamiokande. Super-Kamiokande confirmed earlier suggestions of an atmospheric neutrino anomaly, providing definitive evidence for massive neutrinos and neutrino oscillations due to the quality of the data it provided. Super-Kamiokande has also established the most rigorous bounds on proton decay and produced an exquisitely precise measurement of the $^{8}$B neutrino flux. Recently, the original Kamioka detector was replaced by a kiloton liquid scintillator experiment, KamLAND. By measuring the interactions of antineutrinos produced in distant Japanese power reactors, KamLAND confirmed the solar neutrino oscillation results of SNO, and narrowed the allowed range of neutrino mass-squared differences. Future experiments to be undertaken in Japan include a megaton water Cerenkov detector for long-baseline neutrino and proton decay measurements, and a 10-ton liquid Xe detector for dark matter, double beta decay, and low-energy solar neutrinos.

Smaller laboratories are also making important contributions. In Europe the LSC (Canfranc), LSM (Frejus-Modane), and Gotthard (Mont Blanc) facilities were built off road tunnels. Canfranc (2450 mwe) current houses the IGEX $^{76}$Ge double beta decay experiments as well as developmental efforts on dark matter and solar axions. Frejus (4900 mwe) is the site of NEMO3 double beta decay experiment, which hopes to approach 100 milli-eV neutrino mass limits in measurements on several nuclei (including $^{100}$Mo, $^{82}$Se, and $^{116}$Cd). In Finland cosmic ray studies are being done at the Pyhasala Mine at depths up to 900 m, and the lowest level (1440 m, or 4050 mwe) could be developed for future experiments. There are pending proposals for searches for multi-muon events and for a long-baseline (2288 km) experiment with a CERN neutrino beam. In the United Kingdom the UK Dark Matter Collaboration operates a facility in the Boulby potash mine (3350 mwe). The ton-scale next-generation detectors DRIFT and ZEPLIN will follow the smaller dark matter experiments currently housed there.

The world's deepest active facility is the Sudbury Neutrino Observatory in Sudbury, Ontario, Canada. Located at a depth of 6000 mwe in an operating nickel mine, the SNO solar neutrino experiment recently showed that heavy-flavor neutrinos comprise two-thirds of the solar neutrino flux, thereby resolving the long-standing solar neutrino problem. This was established by measuring charge and neutral current neutrino reactions off deuterium, as well as neutrino-electron elastic scattering: SNO consists of an inner vessel containing a kiloton of heavy water, surrounded by seven kilotons of ordinary water. Very recently the Canadian Foundation for Innovation provided funding for a second hall, with a volume of 15000 m$^3$, for the dark matter experiment PICASSO as well as a future double beta decay detector.

What sites are available in the US? The two operating underground laboratories are Soudan, a former iron mine now operated by the state of Minnesota as a park, and the Waste Isolation Pilot Plant (WIPP), a DOE facility in Carlsbad, New Mexico. Both are relatively shallow, 2080 and 1600 mwe, respectively. Until recently the Homestake Mine was the *de facto* deep facility in the US, housing the chlorine experiment as well as cosmic ray and double beta decay experiments. Because mining has ceased at Homestake, no science experiments are currently active.

US scientists have been prominent in underground science for 50 years. Neutrino astronomy began with the chlorine experiment in 1964. The IMB proton decay experiment, mounted in the Morton salt mine, was the sister experiment to Kamioka, establishing important proton decay limits, providing the first evidence for an atmospheric neutrino anomaly, and seeing neutrinos from Supernova 1987A. US scientists played crucial roles in developing both the $GaCl_3$ (GALLEX/GNO) and metal gallium (SAGE) radiochemical techniques; the pilot experiment for GALLEX was done at Brookhaven. Similarly, Herb Chen and colleagues at UC Irvine developed the conceptual design of SNO.

There have been serious proposals to create a US laboratory similar to Gran Sasso, occurring roughly forty (Luis Alvarez, Aihud Pevsner, and Fred Reines) and twenty (Al Mann and Bob Sharpe; UC Irvine) years ago. Generally the US has had few opportunities, lacking the deep road and railway tunnels common in Europe. Thus the Mann/Sharpe and UC Irvine proposals of the early 1980s were "greenfield" projects. The former proposed sinking a vertical shaft at a site near Yucca Mountain, Nevada, and the latter advocated excavating a long tunnel beneath Mt. San Jacinto, near Palm Springs, California. Department of Energy officials also encouraged an early EarthLab proposal in the 1980s. These proposals were considered seriously, but ultimately were not funded. Today the situation is qualitatively different: there has been an explosion of interest in underground science, many of the world's current facilities are either too shallow or technically inadequate for the next-generation experiments now under discussion, and a superb existing site has become available in the US.

**Why the US needs NUSEL: Collaboration and Community Views.** Here we summarize the key arguments for establishing a dedicated, multipurpose deep underground science and engineering laboratory in the US:

*The science is compelling:* This was summarized in Section I. While the impetus for NUSEL came from the physics community, the laboratory is also important for earth science, for applications such as detector development for homeland security, materials purity, environmental geochemistry, etc., and as an education and outreach center.

*There is a lack of deep sites:* Gran Sasso, at a depth of 3800 mwe, and Kamioka, at 2700 mwe, were proposed twenty years ago. Since that time the background requirements of underground experiments have typically increased by two to three orders of magnitude. While clever chemistry and careful materials science can eliminate many radioactivity backgrounds, often there is no solution to cosmogenic backgrounds other than depth: the long-lived activities associated with cosmic rays as well as associated "punch-through" neutrons can be impossible to veto effectively. One thus concludes that a next-generation Gran Sasso, built today, ought to strive for a factor of 100 lower cosmic ray muon intensity, corresponding to an additional ~ 3000 mwe in cover. The proposed deep level of NUSEL, at 7400 feet or 6500 mwe, very nearly satisfies this rule of thumb. The deepest level of the Homestake mine is 8000 ft, or 7200 mwe.

The need for such depths is apparent from existing experiments. For example, SNO faced difficult challenges, detecting a charged-current signal that exhibits only gentle angular variation (1- cos(θ/3)) and a nonspecific neutral current signal, single neutrons. Delayed βs from cosmic ray muon activation of nuclei were the principal background concern in designing the detector. In Kamioka II, where larger backgrounds could be tolerated because elastic scattering events are sharply forward peaked and thus correlated with the sun's position, ~ 1% of muons penetrating the detector produced spallation-product βs above a ~ 10 MeV threshold. The SNO experimentalists wanted to reduce this background by a factor of 200, cutting the number of spallation events above detector threshold to about 1.5/day. This required the collaboration to develop a new site at 6000 mwe in an operating nickel mine, with the cleanroom and access challenges accompanying such a site. While the experiment is a spectacular success, the difficulty of working in an active mine environment contributed to the four-year delay in obtaining first results. If SNO had been done at Gran Sasso depths with the same stringent background cuts, the resulting detector deadtime (40%) would have seriously diminished the experiment.

Our collaboration, the working groups at the Lead and NESS02 meetings, and several review committees, including the NRC Neutrino Facilities Assessment Committee, have studied the depth requirements of future underground experiments. A consistent picture has emerged, described in the Science Book. A few examples:

- Next-generation dark matter searches. These have as their ten-year goal sensitivity to a Weakly Interacting Massive Particle (WIMP) cross section of $10^{-10}$ pb. Scaling background rates in existing experiments and assuming that an additional factor-of-ten reduction is obtained by vetoing multiple scattering events (by improving detector granularity), a depth of 4500 mwe is required to reduce backgrounds to 50% of the expected signal (10 WIMP events/year in a one-ton experiment). If WIMPS are not observed at $10^{-10}$ pb, a depth of 6000 mwe might be helpful in establishing the tightest possible limit. If such depths are not available, sophisticated shielding and vetoing might gain back a factor of 10 or more: this would include a thick (1-2m) scintillator active veto around the detector (to tag inward penetrating high-energy neutrons) and instrumentation of the cavity's rock walls to catch some part of the shower associated with the initiating muon. As such mitigating steps are expensive, depth is the simpler solution.
- Double beta decay. The most reliable estimate is likely that for the $^{76}$Ge experiment Majorana, as cosmic ray background rates are known for smaller current-generation Ge experiments, such as IGEX. (Approximately one third of the total IGEX count rate in the endpoint region of interest (2038 keV) is correlated with the cosmic muon veto. IGEX is located at 2450 mwe in the Canfranc tunnel.) The principal concern is cosmic ray interactions near but external to the veto shield, producing neutron secondaries which enter the detector. The experimenters have estimated that 4500 mwe is the minimum depth for Majorana, with additional cover being of benefit. The background rates for the Mo double beta decay experiment MOON are more worrisome, with reactions like $^{12}$C(n,3n) occurring in the scintillator producing $^{10}$C, a positron and γ-ray emitter. While the effectiveness of possible cuts will depend on the position resolution achieved in MOON, a safe depth is considered to be 6000 mwe. In contrast, one study indicates that EXO, a liquid Xe TPC, might require an overburden of only 2400 mwe. EXO experimentalists are developing a laser resonance ionization technique to tag the Ba ion daughter produced in the ββ decay, a novel method of background suppression. The trigger rate for this tagging depends on the background rate, and there are potential cosmic ray neutron backgrounds that have not been fully assessed. Thus EXO experimentalists have proposed a pilot experiment to determine background rates and thus the overburden required for a reasonable trigger rate.
- Proposed solar neutrino experiments to measure the low-energy pp neutrinos, thereby better constraining $\theta_{12}$, include HERON and CLEAN, helium- and neon-based detectors for elastic scattering, and the TPC experiment HELLAZ. The minimum depth for HERON and CLEAN is estimated to be 4500 mwe, though the experimenters would prefer to site these detectors at ~ 6000 mwe: especially in the case of CLEAN, there are significant uncertainties in cosmic-ray-induced activity levels due to poorly known spallation yields for neon. HELLAZ is an exception, capable of operating with a 1% deadtime at a depth of only 2200 mwe.

A fair summary is that 4500 mwe is a minimum depth for most of the background-sensitive underground experiments that are now approaching readiness; but generally "deeper is better" because the background estimates are uncertain. Most of these experiments would be problematic at Gran Sasso depths (3800 mwe).

For EarthLab, the issues are simple. The earth science benefits from access to the greatest range of depths. Homestake is the deepest mine in the US and provides access approximately every 150 ft, from surface to 8000 ft. The geomicrobiology goal is to reach the limits of life, expected at 16,500 ft. Thus 8000 ft is a good starting point for the drilling and coring program the geomicrobiologists will initiate.

*There is a lack of space in existing laboratories:* Gran Sasso, which is fully subscribed, has tried, over the past five years, to expand. This generated a lively public debate over environmental concerns connected with the water table, recently complicated by a 50-liter pseudocumene spill that reached a local stream. Baksan, the deepest of the multipurpose laboratories, has suffered because of Russian cutbacks in science. In the US, the cessation of mining in Homestake has (temporarily) stopped science there. Remaining are two relatively shallow US laboratories at Soudan and WIPP.

*The lack of a US laboratory has inhibited the development of underground science here:* Despite the discoveries spawned by the chlorine experiment, no further solar neutrino experiments were sited in the US. Similarly, all active double beta decay experiments are located overseas. Nevertheless, the US community has remained very active, collaborating on GALLEX, SAGE, Kamioka and Super-Kamiokande, SNO, KamLAND, etc.

Given the growing importance of nonaccelerator physics, we are fortunate that a new window of opportunity for creating a US laboratory has opened. The recognition that very long baseline oscillations are crucial for probing neutrino parameters and new phenomena like CP violation favors the US, where we can mount such experiments. The availability of a site that can easily accommodate megaton experiments while providing incomparable depth should attract international experiments here, if the site is properly developed.

*There are important cost advantages in creating a multipurpose laboratory now:* We recognize that underground science is a rapidly growing field, and it is apparent that many next-generation experiments will be large, technically demanding, and costly. Viewed in the long-term, there are significant cost savings in planning a multipurpose facility that can facilitate next-generation physics experiments while also serving earth science and various applied fields well. These include:

- Capital costs savings in having a single site with the capacity to manage several generations of new experiments. The capital costs include the physical plant – the shafts and hoists, the experimental halls, sophisticated HVAC systems, dewatering systems, fiber optics and phone communications, and surface office buildings – as well as ancillary services such as the computing system, libraries, and machine, chemical, and glass shops. In general these facilities can be shared by many users, and reused as early experiments are completed and new ones are mounted. The alternative – experimenters operating in a variety of more limited laboratories, or attempting to develop their own infrastructure at a parasitic site – leads to duplication and, generally, a lower standard of support at each facility. For example, the kind of low-level counting facility that we envision for NUSEL was identified by the Bahcall Committee as a key component of NUSEL infrastructure, required (and requested by) many of the groups planning new experiments. Both the facility and the highly trained staff necessary to operate the facility would be beyond the reach of most individual groups. But as a component of common infrastructure, the facility can advance many projects by allowing experimenters to test materials before they are incorporated into new detectors.
- Human cost savings possible with a dedicated, multipurpose facility. Less than ideal sites imply hidden costs because experiments take longer, in effect shortening the useful lifetimes of the nation's most talented experimentalists. For example, factors contributing to the four-year "stretch-out" of SNO included difficulties with mechanical systems and access limitations to the site. Certain access limitations – SNO shifts have been canceled because of seismic activity – may be endemic to a site, regardless of its operations. But other SNO delays, such as an INCO contractor needing exclusive use of the hoist for an extended period, would not arise in a dedicated facility. Access is also a crucial issue in safety: if a fire occurs when an experiment is off limits, the consequences could be catastrophic.

- Operations cost savings. Hoist maintenance and operations and other activities necessary to provide access to depth as well as dewatering costs are essentially independent of the number of experiments conducted at depth.
- Training costs. We believe every experimental group that has used a new site has gone through a painful period when either scientists have had to be trained in underground operations, or miners have had to be trained to help with experimental operations. Similarly, no facility can operate without an experienced safety crew. An important efficiency in creating a multipurpose laboratory now is the opportunity to train these staff once, then have them available to assist many projects with safety and detector installation.

*NUSEL will encourage synergies that will advance science generally.* One of the great contributions of particle physics has been the creation of laboratories like FermiLab and CERN that have advanced science and technology beyond what was otherwise possible: the sum is greater than the aggregate parts. During this past decade we have seen the consequences of rapid dissemination of new technology in underground science. KamLAND mounted a difficult experiment in record time in part because of the experience with radiopurification gained in SNO and Borexino. A strong laboratory accelerates the pace of science because it creates and maintains a technology base that eases the way for the next experiment – NASA's goal of faster, cheaper, and better. As with FermiLab and CERN, NUSEL is an essential step in allowing the underground science community to tackle experiments of a qualitatively new scale. NUSEL must provide the infrastructure to support ambitious projects, and it must help focus the community. The expertise that will come to NUSEL includes the large-detector project management skills of particle physics, the low-level counting skills of nuclear physicists and chemists, and the geotechnical knowledge of the earth science community. NUSEL will ensure that these groups interact daily.

These and other arguments have been considered by several high-level advisory panels:

***The Nuclear Science Advisory Committee****:* NSAC is the standing committee charged with advising the NSF and DOE on nuclear science policy. In NSAC's recent Long-Range Plan process, NUSEL was one of nine mid-scale new construction projects (defined as costing $300M or less) considered for the next decade. The initial push to create NUSEL came from the LRP preTown Meeting on Neutrinos, attended by some 200 community members: Homestake's closing was announced just two weeks prior to that meeting. Creating a US deep underground science laboratory was the #1 recommendation of the preTown meeting, a ranking that was confirmed later in the larger Town Meeting on Astrophysics, Neutrinos, and Symmetries. The Bahcall Committee, formed after the Neutrinos meeting, submitted its report as a White Paper to NSAC. This formed the basis for presentations and discussions at the final LRP summary meeting in Santa Fe, March 2001. NUSEL became the highest midscale construction priority for the field and the third priority overall. (The first priority was support for the existing program, and the second was the one major new construction project proposed, the Rare Isotope Accelerator.)

The final LRP recommendation reads:
***We strongly recommend immediate construction of the world's deepest underground science laboratory. This laboratory will provide a compelling opportunity for nuclear scientists to explore fundamental questions in neutrino physics and astrophysics.***

*Recent evidence for neutrino mass has led to new insights into the fundamental nature of matter and energy. Future discoveries about the properties of neutrinos will have significant implications for our understanding of the structure of the universe. An outstanding new opportunity to create the world's deepest underground laboratory has emerged. This facility will position the U.S. nuclear science community to lead the next generation of solar neutrino and double beta-decay experiments.*

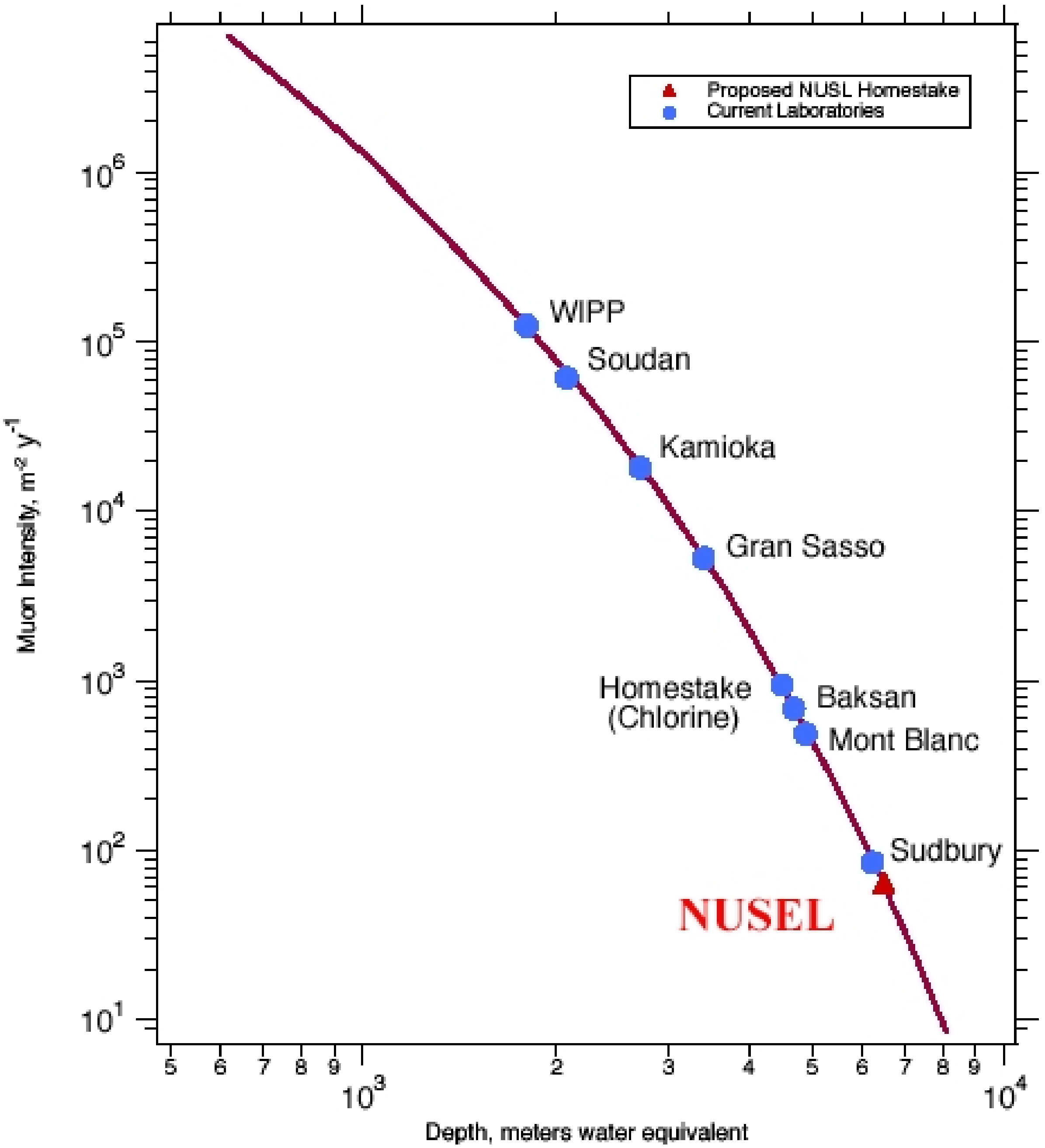

Figure B.1: The depth and muon intensities for existing laboratories and for the produced developments at Homestake. The Homestake main level (shown in the figure as NUSEL) is at 7400 ft, though development will also occur at the 8000 ft (earth science) and 4850 ft (Homestake-chlorine; accelerator physics, megadetector development). Note that the two current US laboratories, Soudan and WIPP, are shallow. The one deep North American site, Sudbury, provides limited space and requires parasitic use, as Sudbury is an active nickel mine.

Following the LRP process NSAC wrote Dr. Robert Eisenstein, then NSF Assistant Director for Mathematical and Physical Sciences:

*This letter is written to express unanimous support by the members of the Nuclear Science Advisory Committee for the creation by the National Science Foundation of a National Underground Science Laboratory at the Homestake Mine site in South Dakota … In view of the compelling nature of the science to be done in a deep underground laboratory, and the potential benefit to the United States of world leadership in this area, we strongly recommend that the NSF undertake an immediate study to assess the feasibility of the Homestake site for a NUSL, and that it invite a detailed proposal and technical design report for serious consideration. NSAC is unanimous that the initiative to create NUSL is very important to further fundamental progress in the physical sciences…*

***The High-Energy Physics Advisory Panel**:* HEPAP started a similar Long-Range Plan in summer 2001. While the recommendations of the LRP Subpanel's January 2002 report focused on one megaproject, a high-energy electron-positron collider sited either in the U.S. or overseas, NUSEL was addressed in the body of the report:

*Worldwide, the program of experiments of interest to particle physicists that require underground locations is broad and technically challenging. Experiments include: searches for neutrinoless double beta decay; searches for weakly interacting dark matter; measurements of solar, atmospheric, reactor, and supernova neutrinos; searches for proton decay; and studies of neutrino properties using beams from distant accelerators.*

The subpanel noted that some experiments could be done in shallow sites like Soudan and WIPP, but that only Homestake and the proposed San Jacinto excavation would be deep enough to provide the very low background required for a variety of experiments. The discussion concludes:

*Construction of a National Underground Science Laboratory at the Homestake Mine has been proposed to the NSF. A proposal for a laboratory under the San Jacinto Mountain has been submitted to DOE and NSF. These proposals are motivated by a very broad science program, from microbiology to geoscience to physics. Construction of a national underground laboratory is a centerpiece of the NSAC Long Range Plan.*

***We believe that experiments requiring very deep underground sites will make important contributions to particle physics for at least the next 20 years, and should be supported by the high-energy physics community. Particle physics would benefit from the creation of a national underground facility.***

Recently, in response to a request from the Director of the DOE Office of Science, HEPAP examined the scientific merit and readiness of major projects proposed for the next two decades. Their report, "High Energy Physics Facilities Recommended for the DOE Office of Science: Twenty-Year Roadmap," was released March 2003. To achieve the highest science rating, *absolutely central,* HEPAP required "that the intrinsic potential of the science be such as to change our view of the universe. This is an extremely high standard, at the level at which Nobel Prizes are awarded." Of the eleven projects recommended, four would impact the science program of NUSEL: liquid Xe double beta decay; a neutrino superbeam; a megadetector for next-generation experiments on nucleon decay, long-baseline oscillations, supernova neutrino detection, and related neutrino physics; and a neutrino factory. The first three were rated absolutely central. The neutrino factory was considered to be too early in the R&D phase to allow a meaningful evaluation of the physics potential. In NSAC's parallel exercise, two of the seven projects considered require NUSEL: a double beta decay experiment (Majorana or MOON) and a low-energy pp solar neutrino detector. Both were rated absolutely central.

***National Research Council Committee on the Physics of the Universe:*** This committee examined the intersection of physics and astronomy in order to identify opportunities for breakthroughs in understanding the birth, evolution, and destiny of the universe, the laws that govern it, and the nature of space and time. The charge included many interdisciplinary problems of concern to NASA, the NSF, and the DOE. The committee organized its work around 11 unresolved grand questions. In its 2003 report, "Connecting Quarks with the Cosmos: Eleven Science Questions for the New Century," the Committee included recommendations for three new projects: probing aspects of the polarization of the cosmic microwave background that would test the inflationary universe; determining the nature of the dark energy; and underground science. The underground science recommendation states in part:

*Three of the committee's 11 questions – the nature of dark matter, the question of neutrino masses, and the possible instability of the proton – must be addressed by carrying out experiments in a deep underground laboratory that is isolated from the constant bombardment of cosmic ray particles. One of the most important discoveries of the past ten years, that neutrinos have mass, was made in an underground laboratory. This discovery has implications for both the universe and the laws that govern it. The mass scale implied by the measurements to date suggests that neutrinos contribute as much mass to the universe as do stars; and neutrino mass points to a grander theory that brings together the forces and particles of nature and may even shed light on the origin of ordinary matter.*

*The committee believes that there are more opportunities for discovery at an underground laboratory. Experiments proposed for the near future to address the fundamental questions it has identified require depths up to 4000 meters of water equivalent (mwe). More visionary experiments, as well as the long-term potential of such a laboratory to make discoveries, require even more shielding, to depths up to 6000 mwe …*

*A North American laboratory with a depth significantly greater than 4000 mwe and adequate infrastructure would be unique in the world and provide the opportunity for the United States to take the lead in "underground science" for decades. Such a laboratory might also be useful for carrying out important science in other disciplines, such as biology and geophysics.*

***Recommendation: Determine the neutrino masses, the constituents of dark matter, and the lifetime of the proton. The committee recommends that the DOE and NSF work together to plan for and fund a new generation of experiments to achieve these goals. It further recommends that an underground laboratory with sufficient infrastructure and depth be built to house and operate the needed experiments.***

***National Research Council Neutrino Facilities Assessment Committee:*** A second NRC committee was charged in 2002 to examine the physics justification for IceCube and NUSEL. The committee's report, "Neutrinos and Beyond: New Windows on Nature" was released in 2003. Its recommendation on underground science states:

*Underground research was pioneered 35 years ago in the United States with the detection of neutrinos from the sun. Development of a new underground facility could restore U.S. leadership in this important research area. Such labs are required to study rare forms of penetrating radiation and rare nuclear processes in a low radiation background environment. Two key attributes of an underground laboratory are required: it must be able to site experiments as deep as 4500 mwe with future capability of siting them down to 6000 mwe, and it must be located more than 1000 kilometers from accelerators capable of producing intense beams of neutrinos. The latter would allow the use of such beams to study the properties of neutrinos as they travel over long distances, helping to measure the small but important neutrino masses. Siting the laboratory in the United States would permit the utilization of its powerful particle accelerators already operating. A lab with sufficient shielding could also be a site for various geophysics and geobiology projects.* ***A deep underground laboratory can house a generation of experiments that will advance our understanding of the fundamental properties of neutrinos and the forces that govern elementary particles, as well as shedding light on the nature of the dark matter that holds the Universe together. Recent discoveries about neutrinos, new ideas and technology, and the scientific leadership that exists in the U.S., make the time ripe to build such a unique facility.***

## III. What are the important attributes of a next-generation underground laboratory?

The following are among the key considerations in planning for the next-generation underground science and engineering laboratory:

*Depth:* If the facility is to accommodate the background-sensitive experiments now being planned (as well as those that will arise in the next two decades) a depth of at least 6000 mwe is needed. Other experiments may prefer intermediate depths for (a) more convenient access, (b) lower lithostatic pressure and thus easier and less costly large-cavern construction, and (c) a non-vanishing muon flux for detector calibration or other purposes. Thus a versatile facility must be able to offer a range of depths.

*Detector Halls:* One key element of the laboratory is the underground halls. There are two models now employed. The Gran Sasso excavation produced three large detector halls, each approximately 20 m wide by 100 m long by 20 m in height. The director, aided by an advisory committee, allocates space in these halls based on the strength of the proposed experiments. Some small detectors at Gran Sasso are sited in various access tunnels and ancillary spaces. This large, multipurpose hall approach has worked well. The planned second SNO hall is in the tradition of those at Gran Sasso (though about 3/8ths as large). The Kamioka and original SNO excavations are quite different. First, the constructed cavities are upright cylinders with hemispherically-domed ceilings and primary access near the top. This is an attractive design for very large water detectors like SNO (8 ktons) and SuperK (50 ktons). Second, the excavation was specialized, not general purpose, designed to optimize a specific experiment.

The "hybrid" model to which SNO is evolving is an attractive one. Certain classes of experiments – such as dark matter searches and double beta decay – have very similar needs in terms of floor space, ventilation, background requirements, utilities, and depth. Thus several smaller halls or one general-purpose hall could be optimized for this class of experiment, and reused for successive efforts. But NUSEL must have the flexibility to construct customized space for detectors with special requirements, including large volumes of water, electrical isolation, and physical isolation and venting for safe use of flammables, cryogens, toxic materials, and suffocating gases. This requires a very special site both physically and in terms of the capacity and flexibility of its ventilation system.

The capacity to customized cavities for challenging experiments implies the retention of mining capability throughout the laboratory's lifetime. As these excavations will occur while other experiments are operating, the ventilation system (and other barriers) must be adequate to keep dust out of occupied halls.

*The capacity to excavate a megadetector at depth:* If a megadetector is to serve multiple purposes – e.g., proton decay, long-baseline neutrino physics, and a supernova neutrino observatory – it must be located at a reasonable depth and be operable for a period ~ 50 years. The Bahcall Committee specifications for such construction were inspired by UNO: a "rural mailbox" hall with gross volume of at least 0.9 million $m^3$ (0.5 million $m^3$ of liquid) at depths of at last 4000 mwe, specifications necessary for a next-generation water Cerenkov detector to be factors of 10 more sensitive and quieter (in terms of background rate) than SuperK. This places very stringent constraints on the host site. As such an experiment is crucial part of the NUSEL program, it must be **known** that rock in the vicinity of the laboratory is sufficiently competent to permit such an excavation. The only accepted technique for doing this is coring of the site, followed by laboratory testing of the core samples. A geotechnical site history, particularly one in which other large cavities have been excavated and monitored for stability, is also helpful.

Another issue is the feasibility of rock disposal. The "rural mailbox" excavation will produce about 2.5 megatons of rock, which is considered solid waste when brought to the surface. Thus, in addition to the civil engineering challenge of mining this volume, NUSEL must solve the transportation and permitting/waste disposal issues.

*Cleanliness.* While depth is the optimal solution to troublesome cosmic ray backgrounds like "punch through" neutrons and delayed spallation-product activities in the detector volume, good ventilation engineering and infrastructure are important in combating natural radioactivity backgrounds. A great deal has been learned about the necessary engineering. For example, radon can be controlled by mineguard (polyurethane sealant) coating of rock walls (which can provide reductions of ~ $10^6$) and proper ventilation. The scrubbed air in contact with experiments should have residual radon levels of no more than 1 $Bq/m^3$. Some detectors have a critical volume needing to be completely purged of radon, which can be accomplished by using boil-off gas from radon-free liquid nitrogen. A radon-free materials storage area is also important. Experiments such as Majorana and MOON are sensitive to activities induced when materials are exposed on the surface, including $^{68}$Ge, $^{60}$Co, $^{93}$Mo, $^{99,100}$Nb, and $^{91}$Mo. For this reason Majorana requires an underground Cu electroforming facility, and is interested in the possibility of underground Ge crystal growth and detector preparation. In experiments like SNO, Borexino, and KamLAND, low-level counting techniques capable of measuring U and Th at levels of $10^{-16}$g/g are an important part of background control. NUSEL must maintain the infrastructure necessary to address these background issues. The low-level counting facility is one example. NUSEL will help raise the overall level of background reduction expertise: it brings experimentalists into contact with one another, so that new background reduction methods are quickly disseminated.

*The importance of 24/7 access:* Observers familiar with how experimental science is really done understand that very few factors are as important as personnel access to detectors, 24 hours a day and seven days a week. Such access may be necessary in construction and operations phases of experiments, and the ability to visit any hall at any time is an important safety issue.

*Safety, special materials, and materials handling:* Next-generation experiments will require safe use and handling of flammables, cryogens, certain toxic chemicals, suffocating gases, and possibly even mega-Curie neutrino sources for detector calibration. NUSEL should have the physical and operational systems necessary to guarantee safe storage and transport of such materials. The required facilities include isolated laboratories, ventilation systems capable of isolating and separately venting certain areas, and effective sensor and suppression systems. Cavities housing large quantities of liquids must be located below grade. The operations requirements include safety teams capable of handling hazards from both conventional mining and construction and from exotic detector materials. Also included are all possible preventative measures: the laboratory must be able to guarantee that each experiment is being conducted responsibly, and that all regulatory and licensing constraints are respected.

*Access for large and heavy equipment:* The cost of detector assembly escalates if one encounters a “ship in the bottle” problem: hoists or other access that requires detectors to be broken down into very small modules, then reassembled at depth. NUSEL must have the capacity to transport standard modules of significant size and weight from a cleanroom in the surface laboratory (or from a loading dock) to an underground hall. It must have a trained staff that can manage such transport in minimal time and effort, without risk of equipment damage. The standardization of module sizes will allow experimentalists in any laboratory to preassemble their equipment in the most efficient way, knowing that NUSEL has the capacity to handle the modules. NUSEL should retain the capacity to handle special loads, such as long steel I beams, oddly shaped sections of tanks, etc. (Mining companies frequently handle large, odd-dimensioned loads by slinging them beneath the hoist cage.) A pumping system with the capacity to empty a water megadetector, either for maintenance or decommissioning, is important.

*Quietness and stability: electrical, mechanical, seismic:* Many underground detectors use sensitive transducers that produce signals at very low levels. While sensitive pre-amplifiers can be protected with electrical shielding and good grounding, experimenters will benefit from well-designed and well-isolated clean power sources, and from electrical isolation and grounding systems that minimize ground loops and cross-talk. Electrically and mechanically noisy equipment should be isolated in specialized underground rooms. The NUSEL power supply should be stable, with a history of infrequent power outages, and uninterruptible power supplies should be in place to keep critical equipment operating during power spikes or failures. The laboratory should be in a seismically quiet region, and rock bursts should be rare, as detectors respond to both.

*Adequate support facilities and a strong scientific environment:* The scientific environments in which many underground experiments have been conducted in the past have been far from ideal. Scientists far from their home universities need shop services. They also benefit from the intellectual atmosphere that accompanies a critical mass of experimentalists and theorists. Therefore, if NUSEL is to be a national and international center, we believe the right model is Gran Sasso, which strives to provide significant services to the experimentalists in a setting not unlike a university physics department. The technical services include machine, chemical, and glass shops, and administrative services include shipping and receiving, computing, maintaining inventories, libraries, and visitor services. Gran Sasso employs approximately 33 staff for these functions. The low-level counting and underground materials facilities envisioned for NUSEL would be additional examples. For below-ground operations, the hybrid model for NUSEL – initial excavation of multipurpose halls coupled with continued development of special-purpose cavities throughout the lifetime of the laboratory – will require the laboratory to maintain adequate engineering and excavation capabilities (permanent or contract staff), in addition to operations/maintenance and safety/environment groups. Gran Sasso has a theory group and a vigorous seminar program. It frequently hosts conferences or workshops, and is playing a leadership role in IUPAP-sponsored underground science activities.

*Quality of life issues:* These begin in the workplace, with efficient layouts, good lighting, well-controlled temperature and humidity, and pleasing aesthetics – in contrast to some of the difficult (and occasionally dangerous) conditions experimentalists have endured in certain parasitic sites. The goal is to minimize fatigue that scientists and staff experience because of long hours of work, and thus accidents connected with fatigue. The underground campus should be designed with the same attention to workplace ergonomics that is now routine in surface laboratories. Support services that contribute to the quality of life include a library, access to electronically-archived literature, a cafeteria and vending machine areas, comfortable accommodations for visitors (laboratory or privately operated), seminar rooms, and reciprocal agreements with area universities. The community context, good housing and schools and cultural and recreational opportunities, affect the quality of life outside the laboratory.

*Outreach and education:* The Kamioka, Soudan, and Gran Sasso laboratories all conduct successful outreach and educational programs. They host regular, ongoing tour programs for K-12 students. The Soudan Laboratory, as part of a state park, provides both historical and scientific tours to the lowest level of the Soudan Mine. The outreach and educational opportunities depend on site-specific factors like regional K-12 student and visitor populations, whether NUSEL is a regional science focal point or one of many, and historical or other connections that help to interest the general public.

*Neutrino beams:* Proposals for next-generation neutrino oscillation experiments to probe phenomena like CP violation favor baselines of 1000-3000 km. Thus the distance between accelerator laboratories and NUSEL is relevant.

**IV. Why is Homestake the preferred site for NUSEL?**

This section is broken into two parts. The first is a rather complete description of the physical plant available at Homestake, at no cost to the public: the mine owner, Barrick Gold Corporation, has agreed in principle to donate the needed portions of the Homestake campus to the state of South Dakota. The second section is a point-by-point discussion of the advantages of Homestake, addressing each of the issues raised in Section III as well as others we feel are pertinent.

As an introduction, it is important to be able to distinguish what is purposed here – a multipurpose, next-generation underground science and engineering laboratory that will facilitate new physics experiments of unprecedented size and complexity, meet the goals of EarthLab proponents, contribute to a variety of important applied-science fields, and conduct a vigorous outreach program -- from past practices in underground science.

In the past, site preparation, access, and operating costs associated with going underground often had the potential to account for a large fraction of an experiment's budget. The cost-benefit analyses have led experimenters to seek shared facilities, even if those facilities are less than ideal. Homestake, Kolar Gold Fields, Kamiokande, Super-Kamiokande, SNO, and WIPP share/shared facilities with mining operations, where the key limitations have been access/capacity restrictions (Homestake, Kolar, SNO), safety (Kolar), or lack of depth. Generally these mine environments have been the locations where specialized cavities were built, however, utilizing the host's in-house mining expertise. Facilities sharing with motor or railways include Gran Sasso, Frejus, Mont Blanc, Canfranc, and Oto. These facilities have offered excellent access, and particularly in the case of Gran Sasso, have further reduced costs per experiment by hosting multiple experiments. The limitations of these facilities include lack of depth (Oto, Canfranc, and perhaps soon Gran Sasso) and size (Frejus, Mont Blanc, Oto), when viewed in the context of experiments planned for the next ten years. To date these laboratories have not tackled the specialized cavity excavations that were central to the key discoveries by Super-Kamiokande and SNO: the tunnel laboratories are of the experimental-hall type, with the space excavated at the time of initial construction. (The possibility of an UNO-like new construction at Frejus, however, has been discussed.)

One exceptional case is Baksan, a "greenfield" project in which a tunnel was excavated to considerable depth (4700 mwe) for the purpose of creating a scientific laboratory.

The motivation for parasitic use is cost. For example, the hard-rock mining costs of sinking new shafts are typically $10,000-20,000 per mwe overburden. Thus the creation of dual access to a site at 6000 mwe represents an initial investment of $120-240M, **after which** one must invest in hoists, utilities, ventilation, hall construction, and other facilities necessary to a laboratory. Thus most experimental groups have elected to share space and facilities with nonscientific hosts. In mine environments parasitic use has generally increased science operations costs due to the resulting longer "time to physics." Capacity on the lifts for personnel and materials is limited and the host's prime function (mining) has priority. Lifts optimized for mining generally are divided in ways that make transport of large experimental modules very difficult. The mine environment is dirty. Whether a mine or tunnel, the need for more exotic detector materials and large volumes of flammables, compressed gases, and asphyxiates may make relations between host and scientists increasingly complicated in future experiments.

This context helps one to appreciate the opportunities Homestake provides:

- The existing physical plant is remarkable. Massive shafts and hoists provide dual access to every level. The ventilation, air conditioning, water supply, dewatering system, phone and radio communications systems, fiber optics network, compressed air supplies, and sensor and safety systems are in place. **No major construction** is required underground to convert Homestake into an unmatched scientific laboratory, apart from hall excavation. Experiments using flammables, cryogens, etc., can be isolated because of the physical extent of the site and of the utilities systems.
- The geotechnical attributes are equally remarkable. Homestake is the **deepest site in the US** (8000 ft, or 7100 mwe). (The density of Homestake rock is 2.91.) It provides a variety of levels and great flexibility is optimizing sites for experiments. The region is seismically quiet and the rock exceptionally competent, with rock bursts rare even at 8000 ft. The areas proposed for experimental halls (6600 mwe and 4300 mwe) can be cored and evaluated before construction. There is an extensive geotechnical database, including studies of the stability of large cavities at depths up to ~ 7000 ft. The site is ideal for earth science, as it provides 3D access to approximately 9 $km^3$ of well-characterized and interesting rock. Thermophilic methanogens have been identified at the 8000 ft level. The **existing mining capacity** of either main shaft to the 4850 ft level is sufficient to excavate an UNO-size cavity in 2.5 years.

- Highly skilled workers – engineers, geologists, miners, administrators – who **know how to operate a complex facility efficiently and economically** are available. They are part of an established operations system. Thus nonscientific operations costs are known. As an operating facility the many permitting obstacles facing greenfield sites – construction, environmental, safety, waste rock disposal – have been addressed. There is an on-site location for waste rock that avoids surface transport (e.g., in the case of UNO).

**Homestake's Physical Attributes:** A summary of the site's present physical characteristics, hoist capacity, utilities, and operations is given below.

*Site physical characteristics (underground):* Levels at Homestake have been developed from the surface to the mine's bottom, at 8000 ft, roughly every 150 feet. Personnel and equipment access to the underground is primarily through the Ross and Yates shafts, both of which terminate at the 4850 ft level (4300 mwe). The shaft cross sections are very large, 19.3 by 14.0 ft and 27.7 by 15.3 ft, respectively. The shafts, separated by about 1000 m, are connected at their bases by major drift, 13-15 ft wide and arching to 12 ft, equipped with an electric trolley line. This drift is one of two main levels that will be developed for NUSEL.

Below this level there are three ways to access lower levels. The No. 6 Winze begins 100m horizontally from the bottom of the Ross shaft and continues to the base of the mine, at 8000 ft. With the installation of a convenient transfer station between the Ross and No. 6, these two shafts will serve (in the development plan we prefer) as the main scientific (clean) access to the mine, serving the 4850 ft, 7400 ft, 8000 ft, and other levels. A series of sloped ramps provides a second access route to levels between the 4850 ft and 8000 ft levels, and the No. 4 Winze a third route between 4850 ft and 7400 ft levels. On the 4850 ft level the base of the Yates, the Ross/No. 6, and the top of the No. 4 shaft form a triangle, with the No. 4 approximately 1900 m from the Yates and 1600 m from the Ross. On the 7400 ft level another major drift, with cross section 9 by 9 ft, connects the base of the No. 4 shaft with the No. 6 shaft. It is along this drift, beginning at the No. 6 shaft end, that the deep level of NUSEL will be developed.

Thus the existing Homestake layout allows development along the main drift at 4850 ft (4300 mwe, 500 mwe below the level of Gran Sasso), with the Yates and Ross providing direct access to the surface. This is the level recommended for megadetector construction (e.g., UNO) and for other experiments wanting to simplify construction and transport, assuming moderate overburden needs. The deep level at 7400 ft (6600 mwe) is the choice for experiments needing great overburdens. No shaft construction is required in this plan, and new drift construction is limited to the immediate areas near the experimental halls. Access to the 8000 ft level, where the geomicrobiology program will be based, is also through the Ross and No. 6 shafts, with the ramp system providing secondary access.

Later, in the **Homestake Facilities Development Plan** section, we will describe how the existing access and other mine infrastructure will be optimized for NUSEL. One goal is to exploit the existing dual access to all levels to sequester mining from scientific access. A second goal is to retain mining capability on all levels, so that specialized cavities can be constructed at any time during the lifetime of NUSEL. We discuss adjustments that can be made to optimize massive detector construction projects. The plan also describes how earth science needs – for instance, the desire to instrument much of the available 9 $km^3$ of rock – will be addressed.

*Capacity and condition of shafts and hoisting equipment:* Currently the Yates hoist is configured with four hoisting compartments, two for people and two skips for rock. The Ross is divided into two skips and a double-decker compartment for people and equipment. The No. 6 is very similar, but is not double-decker.

The hoisting equipment for the Yates and Ross shafts are pre-World War II, manually operated drum hoists with mechanical Lilly speed controls. Both hoists require mechanical motor-generator (MG) sets. The No 6 Winze hoist is approximately 30 years old and is semi-automatic. It also uses a MG set.

The hoists are powered by sets of 1250 hp DC Nordberg motors. The shafts and hoisting systems are serviceable and, with continued proper maintenance, will last indefinitely. The maximum cage load is 6 .7 tons, though the load is normally limited to 6 tons.

The Facilities Development Plan includes a series of important improvements to these hoists to make them suitable for scientific use, to increase the "footprint" (to 11 ft by 12 ft) and load limits, and to reduce operating costs. These include replacing the wooden Yates sets with steel sets; replacing the MG sets with semiconductor power supplies; replacing the manual and semi-automatic controls with fully automatic controls; and an option for upgrading to Kevlar ropes to increase the cage load.

*Dewatering system:* The Homestake Mine dewatering system is extensive, with excess capacity that could be very useful in filling or draining detectors. The Ross pumping system comprises major pumping stations on the 8000 ft and 6800 ft levels of the No. 6 shaft and on the 5000 ft, 3650 ft, 2450 ft, and 1250 ft levels of the Ross shaft. The total capacity of this system to the surface is approximately 2300 gpm through a 12-in-diameter pump column. Most of the water pumped from the mine is industrial water introduced from mining operations. The mine itself is quite dry, with just 470 gpm from groundwater. There is a small additional pump station located at the 1100 level of the B&M No. 2 shaft.

*Ventilation:* Three separate intake and exhaust circuits deliver 860,000 cubic feet per minute (cfm) of ventilation air to the headings of the underground mine. The surface fans include the Oro Hondo fan (3000 hp American-Davidson centrifugal with backward-curved airfoil blades) and the Ellison Fans (two Trane 89" DWDI centrifugal fans with 700 hp motors). The mine has also used five 100 hp section booster fans. Principal intakes include the Yates, Ross, and No. 5 shafts, with capacities of 270,000, 220,000, and 150,000 cfm. All of these intakes will be within the surface "footprint" inherited by NUSEL from the mine owners.

The Oro Hondo fan will be used for ventilating the underground drifts during NUSEL construction. The facilities development plan calls for a new fan system to be installed for permanent operations, while the Oro Hondo fan will be repositioned and reserved for emergency laboratory purging needs.

*Air conditioning:* The mine employed 2480 nominal tons of cooling in the deep levels. Cooling is delivered to the headings via nineteen 30-ton and two 60-ton spot-coolers, 290-ton and 350-ton water chillers with cooling coils, and a 2300-ton vent plant located at the 6950 ft level, which bulk-cools air and circulates it to the headings. This system exceeds considerably the airflow and cooling requirements of NUSEL. As the vent plant is expensive to operate, a replacement air-cooled chilling system will be implemented during NUSEL construction.

*Compressed air:* Current capacity of the Yates compressors is 20,000 cfm at 100 psi. Installation of local filter/dryer units will be required to obtain instrument quality air from this system. While the Yates compressors will be used during initial construction of NUSEL, we intend to install new compressors on both of the main experimental levels, and may use skid-mounted compressors for later underground excavation.

*Water supplies:* The mine has both potable and industrial water supplies. The potable water system is fed from several storage tanks located in Lead and meets all federal Safe Drinking Water Act requirements. A majority of the supply is fed through a 4" pipeline that extends from the surface to the 5000 ft level in the Ross shaft. Each level off the Ross has a 2" pipe branching off the 4" main. The No. 6 and No. 4 shafts are served from the Ross system by a 2" pipeline with outlets on each station. The capacity of this system is up to 400 gpm. This system will satisfy NUSEL potable water needs and will require minimal updating.

The industrial water is very high quality. It comes directly from Homestake's surface water collection system gathered from local streams. The water is stored in a surface 1.2-million-gallon reinforced concrete storage tank. A minimal amount of treatment is provided by a chlorinating system. The industrial water is gravity fed to the underground through a 6" pipeline in the Ross shaft and a 4" pipeline in the Yates shaft. The Ross and Yates pipelines feed 4" pipelines in the No. 6 and No. 4 shafts, respectively. The Ross and Yates pipelines are rated at 1400 and 400 gpm, respectively. The industrial water system will be sufficient for NUSEL needs, with some specialized treatment of water for experiments provided on the laboratory levels.

*Telecommunications and computing:* Homestake has a Bell-based telecommunications touch-tone phone system that includes voice mail, a surface radio system, and a pager system. The mine has state-of-the-art communications and data transmission systems throughout the Ross and Yates shafts. Barrick intends to pass all of these systems on to NUSEL. Much of the wiring for phone, control, security, and intercom communications surface system was upgraded in 1999.

Voice: The current Bell system can support in excess of 600 lines. The surface and underground voice communications (telephone) use copper Category 3 cabling. The subsurface cabling consists of two 100-pair phone lines (one serving the 4850 ft level and No. 4 Winze and the second serving the No. 6 Winze area) and one 50-pair phone line (serving the Yates shaft stations). All are routed down the Yates shaft. During the 1999 surface upgrade new Category 3 cables were installed, primarily as buried cables or as bundles in the tunnels. These new cables replaced older wires that had been hung overhead.

Voice/radio: An existing leaky-feeder system extends the entire length of the Yates and Ross shafts for redundancy and for emergency communications. This system is just a few years old and is one of the most flexible and reliable mine-ready radio communications systems on the market.

Internal data: The mine has internal and external data communications systems. Both systems have hardware and connectivity that can be expanded to increase redundancy and speed to more remote locations in the mine. Mine computers are linked by fiber optics, with the subsurface serviced from a vertical fiber down the Yates shaft. This fiber is routed to a concentrator outside the 4850 ft electrical shop and then down the No. 6 shaft to the 6950 ft vent and across to the 7400 ft LHD shop. Currently all pumping controls and carbon monoxide monitoring between the 4850 ft level and the surface interfaces locally to the fiber system by copper line, so that these functions can be monitored from the surface. PC interfaces are located on the 6500 ft, 6950 ft, and 7400 ft levels. NUSEL plans include complete redundancy for data communications between internal facilities, including redundant fibers in the primary shafts.

External data: Black Hills FiberCom provides connectivity to the mine with a direct fiber connection. The current speed is OC-1 (51.84 Mbps) supported by a SONET (Synchronous Optical Network) hierarchy. Black Hills FiberCom ensures redundancy through alternate fiber paths should the direct fiber be severed in any location. Future expansion to OC-3 (155 Mbps) speed is possible. This would allow outside users to obtain speeds equal to that possible for inside users operating over the internal data network.

*Controls and monitoring:* The mine's infrastructure control and monitoring system is well-tested and thoroughly reliable. The system has four nodes, with one operator interface in the surface mine office and another underground. The existing operations staff installed and programmed the system, so that all necessary expertise is available locally.

FIXDMACS system: The FIXDMACS system is used to monitor, throughout the mine, the control pumps, fans, and other essential equipment, as well as carbon monoxide levels. The system operates on uninterruptible power. The control area where facility operators monitor the output includes 50 process control screens, as well as several equipment status screens and control interface templates. The graphics displays are high quality, based on the operator interface software Intellution.

Subsurface gas detection: There are approximately 20 carbon monoxide monitoring stations located throughout the mine. As new ventilation paths will be set up to optimize NUSEL, many of these monitors will be relocated.

*Personnel and operations:* As we will argue in the next section, there is no mine attribute more important to NUSEL than the existence of an efficient operations infrastructure manned by highly trained personnel, many of whom desire to remain with NUSEL. Mine operations are carried out by five groups:

- Operations/maintenance group: This group is responsible for the operations and maintenance of plant infrastructure including the hoisting facilities, crushing operations, ventilation and pumping systems, power, compressed air, water and communications services, surface grounds and buildings, and the main underground haulage and access drifts. This group supports underground excavation and construction activities by transporting supplies, materials, waste rock, and personnel. At NUSEL this group would help with the transport of experimental equipment between the surface and the experimental halls. This group is responsible for the installation of new services and equipment, such as power and communications, air handlers, fans, and other hardware.

- Underground excavations/construction group: This group is responsible for the drill, blast, muck, and ground control functions for all underground excavations. Construction activities include track, concrete, and shotcrete placement, wall coatings and paint finishes, wall and barricade construction, and excavation support installations (e.g., construction of rock dumps and bypass gates).
- Engineering group: This group is responsible for the evaluation and design of all underground excavations including technical support for ongoing excavation and construction activities. Technical support includes rock mechanics analysis, design and monitoring, surveying, scheduling, blast design and materials, and equipment testing. At NUSEL these responsibilities would include the scheduling of facilities used in transport of experimental equipment and the oversight of transport and underground installation. Plant engineering responsibilities include preventative maintenance, systems engineering, equipment replacement and procurement, and plant upgrades.
- Safety/environmental group: This group evaluates all work done on the site to ensure safety and environmental protection. It has specific reporting, database, permit, inspection, monitoring, and compliance responsibilities. It formulates emergence response plans, policy books, and training materials, and is responsible for training site personnel, contractors, the mine rescue team, and for NUSEL, the visiting scientists. The group is responsible for security functions and the occupational health program, including documentation, operation, and maintenance of the mine operating system.
- Administrative group: This group is responsible for all personnel functions, public relations, legal review, accounting, and purchasing. Purchasing functions include competitive bid preparation, contract agreements, legal review, the documentation and accounting interface, inventory management, supply ordering, and warehouse and yard supervision. Thus the NUSEL shipping and receiving, computing, inventory, library, and visitor services would fit within this group, as would all funding agency reporting requirements.

**Note:** This section and the remainder of the Reference Design were written and costed under the assumption that the site agreement for Homestake will allow the scientists to make use and build upon the Homestake's existing underground infrastructure. See section G for a discussion of recent flooding and site liability issues that invalidate this assumption. When a site agreement is completed, our collaboration intends to work with South Dakota officials to submit an addendum to this Reference Design Project Book.

**<u>Why Homestake is the optimal site for NUSEL</u>**: Later, in the Facilities Development Plan section of this version #2 proposal, we present a detail plan for exploiting the existing site and infrastructure of Homestake to produce an underground science laboratory unique in its potential and flexibility. The plan is more efficient than that offered in the version #1 proposal, a result of a year of discussions between the collaboration's scientists and engineers (including two with a total of 40 years of experience as head engineers at Homestake). It achieves the following goals:

- A two-level laboratory divided between intermediate (4850 ft) and deep (7400 ft) levels, which is attractive scientifically (allowing experimentalists to balance great depth and easy of access) and from the engineering perspective (distributing ventilation loads). Some activities will use other parts of the mine, such as the geomicrobiology center at 8000 ft.
- Clean, efficient scientific access with large-load capabilities (module weights of at least 8 tons and footprints of at least 11 ft by 12 ft) from surface to 8000 ft.
- All hall excavations occur in rock that is immediately accessible, and thus where rock quality is known.
- Dual access is maintained for all levels, enhancing safety. Because of this access, future expansion of both the 4850 and 7400 ft levels is possible while keeping all excavation sequestered from all science operations.
- The plan optimizes both excavation and detector construction possibilities for a megadetector on the 4850 ft level.

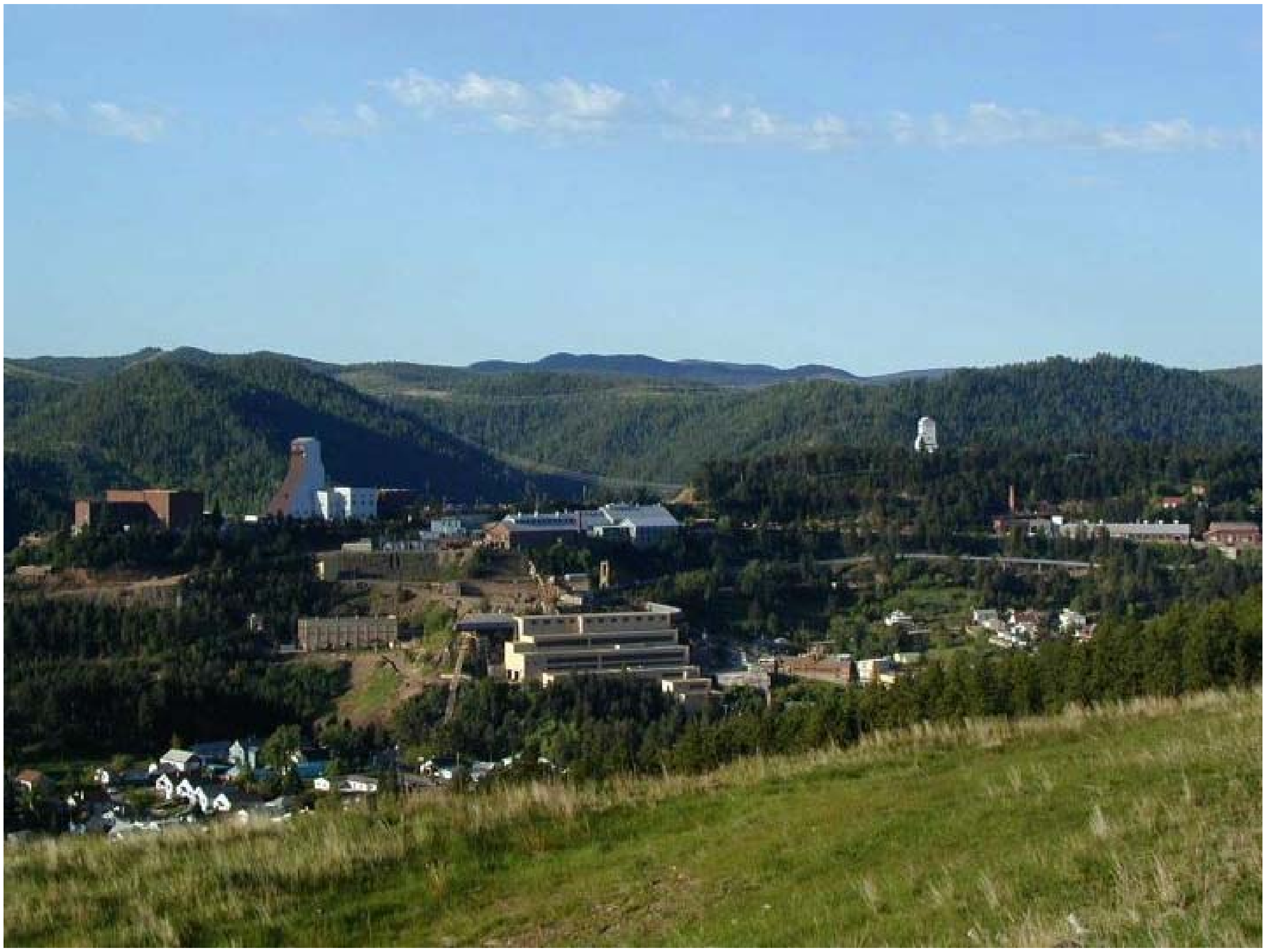

Figure B.2: A view of the Yates (left) and Ross (right) headframes. The distance between the two shafts is approximately one kilometer. These shafts/hoists provide the principal access to the 4850 ft level. In the facilities development plan, the Ross and the No. 6 Winze provide scientific access to all laboratory levels. The Yates would normally be used as the excavation shaft (together with the No. 4 Winze to reach the 7400 ft level), but could also be dedicated to megadetector construction, to give this experiment maximum access. See Section E, Figure 3, for the corresponding underground view, showing planned laboratory excavations.

The Bahcall Committee recognized the potential of Homestake long ago:

- The **existing infrastructure** at Homestake will allow us to produce a facility at very **low cost** that has **capabilities beyond any other envisioned laboratory**. The important attributes include the multiple shafts and thus the multiple access, the ability to sequester potentially hazardous experiments, the depth and range of depths available, the existing mining capacity (sufficient to excavate the UNO cavity in 2.5 years), and the reliable facilities (hoists, ventilation, air conditioning, water, dewatering, data and communications, controls and monitoring systems, and local hydroelectric capacity).
- The Homestake site offers **low risk**: a) the overall site geology and hydrology are very well understood and very favorable; b) the rock structures designated for hall development are directly accessible; c) Homestake is an operating, efficient facility with a highly trained staff, many of whom would remain at NUSEL; d) mining and operations costs are known; e) the area is seismically quite; f) the area has a long history of support for mining; and g) as an operating facility, Homestake does not face the substantial permitting obstacles confronting proposed greenfield sites.
- The Homestake site requires very little development, greatly **shortening the time to first physics**. The facilities development plan envisions the low-level counting facility operating at 7400 ft within one year of initial funding. The earth science program will commence immediately.
- Homestake is an **ideal site for EarthLab and for geomicrobiology,** and is likely the only site that could serve both the physics and earth science communities well.
- Homestake presents **exceptional opportunities for education and outreach**. The mine is located in a major tourist area and in a region of the US that currently lacks a major science focus. NUSEL will be an important partner of the universities and tribal colleges in South Dakota and in surrounding EPSCoR states, working to strengthen science and education in the region. Distance education and other K-12 outreach activities are crucial to this region because of the demographic problems facing school districts.

Below we expand on the first two bullets above by relating Homestake's physical attributes to the discussion in Section II:

*Depth:* Both of the NRC committees noted that proposed next-generation experiments require overburdens of 4000-4500 mwe, and that NUSEL must be able to provide depths up to 6000 mwe to meet scientific needs over the next two decades. The proposed Homestake development plan provides main levels at 4400 mwe and 6500 mwe, and potential access to any depth from surface to 7200 mwe. The rock in Homestake is of excellent quality (the results of geotechnical studies are cited below), believed by our engineers to be superior to any of the 50 hard-rock North American mines with which they are familiar. All areas planned for hall development are immediately accessible for coring. The site is seismically quiet and very dry.

Unlike any other proposed NUSEL sites, Homestake meets the needs of EarthLab proponents and of the geomicrobiology community, as well. Important attributes include its 3D extent (9 $km^3$ are accessible through 600 km of drifts), a massive geotechnical data base including 3D computerized models of the site, and complex, highly-folded preCambrian rock with an interesting geothermal history. The microbiologists have found interesting extremophiles at the 8000 ft level and propose to study deeper layers, to 16500 ft.

*Detector halls:* Because of the excellent access provided underground by the existing system of hoists and ramps, a uniquely flexible laboratory is possible, one that merges the best elements of Gran Sasso and Kamioka. Initial construction will produce a series of halls similar to those of Gran Sasso, though specialized somewhat to meet the needs of the low-level counting facility and other experiments described in the Science Book. But because the existing hoisting system provides dual access to all levels, NUSEL-Homestake will be a **living** facility, one where Super-Kamiokande-style specialized cavities can be constructed in future years, without interfering with ongoing science operations. As noted below, this includes the construction of megacavities much deeper and much larger than any attempted to date.

Homestake is unique among the sites proposed because experiments using large quantities of cryogens and flammables can be sequestered and separately vented. Provisions will be made in NUSEL construction to reserve such sites, and to anticipate how the existing ventilation system can be adapted for those sites.

*Megadetector capabilities:* This version #2 proposal was profoundly influenced by very recent neutrino physics developments. With the realization that very long baseline experiments might be able to probe new sources of CP violation relevant to leptogenesis, great attention has been drawn to neutrino superbeams and to megadetectors designed as distant neutrino detectors, but also serving as proton decay, supernova physics, and precision atmospheric neutrino detectors. There is growing recognition that such a detector, located at depth, will support a much richer science program than one located just below the surface. While such a detector is far outside the scope of this proposal, the new NUSEL-Homestake configuration prepares the way for such a project by preserving an ideal site for the construction.

Proposed megadetectors include water Cerenkov detectors like UNO and Hyper-Kamiokande, variations where an array of Super-Kamiokande-like modules is excavated, and new-technology detectors like SuperIcarus. As the rural mailbox design of UNO may be the most difficult geotechnically, we focus on this.

Among the challenges posed by such a detector are:

- Is it demonstrated that the excavation is geotechnically practical in the proposed site?
- What are the permitting or practical obstacles to disposing of the associated 2.5 megatons of waste rock?
- Is the excavation cost known reliably?
- After excavation, can the detector construction be done efficiently?
- Can one efficiently fill and drain (for periodic maintenance/decommissioning) the detector?

Stability: The geotechnical stability of large caverns at Homestake has been established through construction and measurement as well as through laboratory testing and modeling. Among the "permanent" deep large chambers the Homestake Mining Company constructed are an equipment repair shop at 7400 ft and an air conditioning plant at 6950 ft. In addition, HMC has carried out extensive Vertical Crater Retreat (VCR) mining at depths below 7000 ft. VCR involves the excavation of 150 to 200 ft high cylindrical regions by drilling vertical holes for explosives into the volume from above, then blasting out progressive horizontal sections from the bottom of the region.

In 1985, as VCR mining was being developed at Homestake, HMC arranged to have the Bureau of Mines evaluate a test VCR excavation at 7100 ft. This involved premining stress tests of the site to fix parameters in a rock mechanics model, followed by postmining measurements of rock shifts. The model predicted the postmining behavior accurately (W. G. Pariseau, "Research Study on Pillar Design for Vertical Retreat Mining," Bureau of Mines Report J0215043, October 1985). The rock stress and shear characteristics utilized in these calculations were determined by Z. Hladysz (now a NUSEL-Homestake collaboration member). Additional rock mechanics studies of Homestake shaft stability were carried out by W. Pariseau, J. Johnson, M. McDonald, and M. Poad in 1995 (Bureau of Mines Report of Investigations 9531, 9576, and 9618). These involved both 2D and 3D models.

In fall 2000 Johnson and Tesarik of the National Institute of Occupational Safety and Health (NIOSH), employed numerical models similar to those of Pariseau to evaluate the stability of a 200 ft diameter by 400 ft cylindrical excavation at various Homestake depths. They concluded that such an excavation would be stable with stresses below the rock limit at both the 4850 and 6800 ft levels, but could be problematic at the 8000 ft level. In February 2001, in response to the Bahcall Committee interest in rural mailbox excavations for experiments like UNO, Callahan, Osnes, and Vlankenship of RESPEC (Rapid City, SD) carried out stability analyses for various deep chambers, including a rectangular configuration with a 50 by 50 m cross section at 4850 ft. Their conclusions were consistent with those of Johnson and Tesarik.

The summary from these studies is that, with normal ground support, megacavities of the type proposed for UNO can be constructed at depths up to 6500 ft, with the expectation that they will be stable for several decades. However, as there is no strong science justification for going below 4850 ft, this is optimal depth for an UNO-style cavity in Homestake. (The Callahan et al. study concluded that the construction would then have a safety factor of 10, where 3 is considered acceptable.) Both excavation and detector construction can be optimized by locating the detector near the base of the Yates shaft. No cost savings results from siting UNO at a shallower depth – any additional hoisting costs at 4850 ft are offset by the improved haulage and support infrastructure on this level.

There have also been studies of the stability of an alternative design, an array of VCR cylinders 50 m in diameter and 50 m long. With a center-to-center separation of the cylinders of 100 m, the excavation is within Homestake rock limits. Increasing this separation to 150 m increases the safety factor significantly.

Much less is known about the suitability of the rock in competing proposals. In one case the rock cannot be tested prior to constructing the 6 km tunnel to the proposed laboratory location. (The national monument status of the site prevents coring prior to construction.)

Rock waste disposal: The excavation of a 0.9 million $m^3$ volume for a 0.5 megaton water detector will produce approximately 2.5 megatons of displaced rock, a solid waste. A disposal site must be found, and the noise and dust accompanying trucking of the rock can lead to public opposition and permitting difficulties. Assuming 20 ton loads, the task corresponds to 100 truck round trips per day, 365 days a year, for 3.5 years.

This very difficult problem is avoided at Homestake. A large open pit exists on site where the waste rock can be deposited. This can be accomplished without any surface transport, by installing an underground conveyor on the 600 ft level, connecting the Yates shaft to the open cut, a distance of about 800 m. The anticipated cost of the conveyor system is approximately $5M.

Excavation costs: The track record in physics for correctly predicting the cost of large excavations is not a good one. In contrast, at NUSEL the excavation would be done with the same equipment, personnel, and techniques that Homestake used for mining, and in familiar rock. The Homestake rough mining cost for the 4850 ft level is $34.46/ton.

Detector construction plan: The facilities development plan anticipates that work on a megadetector would commence after scientific halls on the 7400 and 4850 ft levels have been completed, and after the Ross and No. 6 hoists have been modernized and upgraded for science access to these levels. At this point the Yates hoist could be dedicated to the excavation, which could be completed in as little as 2.5 years if mining is done at capacity. After excavation is completed, the Yates cage could be modified – very much as the Ross and No. 6 were to enhance science access – to make detector construction as easy as possible. Because of the very large Yates shaft cross section, it is possible to have a single-cage configuration as large as 5.5 m by 4.5 m, so that a large cargo container could be lowered to the detector site. Dedicating the Yates in this way to megadetector excavation and construction would give this project maximum flexibility, and avoid any interference with other laboratories activities, which would be conducted through the Ross and No. 6.

The cost of megadetector construction is not part of the NUSEL proposal, though our proposal is designed to make such a project feasible at a later date.

Filling and draining the detector: The mine's dewatering system may be very helpful in filling and draining the detector. The excess pumping capacity of the system (that in excess of the 470 gpm from groundwater) is about 2000 gpm, or 11.5 ktons/day. Thus a detector comparable to SuperK could be drained in 5 days, while UNO could be emptied in about three months. As the ground water is of good quality, it may be possible to filter it for use in the filling stage, as well. This source could fill SuperK in 20 days and UNO in one year, timescales that might well be adequate.

*Cleanliness:* The cleanroom environment – a variety of rooms varying from Class 100 to 1000 would be useful – is important on both the detector assembly and detector operations ends. For example, some groups will use NUSEL's surface laboratory cleanroom facilities to prepare detector modules. Transport of large, precleaned, and carefully wrapped instrumentation assemblies from the surface to depth will be via the Ross and No. 6 hoists, which will be fitted with enclosed lift cages resembling an elevator. While not a cleanroom environment, the lifts will be dedicated to science. There will be facilities for washing and cleaning the assemblies before re-entering the cleanroom environment at depth. This is the SNO model, with an important improvement: SNO transport was through an active and thus very dirty mine environment. At NUSEL the transport leg, while not a cleanroom environment, will be isolated from the mining environment to minimize contamination during transport.

The standard facilities described in Section III – underground materials storage areas, the low-level counting facility,

necessary underground fabrication facilities – will be part of NUSEL-Homestake infrastructure.

*24/7 access:* During major excavations and detector construction NUSEL hoists would likely operate 24 hours and seven days a week. At other times access might be limited at night. To provide access NUSEL will install an automatic man-hoist – a small elevator – on the Ross and No. 6 that can be operated by an individual and controlled by a keycard. There must be an operator, however, who can be called in an emergency. This might be addressed by combining this job with others (e.g., night security).

*Safety, special materials, materials handling:* The FIXDMACS system for mine monitoring and the safety experience accumulated by Homestake operators form an important baseline for safety oversight and procedures at NUSEL. The Homestake Mine fully complies with all MSHA regulations (one provision of these regulations being two means of escape from each level). All operating shafts have fresh, incoming air (fumes will not follow workers attempting to exit the mine).

The sensor system described earlier permits rapid detection of carbon monoxide excesses, fires, etc. We plan certain control and monitoring upgrades for NUSEL. One of these is a new addressable fire detection and alarm system that is fiber linked to the main fire alarm panel at the surface, which serves as the fire alarm command center for the entire Homestake facility. The subsurface fire detection system will also include a remote fire alarm annunciator, door holders/closers, manual pull stations, and smoke detectors and audio/visual notification in the corridors, labs, and shops. The carbon monoxide detection system will be upgraded to a multi-gas system in the detector areas. The subsurface PA system will provide paging coverage for all normally occupied subsurface science areas.

Surface security and access will be controlled by a card reader system, with card readers installed at all new-structure exterior doors. Access to the facility will be monitored by relocated video cameras. We have been asked about the possibility of a higher-security area for national security work. There are possibilities for developing a separate drift for such work, with tightly controlled access, should the agencies decide a need exists.

MSHA oversees all mines that produce commercial materials, and thus was the responsible agency during mine operations. OSHA oversees all other underground construction as well as most surface facilities. OSHA is familiar with the activities in standard university, industrial, and national laboratories, while MSHA is not. However, MSHA regulations governing hoist operations, ventilation, and mine rescue teams are well tested and successful. A reasonable approach would be to place NUSL under OSHA oversight, but have the laboratory follow MSHA regulations with respect to mine rescue and hoist operations.

Homestake's mine rescue team is outstanding, generally regarded as the best in the US. All team members are regular employees, but they donate their time when training for mine rescue. The only facility cost is their equipment and training materials. NUSEL would like to retain this team and expand their training to include potential problems in the scientific areas.

NUSEL will establish a safety office to oversee all safety and security issues, and to guarantee that experimental operations are carried out in a responsible way. In addition to its internal responsibilities, the safety office will be responsible for coordination with local police, fire, and medical personnel. It will also be responsible for a safety review of new experiments before they are formally considered, and for control of all flammable materials, radioactive materials, noxious gases and liquids, and asphyxiating gasses. Before any new experiment is approved, the report of the safety office will be reviewed by an independent operations safety oversight committee that includes mine safety experts, experimentalists, and a representative of the South Dakota Department of Environment and Natural Resources. This committee will report to the Director.

*Access for large and heavy equipment:* The key element in the Ross and No. 6 hoist upgrades is to enable efficient transport of standardized detector modules, 11 by 12 feet and up to 8 tons, from surface to 8000 ft. In addition, Homestake personnel have experience transporting very large or awkwardly shaped cargo underground by slinging the load beneath the cage. We noted, in the discussion of UNO, that the Yates could be adapted for a module footprint of about 14 ft by 18 ft, should this be helpful to that experiment.

*Quietness and stability: electrical, mechanical, seismic:* There are two issues specific to Homestake. First, the area is very quiet seismically, and the mine has very few rockbursts. Second, there are several hydroelectric plants on the site that could provide power to NUSEL at very low cost. This would give NUSEL a backup power supply should Black Hills Power suffer an outage. One plan would have Barrick Gold retaining ownership of the hydroelectric facilities, but selling the power to NUSEL at rates substantially below market. There is about 2.5 MW of generating capacity that interests NUSEL. The cost of power generation is $0.01/kW-hour.

*Adequate support facilities:* We feel the Gran Sasso model of strong in-house support facilities – library, shops, visitor support, etc. – is the right model for any underground laboratory that aspires to be a national and international center. For Homestake the argument for following this model is even stronger, as the universities within 40 miles of the site, Black Hills State College and the South Dakota School of Mines and Technology, do not have the research infrastructure typical of large research universities.

*Quality of life:* There are several site-specific issues affecting the quality of life. The town of Lead (population 3500, elevation one mile) has a sunny, semi-arid, high-plains climate, warmed by the compression of air down-drafting from the Front Range of the Rocky Mountains. Mean January and July temperatures (high/low) are 25°/10°F and 75°/69°F, respectively. Lead retains much of its mining heritage. A major civic project now underway is the reconstruction of the 1914 Lead Opera House, following faithfully the original architectural drawings. The city webpage can be found at http://www.leadmethere.org.

The nearest city is Rapid City (population 57,500), a 45 minute drive on very good roads. Rapid City has a very convenient modern airport with jet and air commuter service to Denver, Minneapolis, etc. The Black Hills offers exceptional outdoor recreational opportunities, including skiing, hiking, boating, trout fishing, horseback riding, and mountain biking (including on the 114 mile long Mickelson Trail): this is the main reason for the area's strong tourism industry. The area offers many campgrounds. Spearfish Canyon, an area preserved for the past 100 years by the Homestake Mining Company, is just a few miles south of Lead. This US Forestry Service Scenic Byway is the location where the winter scenes in "Dances with Wolves" were filmed.

Various services give a cost-of-living index for Rapid City of 89.3-100.2, depending on the choice of weighting factors.

*Outreach and education:* This is addressed in Section IV of the Science Book.

*Neutrino beams:* Homestake is 1290 km from FermiLab and 2530 km from Brookhaven.

## D. SCIENCE TIMELINE

The NUSEL proposal under discussion here is tied in detail to the science projects that the broad science community will approve through existing mechanisms, such as SAGENAP. While one cannot predict precisely either the R&D progress of developing experiments or future funding decisions, the Science Book nevertheless provides a useful guide to the number and variety of experiments NUSEL may be asked to accommodate. We summarize this information in the timeline given below. The Reference Design, in consideration of this timeline, then answers the following engineering questions:

- Given the science needs, what are the appropriate parameters and design goals for NUSEL-Homestake?
- How does one meet those goals while minimizing costs and construction risks?
- How does one ensure safety?
- What can be done to preserve future options for mounting next-generation experiments, especially as future project needs – such as size and safety requirements – cannot be fully anticipated now?
- What are the major uncertainties or options in the reference design, and what is the plan for reaching decisions on the way to the baseline definition?

**Science Timeline: Physics**. Here we summarize the conclusions we have drawn from the Science Book:

***Experiments mounted at the start of construction***: There are several physics experiments and R&D efforts that would like space, even on a temporary basis, at the earliest possible time. Examples include the development of Ge detectors for national security (MEGA/SEGA), the early-stage efforts on Majorana (the Majorana proposal, which is being submitted approximately now, proposes a staged development of detector arrays), and nonproliferation projects similar to the PIsCES effort. These experiments have modest footprints and utility requirements and could be set up in existing space, using prefabricated cleanrooms of low cost.

Several groups involved in R&D have materials that should be "cooled," stored underground to allow cosmogenic activities to decay away. Storage space at modest depth is needed.

EarthLab scientists desire access to as much of the 600 km of drifts as possible, to install sensors that will continue to monitor humidity, stresses, and rock movement after the drifts are closed. The closing of unused drifts will be one of the first construction activities. Thus EarthLab scientists need to be integrated into initial construction activites.

For definiteness we will take FY06 as the start of construction. (This follows agency advise. We believe the Baseline Definition Project Book could be completed in time to allow FY05 consideration, however.)

The program development plan calls for the first new-construction finished rooms to be available on the 7400-ft level at the beginning of FY07, with all rooms completed four months into FY08. Therefore to meet the early needs described above, suitable existing space must be provided for approximately two years. That space must be accessible during construction, and must have adequate utilities. This presumes that maintenance continues on the Homestake site, from now until construction, so that at least moderately deep drifts can be reached safely in FY06.

***Initial experiments and R&D efforts:*** There are a number of efforts that are clearly either ready to go now, or should be with an additional four years of R&D and preparation. First among these is the low-level counting facility, crucial not only to the first experiments to be constructed at NUSEL, but also to most R&D efforts that will want space there. The low-level counting facility will be central to many of the detector development efforts undertaken because of national security concerns. It will also serve as a national user facility. The importance of this facility became clear at the Lead and NESS workshops, leading to the world-class proposal incorporated into this Reference Design document. The low-level counting facility is under active development by physicists from Los Alamos, Princeton, NIST, Alabama, and other institutions.

Majorana and likely one additional one-ton-scale double beta decay experiment will ready by FY07. Majorana requires a deep site, at least 4500 mwe, though the experimenters feel additional depth is important to provide a safety margin.

The geomicrobiology program and many other components of the earth science program would be underway by FY07. It is likely that boreholes for the geomicrobiology program could be done in FY06. Required supporting facilities include a small room at the 8000 ft level, a clean laboratory in the complex at 7400 ft, and eventual access to chemistry and other surface laboratory facilities.

There are several dark matter experiments that will be seeking deep locations at about FY07. In particular, CDMS II, currently being deployed in Gran Sasso, will run through 2006. Its one-ton successor CryoArray shares much of the same technology, though some materials and background R&D remains to be done. CryoArray is likely to be ready by FY07. The depth requirements are similar to those for double beta decay.

The Science Book describes many R&D projects – solar neutrino detectors, new technologies for supernova detectors, applied science detector developments, etc. – that will require space at depth as soon as it is available.

***Next experiments***: Although some of the solar neutrino experiments have more ambitious R&D schedules, we have placed the mounting of a full-scale experiment no earlier than FY08. Most of the proposed experiments require construction of a specialized cavity. As selection of a pp solar neutrino experiment is unlikely to occur before initial construction on the 7400-ft level, this requires NUSEL to build such a cavity after the deep level is already an operating scientific laboratory. Thus a scheme is needed to facilitate this and future construction while maintaining the cleanliness of existing, operating experiments.

A second facility that will be installed in FY08 is the low-energy accelerator for nuclear astrophysics. This is a staged facility for which the planning has already commenced (a national coordinating committee has been meeting regularly). This group's schedule calls for submission of a proposal by FY06. The facility requires moderate depth, does not require exceptional cleanliness, and could be conveniently located on the 4850-ft level, away from most other experiments. This would minimize concerns about electric interference with other experiments.

***The megadetector:*** The excavation requirements of the megadetector dwarf those of other experiments discussed here. This excavation is not part of the NUSEL proposal – yet it is crucial for NUSEL to be able to accommodate this detector, given the intense interest in next-generation proton decay searches and very long baseline experiments to measure CP violation and other new neutrino properties. HEPAP, in its recent response to the DOE Office of Science call for comments on future megaprojects, called for a depth of at least 4000 mwe. It also stressed that a rigorous professional civil and mechanical engineering design for the detector depends on the NUSEL site choice. NUSEL-Homestake's first option would be to place this detector near the base of theYates hoist (4400 mwe), dedicating the hoist to the excavation. The needed conveyor for rock disposal in the open cut could be built in FY08 (or earlier, if there is need), so that megadetector construction could begin in FY09.

This **proposed initial NUSEL physics program** is displayed in the **Timeline**.

## TIMELINE: NUSEL-Homestake Initial Science

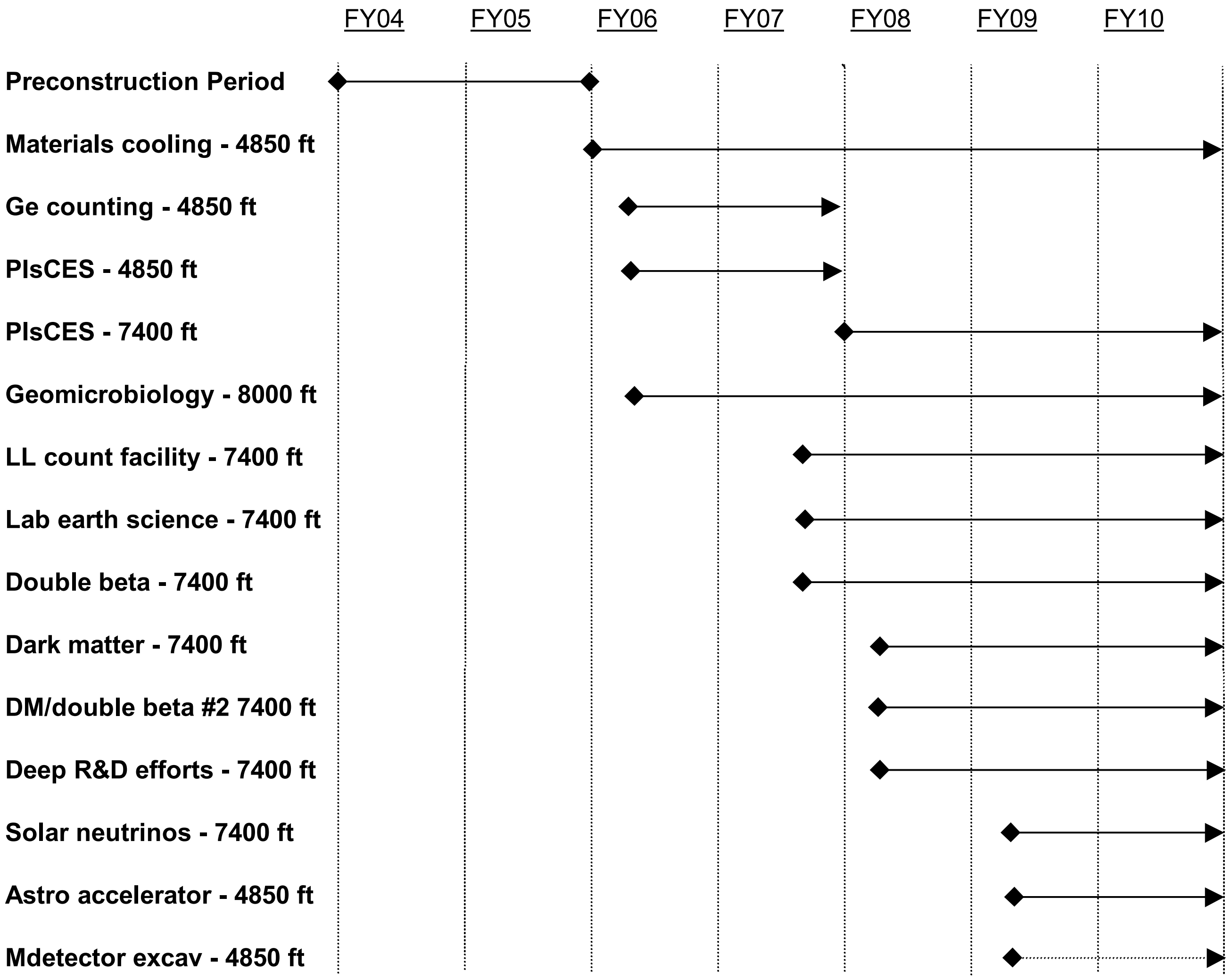

Assumptions:

1) No. 6 Winze renovations completed by fourth quarter, FY07
2) Some 7400-ft rooms finished by first quarter, FY07, remainder by second quarter, FY08

**Science Timeline: Earth Science**. The EarthLab community also proposed a plan for its activities over the first five years of NUSEL:

*First year:* The earth scientists will undertake two surveys, the mining of existing Homestake data bases and direct reconnaissance of all accessible mine areas in order to compile relevant geochemical, hydrologic, and rock mechanics data. This will help the community identify optimal candidate sites for subsurface experiments. As many unneeded drifts will be closed in the first year, the positioning of various sensor arrays must be connected with this work: fluid pressure transducer arrays, tensiometer arrays, and temperature probes will be installed in areas where future access may be impossible. Initial installation of the Deep Seismic Observatory will begin. The data obtained in the initial surveys will be used to plan a fracture characterization, geophysical imaging, hydrologic measurement, and stress measurement program. The EarthLab steering group will begin development of a web-based system for data distribution, and of a preliminary mechanical, thermal, and hydrological model based on drilling and excavation history.

*Second year:* The plan includes completion of hydraulic fracturing measurements for the stress field in key areas of the mine and the performance of tracer tests on candidate sites for coupled experiments, with the goal of picking shallow, intermediate, and deep candidate sites. Borehole drilling and coring from 8000 to 14000 ft will be done for the ultradeep biology experiment and associated experiments, and stress and permeability measurements will be made. The Deep Seismic Observatory installation will be completed. Boreholes will be excavated for the deep heater experiment and temperature sensors installed for the coupled thermal-hydrologic-chemical-biological experiment. Sites will be selected for hydrology stations, deep percolation, and paleohydrology boreholes.

*Third year:* By the end of the second year the anaerobic room for geomicrobiology will have been excavated and early in the third year the geolab will be available in the 7400-ft complex. Data collection will begin for the deep flow and paleoclimate experiment, after coring and borehole installation, and the ultradeep biology facility will be readied for experiments. The heater experiment will be initiated, obtaining temperature, geophysical, geochemical, microbial, and hydrologic data. The fracture propagation experiment will be conducted, with monitoring of gas fluctuations, microseismic data, and tracer transport. The borehole arrays for the deep-coupled-processes facility sites will be completed, and cross borehole tomography and tracer tests performed for each coupled-processes site.

*Fourth year:* Goals include completion of the microseismic analysis of the induced fracture experiment, of the analysis on changes in fluid chemistry and microbial populations to temperature changes, and of the 3D fracture transport models for coupled experiment sites. The preliminary thermal, hydrologic, and mechanical model will be revised to take into account new data from the heater experiment in progress. An analysis of first results from the deep flow and paleoclimate experiment will be completed, including a preliminary hydrology model, readying the facility for more detailed analyses and in situ experiments.

*Fifth year:* The first heater experiment will be completed and the facility readied for further experiments. It is expected that the first set of funded experiments at EarthLab facilities will be initiated.

The earth science community has recently focused on developing the EarthLab science case, and thus has not played a very active role in the detailed NUSEL facilities planning that is described in this project book. (Efforts have been made to address the needs this community has identified, such as the rooms on the 8000- and 7400-ft level and the scheduled excavation for the earth science drilling rig). **One main goal in the next Baseline stage of the project book will be to better integrate the earth-science timeline into the NUSEL facilities development plan.**

## E. REFERENCE DESIGN AND PROJECT PLAN

In the following we consider the elements of the Reference Design for the Homestake National Underground Science and Engineering Laboratory:

- The Facilities Development Plan: This describes the plan for subsurface access; development for the 4850-, 7400-, and 8000-ft levels of NUSEL; subsurface infrastructure; the surface campus, land acquisition, and related permitting issues; and preconstruction maintenance and operations issues.
- The Facilities Operation Plan: This describes the facility and site operations plan.
- The Outreach/Education Plan: This describes the visitor center and museum/archive plans, as well as associated operations.
- The Program and Management Plans: This describes our plan for carrying out the NUSEL-Homestake project, including roles, responsibilities, and methods for accomplishing our goals. Pending guidance from the NSF, we present a possible management scheme.

**I. The Facilities Development Plan: Subsurface.** The current facilities development plan differs from that of our June 2001 conceptual proposal, reflecting important advances in our understanding of the Homestake site and of its potential for science. We propose a laboratory with two main levels, 4850 ft and 7400 ft. This furthers several of our goals:

- Various science reviews have concluded that a new multipurpose laboratory must provide at least 4500 mwe of overburden and should strive for 6000 mwe, to accommodate future experiments. The 7400-ft level (6600 mwe) of Homestake has a well maintained (and relatively recently constructed) 9 ft by 9 ft drift. It is the deepest level reached by the No. 4 Winze, which will serve as the mining shaft for future expansion of 7400-ft laboratory facilities as well as a secondary escape route. This level is also reached by the No. 6 Winze, which will serve as primary access. The No. 6 end of the main drift at 7400 ft reaches the Yates formation, generally regarded as the most competent rock within the site. The combination of excellent rock, dual hoists, and convenient access argues for locating the deep NUSEL rooms on this level, in the vicinity of the No. 6 Winze.

- The two main shafts from the surface, the Ross and Yates, terminate at the 4850-ft level (4300 mwe). They are connected by a one-kilometer major drift, in excellent condition, with a 13 ft × 13 ft cross section, off of which several serviceable rooms with good utilities exist. There are important reasons for developing a second laboratory level on this level. First, several proposed experiments that were discussed in the Science Book require 4000 mwe of overburden or less, and thus could be located at the 4850 ft level, providing more convenient direct access from the surface (avoiding transfer to the No. 6 Winze) -- science access would be through the Ross. Second, an additional laboratory level allows us to separate experiments that might interfere. (The nuclear astrophysics accelerator will be placed on the 4850-ft level because of electrical interference and machine-associated background concerns.) Third, a split-level laboratory helps to distribute the ventilation load. Fourth, the existing rooms on this level can be used for materials storage ("cooling" activated materials) and for mounting certain experiments at the beginning of construction.

  But by far the most important consideration is reserving an ideal site for the megadetector. The base of the Yates shaft is in the middle of the Yates formation, the optimal rock for large excavations. According to a recent HEPAP report, a depth of 4000 mwe is desirable. By placing the megadetector near the base of the Yates, this hoist could be dedicated to this experiment during both the excavation and construction stages. Mined rock can be removed from the Yates hoist at the 600-ft level and transported by conveyor to the open cut, solving the megadetector's substantial waste rock disposal problem.

- In addition, some development will be done on the 8000-ft level in support of the geomicrobiology program. This includes a small room and excavations to allow headroom for the drilling rig to be used in boring to 16,500 ft.

The earlier conceptual proposal advocated an extension of the Yates shaft to the 7400-ft level, connecting the new base to the No. 6 Winze by constructing a one-kilometer major drift. The Yates extension was the most technically risky aspect of the original proposal, as a single shaft to 7400 ft would be near the limit of what has been accomplished in mining. Both the Yates extension and the new drift are expensive: the hoisting goals of the original proposal would have required replacing the Yates head frame. Most important, the base of the Yates would then be in the Poorman formation, generally considered to be the weakest rock in the site.

The current proposal is far less risky, more cost effective, and places all major construction in the most competent rock. It leads to a somewhat smaller hoist "footprint" for science access: the Ross and No. 6 can support a cage footprint of slightly more than 12 ft × 11 ft. However, given the existing hoist engines, this tradeoff allows larger loads, which virtually all of the experimental groups we consulted preferred. In contrast, a single hoist reaching 7400 ft would pull so much tonnage in rope that the load limit would be reduced substantially.

***I.1 General Subsurface: Rock and Site Characterization:*** This section briefly summarizes general characteristics of the underground site relevant to underground construction.

*Earthquake potential:* South Dakota is rated as a 5 on the 1 to 5 scale of the National Fire code, with 1 being "maximum potential for earthquake damage" and 5 denoting the least potential.

*Regional geologic setting:* The Precambrian core of the Black Hills uplift in South Dakota is exposed in an elongated dome approximately 85 km long and 35 km wide. Archean igneous and early proterozoic sedimentary and volcanic rocks were deformed in a regional metamorphic event. These rocks were then intruded by Harney Peak granite approximately 1.7 Ga ago. The Precambrian core of the Black Hills was uplifted and eroded. Paleozoic and Mesozoic sediments were uncomformably deposited on this uncomformity surface.

*Mine geology:* Early Proterozoic statigraphic units within the Homestake Mine are the Poorman (oldest), Homestake, Yates, and Ellison (youngest) formations. The Poorman is characterized by a gray to black, banded to laminated micaceous phyllite, and contains minor amounts of slate and schist. The Homestake is a carbonate facies banded iron formation. The Ellison consists of phyllite, quartz, and mica schist, with a considerable amount of dark quartzites. The Yates formation is hornblende-plagioclase schist. All of the proterozoic rocks have been subjected to multiple deformation events leaving complex isoclinal folds. Generally the rocks strike northwest and dip northeast. The isoclinally folded synclinorium/anticlinorium couplet plunges at various attitudes to the southeast. Tertiary intrusive rocks, mainly rhyolite-phonolite dikes, are fairly abundant in the district.

There are differences in the ability of these rocks to support large excavations. The most competent rock is that of the Yates formation, and the least competent is Poorman rock. One parameter that illustrates the variation of rock strength in Homestake is the uniaxial compressive strength Co. The rock is characterized by Co(1), Co(2), and Co(3), defined as the strengths parallel to the strike of the schistosity, perpendicular to the schistosity, and parallel to the dip of schistosity, respectively. The results (Co(1), Co(2), Co(3)) for Homestake, Poorman, and Ellison were measured by Hladysz: (20.15, 11.15, 13.27), (13.63, 10.00, 12.27), and (11.34, 11.41, 8.15), respectively, in units of 1000 psi. Thus there are factors of two differences that can be gained by picking the rock judiciously. Hladysz states that the Yates Formation values will be much higher than those for Homestake Formation, the best of the three above. **Measurements in the regions planned for development, of course, will be part of the baseline definition effort**. We have placed all major construction – the large halls planned for the 7400-ft level and the megadetector on the 4850-ft level -- in Yates rock. The figures below show a cross section of Homestake, where the Yates Formation is identified, as well as the views from above on the 7400-ft and 4850-ft levels. Note that Yates rock is immediately accessible on the 7400-ft level from the No. 6 Winze. Also, the base of the Yates is in Yates rock, as is 60% of the vent drift connecting the Yates and Ross shafts.

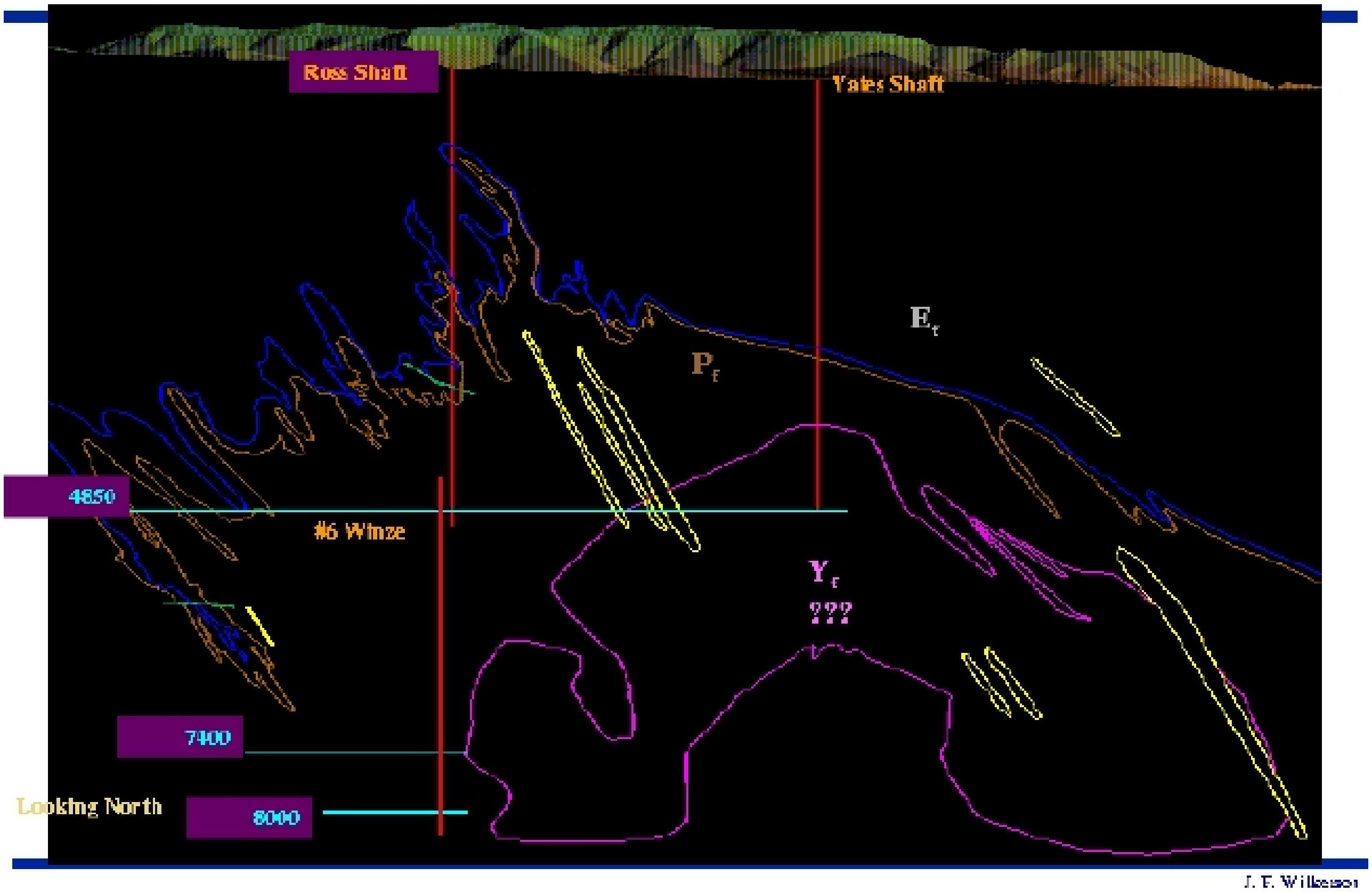

Figure E.1: Cross sectional view of Homestake showing the Yates formation, denoted $Y_f$. (Note that this view is 180° from the cross sections shown early in the Project Book, where the Yates shaft was on the right.) The base of the Yates – the optimal position for the megadetector in terms of depth ease of mining and construction, is in the middle of the Yates formation. Similarly, the drift at the 7400-ft level near the No. 6 Winze extends to the Yates formation, allowing one to develop into that area. The strongest direction is perpendicular to the rock face. Thus one should orient large halls with their longest dimension in this direction.

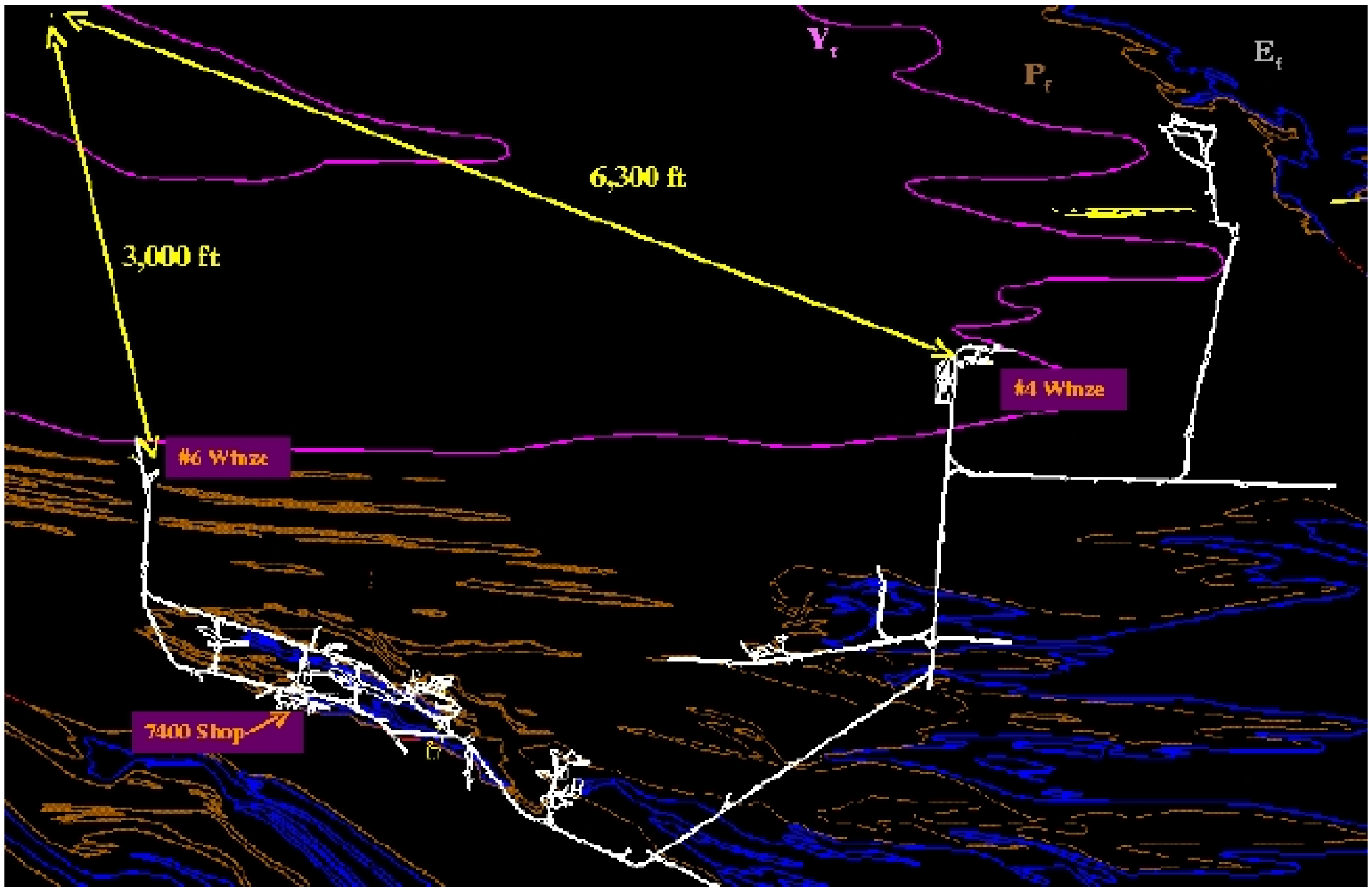

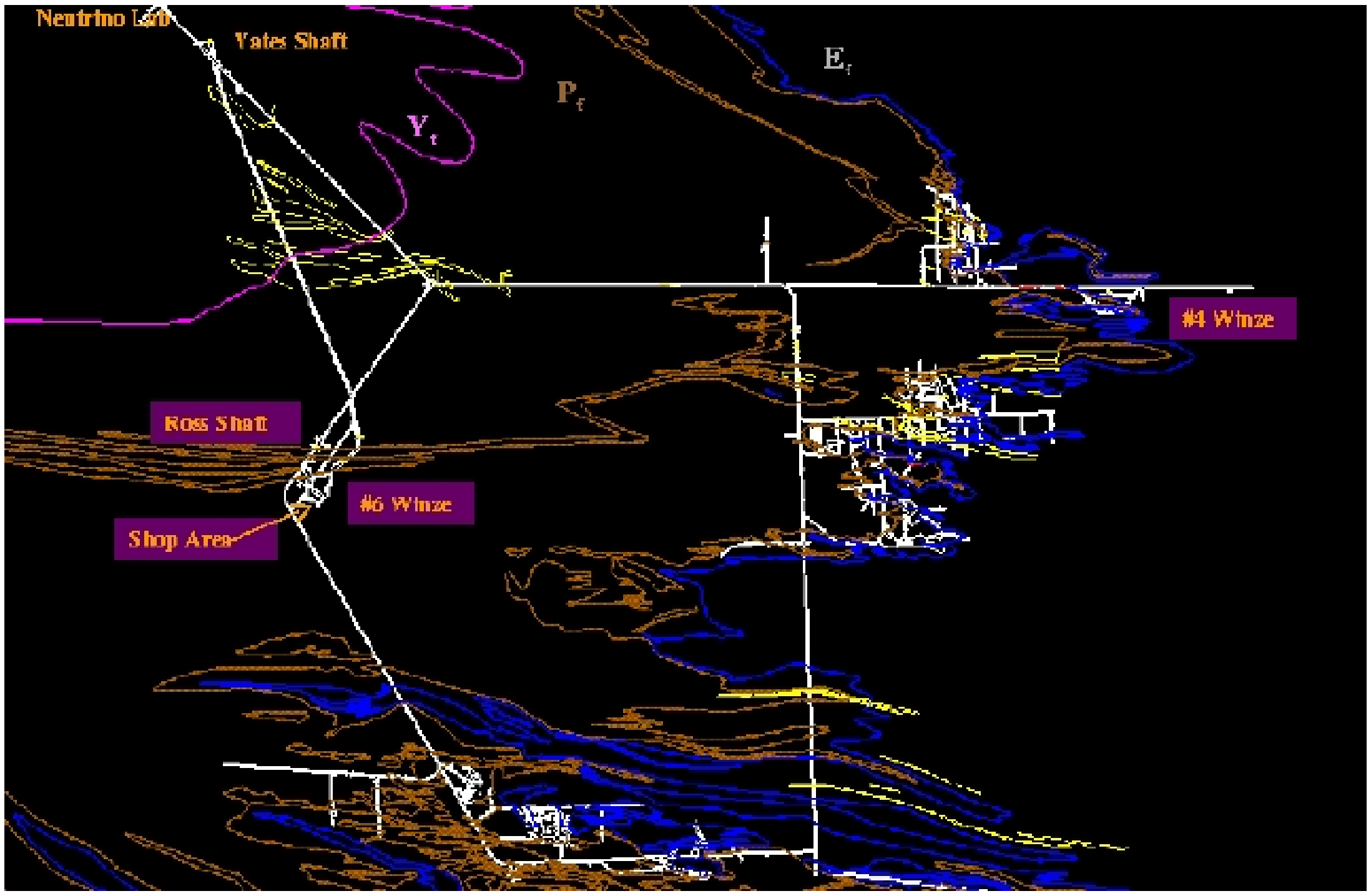

Figure E.2: Top views showing formation boundaries on the 7400-ft (top) and 4850-ft (bottom) levels. The Yates formation is outlined in purple.

*Ground support:* The rock formations described above are stable. Regional stress fields are minimal and regional seismic events are rare. In instances when stress has exceeded compressive strength, as has occurred in some stope pillars, the rocks generally deform in an elastic rather than brittle manner. However underground openings require ground support to assure safety. The different rock bolts used at Homestake are friction stabilizers (splitset and swellex), point anchor bolts (pattin), grouted or resin bolts (rebar or fiberglass), and cable bolts. Mats, screens, and shotcrete are also used. Spans up to 40 ft are routinely excavated using the standard ground supports mentioned above. Larger openings are entirely possible (and have been excavated, as described in the Overview), given appropriate rock mechanics assessments to determine support requirements.

*Typical drift sizes:* Most tracked drifts above the 4100-ft level are 7 ft by 7 ft. Tracked drifts between 4100 and 6800 ft are typically 8 ft by 8 ft. The 4850-ft level vent drift from the Yates shaft to the Ross/No. 6 shaft is about 13-15 ft wide and arched to 12 ft. Rail access drifts between the 6800-ft and 8000-ft levels are typically 9 ft wide and arched to 9 ft. Trackless drifts throughout the mine are typically 10 ft in width and arched to 10 ft.

***I.2 Subsurface access (WBS category 3.1):*** Guiding principles for the proposed underground access include:

- For safety, all levels should have dual access.
- The scientific access should be clean – not necessarily a cleanroom environment, but far cleaner than typical mine access.
- The scientific access should be designed around a standardized module, the dimensions and size of which will determine how most equipment is transported underground.
- Some provision to transport oversized materials is helpful.
- As 24/7 access is important, a single-person automated man-hoist is important.
- As specialized excavations will be necessary for many future experiments, NUSEL must retain the capacity to excavate new rooms on the 7400- and 4850-ft levels, as needs arise in future years. This must be done without disrupting existing experiments or compromising the cleanliness of the scientific access.
- Megadetector construction should be optimized, and interference of this project with the ongoing science minimized.

Figure E.3 shows a two-dimensional projection of Homestake's main drifts and hoists. The relevant hoists are the Yates and Ross (from surface to 4850 ft), No. 6 Winze (from very near the base of the Ross to the mine bottom, at 8000 ft), and the No. 4 Winze (from 4850 ft and 7400 ft). These hoists form a triangle, viewed from above: the Yates – Ross/No. 6 separation is ~ 1 km, the Ross/No. 6 – No. 4 separation is ~ 1.6 km, and the No. 4 – Yates separation is ~1.9 km.

The proposed chronology for optimizing the access for NUSEL is as follows (assuming a FY06 start of construction):

- For the 2.3 years of the preconstruction period, the site must be maintained (pumping, ventilation, etc.), which requires safe access to depth. Because the mine owner, Barrick Gold, has deferred maintenance in anticipation of abandoning the site in 2003, substantial shaft and hoist maintenance will have to be funded by the science community during preconstruction, in order to guarantee safe access. This is an unresolved issue for the proposers and NSF.
- With the start of construction in FY06, the Ross and No. 6 Winze will be used to mine the 7400-ft laboratory level and to clear any areas on 8000-ft needed for the borehole program. This should take about one year. Following this the No. 6 Winze and Ross will be upgraded and modernized, and a major rebuild of the Ross shaft undertaken. When the No. 6 Winze worked is completed, estimated to be 8 months into FY07, some early science can move into the 7400-ft level. The planned work on the Ross shaft and hoist is more extensive, and will not be completed until the start of FY08. At that point the Ross/No. 6 system is fully converted to clean science access, surface to 8000 ft. All future excavation must be done through other shafts.
- The Yates and No. 4 will be retained as mining hoists. Important maintenance and modernization of these hoists will have been done in parallel with the work on the Ross/No. 6. Initial development on the 4850-ft level will proceed from the Ross side of the Yates-Ross drift. Future development can be done by moving rock toward the Yates. Such excavations can be done cleanly, without affecting operating experiments, by arranging ventilation circuits properly, as will be described later.

- Future expansion of the 7400-ft level can be done similarly, routing the waste rock to the No. 4 Winze, along the separate 4850-level drift connecting the top of the No. 4 to the base of the Yates, then up the Yates. Again, the hauling never protrudes into areas being used for science.
- Access for geomicrobiology and earth science to 8000 ft is provided by the Ross/No. 6 and by a ramp that connects the 8000- and 7400-ft levels (thereby reaching the No. 4).
- The most likely site for future excavation of the megadetector is in the Yates formation, near the base of the Yates shaft. As noted before, the rock would be transported to the open cut on a conveyor system built on the 600-ft level.
- It is quite possible that the megadetector experimentalists would chose, on completing excavation, to reconfigure the Yates hoist and cage to more efficiently lower scientific loads. A cage footprint of at least 18 ft × 13 ft could be achieved. A dedicated clean hoist could make construction more efficient. If this option were pursued, then it would be important, during megadetector excavation, to complete any anticipated excavation on the 7400-ft level in parallel. (That is, the Yates would be available for mining during megadetector excavation, but not during later construction.) Parallel excavation on the 4850- and 7400-ft levels is feasible because megadetector excavation requires only 50% of the Yates mining capacity, assuming a five-year excavation period. (Yates hoist upgrades for the megadetector are not included in this proposal.)

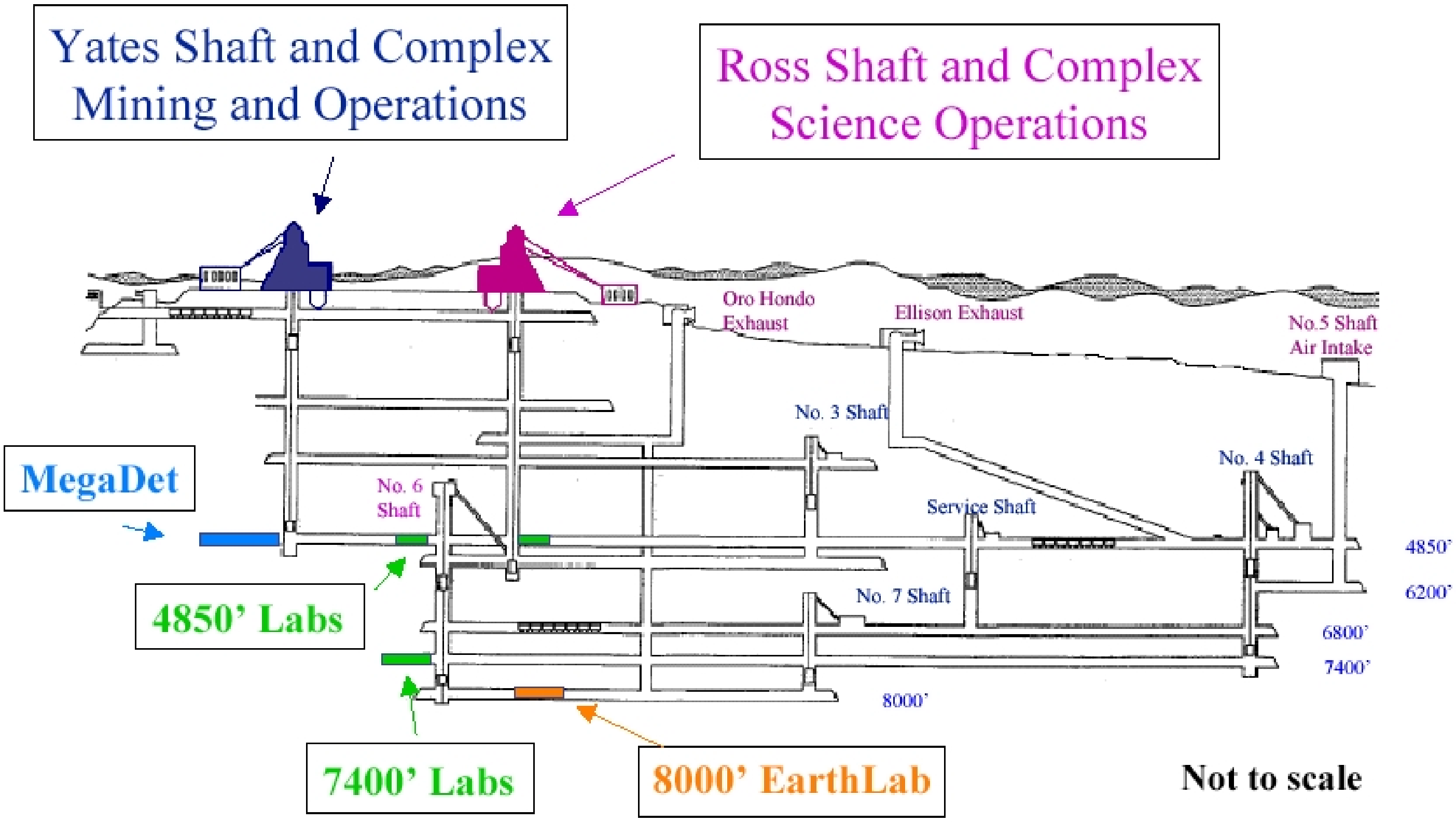

Figure E.3: The cross section of the mine, inserted here to show the shafts relevant to access. This is a 2D projection: see Figures E.2 for a top view showing relative positions.

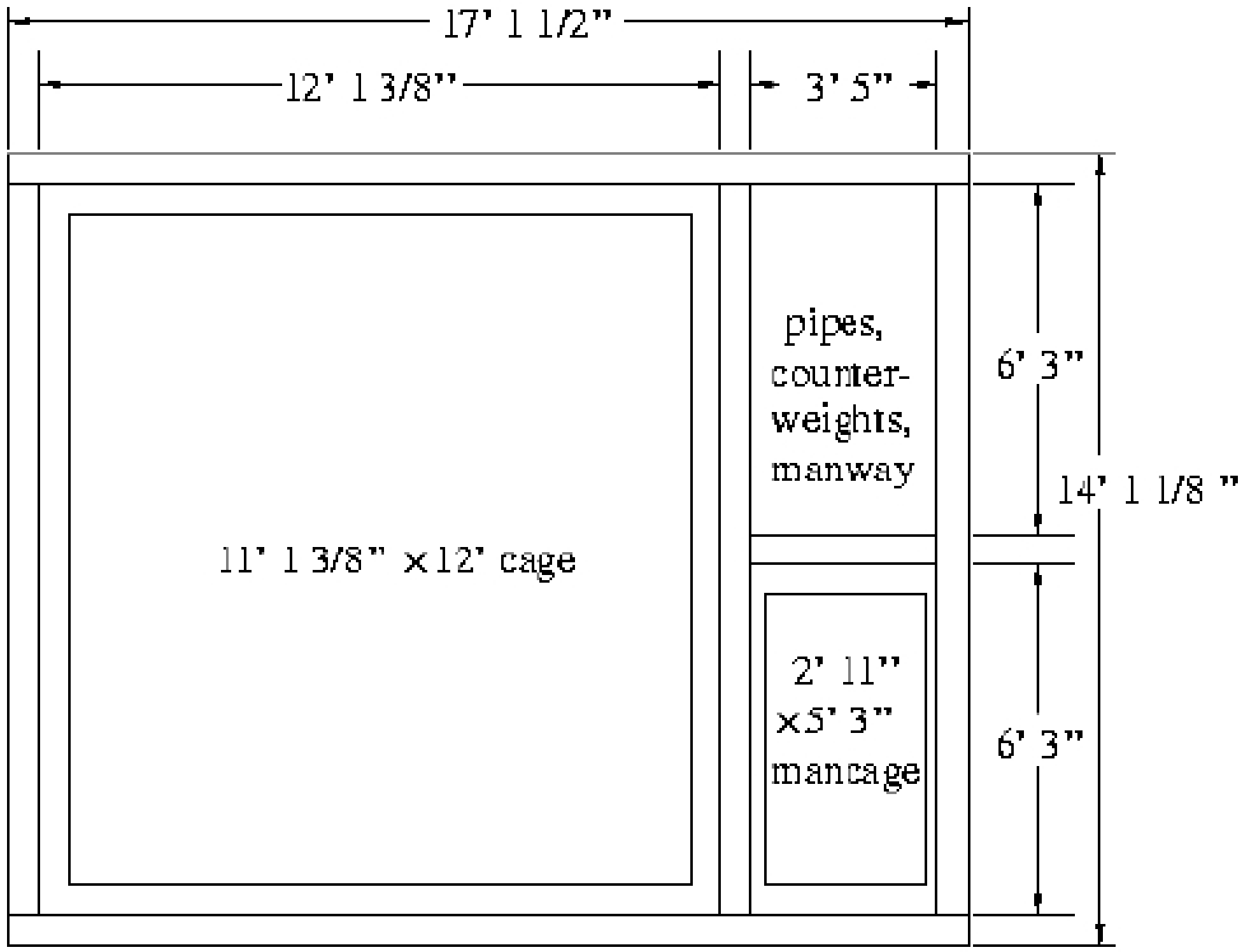

Figure E.4: Schematic showing the planned No. 6 Winze cage reconfiguration and 24/7 mancage.

The envisioned Ross and No. 6 shaft/hoist upgrades represent a substantial investment, but the return is decades of enhanced efficiency in mounting experiments. The improvements on the Yates and No. 4 are focused on maintenance and automation, with the goal of reducing NUSEL operating costs. Shaft improvements include:

- Realign and replace steel and timber as needed
- Realign and replace guides
- Repair and replace utilities, pump column as needed
- Convert the Ross and No. 6 Winze ore hoisting compartments to dedicated personnel lifts and counterweights
- Clean and dewater the Ross and No. 6 Winze
- Rebuild the Ross and No. 6 Winze level station areas at 4850, 7400, and 8000 ft

The plant and equipment improvements include:

- Conduct all deferred maintenance and repairs
- Replace the Yates and Ross MG sets
- Automate the Yates, Ross, No. 4 Winze, and No. 6 Winze hoists
- Convert the Ross and No. 6 Winze ore-hoists to second man-hoists
- Replace the No. 4 Winze ore-hoist and skip pocket
- Investigate high strength ropes and new drum liners to upgrade the hoisting capacity of the Ross and No. 6 Winze

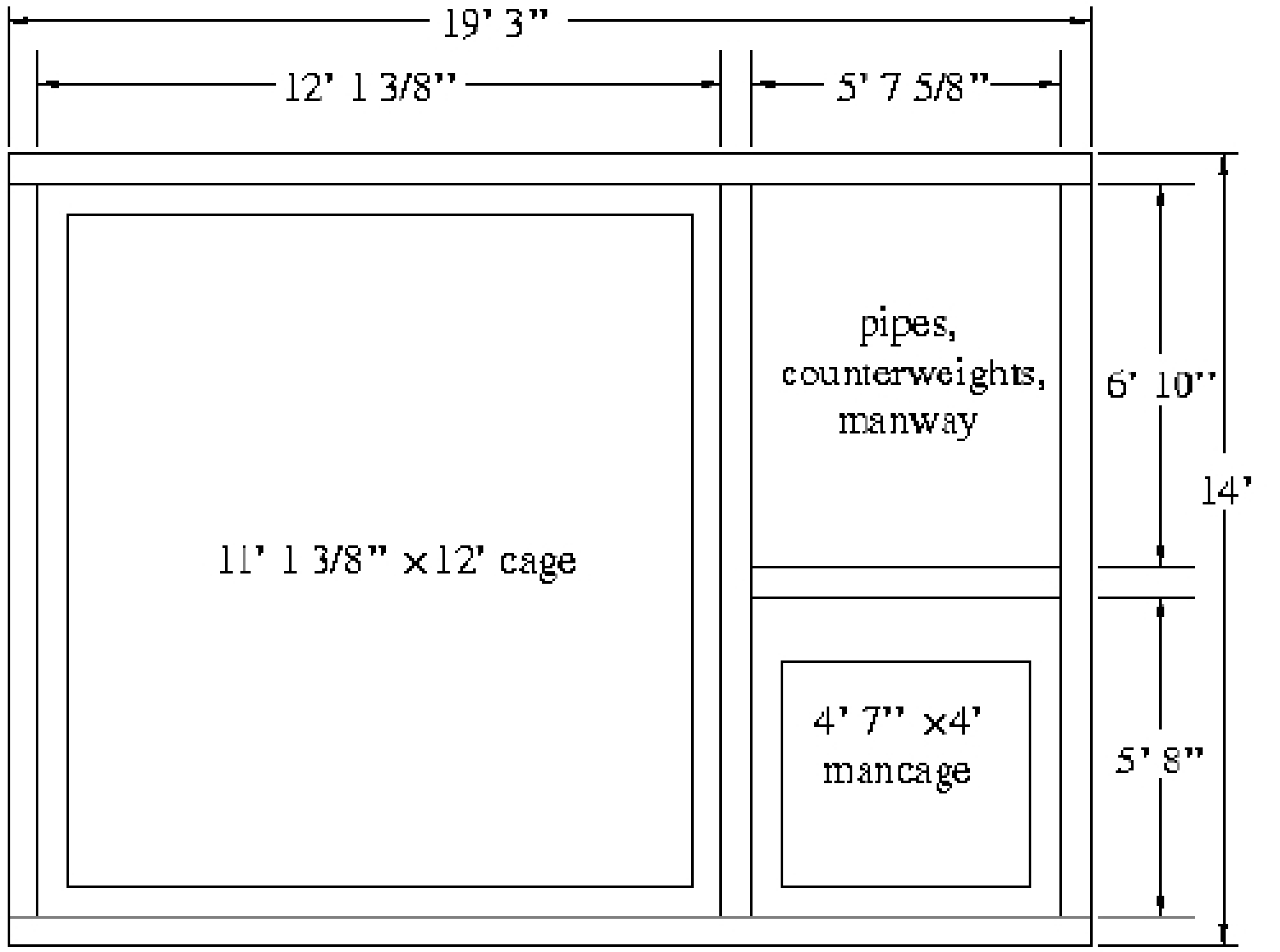

Figure E.5: As in Figure E.4, but for the Ross hoist.

The engineering schedule and costing for these upgrades is presented in Appendix A.

The Ross and No. 6 shafts have cross sections of 19.25 ft × 14.0 ft and 17.13 ft × 14.1 ft. This permits a cage footprint of 12 ft × 11.1 ft. The envisioned cage is light in weight and enclosed, resembling a standard elevator compartment. The vertical dimension is not yet fixed. The load capacity of the hoist is at least 8 tons. In the engineering studies that will be done to fix the final design, one option that will be explored is the use of light-weight ropes and new drum liners, which could substantially increase the hoist capacity.

Both the Ross and No. 6 Winze will include small man hoists, approximately 3 ft × 5 ft, to provide 24/7 access to depth. The after-hours use of these small cages would be via card keys: a system could be put in place where the nightwatchperson could serve as backup operator.

All of the hoist and shaft work is detailed in the appropriate section of the appendix, where the spreadsheets for the proposed work are presented.

*I.3* ***Laboratory developments on the 7400-ft level (WBS category 3.2)****:* The design of space on the 7400-ft level was based on the following conclusions, drawn from the science book and from discussions at Lead, NESS02, and elsewhere:

- The laboratory should be a class 10,000 clean room, with provision for achieving higher levels of cleanliness in selected rooms.
- There is a preference for small, individualized rooms for many experiments (in contrast to the large-hall format of Gran Sasso).
- The low-level counting facility should be state-of-the-art, capable of serving the broader community in addition

to NUSEL's experiments and R&D efforts.
- There is need for some high-bay general-purpose space that can accommodate a large experiment or several R&D efforts.
- A hallmark of NUSEL-Homestake should be the excavation of special-purpose cavities to enable specific experiments. The site should be engineered to allow such excavations without unduly interfering with ongoing experiments.
- The Homestake site presents unique opportunities for guaranteeing safe venting of experiments, e.g., by routing exhausts to higher drifts. These opportunities should be exploited.
- While the Gran Sasso model of a surface laboratory is used, there should be significant facilities underground (cafeteria, clean and outside machine shops, some office space, a meeting room) to eliminate unnecessary use of the hoist.

These considerations – folded with the requirements specified in the Science Book – leads to the following set of rooms on the 7400-ft level. All are initial construction apart for the solar neutrino facility, which we discuss below as a special-purpose excavation coming perhaps two years after initial construction (when a decision is made among pp neutrino detectors now under development). Thus the solar neutrino cavity poses the interesting issue of excavating an additional room while maintaining the cleanliness of the operating laboratory.

| | Dimensions*<br>W x L x H (m) | Volume<br>($m^3$) | Floor Area<br>($m^2$) | Wall Surface<br>($m^2$) |
|---|---|---|---|---|
| 7400' Level | | | | |
| Car wash/change | 10 x 8 x 8+<br>10 x 8 x 5 | 675<br>400 | 80<br>80 | 400<br>260 |
| Lunch/refuge | 15 x 20 x 5 | 1500 | 300 | 700 |
| Utilities | 10 x 50 x 8 | 4400 | 500 | 1700 |
| Geo Lab | 15 x 28.5 x 5 | 2140 | 430 | 860 |
| LLCF | 29 x 14 x 18 +<br>21 x 14 x 9 | 10,000 | 1000 | 3000 |
| LLCF Utilities | 20 x 14 x 6 | 2000 | 280 | 900 |
| Secure Counting Lab | 10 x 15 x 5 | 750 | 150 | 475 |
| General Purpose Lab | 20 x 80 x 20 | 35,000 | 1600 | 7200 |
| Dark Matter #1 | 20 x 15 x 8 | 2700 | 300 | 1100 |
| Double Beta decay | 30 x 15 x 8 | 4000 | 450 | 1300 |
| Dark Matter #2 | 20 x 15 x 8 | 2700 | 300 | 1100 |
| Future Solar Nu Lab | 12 dia x 24 +<br>8 x 40 x 4 | 4500 | 430 | 2200 |
| Central Hall, Entrance | 5 x 265 x 5 | 7950 | 1590 | 5930 |
| Connecting Halls | 5 x 80 x 5 | 2400 | 480 | 1800 |
| Exhaust/utility drift | 3.5 x 400 x 3.5 | 4900 | 1400 | 4200 |
| Seminar room/offices | 20 x 25 x 4 | 2000 | 500 | 860 |
| Interior machine shop | 10 x 30 x 5 | 1000 | 200 | 500 |
| Ext. machine shop | 10 x 20 x 5 | 1000 | 200 | 500 |
| **Subtotal** | | *90,015* | *10,270* | *34,985* |

The main drift on the 7400-ft level extends somewhat beyond the No. 6 Winze station area, ending at the Yates formation, the rock best suited for large excavations. An entrance drift, 5.5m × 5.5m, will be driven into this formation, perpendicular to the edge of the formation. Rooms are oriented in that perpendicular direction, the direction of maximum stability.

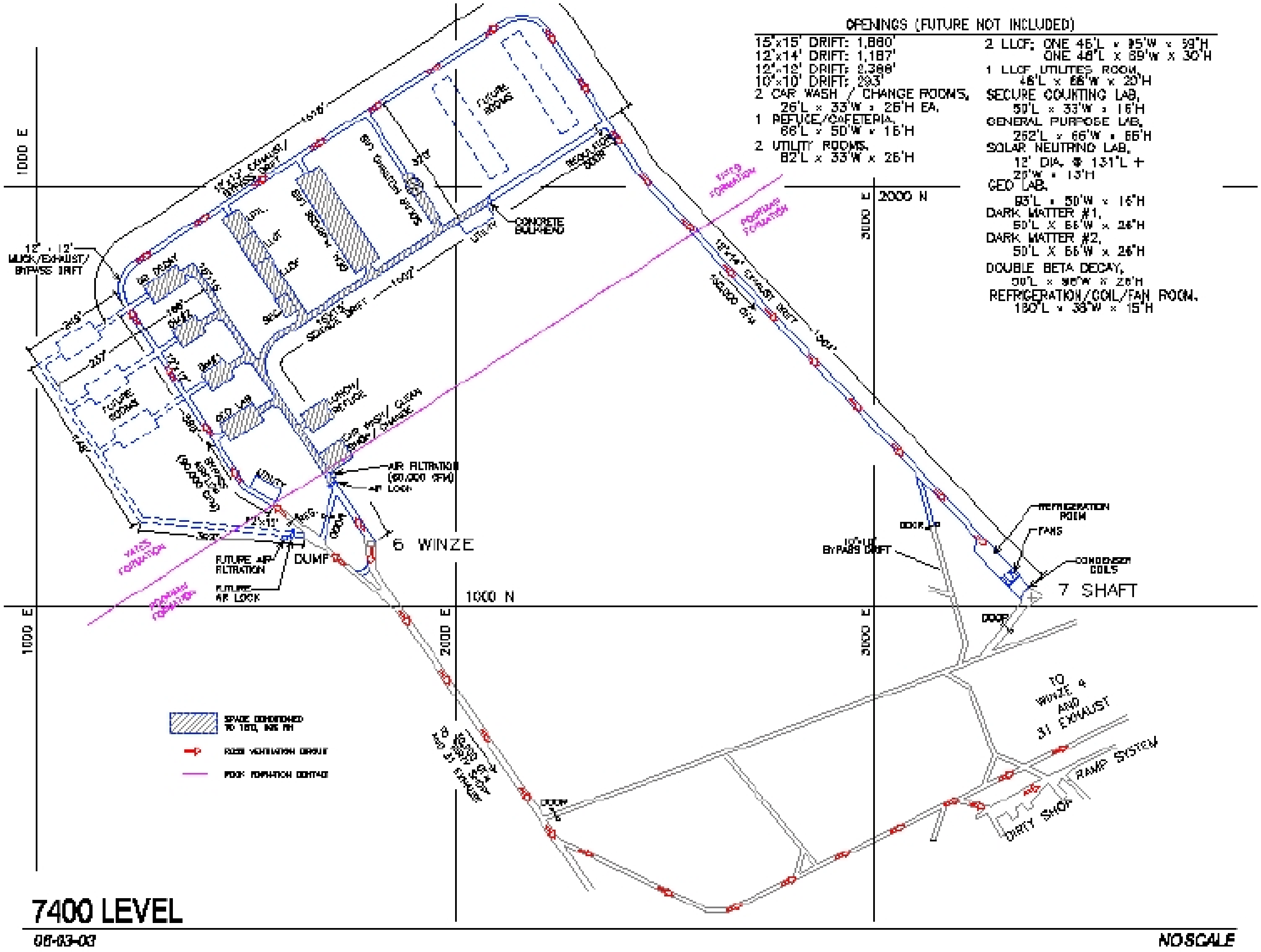

Figure E.6: The proposed layout of the main laboratory level at 7400 ft. The design allows additional rooms to be excavated on the right, through the exhaust drift. Similarly, a major laboratory expansion could be done on the left, again through exhaust drifts so that the existing laboratory remains isolated from the work. Service spaces (offices and the clean machine shop) are not shown. Note that all major construction is well within the Yates Formation (boundary indicated in red).

Figure E.6 shows the laboratory layout. The design was based on the following considerations:

- Positioning of rooms in the Yates formation.
- Alignment of the longest dimension of the major excavations perpendicular to the edge of the formation (and thus to the rock structure), the most stable direction.
- Spacing of major excavations by two times the width of the excavation. It is generally accepted that two parallel excavations should be separated by twice their width. However multiple parallel excavations may require additional spacing, as discussed below.
- The development should have space for future excavations, in that specialized cavities are required for several experiments now in the R&D phase. Expansion should not unduly affect the cleanliness or seismic quietness of operating experiments.

The entrance to the clean area includes a changing room and a car wash, an arrangement that has worked well at SNO. Four smaller rooms for dark matter, double beta decay, and laboratory earth science are arranged along the left side of the main entrance corridor. (The dimensions of these rooms were increased over those specified in the Science Book to allow for unanticipated needs, such as a water shield.) Each has a 5m portal area where a higher-level clean room barrier or addition radon filtration could be installed. Two machine-shop areas (one clean, one intended for equipment not compatible with the clean area and thus accessible only from the outside) are provided. The latter will be located in existing space in the 7400-level shop area. A cafeteria/refuge and a utilities room are positioned immediately to the right of the entrance corridor. Further to the right is the low-level counting facility (described separately below) and a Gran-Sasso-style high bay area, an arched hall with floor dimensions of 20m × 80m and with a minimum ceiling height of 20m.

The third major development, farthest to the right, illustrates how the layout allows future expansion. This is designated as the solar neutrino area, representing a special-purpose excavation that would likely be undertaken one or two years after the science program commences on the 7400-ft level. Approximately 900 feet of expansion space has been left for this and other future excavations. Development will proceed to the right, with

Common areas include the cafeteria/refuge, a modest office area (10 offices and a 30’ by 40’ seminar/discussion room), a machine shop within the clean area, and a second machining area outside the clean area.

Accepted practice calls for the rooms to be separated by about two times their width at this depth. However, one of the issues that must be studied carefully is whether such separation is sufficient when there are a series of parallel excavations, which can increase the net stress on each. This question requires coring of the area, numerical modeling of the rock mechanics, verification of the model by further tests, and, of course, conservative ground support once the answers are known. If the spacing is increased to 3 time the widths, a possible outcome of such a study, this would require us to extend the length of 5m × 5m science drifts (for accessing the rooms) and 3.5m × 3.5m exhaust drifts, a modest alteration. A second point requiring further study is radon. While the entire clean area will use scrubbed air, **additional radon remediation will be studied during the baseline definition process**, as we want to understand better the consequences of ventilating along long drifts providing substantial rock surface area.

The floor space and volume occupied by the experimental halls is 4660 $m^2$ and 66,190 $m^3$, respectively, while the common areas (car wash/change, lunch/refuge, seminar room, offices, machine shops) account for 1360 $m^2$ and 6575 $m^3$. The utility areas (general and within the low-level counting facility) require 780 $m^2$ and 6400 $m^3$.

The laboratory ventilation and cooling system is shown in Figure E.7. As will be shown later, the gross ventilation pattern is down the Ross and No. 6 shafts, into the laboratory complex, and exhausted through a vent drift to the No. 7 shaft.

Similarly, the process diagram – industrial, domestic, and chilled water, the mine dewatering system, sewage, and fire protection – is diagrammed in Figure E.8.

Figure E.7: The 7400-ft main laboratory level ventilation and cooling scheme.

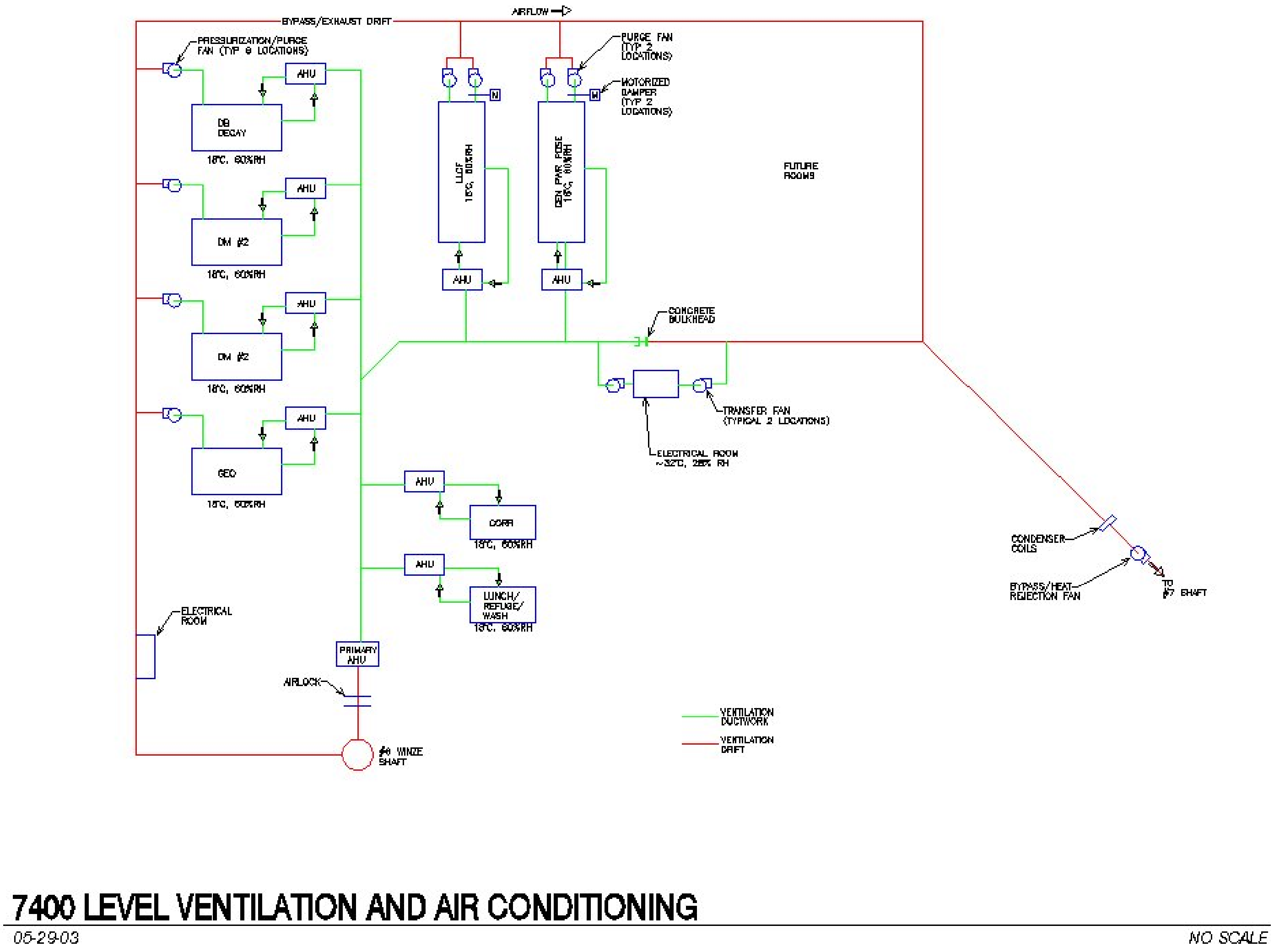

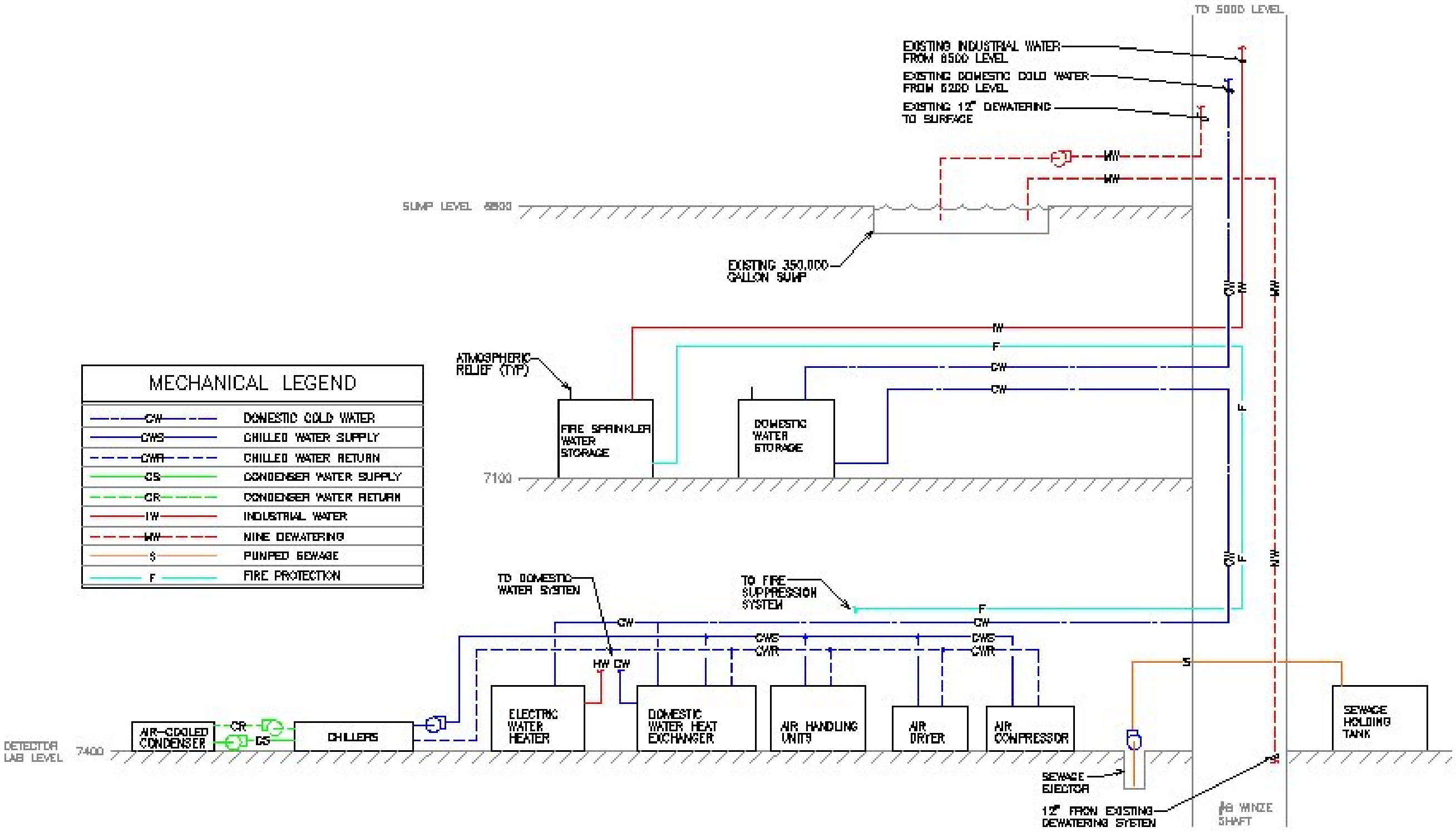

## 7400 CHILLER AND PROCESS PIPING SCHEMATIC

*05-29-03* *NO SCALE*

Figure E.8: The 7400-ft main laboratory level process diagram.

***I.4 Laboratory development on the 4850 and 8000 ft levels (WBS categories 3.3, 3.4).*** Additional developments are planned for the 4850-ft level and for the 8000-ft level. The former, of course, will likely at some point host the megadetector near the base of the Yates. The excavation costs of that major experiment is not part of this proposal. (However, in the spreadsheets presented in the appendix, we include the conveyor that would be needed to transport rock from the megadetector excavation to the Open Cut. This is a major item, approaching $5M. However, if implemented early in the NUSEL-Homestake development plan, this facility would substantially reduce haulage costs. Thus we have included it in our costs and schedules to leave open this option.) Apart from the megadetector, an ideal facility for the 4850 level is the accelerator for nuclear astrophysics. This experiment should be isolated electrically from some of the sensitive experiments that will be done on the 7400-ft level. The 4300 mwe provided by the 4850-ft level is also more than adequate. In addition, storage areas for materials cooling will be placed on the 4850-ft level. This level will be accessible continuously, starting approximately six months after the start of construction. There is excellent excavated space on this level (as Figure E.7 shows), which would allow us to mount moderate scale experiments there very early, using prefabricated clean rooms. Two groups have made inquiries about this possibility.

Also shown in the table below is a room on the 8000-ft level for geomicrobiology and an estimate of the excavation needed to provide space for the drilling rig this group will use (depending on the location they choose). The 8000-ft room is the laboratory for the geomicrobiology anaerobic glove box. There is a larger earth science laboratory in the 7400-ft complex.

| | Dimensions* W x L x H (m) | Volume ($m^3$) | Floor Area ($m^2$) | Wall Surface ($m^2$) |
|---|---|---|---|---|
| 4850' Level | | | | |
| Accelerator | 20 x 35 x 10 + 5 x 10 x 4 | 8000 | 750 | 2700 |
| Clean Room | 8 x 12 x 6 | *600* | *96* | *220* |
| Materials Storage | 10 x 20 x 4 | *800* | *200* | *300* |
| Lunch/refuge room | 10 x 6 x 4 | *300* | *60* | *120* |
| *Subtotal* | | 9700 | 1106 | 3340 |
| | | | | |
| | | | | |
| 8000' Level | | | | |
| Anaerobic Room | 10 x 3.5 x 10 | 350 | 35 | 305 |
| Drilling Area | 7 x 16 x 10 | 1120 | | |
| *Subtotal* | | *1470* | *35* | *305* |
| | | | | |
| *Total* | | *101,185* | *11,411* | *38,630* |

Note that the 4850-ft development is outside the Yates formation, in the weaker Poorman formation. It is likely that the area chosen is quite suitable for excavations of the modest size proposed for this level. In this immediate vicinity, as one can see from the Figure E.7, there are a number of shops and other excavations that have been in existence for a long time. There are two reasons for wanting this development very near the base of the Ross shaft. The first is the obvious convenience of reaching this area quickly. The second is good ventilation schemes off either the Yates or Ross circuits (see below). However, safety is paramount. Thus this area will be cored and carefully studied, to verify its suitability. If the decision is negative (highly unlikely), then this development could be moved approximately 450m along the main vent drift, toward the Yates shaft, where the Yates formation begins.

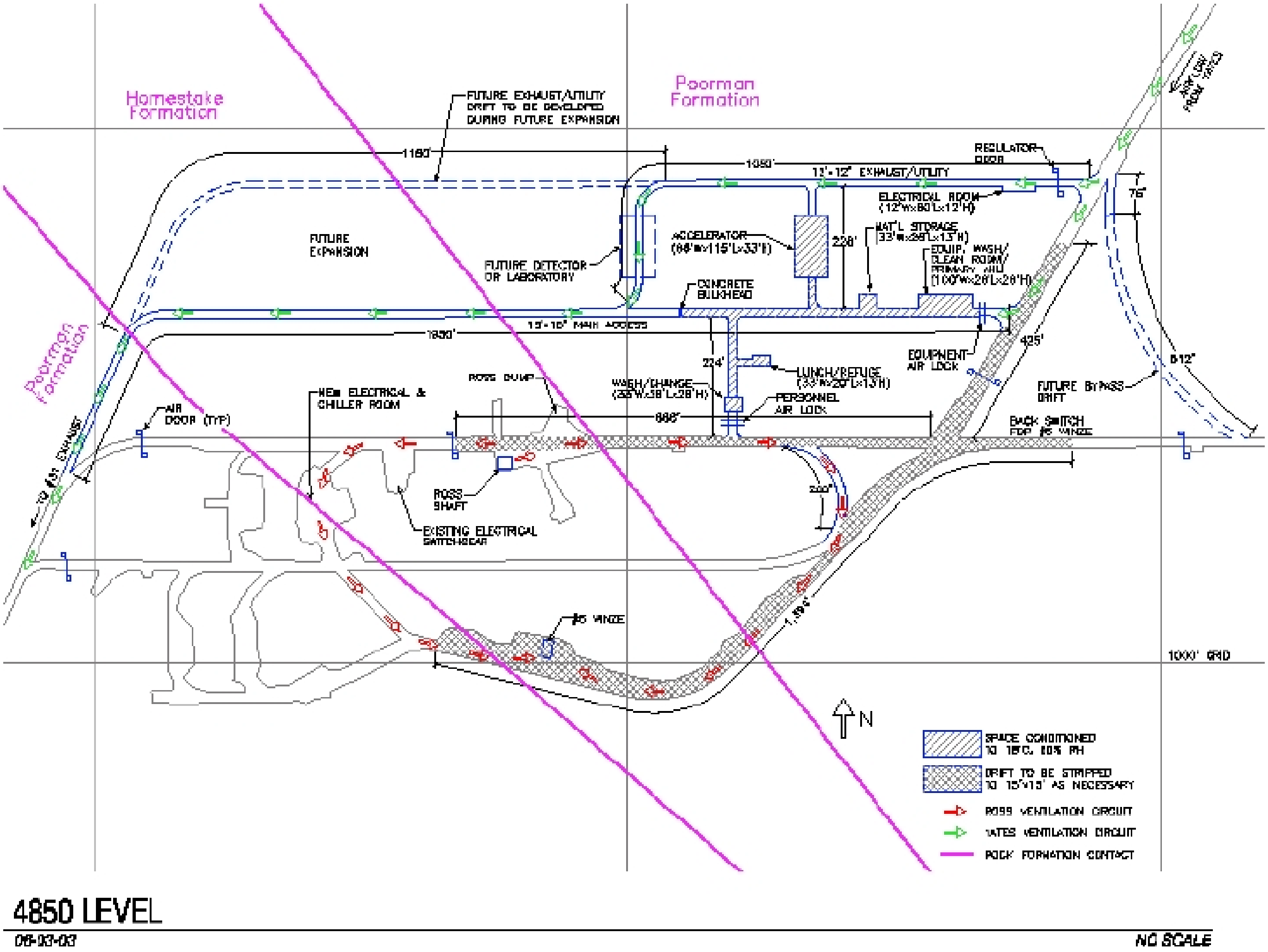

Figure E.9: The proposed layout of the 4850-ft intermediate laboratory level.

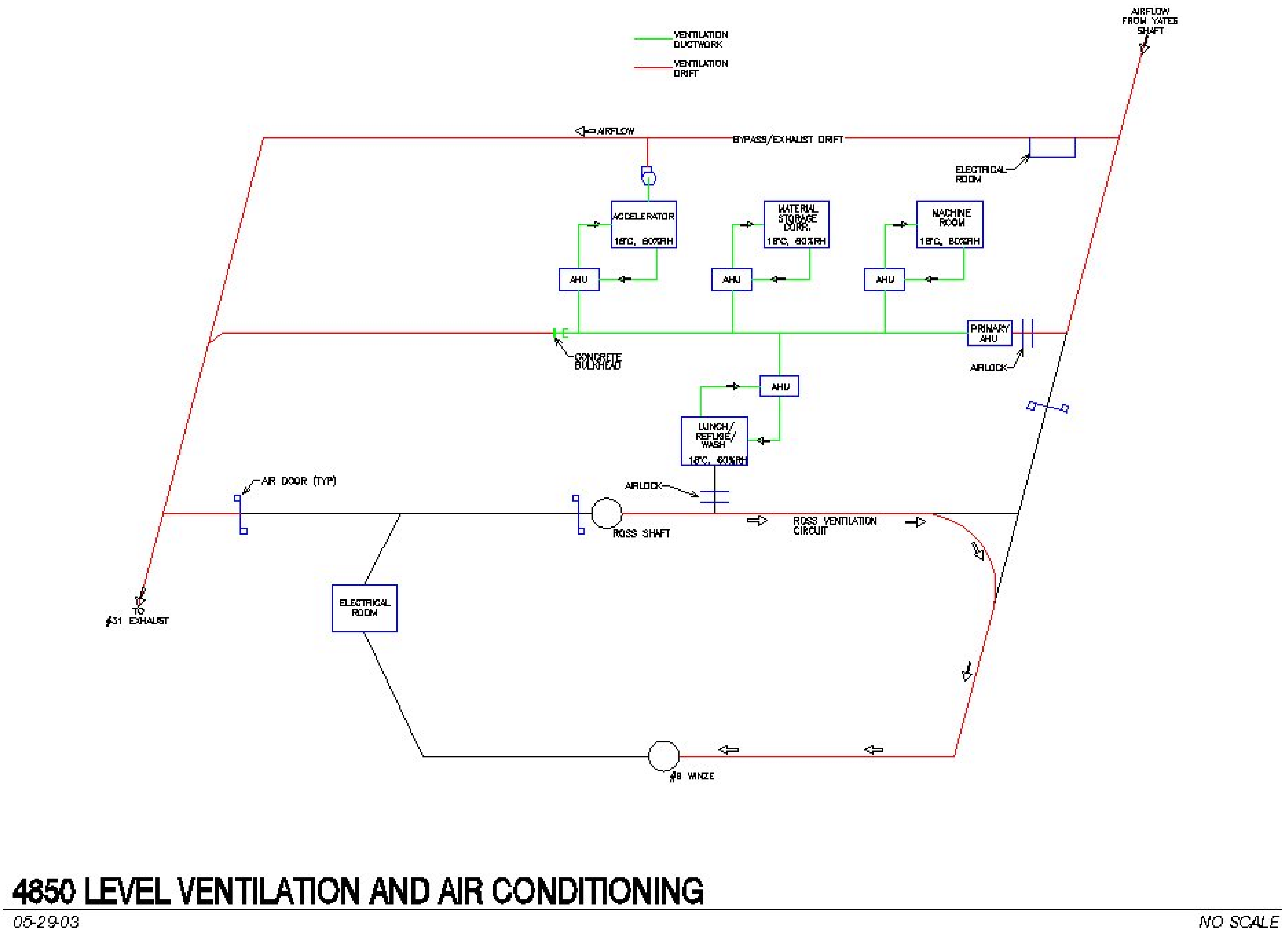

Figure E.10: The 4850-ft level ventilation and cooling scheme.

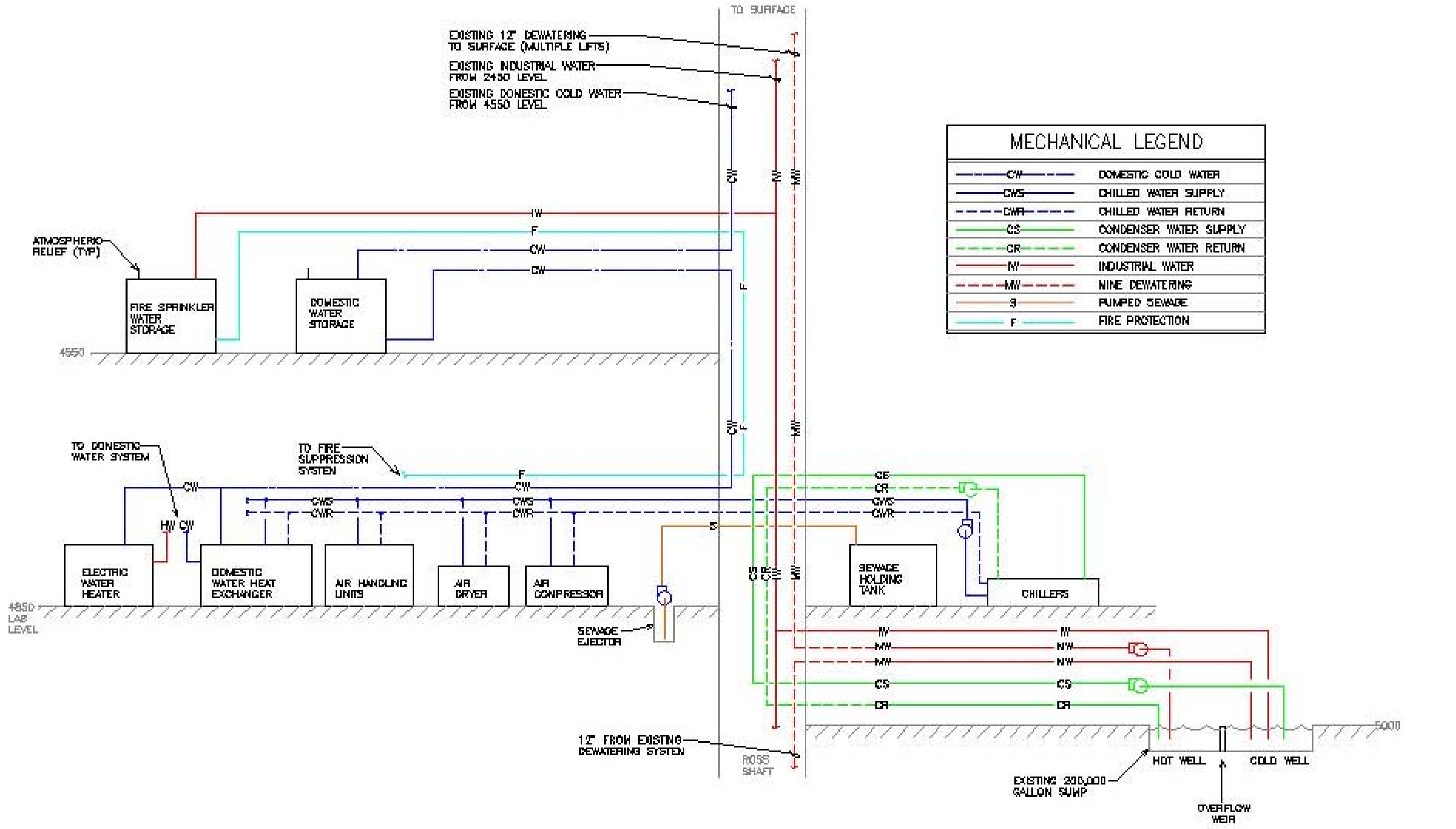

## 4850 CHILLER AND PROCESS PIPING SCHEMATIC

*05-29-03* *NO SCALE*

Figure E.11: The 4850-ft level process diagram.

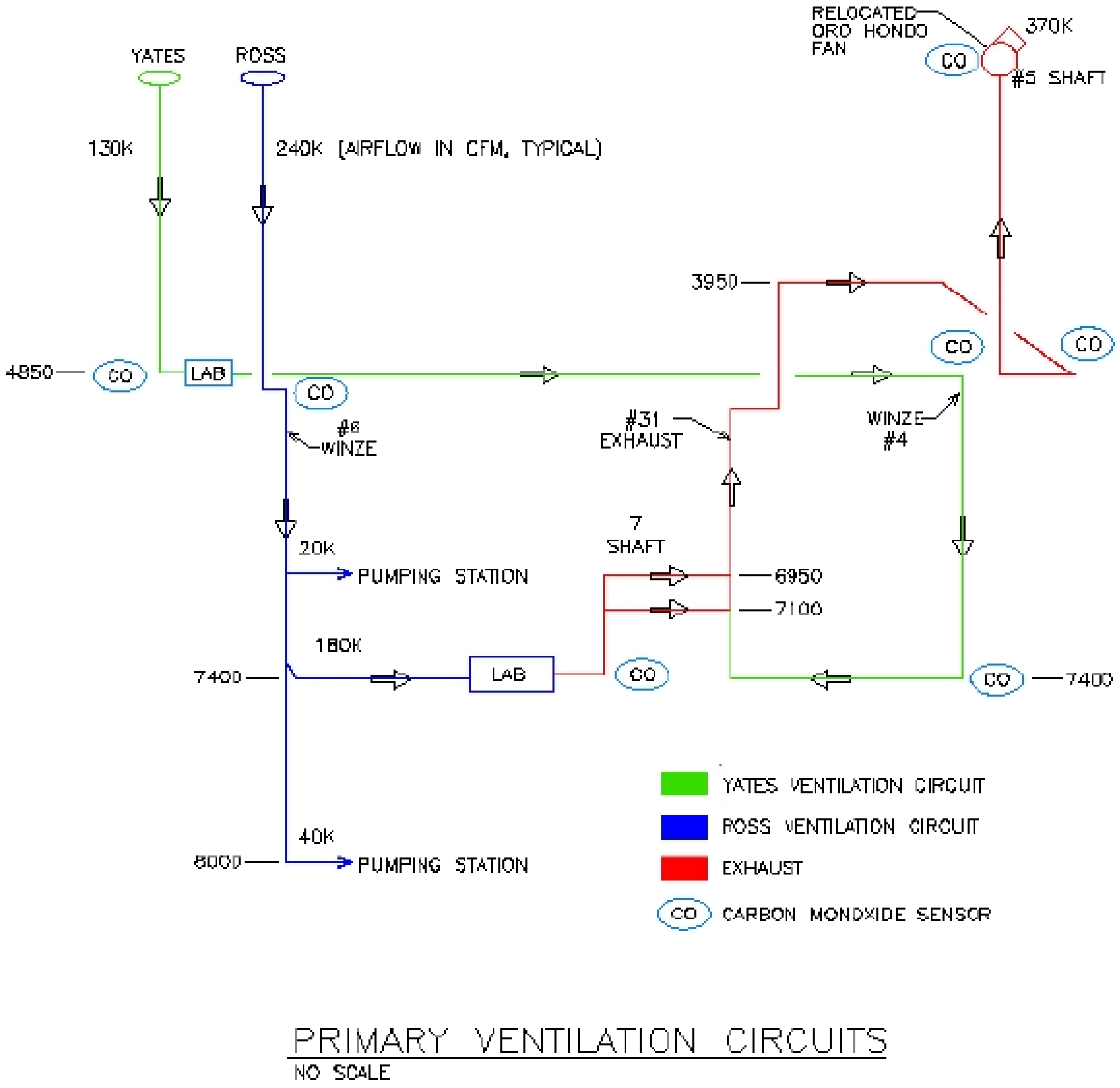

Figure E.12: The primary ventilation diagram for the laboratory, showing the Yates and Ross circuits and the repositioned gas monitors and Oro Hondo fan.

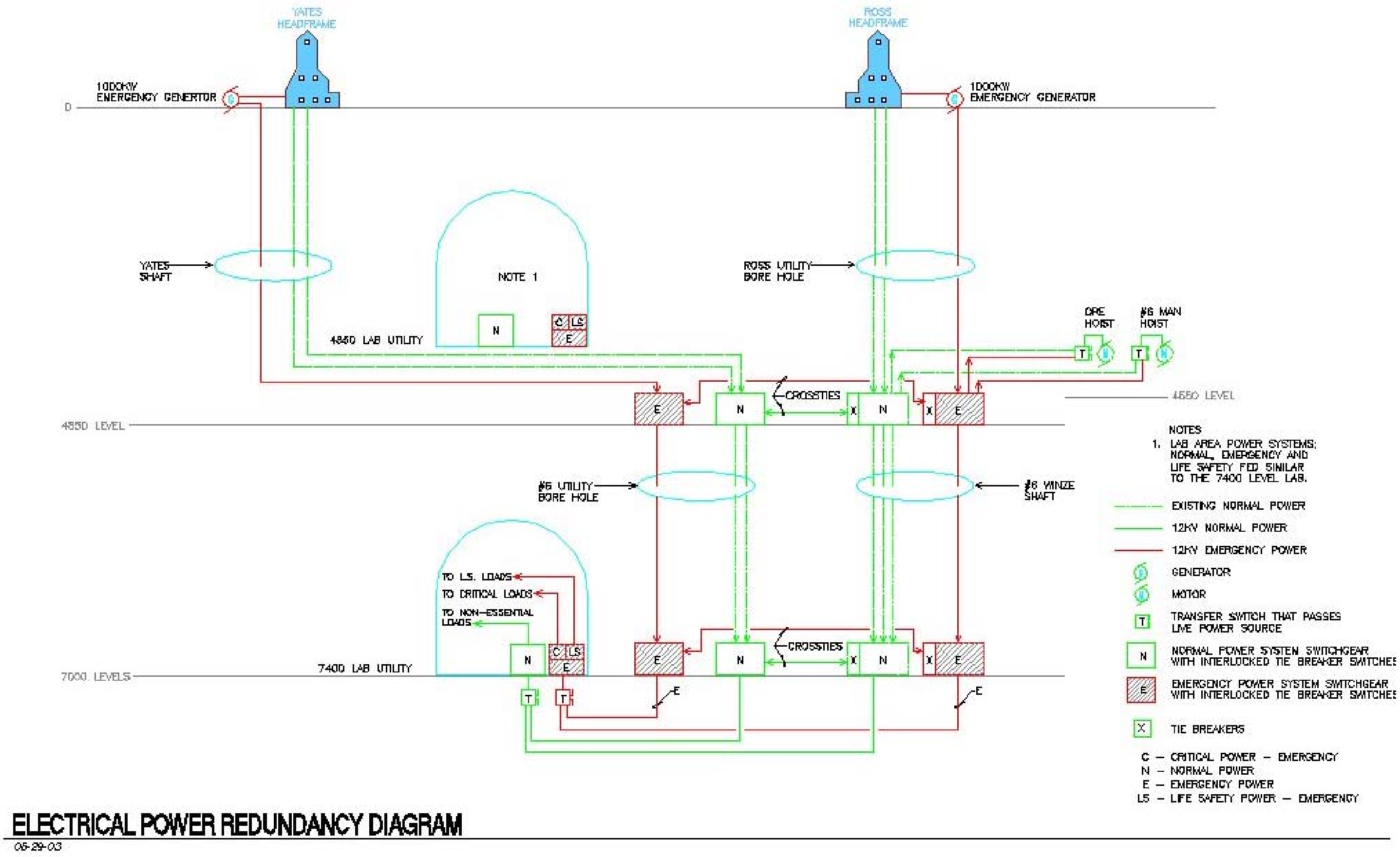

Figure E.13: The overall power redundancy scheme planned for NUSEL-Homestake. See discussion in section 3.9 of the WBS.

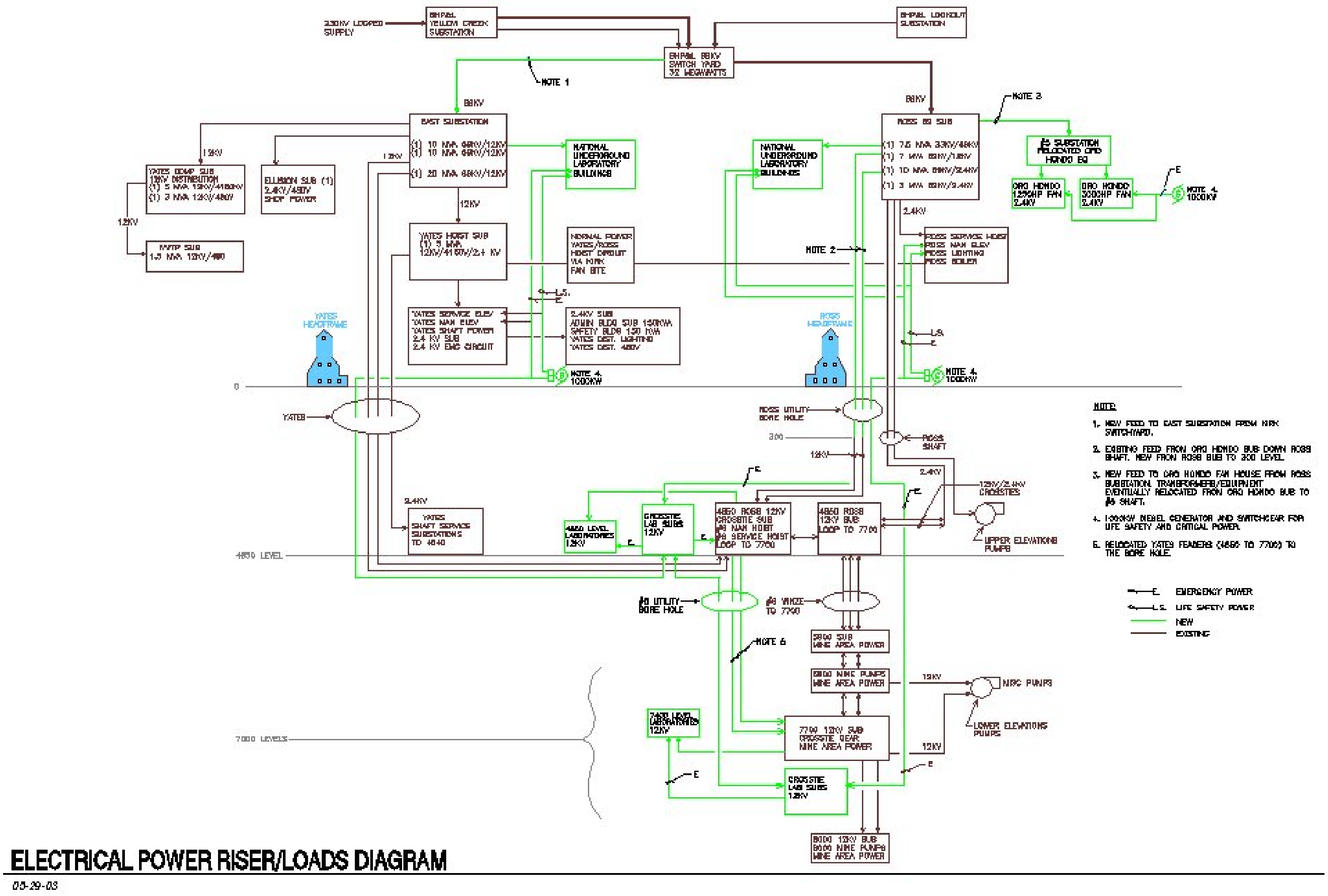

Figure E.14: The power riser-loads diagram proposed for NUSEL-Homestake showing the distribution system including new electrical boreholes. Much of this is detailed in section 3.9 of the WBS, and the associated excavation schedule and costs appear in the appendices.

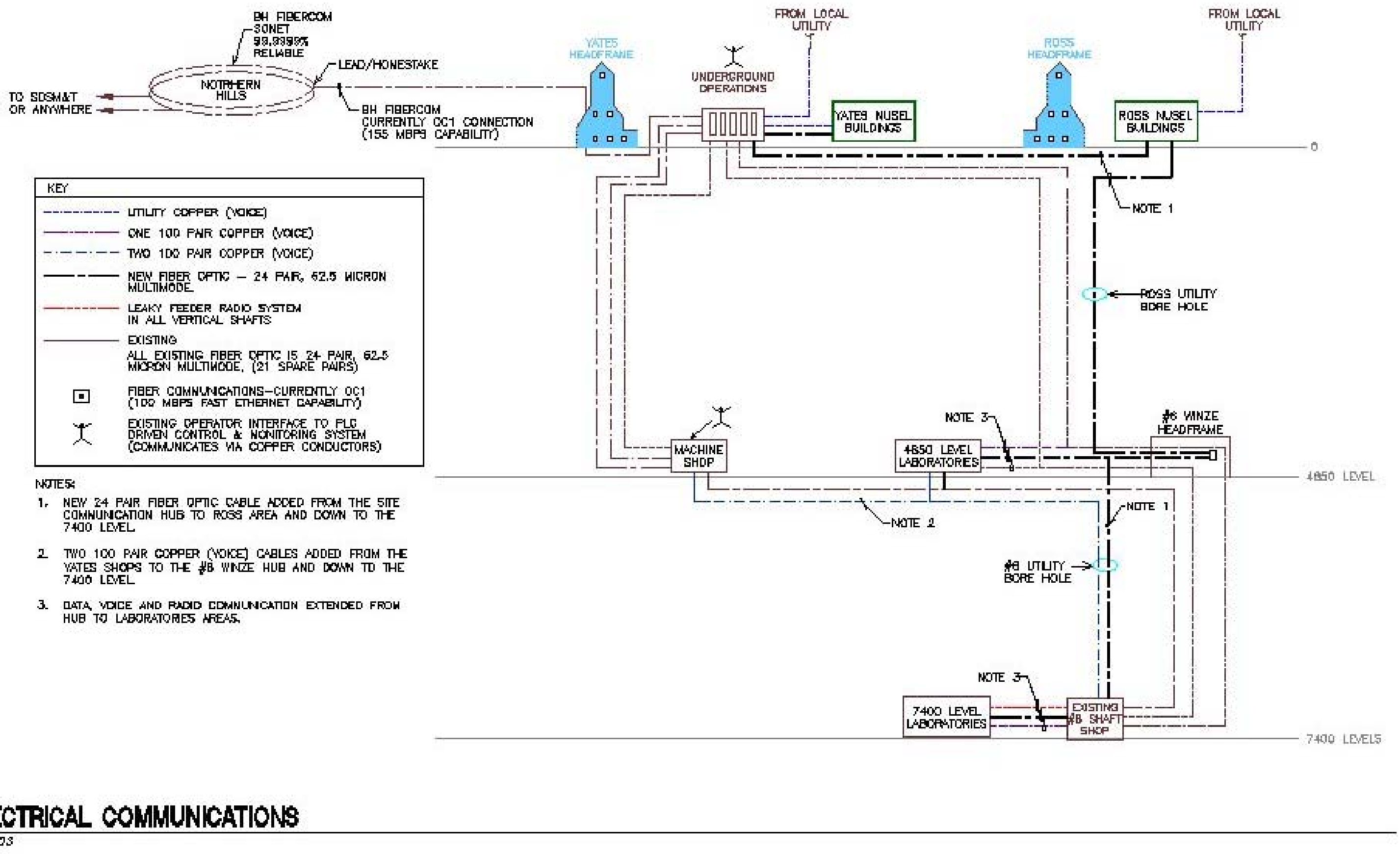

Figure E.15: Proposed NUSEL-Homestake communications plan, adapting current Homestake facilities. See the discussion in Section 3.9 of the WBS.

***I.5 Ventilation circuits.*** An important attribute of the Homestake mine is its sophisticated and flexible ventilation scheme. NUSEL will use two ventilation circuits, corresponding to air flowing down the Yates and the Ross/No. 6. Note that the flow is downward on all access shafts (the Ross, Yates, No. 6, and No. 4). Also note that the main laboratory level, which is on the Ross circuit, is vented through the No 7 shaft to drifts on the 7100 and 6950 levels, up the #31 exhaust, and out the #5 shaft. Thus if an accident produced a large quantity of suffocating gas, for example, it would be vented away from the occupied laboratory level. Workers would be able to exit via either the No. 6/Ross or the No. 4.

The 4850-ft laboratory area is on the Yates circuit. The intake is located near the equipment wash at the eastern end of the main corridor (see Figures E.7 and E.8). This circuit and bypass exhaust to the #31 exhaust vent at the lower right on these figures. The overall scheme is shown in Figure E.10. The red arrows in Figure E.7 show the Ross circuit airflow on the 4850-ft level between the Ross and No. 6 Winze. This crossover is sequestered from the Yates circuit (and thus the laboratory areas) by various air doors. Thus access to the laboratory is achieved by coming down the Ross, walking a short distance in the Ross ventilation circuit, and entering an airlock.

However it is clear that the 4850-ft level ventilation scheme could be reconfigured easily to run off the Ross circuit. This flexibility could be important when megadetector excavation begins "upstream" on the Yates ventilation circuit. By switching to the Ross circuit, the laboratory areas illustrated in Figure E.7 could be kept clean.

***I.6 Utilities and finishing of the underground halls (WBS category 3.9)***: An extensive discussion of utilities – power, communications, HVAC, sensors – appears in the WBS. That narrative walks one through the ventilation, power, and communications diagrams presented in this section (E.12 through E.15). We are concerned that Homestake flooding (see Section G) could alter these plans, which are largely based on cost-effective adaptations of current Homestake utilities.

A second category included in "underground systems" is the finishing of the underground halls, e.g., the transforming of the rough halls (excavated, shotcreted, concrete floors, mineguard) into finished labs. This includes bringing basic utilities into each hall. It also includes the detailed technical design specific to experiments: the clean room requirements, radon mitigation, the quantities and purities of experimental "utilities" such as liquid nitrogen, nitrogen gas, ultrapure water, accommodations for specialized electronics or chemistry needs, mechanical capabilities such as Cu electroforming, etc. In a process that began almost 20 months ago, during the Lead meeting, our collaboration has helped to organize the community into "interest groups" focused around science areas, such as dark matter or double beta decay, in order to move toward a consensus on the needs of each area. (This was a "deliverable" for the Lead convenors: their responses have been incorporated into the Science Book.) This information was essential in locating rooms and in defining their dimensions. It stimulated the further development of the low level counting facility and the specialized utilities it will provide. **We believe strongly that this kind of iterative process must continue and must be augmented: our engineers and the interest groups must check and cross-check each other in the Baseline Definition process.** We do not believe the NSF should be a spectator to this process: it must provide adequate funding to us so that we can continue this process. **The absence of agency support during the Reference Design has already limited this iterative process, which we believe is central to any successful large project**. During the Baseline Definition science-engineering teams will again review the configuration of each hall, to make sure that all utility needs have been adequately met. We have already "flagged" certain issues, such as radon control, that must be examined because of Reference Design changes within the last three months. The use of extended ventilation drifts (necessary in placing the 7400-ft development in Yates rock while utilizing the No. 7 shaft for ventilation) could increase radon levels. Certain rooms may require additional air scrubbing.

Because of these issues – the uncertainties of flooding and our desire to present cost breakdowns only after adequate interation between the science groups and the engineers have taken place – we have taken the following approach:

- We present a reasonably detailed plan in the WBS describing how we will use the existing Homestake utilities (some of which are virtually new) in our engineering.
- But in the WBS budget, costs are not derived in detail (as for underground construction), but by comparing to SNOLab, which kindly provided a detailed breakdown of underground systems costs and which is located at a depth comparable to Homestake's 7400-ft level. We have made adjustments of the SNOLab numbers for use

on 4850- and 8000-ft levels.

- We assign a contingency of 40% to account for any uncertainties.

***I.7 The low-level counting facility (WBS category 6.1)***: This proposal contains funding for one science facility, the Low-Level Counting Facility (LLCF). It is viewed in this proposal as a general facility for users, important to most of the basic and applied science projects that NUSEL will host. The LLCF integrates many functions into three primary areas: ultra-low-level counting and sample preparation, low-level counting and analysis, and utilities.

The ultra-low-level counting chamber consists of two large ultra-pure water pools and associated sample preparation and data acquisition/analysis areas. The water pools are separated by a low-activity steel and sulfurcrete retaining wall and are lined with urylon, a water- and radon-impermeable plastic coating. Each pool is a 12m cube filled with ultrapure water (~$10^{-14}$ g/g U and Th), covered by a hermetic deck, with a 4m high chamber above for handling and insertion of counters and samples into the water. The chamber is supplied with radon-free air, while the air between the water surface and the deck is purged with radon-free nitrogen.

In the first pool several reentrant ports extend into the water. Six of the ports house large acrylic thimbles, each containing an annulus of liquid scintillator surrounding a central sample chamber 0.5m in diameter and 0.5m in length. The light from the scintillator is channeled through light pipes to an array of photomultipliers on the deck above the water. It will be possible to measure U and Th in samples at sensitivities of $10^{-13}$ – $10^{-14}$ g/g. The samples to be counted must be first enclosed in a water- and Rn-impermeable plastic bag that is also highly reflective on the outer surface. VM2000 (manufactured by 3M Corporation) appears quite suitable. LLCF plans provide a sample preparation laboratory and data acquisition (DAQ) room for the pool. Three other reentrant ports are provided for prototyping, important for experiments requiring extremely low levels of external background.

Bulk assay of large amounts of materials for U and Th at the $10^{-13}$ – $10^{-14}$ g/g will be possible in this pool. As the figure shows, two of the ports are in the central region of the pool, where the highest level of sensitivity can be achieved, while four others are near the periphery, allowing assay at the $10^{-13}$ g/g level. Although the energy resolution of the modules is relatively poor (~ 20%/√E/MeV), Monte Carlo simulations do indicated some limited ability to distinguish U from Th, based on the measured energy spectrum. The primary justification for the facility is the need to assay large representative samples of materials to be used in shielding, supports, and front-end electronics in the outer regions of dark matter, double beta decay, and solar neutrino detectors. As the measurement time per sample is two ($10^{-13}$ g/g) to four ($10^{-14}$ g/g) weeks, the first pool will be able to process about 100 samples/y at the $10^{-13}$ g/g level and about 24/y at the $10^{-14}$ g/g level. This should be adequate for NUSEL and national needs, according to our surveys.

Three reentrant ports for prototyping detectors are installed in the first pool. All can accommodate detectors up to 1m in diameter and 1m in length. One of the ports is located in the pool's central region, minimizing external backgrounds, while the other two are closer to the periphery. Separate DAQ rooms are provided for each test port.

As the deck of the first pool is modular, it is possible to reconfigure the area to handle larger samples or a larger number of samples.

The second pool is reserved for future developments, possibly a "mini-CTF" similar to the Counting Test Facility (CTF) built by the Borexino collaboration at Gran Sasso. This would extend sensitivities to U and Th by approximately two orders of magnitude beyond that possible with the modular system pool. Such a facility could also be filled with Gd-loaded liquid scintillator to provide unparalleled sensitivity to neutrons. While the second pool awaits development, experimentalists will use the facility to prototype detectors in a very low background environment.

The ultra-low level hall also houses a leaching/emanation laboratory. It will allow experimentalists to measure U and Th (and other radioisotopes) leaching out of samples placed in ultra-pure water. SNO and other experimentalists have developed the extraction and counting techniques (vacuum degassing and counting in Lucas cells) required for $10^{-12}$ g/g sensitivity to U and Th. The leaching/emanation laboratory contains three large (1m × 3m × 1m high) leaching tanks filled with ultrapure water, associated pumping and degassing systems, and counting and DAQ systems.

The ultra-low-level counting hall contains three staff offices to allow on-site analysis and reporting of measurements made in the ultra-low- and low-level counting halls.

The LLCF will house a full array of alpha, beta, and gamma counting systems. An array of eight well-shielded single, segmented, and multiple hyper-pure Ge detectors will be available. Eight novel ultra-low background tracking detectors sensitive to surface alpha and beta contaminants will be deployed. Small proportional counters (similar to those originally developed for the chlorine solar neutrino experiment) will be placed inside shields to allow single-atom counting of gas samples containing argon, krypton, xenon, radon, and various molecular gases. Sample handle and preparation laboratories will be in close proximity to the counting rooms. As each sample requires one to four weeks for counting (depending on sample size and sensitivity desired), several hundred samples can be processed per year per array. This corresponds to projected needs, based on our survey of groups planning ultra-low-level experiments.

Sensitivities of $10^{-9}$ g/g U and Th can be achieved with the hyper-pure Ge detectors. This can be extended to $10^{-12}$ using neutron activation analysis (NAA). As SNO has demonstrated in background measurements on neutral current detectors, surface activities of 1/day/m$^2$ can be counted. Single-atom counting of Rn and other radio-isotopic gases has also been demonstrated. These capabilities complement those of the ultra-low-level hall: the hall provides unprecedented sensitivity at the cost of lower throughput.

An important aspect of the LLCF is the separate area provided for national security samples. The secure counting laboratory will provide strict access control, and will include electrical, mechanical, and personnel isolation consistent with classified operations. The counting capabilities will be similar to those of the open facility, including high-purity Ge counters; alpha, beta, and Rn counters; and shields. The area is designed for approximately half the throughput of the open area. We expect the MEGA detector (18 high-purity Ge detectors currently being built at the Pacific Northwest National Laboratory, with NNSA support) to be sited in the secure counting area. Sample preparation and DAQ/analysis rooms are included, so that the secure area is selfcontained. When not in full use, unclassified samples could be counted in the secure area.

The LLCF will also provide two passive shields (2-3 m$^3$ inner volume) for measuring internal backgrounds in prototype detectors. Measurements with a similar shield at Gran Sasso have demonstrated remarkably low backgrounds, ~ 1 mHz/liter in scintillator with a threshold of 50 keV. In our survey many experimentalists wanted access to a well-shielded environment for detector development. The two LLCF shields complement the three reentrant ports in the ultra-low-level counting hall, which provide a smaller, better-shielded environment for long-term determinations of backgrounds in detectors that do not require frequent access (access would be disruptive to other detectors operating in the pool). The LLCF shields will provide good shielding of gammas and neutrons while providing experimentalists with access: they can be opened and closed on demand. Separate DAQ rooms are provided for the electronics associated with detectors placed in the shields.

These activities require specialized utilities and/or process support, including radon-free nitrogen gas and liquid nitrogen, radon-free air, ultra-pure water, and ultra-pure liquid scintillator. Due to SNO and Borexino, the required techniques are well developed. The capacity will be sufficient to meet the needs not only of the LLCF, but other 7400-ft level experiments needing ultra-pure gases and liquids. Specifically, 400 m$^3$/h of Rn-free air will be produced, which is sufficient (at one room change per hour of air) to provide Rn-free air to the deck rooms of the ultra-low-level counting laboratory, the leaching/emanation laboratory, the shields of the low-level counting laboratory, the sample preparation rooms of the LLCF, and the electroforming room. The capacity is sufficient to encapsulate four dark-matter experiments, for example, with a shell of Rn-free air while personnel work on the detectors (assuming experimental volumes of 500 m$^3$). Ultrapure liquid scintillator could be provided (and repurified) at a rate of about 2 m$^3$/d. This is sufficient to meet the needs of the ultra-low-level counting hall, including repurification about once a week. Excess capacity of about 1 m$^3$/d would be available to the experimental halls. Rn-free liquid nitrogen and nitrogen gas will be produced in sufficient quantities to meet the needs of the LLCF and several detectors in the 7400-ft halls. (It would not be sufficient for large-volume cryogenic detectors like HERON or Icarus, however.) Finally, ultra-pure water is used in the water shields and in the pools of the ultra-low-level counting facility. The LLCF will have a capacity of about 65 tons/d and a purity of $10^{-14}$—$10^{-15}$ g/g U and Th. This production rate corresponds to three weeks for filling one of the pools in the ultra-low-level counting hall. As water repurification will consume about 20% of the system's operating time, there is sufficient excess capacity to produce and repurify 5-10 ktons/y of ultra-pure water for use in the experimental halls.

Other capabilities offered in the utility area of the LLCF are electroforming of Cu (and possibly other metals), machining of the electroformed Cu, and a detector construction area. The electroforming area is isolated and exhausts directly into the ventilation drift, for safety reasons. The electroforming systems would be provided with Rn-free air to ensure that Rn daughters are not incorporated into materials. The procedures have been worked out with the Majorana collaboration. Production capacity is expected to be several kg/week. The machining area will include a clean lathe, mill, and small tools, and will be dedicated to electroformed materials, with specially selected tool bits, etc., to ensure cleanliness and radiopurity. An assembly area (4m × 8m × 4m high) is provided on the second floor of the utilities room.

Supporting surface facilities include a chemistry laboratory to handle Neutron Activation Analysis (NAA) samples, as irradiated samples will not be allowed directly into the LLCF because of contamination worries. Additional needed surface chemistry capabilities include optical emission spectroscopy, chromatography, inductively coupled plasma mass spectrometry (ICPMS), and standard chemical analysis. Clean rooms will be provided for high-sensitivity measurements, such as those with ICPMS. A surface contamination laboratory is needed for optical and x-ray fluorescence measurements of surfaceswipes from underground experiments. None of these facilities requires an underground location. The surface chemistry laboratory is an essential complement to the LLCF.

Space for detector development – important to many NUSEL activities – is also crucial to the LLCF, providing a dedicated area for developing new counting techniques that will later be incorporated into the LLCF. This is an essential part of the proposed facility: we envision the LLCF as an evolving, dynamic facility that not only serves the community, but also makes innovative advances in experimental technologies. In this way the diagnostic capabilities of NUSEL will progress as new experiments make increased demands.

We have budgeted two full-time staff members for the LLCF. We anticipate that the staff will in fact be larger because the counting specialists will share their time between the LLCF and experiments needing counting expertise. This would be helpful in keeping the LLCF responsive to the needs of the experimental community.

***I.8 Costs and schedules for underground development:*** The detailed costing for this Reference Design Project Book have been completed for underground developments, not yet including costs for finishing the experimental halls or for instrumenting the low-leveling counting facility, as noted above. Thus the spreadsheets given in the appendices will be updated when these engineering studies are completed.

The master schedule spreadsheets and summary of costs are the synthesis of the engineering plans given in the various appendices. These cover the specifics of the upgrades on the Ross shaft/hoist, the No. 6 shaft/Winze, and No. 4 shaft/Winze, and the Yates shaft/hoist; pump column and ventilation conversion; and the various underground excavations, including rough excavation, shotcreting, concrete floors, and rock bolting.

The chart following Fig. E.16 summarizes the schedule implicit in the master sheets. The schedule assumes an FY06 start.

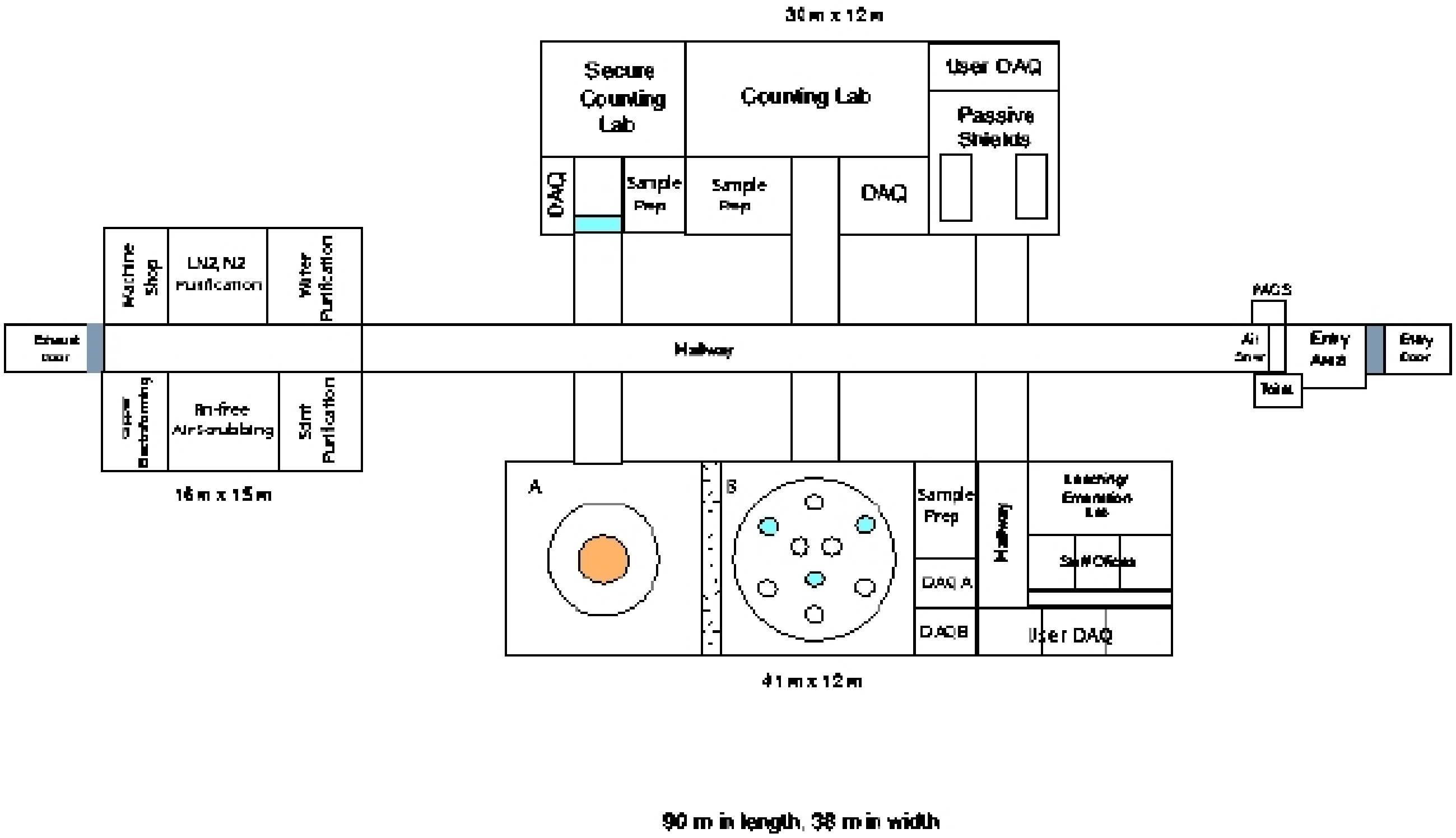

Figure E.16: One of two configurations now under consideration for the low-level counting facility.

## ANTICIPATED UNDERGROUND CONSTRUCTION SCHEDULE

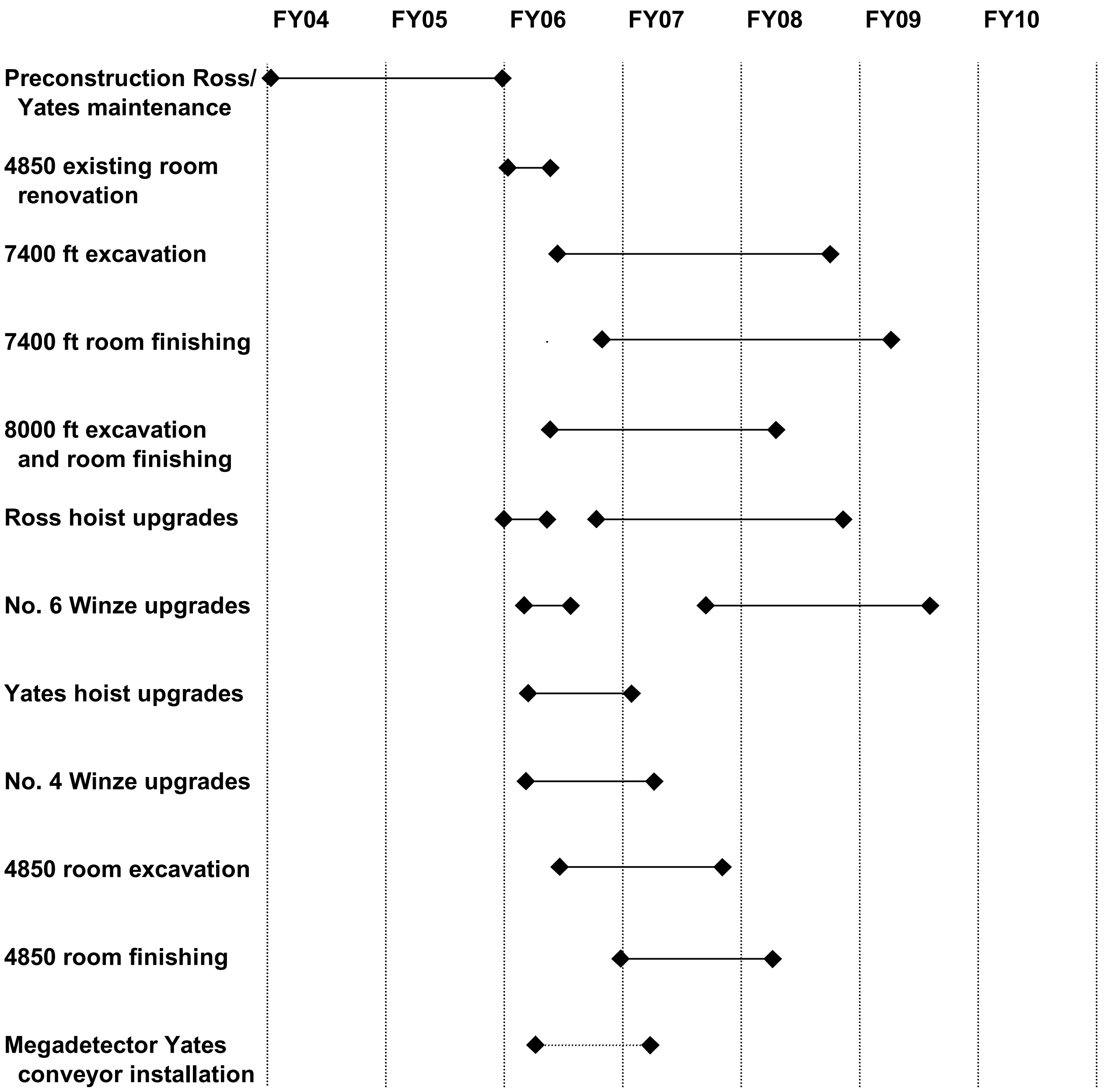

**II.Facilities Development Plan: Surface Campus**.

***II.1 General issues:*** As discussed above, NUSEL underground activities will be centered on the Ross and Yates shafts. Early excavation will be done through the Ross, then terminated. Future excavations (including the megadetector) will be done through the Yates. Mine maintenance activities will be carried out through the excavation shafts. This proposal advocates a Gran Sasso-caliber surface laboratory building, so that NUSEL-Homestake can host major experimental activities and collaborations on site. (This departs from our conceptual proposal where we advocated a five-story underground office building approximately equal in square footage to the surface science campus. We no longer believe this is a cost effective strategy.) Ideally that building would be placed near the scientific shaft (the Ross shaft). It is anticipated that the site will include a Visitor Experience Center as well as a museum/archive. These efforts require coordination and partnership with the city of Lead and the state. Several office building and other usable structures occupy parts of the site – preserved at our request.

Discussions with Homestake/Barrick on proposed site boundaries have occurred, though no resolution is possible before an agreement for transferring the site is obtained. We would ask for only a fraction of the available surface land (though all of the subsurface workings). Some of the important surface site boundary issues include:

- NUSEL would like to control all shaft accesses to the subsurface. This includes the Yates, Ross, Oro Hondo, B&M, and No. 5 shafts, and the portal entrances for the Kirk Power Tunnel, the Kirk Powder Tunnel, and the Ellison Fan Tunnel.
- It is quite likely that some future experiments, e.g., ones using cryogenic detectors, will require additional ventilation shafts with surface access. Thus it would be desirable for NUSEL to own all surface areas immediately above the usable underground workings.
- NUSEL underground excavations will be for civil purposes, not mining. It is not yet clear to us what this implies about the surface rights required about these excavations. While the simplest solution would be to contain all such excavation within the site boundary (e.g., the point made above), if this is not possible, some transfer of rights might be needed.
- Experiences of existing national laboratories demonstrate, over laboratory lifetimes, that surface space often becomes confining. Thus a surface footprint larger than that needed for the initial program would be prudent.

The conclusion is that the surface footprint should include all land directly about the underground workings, with the southern boundary extending south to the No. 5 shaft. Two isolated areas that would be desirable to acquire are the Ross parking lot and an area around the B&M shaft. The proposed boundary is shown in the Figure. Note that the northern boundary is constrained by the town and by the area that formerly housed the mill, while the western boundary is constrained by the town. Thus these boundaries, apart from possible minor adjustments, are fixed. The southern boundary extends beyond Whitewood Creek in order to include are surface areas above the underground drifts. Within this site, steep slopes and other topographic challenges render much of the requested land of limited use.

This proposal excludes the Open Cut, the proposed dump for waste rock from the megadetector excavation. Thus the agreement with Barrick must provide permission for this use. Other items we have marked for future discussions with Barrick include the inclusion/exclusion of the Yates rock dump, waste facility, and Mill reservoir within/from the NUSEL boundary; the inclusion of private or leased buildings within the property boundary; and possible steps to enhance public access to the site. These issues will require detailed discussions between Barrick, the new landlord (the state of South Dakota), and the collaboration at the time a site agreement is reached.

***II.2 NUSEL boundary and development possibilities:*** Below we present a series of maps of the Homestake complex, with notes on building sites and other issues. These are referenced several times in the WBS narrative.

Figure E.17: This large-scale map of the Homestake site and surrounding areas gives the proposed NUSEL boundary that was under discussion with the Homestake Corporation at the time it merged with Barrick Gold Corporation. Since that time, site discussions have been on hold while site transfer issues are being negotiated. In the first figure three areas are outlined in red, corresponding to the larger-scale Figures E.14-16.

Figure E.18: (Marked NUSEL Figure #1) This figure shows the area around the Ross head frame that would be the first choice site for developing the main surface science laboratory. A parcel covering approximately two acres is available at the top of Mill Street near the Ross head frame. This site could be prepared for building by relocating or demolishing some homes, and by cutting and grading this hill site to create a level building site. The rock and soil from the excavation could be used to fill a nearby depression, creating an additional level building site for other facilities or parking. The existing Ross parking lot is shown along with the extension that could be created by regrading. Several of the buildings in the Ross development will be retained for NUSEL. For example, the Ross dry facility has underground access to the Ross head frame and could easily be converted to additional laboratory or office space.

Figure E.19: (Marked NUSEL Figure #2) This figure shows an area between the Ross and Yates head frames and very near the open cut – the site of the original Homestake claim. Immediately next to the open cut is the site of the Homestake visitor center, which runs surface tours much of the year, and its parking lot. There is a substantial area outlined in green, outside the NUSEL property boundary, that could be used for the NUSEL Outreach Center. This location, a 1.5 acre site adjacent to Highway 85 and Mill Street, has good exposure along the highway, ample parking, and is located near the road that will serve the main Ross facility. The site is currently vacant. Placing the NUSEL Visitor Experience Center near the Homestake visitor center would likely be viewed very favorably by the Lead development officials. Also outlined in green is an additional six acres across from Mill Street that is available for new facilities (NUSEL or commercial) and for parking. Before such use demolition of some existing structures and grading would be required.

Figure E.20: (Marked NUSEL Figure #3) This figure shows the area around the Yates complex, which will serve as the main mining hoist for NUSEL. The large area outlined in green was the former Homestake mill area, which we propose leaving outside the NUSEL boundary. All buildings in this area have been demolished during the Barrick reclamation. There is a second area in green, located within the NUSEL boundary, occupied by homes that have been purchased by Barrick. Very near the Yates head frame is the mine office and the Yates dry facility. The mine office building is very serviceable. It will likely be used initially as temporary office space, and it and the Yates dry facility might prove valuable as an assembly area during megadetector assembly, particularly if the Yates is the hoist used by the experiments during construction. (We discussed elsewhere the possibility that the Yates would be adapted to this use, giving this experiment a dedicated hoist.) The foundry building is in good shape and is interesting architecturally. The renovation of this building for the museum/archive has been discussed.

***II.3 NUSEL surface facilities: renovations.*** Discussions were conducted with the Homestake Corporation about the extent to which existing buildings might be preserved and adapted to NUSEL needs. However, the surface campus is the least well developed part of the current Reference Design because, in the absence of a site agreement, we were unable to proceed to the next step, an inspection by a civil engineer. That is, we do not have a professional assessment of the state of repair and potential renovation costs of existing buildings. **Thus the Baseline Definition must include a reassessment of this section based on such an engineering inspection: this reassessment could in turn impact the new space we will build.**

NUSEL activities will be concentrated in two areas, around the Ross and Yates complexes depicted in Figs. E.18 and E.20, respectively. Most science operations will be carried out in the Ross complex and most excavation activities will focus on the Yates. The Ross complex will likely undergo major renovations, including the addition of a new surface laboratory. The changes in the Yates complex will be less extensive, as its use will continue to be mining. As we have noted in connection with Figure E.19, additional areas will be used for public outreach, and we anticipate that the operations of the outreach areas will involve state and local partnerships.

There are many potentially serviceable buildings that Barrick intends to make available to NUSEL. As noted above, none of these has been fully evaluated, nor have anticipated maintenance costs been estimated (often a key factor is deciding between renovation and new construction). Renovation would include new HVAC systems for all needed buildings along with structural reconfiguration, as needed. Depending on the future cost analyses, it is possible that existing buildings could meet many (and perhaps most) NUSEL needs. A partial list of existing structures potentially important to NUSEL includes:

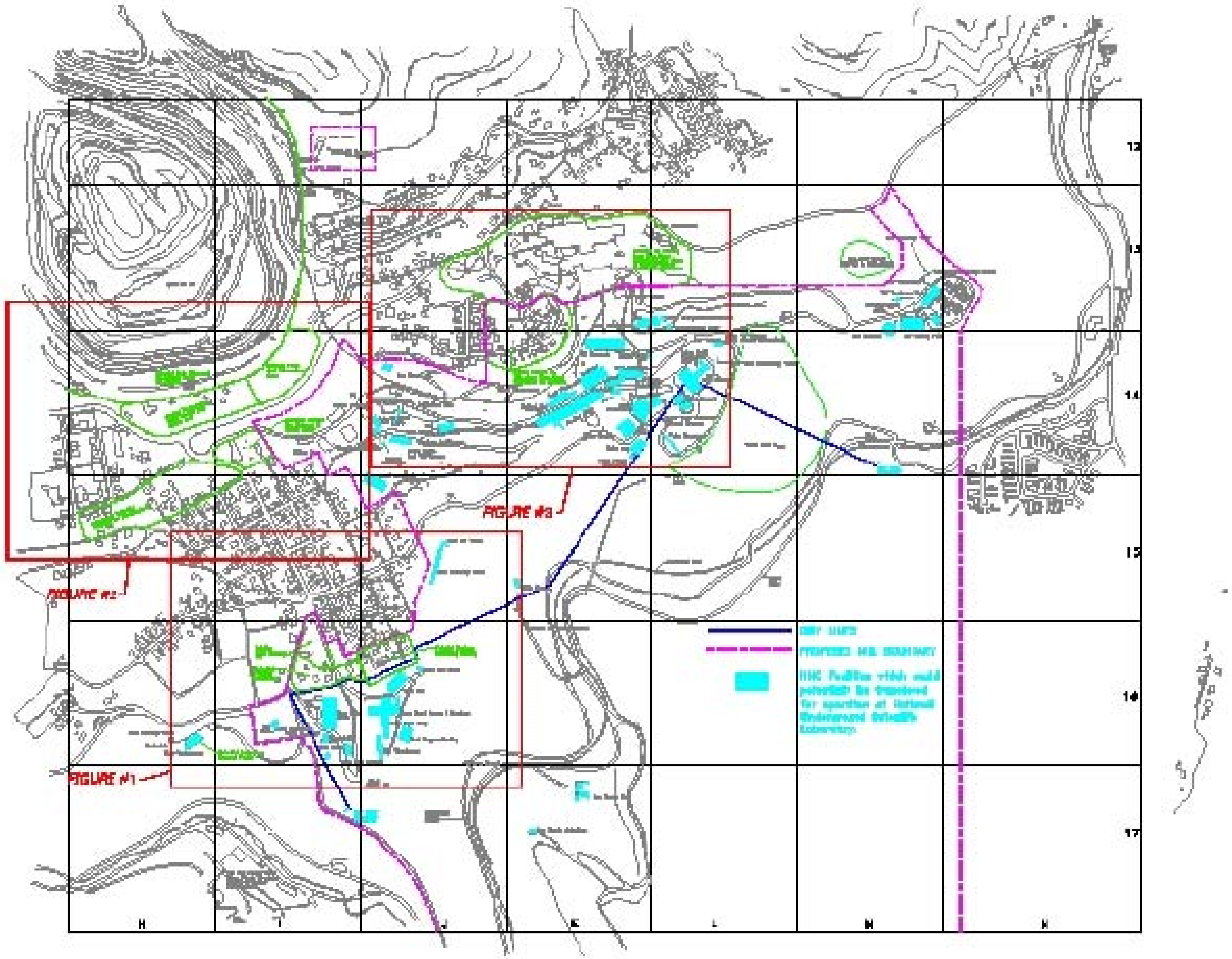

# NATIONAL UNDERGROUND SCIENCE AND ENGINEERING LABORATORY SITE PLAN

05-08-03

1"=1000'

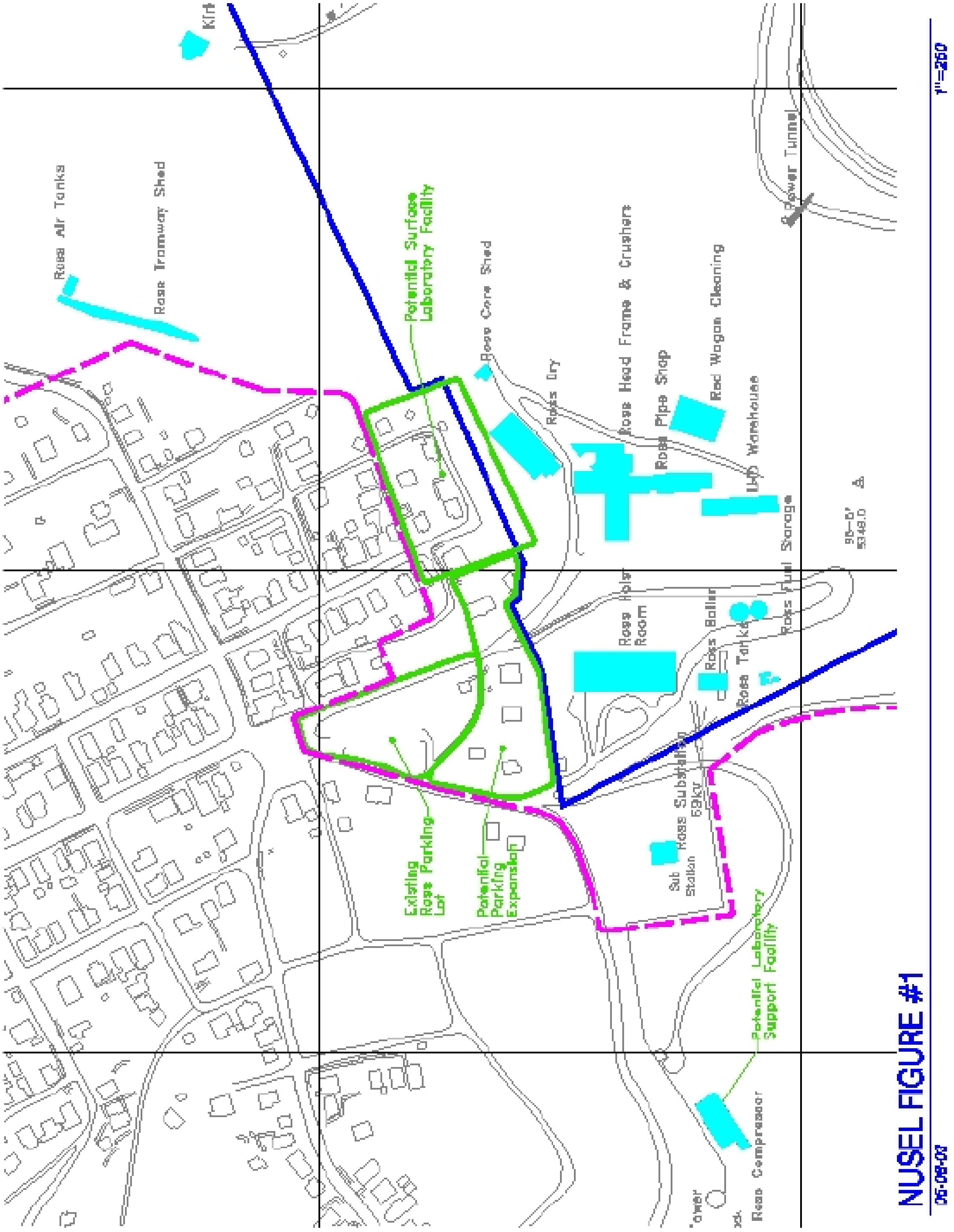

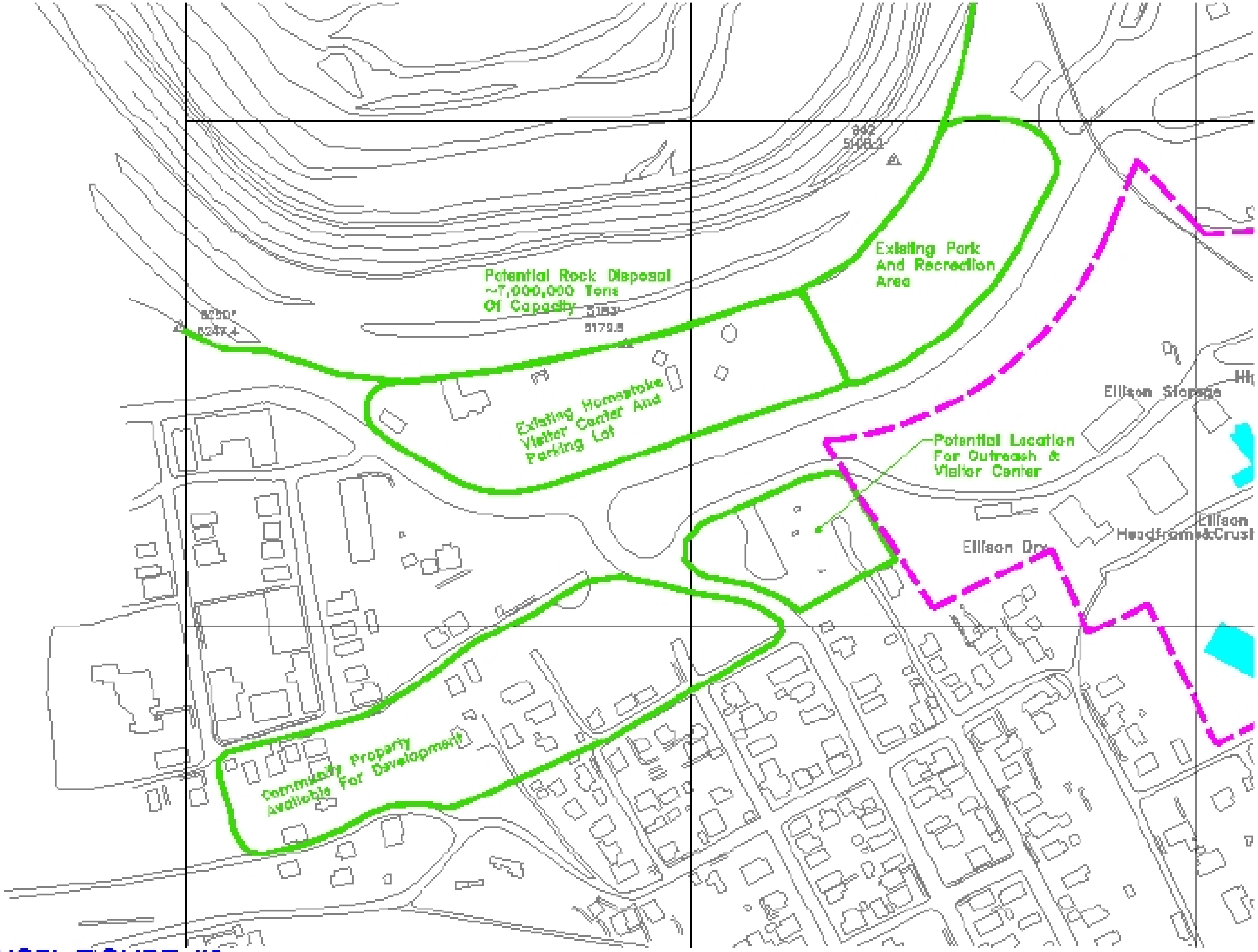

**NUSEL FIGURE #2**

05-08-03 1"=250'

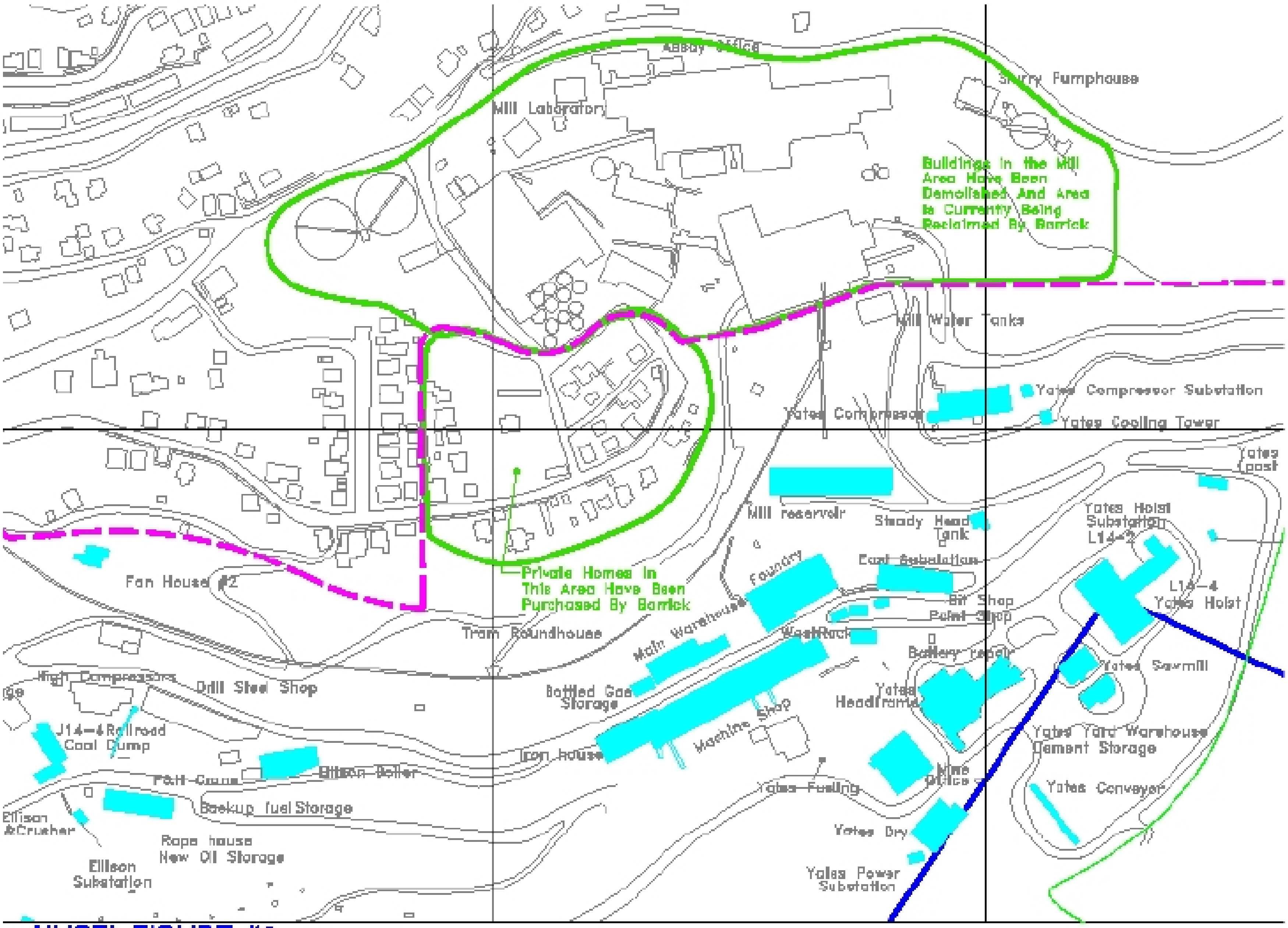

NUSEL FIGURE #3

05-08-03

1"=250'

- Ross headframe and crushers: required for underground operations and rock handling during initial excavations; after the Ross becomes the scientific access, the crushers will be removed.
- Ross hoist room: required for underground operations.
- Ross dry facility: potentially convertible to additional laboratory or office space – has underground access to the Ross head frame.
- Ross LHD warehouse: usable as a facility maintenance warehouse.
- Ross core shed: candidate site for a NUSEL operations office.
- Ross boiler: used to heat all buildings at the Ross complex.
- Ross electrical substation.
- Ross compressor: most of the underground civil excavation will utilize skid-mounted compressors. A dedicated compressor for science activities will be installed at the subsurface facility. Thus the Ross compressor facility can be utilized for other purposes. As it has an overhead crane, a warehouse is one potential use.
- Yates head frame and crushers: required for underground operations.
- Yates hoist room: required for underground operations
- Yates dry facility: potentially convertible to additional laboratory or office space – has underground access to the Yates head frame.
- Yates east electrical substation.
- Mine office (Yates): usable as initial office space; could be used as assembly or office space during megadetector construction.
- Foundry: candidate building for museum/archive. Brick construction requiring major renovations.
- Yates machine shop: candidate building for museum/archive or other outreach. Wooden building requiring major renovations.
- Oro Hondo fan: will be adapted to NUSEL ventilation needs.
- Oro Hondo electrical substation.

The Ross pipe shop is also very convenient to the Ross head frame. We expect to demolish this building and replace it with a new receiving and assembly facility. This new facility would incorporate clean assembly space, convenient to the Ross, the science access to the underground campus.

***II.4 NUSEL surface facilities for science.*** The discussion above establishes the existence of convenient, buildable sites on the proposed campus and of a variety of existing structures that could prove useful to NUSEL. This section deals with the surface campus needs of NUSEL, which we have based on Gran Sasso and similar laboratories. Gran Sasso is an excellent model for NUSEL-Homestake because this laboratory was built in a region of Italy without strong research universities. Because of this, and because the INFN wanted Gran Sasso to serve as an international center for underground science, the laboratory included from the outset significant infrastructure to support visiting scientists. Generally Gran Sasso gets very high marks from its user community for the support it provides. The present proposers feel that the Gran Sasso model is the correct one to follow in the US, particularly as the Homestake site is in a region of the country where the existing scientific infrastructure is somewhat underdeveloped.

Below we describe the procedure by which we estimated the facilities and space the upper campus should provide. The needs are better defined than the cost, we believe. The lack of site access is a general source of uncertainty in defining costs for the surface campus (and is the reason we have employed a 40% contingency for this development in the WBS). Until we can access the site to inspect existing buildings and construction sites, costs will remain somewhat uncertain.

While our group used Gran Sasso, the Sudbury Neutrino Observatory, and KamLAND as benchmarks for infrastructure, space, and manpower needs, there are obvious assumptions that must be made about the size of the NUSEL-Homestake facility to scale to these other projects. Key factors are the number of simultaneous experiments and the average number of scientists on campus – we will return to these assumptions at the end of this section, when our projections for Homestake are made.

Our information on Gran Sasso came from the Bahcall Committee Technical Subcommittee site visit, presentations at NESS02, and several public presentations by Gran Sasso Director Sandro Bettini. Gran Sasso maintains a web site that provides considerable detail on space and buildings.
Information on SNO and KamLAND comes directly from the collaborations, as several of the scientists belong to the NUSEL-Homestake Executive Committee.

Using Gran Sasso as a model, but taking into account certain infrastructure needs of SNO and KamLAND, we conclude that the following functions would be highly desirable and perhaps essential for a national facility:

- Surface and underground machine shops. (Note our underground laboratory plans include both a clean machine shop and an outside machine shop, with the latter located in the existing shop area on the 7400-ft level.) The underground facilities will help scientists avoid unnecessary trips to the surface for machining. But the surface shop will be the main facility, critical to detector module assembly and other preparatory activities.
- Surface electronics laboratory.
- Surface and underground assembly facilities, including clean room facilities.
- Plating, glass, and high vacuum technology shop.
- Stores and receiving.
- Chemistry facilities.
- Surface support for the low-background counting facility, as described elsewhere in this document.
- Cryogenics support.
- Computing and data acquisition support facility.

While Gran Sasso is generally praised for its support and organization, the most frequent complaint we heard was inadequate underground support, which then costs time by requiring travel back to the surface laboratory**.** (The Gran Sasso surface laboratory is about a 10 km drive from the underground halls.) This complaint is common to most of underground laboratories. For example, a simple mechanical adjustment to a steel member typically involves a three-day delay at SNO: 1) discovery of the problem and shipment to the surface; 2) locating machine facilities and alteration of the component; and 3) cleaning, shipment underground, and installation. **An issue flagged for the Baseline Definition is the adequacy of the underground support we have provided in this proposal.** Note that chemistry and copper electroforming facilities are included in the Low Level Counting Facility (to be discussed below). We want to reexamine the need for improved underground electronics capabilities.

In addition to the specific facilities mentioned above, NUSEL administration would provide the usual functions of safety, security, library services, shipping and receiving, and visiting scientist liaison and support functions. Some of these activities are discussed in the project management section (in connection with the NUSEL offices that will shoulder these responsibilities).

What is the size of the community NUSEL will serve? Gran Sasso supports between 450 and 600 scientists each year: the lower number is the visitor count, while the higher corresponds to all of the collaborators listed on experiments that are underway. On a typical day there are approximately 200 individuals on site, including contractors. (This number is consistent with the number of lunches the cafeteria serves each day.) There are on the order of 10,000 visitors/year to Gran Sasso, with associated tours and public outreach being a significant focus for the local administration. We expect the NUSEL outreach program to be at least a factor of ten larger.

The Gran Sasso surface facilities total approximately 110,000 gross square feet (gsf). The laboratory has had plans to expand by about 60%, as current facilities are oversubscribed, but these plans are on hold due to recent environmental issues that have affected laboratory operations.

The space estimates we have made are summarized in the table below. The gsf/asf is the approximate gross/assigned square feet ratio appropriate for the designated types of construction. The Gran Sasso numbers are gsf.

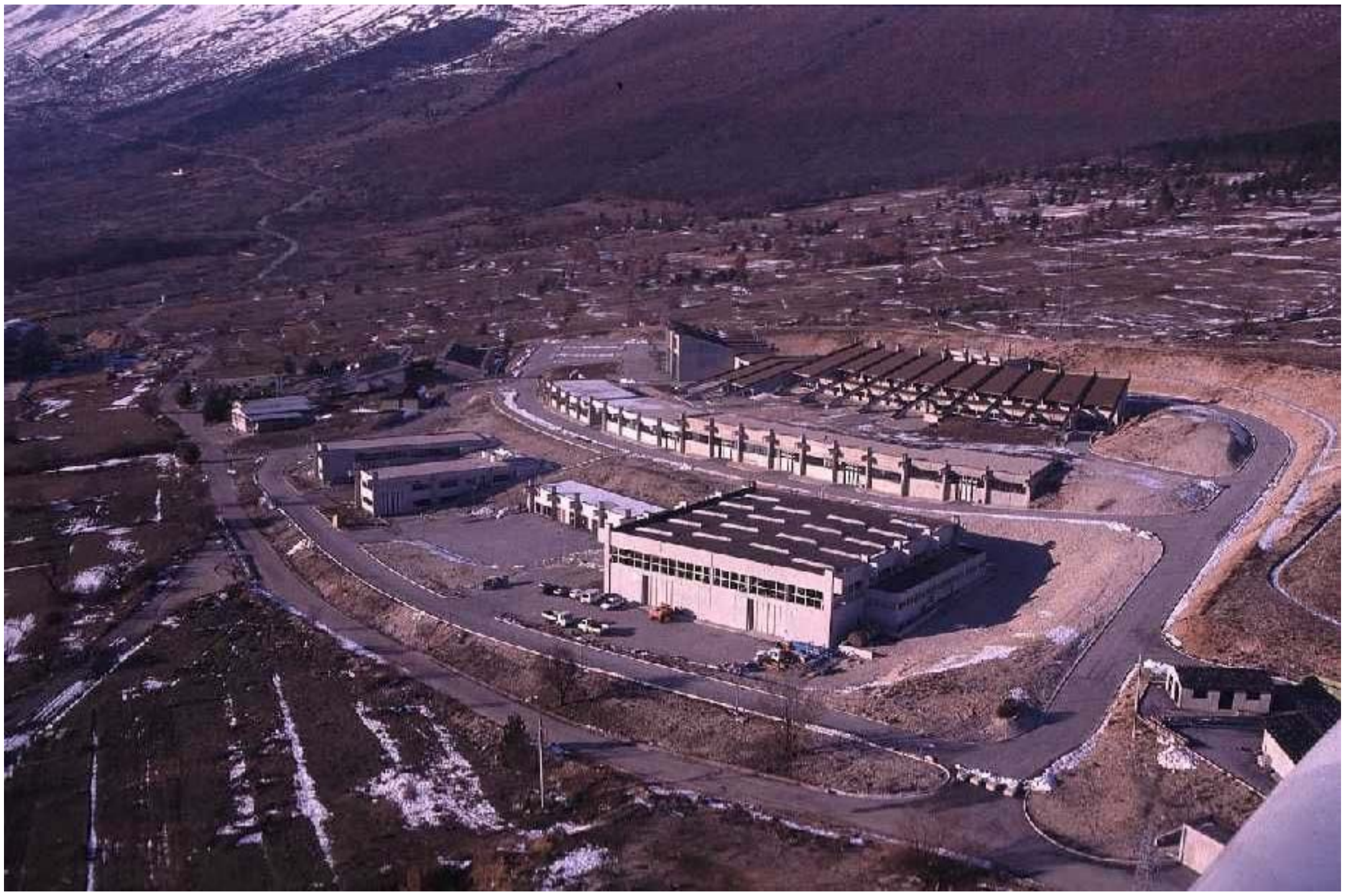

Figure E.17: Aerial view of the Gran Sasso facilities.

| Use | Assumptions | gsf/asf | gsf | Gran Sasso (est) |
|---|---|---|---|---|
| Visiting scientist office space | 100 asf/visitor; 100 visitor spaces | 2.0 | 20,000 | 12,000 (building 3 story addition, too) |
| Scientist and postdoc office space | 30 scientists; 160 asf/scientist | 2.0 | 9,600 | 8,500 |
| Administrative and support staff space | 60 personnel; 130 asf/staff member | 2.0 | 15,600 | 22,200 |
| Auditorium, meeting rooms, library, cafeteria | 14,500 asf | 1.6 | 23,200 | 21,700 |
| Labs; stores; chem/glass/electronics labs; computing | 18,000 asf | 2.0 | 36,000 | 28,500 |
| Assembly, receiving, staging, materials storage | 45,000 asf | 1.25 | 56,250 | 17,000 (GS says their existing space much too small) |
| *Subtotal science* | | | 160,650 | 109,900 |
| Visitor Experience Center | 40,000 asf | 2.0 | 80,000 | |
| Museum/archive | 10,000 asf | 2.0 | (20,000) | |
| *Subtotal outreach* | | | 80,000 + (20,000) | |
| **Total** | | | 240,650+ (20,000) | |

Table: Projected space needs for NUSEL-Homestake and estimated Gran Sasso usage.

NUSEL-Homestake would provide 40 shared visitor offices, assigned flexibly to collaborations doing work on sight, and 30 offices for the permanent scientific staff and postdocs. We have planned for 15 scientists and 10 postdocs, estimating that perhaps two-thirds of these will be in physics and the remainder in earth science/engineering. The estimated auditorium, meeting rooms, library, and cafeteria space matches closely that of Gran Sasso. Gran Sasso's large auditorium accommodates approximately 200, allowing the laboratory to host mid-sized conferences like TAUP (the IUPAP meeting in underground science). (The city of Lead is currently rebuilding its historic opera house, an off-campus venue that could be used by the laboratory for larger conferences.) Gran Sasso has a few small seminar and conference rooms, and about 10,000 gsf of cafeteria space. The NUSEL-Homestake estimate of special laboratory space, electrical/chemical/glass laboratories, stores, and computing/data acquisition labs is somewhat larger than Gran Sasso's, but reflects the earth science needs and the special chemistry facilities that will be needed to support the advanced NUSEL low-level counting effort. NUSEL detector assembly, staging, and receiving space is much greater than that provided by Gran Sasso. However, we anticipate, given the scale of some of the experiments NUSEL will tackle and the distances collaborations will travel, that a great deal of assembly will be done on site. Gran Sasso's modest assembly space is also partially attributable to the horizontal access, which makes assembly at a remote location and transportation by truck to the laboratory more convenient. We consider the 56,000 gsf estimate to be a realistic estimate for a facility like Homestake. There is a good possibility that much of the detector assembly space will come from renovated Homestake buildings. But our inability at the present time to assess the quality of the available space makes it difficult to guarantee that this is the case. **In the WBS we assume all space is new, to guard against underestimating costs.**

***II.5 NUSEL surface facilities for outreach.*** The Table also provides an estimate of 40,000 asf (80,000 gsf) for the space needed for the Visitor Experience Center described in the Science Book. This space includes exhibit space, flexible "classroom-type" space, workspace for the staff, auditorium space for multimedia, and the center's store. The space estimate is based on a projected visitor load that we anticipate could grow to 400,000/year, with the summer peak tourist season accounting for more than half of the total. Space/visitor ratios for existing centers vary somewhat. For example, the Museum of Science and Technology in Tampa provides 210,000 gsf for 645,000 visitors/year; the new addition of the McDonald Science Center of just 12,000 gsf is planned to allow visitor growth from the current 130,000/year to 250,000/year. A science center with a size and focus similar to that planned for Homestake (a mix of tourists and school groups) is the Lawrence Hall of Science in Berkeley, which handles 250,000 visitors/year in a facility that one administrator estimates to be 60,000 gsf. The 80,000 gsf used in this proposal is typical of centers with capacities up to 400,000 visitors/year.

The table also shows 20,000 gsf provided for the museum/archive. This construction is not currently funded in this proposal (which is why this item is set off by parentheses in the Table). Early discussion between the Homestake Corporation and us focused on private arrangements that might finance renovation of space for this facility and subsequent operations. The foundry building is one possible building of the requisite size. The archive function of the museum/archive is important to NUSEL earth science because Homestake maps and mining records will be valuable in site characterization: the archive has an important science function.

**III.Operations.** This section deals primarily with the personnel required for NUSEL operations. Most nonpersonnel operations costs are addressed in the Work Breakdown Structure, not here.

***III.1 Site operations and maintenance.*** Facility operations costs at Homestake are very well known because of the long history of successful mining operations.

The facility operating costs divide into personnel and other costs, as shown in the tables. The first 15 operationsstaff positions are technical, supervisorial, or administrative. The remaining workers operate the hoists, shafts, crusher, yard/warehouse, and construction. The effect of the hoist modernization described in I will be to reduce operator needs by more than a factor of two, from 48 to 22 FTEs, by year five. The needs in year five are six hoist operators, 12 shaft maintenance workers, two yard/warehouse workers, and two mine construction workers. The resulting year-five FTEs and associated personnel costs for operations are 37 FTEs and $2.34M (fully loaded). The detailed spreadsheets giving the costs over the five-year grant period are presented in the appendix. **These personnel are under the Associate Director for Construction, Facilities, and Operations,** as described in the Management Plan.

| *Operations personnel* | *Year-five FTEs required* |
| --- | --- |
| Operations/maintenance manager | 1 |
| Safety/environment manager | 1 |
| Administrative manager | 1 |
| Mine engineer/geologist | 2 |
| Environmental engineer | 1 |
| Safety technician | 2 |
| Surveyor/engineering technician | 1 |
| Operations/shaft supervisors | 2 |
| Secretary/clerical | 2 |
| Accounting/purchasing | 2 |
| Shaft maintenance staff | 12 |
| Hoist operators | 6 |
| Yard/warehouse staff | 2 |
| Mine construction staff | 2 |
| ***Total*** | ***37*** |

The other operating costs are based on very recent operating experience at Homestake. In general these costs may somewhat overestimate future operating costs because the NUSEL footprint will be reduced. The table shows these costs total $2.479M/y, yielding a total year-five operations cost of $4.82M. The personnel site operations and administration spreadsheets included in the appendix show how the personnel portion of these costs evolves as the hoist modernization is implemented.

| *Operations activity* | *Overhead and fixed (annual)* |
| --- | --- |
| Haulage | $150,000 |
| Other mine operating | $147,000 |
| Mine general | $279,000 |
| Ventilation/cooling | $691,000 |
| Hoists/shafts | $507,000 |
| Waste water | $88,000 |
| Administration/general | $617,000 |
| ***Total*** | ***$2,479,000*** |

| *Site maintenance activity* | *Required FTEs* |
| --- | --- |
| Mechanical plant engineer | 1 |
| Electrical planner/supervisor | 2 |
| Field maintenance supervisor | 1 |
| Maintenance planner | 1 |
| Surface electrician | 1 |
| Underground electrician | 3 |
| Maintenance technician | 4 |
| Field maintenance | 4 |
| Mine maintenance (ground control/track) | 4 |
| ***Total*** | ***21*** |

As was done in Homestake operations, mine safety will be handled by a crew selected from the workers listed above, and especially trained. Homestake's crew has won many national awards.

In year five, site maintenance requires 21 FTEs, as listed in the table. The year-five loaded costs are $1.34M/y. The detailed spreadsheets are presented in the appendix**. These personnel will also be the under the Associate Director for Construction, Facilities, and Operations**.

***III.2 Science operations.*** Science operations needs were evaluated by examining what has been needed at Gran Sasso, KamLAND, and SNO, with adjustments to reflect the special requirements or goals of NUSEL-Homestake. The basic information is summarized in the table:

| ***Category*** | ***NUSEL FTE needs*** | ***Baseline FTEs*** | ***Comments*** |
|---|---|---|---|
| Director (0.7), Assoc. Directors (2) | 2.7 | | |
| Senior, associate, and assistant scientists | 12 | 10 (Gran Sasso) | Homestake includes EarthLab effort |
| Postdocs | 10 | 30 (Gran Sasso) | |
| Graduate students | 10 | ? | |
| Undergraduates | 5 | ? | |
| ***Subtotal: scientists*** | ***40*** | | |
| Librarian/publication | 2 | 2 (Gran Sasso) | |
| Machine shop | 3 | 3 (Gran Sasso) | |
| Undergrd. machinist | 1 | | New to NUSEL |
| Draftsman | 1 | 1 (Gran Sasso) | |
| Electronics | 3 | 1 (SNO) | |
| Glass | 1 | | |
| Low-level counting | 2 | 1 (SNO) | |
| Chemistry | 5 | 5 (SNO) | |
| Computing | 5 | 5 (SNO) | |
| Receiving | 2 | 1 (SNO) | |
| Administration | 4 | 30 (Gran Sasso) | Gran Sasso number is total lab administration |
| ***Subtotal: support*** | ***29*** | | |

***III.3 Detector operations.*** The following personnel are required for detector operations.

| ***Category*** | ***NUSEL FTE needs*** | ***Baseline FTEs*** | ***Comments*** |
|---|---|---|---|
| Assistant manager | 1 | | |
| Large assembly/staging | 3 | 1 (Gran Sasso warehouse) | Vertical access → additional needs |
| Underground assembly/staging | 3 | 3 (SNO carwash cleaners) | |
| Transportation | 1 | 0.5 (SNO) | |
| Maintenance machinist | 2 | 0.5 (SNO) | |
| Professional operators | 3 | 2 (SNO) | |
| Mechanical engineer | 1 | 1 (SNO) | |
| Process engineer | 1 | | |
| Secretary/clerical | 1 | | |
| ***Total*** | ***16*** | | |

**Science and detector personnel are under the Associate Director for Research.**

***III.4 Administrative operations.*** Administrative operations describe those administrative activities within the **Administrative Office** (see the project management organizational chart) within the Director's office.

| *Category* | *NUSEL FTE needs* |
|---|---|
| Administrative Office director | 1 |
| Personnel/benefits officer | 1 |
| Financial officer | 1 |
| Visitor liaison officer | 1 |
| Public information officer | 1 |
| Secretary/clerical | 4 |
| Security guards | 8 |
| ***Total*** | ***17*** |

This office will also oversee various activities that are handled by outside contracts, including the cafeteria, custodial services, underground (MSHA) and other safety training, and possibly administering Lead housing for visitors. This office would also support the Associate Director for Construction, Facilities, and Operations in outside contract activities, including contracts for mining and for facilities construction.

***III.5 Outreach/education operations.*** Outreach and education are also positioned as a **separate office within the NUSEL directorship**, which will allow the Director to take a personal role in supporting this office and its director. This could be helpful in the formation of collaborative outreach and education agreements with South Dakota, regional, and national organizations.

| *Category* | *NUSEL FTE needs* |
|---|---|
| Laboratory Director | 0.2 |
| Director, Education/Outreach Office | 1 |
| Manager, K-12 Education | 1 |
| Manager, Visitor Experience Center | 1 |
| Manager, Computing and Networking | 1 |
| Display design staff | 3 |
| K-12 education staff | 3 |
| Visitor experience center staff | 6 |
| Web and interactive display staff | 4 |
| Secretarial/administrative | 2 |
| ***Total*** | ***22.2*** |

***III.6 Management operations.*** Our working assumption is that a nationally recognized nonprofit university management group, such as URA, UCAR, or SURA, will manage NUSEL-Homestake. (AUI is another possibility.) This group will maintain an office on site to work with the laboratory Director's office and with the NUSEL landlord (the state of South Dakota). It will be responsible for organizing periodic construction and operations reviews, for guaranteeing that all agency reporting requirements are satisfactorily addressed, and for addressing state reporting requirements (particularly those associated with land stewardship). Minimum personnel needs are:

| *Category* | *Management FTE needs* |
|---|---|
| Laboratory director | 0.1 |
| Site Office director | 1 |
| Site office adm. assistant | 1 |
| Off-site administration | 0.5 |
| ***Total*** | ***2.5*** |

**IV. Site Development:** A small staff for site development – science and education/outreach – will reside within the Director's office. This group will also work closely with Lead and other regional officials on issues such as site access, traffic impacts, etc., that affect NUSEL science or outreach.

| *Category* | *NUSEL Science FTEs* | *NUSEL Outreach FTEs* |
|---|---|---|
| Manager | 0.8 | 0.2 |
| Assistants | 1.6 | 0.4 |
| Secretarial/clerical | 0.8 | 0.2 |
| ***Total*** | ***3.2*** | ***0.8*** |

**V. Proposed Management Plan*:*** The collaboration looks forward to working with the National Science Foundation in formulating a management plan for NUSEL-Homestake. To start this process, the scientists in our collaboration offer a plan we believe has considerable merit.

*General structure and overall principles:* We believe that a recognized, stable, nationally representative management body is important to the success of this project:

- The underground science community has not previously united behind a national initiative of this sort. The current competition for NUSEL has produced four proposals. We believe a nationally prominent, inclusive management body could help unite the community behind the winning proposal. The current proposers would be comfortable stepping aside, in favor of such an organization.
- NUSEL will be a challenging project to manage successfully. While submitted as an MRE, the laboratory probably will operate for decades. As the NRC Quarks and the Cosmos report stressed, the NSF and DOE will each support major NUSEL projects. NUSEL will also host international projects, important national security work, and industry scientists. A management group with experience in dealing with a spectrum of funding sources is needed. The project is unique in combining two rather distinct disciplines, physics and earth science.
- An agreement with the site owners, Barrick Gold, is essential. Barrick has agreed to donate the site, but has placed conditions on the transfer, some connected with potential liabilities. The state of South Dakota has made a major commitment by agreeing to accept ownership of the needed portions of the mine (the underground workings and certain surface areas and buildings). Despite these steps, a final agreement is not yet in hand. We believe an experienced management group could help the state and the NSF put into place the insurance and trust fund guarantees that would address Barrick's concerns about government use of its property, while also assuring South Dakota that science uses of the mine will not adversely affect its new property. An established management group would address a key concern in the current negotiations, how to guarantee the longevity of any agreement.

A cooperative agreement between the NSF and an experienced nonprofit management group would provide the stability and experience necessary for NUSEL's success. There are several groups that might satisfy the requirements, including AUI, AURA, and URA. We concluded that each of these organizations is capable of providing high-level management for NUSEL, but found that URA offers special advantages:

- URA has a 38-year record of successful management. Its broad charter to "…acquire, plan, construct, and operate machines, laboratories, and other facilities, under contract with the Government of the United States or otherwise, for research, development, and education in the physical and biological sciences… and to educate and train technical, research, and student personnel in said sciences" encompasses NUSEL's activities. Recently URA has become increasingly involved with particle astrophysics and astronomical sciences relevant to NUSEL, through participation in the Sloan Digital Sky Survey, the Pierre Auger Cosmic Ray Observatory Project, and the dark matter experiment CDMS II at the Soudan Underground Laboratory.
- Almost all of the institutions prominent in NUSEL physics, including those interested in other NUSEL proposals (e.g., UC Irvine, the University of Minnesota, and New Mexico State University), can be found among URA's 90 US member universities. Thus URA is naturally positioned to unify the community behind the proposal NSF selects.

- URA includes member universities from three other nations, Canada, Italy, and Japan. These nations are host to the three most prominent underground laboratories, SNO, Gran Sasso, and Kamioka. (Also, the NUSEL-Homestake collaboration includes researchers from Japan (Osaka University) and Canada (University of Toronto). The University of Toronto, a URA member, has close ties to Barrick Gold.) Thus URA addresses the collaboration's concerns for strong relations with laboratories and scientists outside the US.
- URA will create a separate governing board to oversee NUSEL.
- A major goal of our proposal is the stimulation of new research and education endeavors in South Dakota and neighboring Northern Great Plains states. URA has expressed its desire to help in these efforts. URA will welcome regional research universities as new URA associate members. The management plan calls for South Dakota and regional universities to have a significant presence on URA's NUSEL Board of Overseers. The Board of Overseers' Environment, Safety, and Health Committee would meet regularly with the cognizant state officials, so that the state can verify that all aspects of the site's use agreement are respected. The management plan calls for state and regional organizations to play the leading roles in education and outreach, proposing Memoranda of Understanding to delegate these responsibilities. Finally, URA will be supportive of NUSEL plans to partner with regional universities, through NUSEL's university associates, in efforts to enhance research and education.

The Organizational Chart below shows a possible management structure. Below we describe its basic elements:

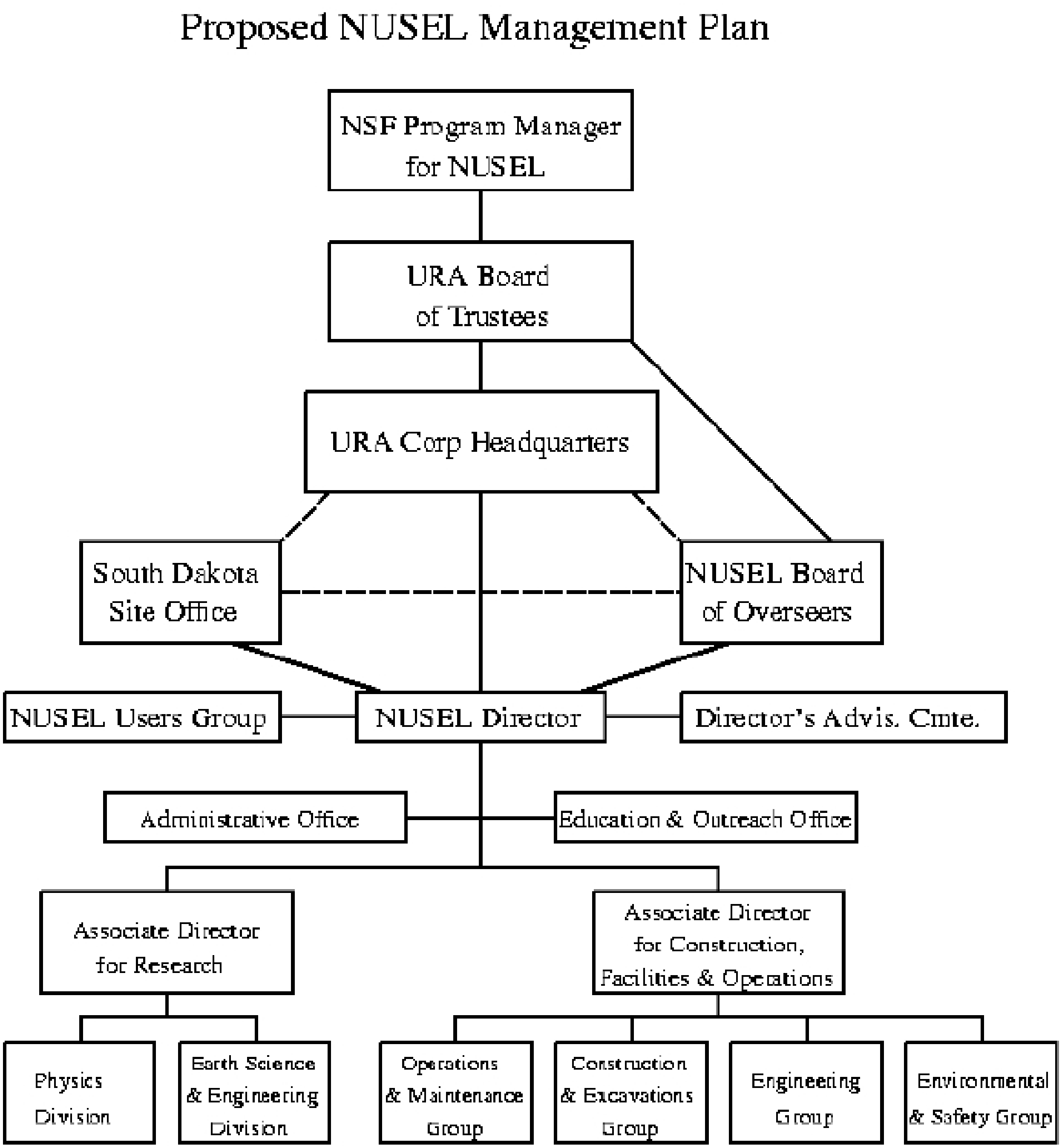

Figure E.21: A possible management plan for NUSEL-Homestake.

- NSF would administer NUSEL through a cooperative agreement with URA. URA has suggested an internal management scheme similar to that for FermiLab with an independent Board of Overseers. The Board structure would include Science; Administrative and Audit; and Environment, Safety, and Health Committees, as well as external and visiting committees for program review. The Board will appoint the NUSEL Director, with NSF concurrence, after conducting an international search. The Board would also schedule periodic project management ("Lehman"-style) reviews of NUSEL during and after construction, to ensure that the project remains on schedule and on budget. If a management agreement were put in place soon after site selection, URA would like to be active in helping the current collaboration join with others in the underground science community in completing the final proposal update.

  Given the breadth of NUSEL science and the need for both science and large-project advice, an appropriate size for the Board of Overseers would be about 15. Of these, URA would reserve four places for regional scientists, two from South Dakota and two from the neighboring EPSCoR states. The remaining appointments would also take into account geographic balance and any need for international representation. In addition, URA would welcome research universities from the Northern Great Plains into the Association. (Among South Dakota and its EPSCoR neighbors, the University of Nebraska is currently the one URA member institution.)

  The University of Washington submitted the NUSEL-Homestake proposal. In this update the UW has designated URA as the performing institution. If URA were designated by the NSF as the managing entity for NUSEL, the UW intends to turn over all responsibilities to URA. It would cease to have any role other than as one of the member universities of URA.

  URA will create a local office in South Dakota. It will provide any needed assistance to the Director during NUSEL's first year, when Laboratory budget and personnel procedures are being formulated.

- We assume South Dakota will act as custodian of the Homestake site, designated the Site Office in the organizational chart. URA and the State of South Dakota, in full consultation with the NSF, will negotiate a mutually satisfactory site use agreement. It will work with the NSF and other agencies supporting NUSEL science projects to put into place use and decommissioning agreements addressing all state concerns connected with the scientific use of the site. The state Site Office and/or Department of Environment and Natural Resources will consult closely with the URA Board's Environment, Safety, and Health Committee. URA and the State of South Dakota would agree on the level of funding needed to support the State's Site Office, and that funding would be provided to to the State through the site agreement.

- URA, after an international search and with the concurrence of the NSF, will appoint the NUSEL Director. The Director will be responsible for all day-to-day decisions affecting NUSEL operations. TheDirector will appoint an outside, independent Advisory Committee, representing the broad community that NUSEL serves (including university, national laboratory, and international scientists). As with the URA Board of Overseers, four members of the Advisory Committee will be chosen from South Dakota and regional EPSCoR institutions. The NSF, URA, and Director, by mutual agreement, will choose the initial group of Committee members. Members will serve three-year staggered terms. The Committee is intended as an independent community body that provides advice to and otherwise assists the Director.

  A NUSEL's Users Group will be formed to provide advice to the Director on user needs. As in other national laboratories, the Users Group will have its own governance and select its representatives. The chair of the Users Group will be an ex officio member of the Director's Advisory Committee.

- Two Offices and two Associate Directorships are positioned under the Director. The Administrative Office will be responsible for all personnel functions, purchasing and procurement, accounting, legal review, and public relations. Purchasing functions include competitive bid preparation, contract agreements, legal review, the documentation and accounting interface, inventory and supply management, and warehouse and yard supervision. This office will coordinate NUSEL shipping/receiving and visitor and user services. The office will assist the Director's Office with funding reporting requirements for the NSF and other agencies and for URA.

The Education and Outreach Office will operate the Visitor Experience Center and the Museum/Archive, and will direct all on-site programs connected with education and outreach, such as the REU and UROP efforts. As noted in the Science Book Education and Outreach chapter, our South Dakota and regional partners would like to play the lead role in education and outreach activities involving South Dakota and its EPSCoR neighbors. We envision these responsibilities being transferred to appropriate state and regional organizations through Memoranda of Understanding incorporated into the Cooperative Agreement. For example, we noted previously that the South Dakota Space Grant Consortium includes most of the partners we have identified as important to outreach and education in South Dakota: its research and teaching universities, the tribal colleges, affiliates in industry and government, and community Science Centers. The NUSEL Education and Outreach Office would then be responsible for coordinating on-site activities with those of the South Dakota Space Grant Consortium (or whatever organization shoulders these responsibilities). Similarly, it could coordinate regional outreach with the organization holding those responsibilities. The Education and Outreach Office would also work closely with the city of Lead and with state tourism officials, so that traffic, services, and other visitor requirements are adequately addressed.

The Administrative Office and the Education and Outreach Office are both viewed as extensions of the Director's Office, and thus directly accountable to him/her.

The Associate Director for Research will oversee all research activities of the laboratory, and thus will be accountable for the wise use of laboratory resources in support of the science program. He/she will be responsible for building NUSEL's permanent scientific staff and for attracting a strong user community to the laboratory. We envision the Directorate being divided into two Divisions, Physics and Earth Science/Engineering, with separate program advisory committees. All technical services and personnel that primarily support science – the chemistry, glass, and machine shops, the low-level counting facility, the NUSEL library, computing service, etc. – will be under the Associate Director for Research. This office also handles visitor and scientific staff office and laboratory space needs in the surface campus.

The Associate Director for Construction, Facilities, and Operations is the individual immediately under the Director that is accountable for on-time and on-budget execution of the construction plan, as well as the operations of the resulting facilities. The structure of the Directorate follows that used successfully at Homestake for many years. As many of the Directorate's employees will be former Homestake engineers, operators, and geologists, this will allow NUSEL to begin with a well-tested operations structure. The tasks assigned to the four operations groups (Operations/Maintenance, Underground Excavations/Construction, Engineering, Safety/Environmental) are described in detail in the Overview section of this Project Book. The employees will be very experienced in underground operations, but will need to shoulder additional, unfamiliar tasks associated with science operations, and thus will undergo additional training. For example, the Operations/Maintenance Group would help with transport of experimental equipment between the surface laboratory and the underground halls; the Engineering Group would be responsible for scheduling of facilities used in transport of experimental equipment and in oversight of transport and underground installation; and the Safety/Environmental Group would be responsible for the safe handling of scientific materials like cryogens, compressed gases, and hazardous liquids.

*Other project management issues:* There are several points relevant to project management during the current interim period:

- The Homestake Collaboration has an interim structure that includes a PI (Haxton, Washington) and eleven other members of the Executive Committee (Balantekin, Wisconsin; Bowles, Los Alamos; Conrad, Columbia; Farwell, South Dakota School of Mines & Technology; Lande, Pennsylvania; Lesko, Lawrence Berkeley; Marciano, Brookhaven; Marshak, Minnesota; Onstott, Princeton; Shaevitz, Columbia; Wilkerson, Washington.) The Collaboration has a set of bylaws and is open, with a posted procedure for including new members. It is divided into interest groups, which have shouldered responsibilities for preparing Science Book materials, for recommending hall sizes, utility and cleanliness requirements, etc. The recent NRC Barish panel report noted the strong leadership behind the NUSEL effort.

- The Executive Committee has frequent conference calls and meets several times a year. On several occasions meetings have been scheduled in Lead, to encourage local supporters of NUSEL. The full collaboration meets during major conferences. The next such meeting is schedule for September (TAUP 2003). The PI recently began monthly newsletters to the collaboration with the goal of improving information flow. The newsletter and all other NUSEL-Homestake information is kept on the collaboration's web site, http://int.phys.washington.edy/nusl/.
- We envision this organization giving way to a Director and Users Group some time during the next phase of the project, in which the Baseline Design Project Book is created. The Director will be the new PI. The process would be expedited by early decisions by NSF on site selection and project approval.

## F. WORK BREAKDOWN STRUCTURE: BUDGET AND EXPLANATION

The Work Breakdown Structure presented here describes the costs and contingencies associated with laboratory construction and operations. It is divided into three sections:

WBS-1: Laboratory Science Construction and Operations
- Property
- Insurance
- Underground Development for Science
- Surface Development for Science
- Site Operations and Maintenance
- Science Operations
- Detector Operations
- Director's Operations

WBS-2: Laboratory Education and Outreach Construction and Operations
- Property
- Insurance
- Underground (near-surface) Facility
- Surface Development for Education and Outreach
- Education and Outreach Operations
- Director's Operations

WBS-3: Laboratory Management
- Management Site Office
- Landlord (South Dakota) Site Office

These WBS spreadsheets are supported by two other sets of documents:
- The WBS explanations, which set the stage for going beyond the Reference Design to the Baseline Definition by explaining the costing procedures here, and the factors that limit the current Reference Design. These limitations are almost exclusive connected with the absence of a site agreement and property transfer, and the resulting lack of access necessary for some of the detailed planning.
- The appendices, which present far more detailed spreadsheets for all of the planned underground work (the most important cost item), for site operations and maintenance, and for science, outreach, detector, and administrative operations.

An effort has been made to assign reasonable contingencies. The scheme adopted is:
- 50% contingencies have assigned to items like property issues (acquisition, environmental, permitting, insurance) where substantial uncertainties will exist until a site agreement is finalized and a site boundary defined.
- 50% contingency was used for sealing unused areas: we lack an adequate survey of the sealing needs.
- 40% contingencies have been used for surface construction and for underground systems. We have good benchmarks for these costs from other laboratory developments, but a 40% contingency is prudent until we have a design specific to Homestake.
- 40% contingency has been used for underground shaft and hoist upgrades. Here we have detailed engineering estimates that justify a smaller contingency, but the flooding now occurring (see Section G) will affect at least one shaft/hoist system needed by NUSEL, the No. 6 Winze. Thus a 40% contingency was considered prudent.
- 40% contingency was used for surface demolition
- 40% contingency was used for certain non-personnel operating costs and for contract services (e.g., custodial and cafeteria) for which estimates specific to Homestake cannot be made at this time.
- 40% contingency was used for South Dakota site office costs, while 25% contingency was used for the management group.
- 25% contingency has been used for underground rough construction (excavation, rock bolting, shotcreting). Detailed engineering estimates of these costs have been made and are presented in the appendices.

- 25% contingency has been used for administrative personnel costs, as there is some uncertainty in the adequacy of staffing proposed.
- 10% contingency was used for personnel areas where we felt it was feasible to assign a definite manpower level. As discussed in the previous section, we made such assignments by using existing laboratories as models, generally choosing as our baseline the laboratory providing the best level of service (as we believe service needs of future experiments will be greater, not lesser).

Note that our conceptual proposal, submitted in June 2001, used a contingency of 25%.

The total cost of the project at the Reference Design stage is $331M, an increase over the conceptual proposal of approximately $50M. This results from the following combination of factors:

- Inclusion of a much-enhanced low-level-counting facility, adding about $16M.
- Inclusion of a waste conveyor for disposing of rock into the open cut. We feel it is important to establish this facility at the outset, as it is essential to future megadetector construction. Early installation will, in the end, lower rock disposal costs, including those associated with the developments of this proposal. The cost is about $7.3M.
- More cautious use of contingencies.

During the Reference Design effort we found several items that had been overlooked earlier, when we had a less complete understanding of the site. But our increased site knowledge also led to a far more efficient design, one that avoided the costly Yates shaft extension and major drift construction. There was a net savings in this process.

Because sites are often compared, we wanted to characterize Homestake costs, distinguishing costs that are site-dependent (and thus form a basis for selecting a site) from those that are less site dependent (and thus would be comparable, if each site provides the same level of service):

| *Category of expenditure* | *Homestake cost* | *Dependent on site?* |
|---|---|---|
| Access (cost of going deep) | $44.3M | Very |
| Site operations/maintenance | $44.0M (5 years) | Very |
| Property | $1.8M | Very |
| Underground systems | $45.8M | Somewhat |
| Waste rock/coring/sealing | $11.6M | Somewhat |
| Underground room excavation | $37.6M | Somewhat-weakly |
| Surface development | $55.4M | Weakly |
| Science operations+LLCF | $35.2M (5-years) | Weakly |
| Director's operations | $11.8M (5 years) | Weakly |
| Detector operations | $4.3M (5 years) | Weakly |
| *Total Science* | *$291.7M* | |
| *Total outreach/education* | *$30.5M* | Weakly |
| *Total management* | *$8.9M* | Somewhat |

The year-five site operations/maintenance costs for Homestake are $8.2M/y.

## WBS-1: LABORATORY SCIENCE CONSTRUCTION AND OPERATIONS

### 1 Property

1.1 ***Property acquisition***: The "landlord" agreement between Barrick Gold and the state of South Dakota remains under discussion. When resolved, the laboratory must verify that the agreement meets its legal needs; negotiate a use agreement with the state of South Dakota; resolve ownership issues for remaining Barrick buildings and for buildings the Laboratory constructs; address issues associated with private ownership or leases on buildings within the laboratory boundary; negotiate shared use agreements with Barrick for roads, fences, and other facilities that cross site boundary lines; and understand responsibility for and disposition of properties at the time the laboratory ceases to operate. An estimate of $500,000 is given to cover the associated legal fees.

1.2 ***Environmental assessment and mitigation***: The laboratory will conduct an environmental assessment of the property, to make sure that all laboratory and funding agency environmental requirements can be met.

1.3 ***Easements***: Easement issues include access roads and utilities, including civil (e.g., drainage), mechanical (gas, industrial water, potable water), and electrical. For electrical power, Homestake Mine is currently a "primary customer" which means that it owns, operates and controls the existing on-site power distribution required for all mine facilities. All land anticipated to be required for future on-site overland or underground transmission is currently owned by Barrick Gold Corporation. Transference of the land on which the power distribution network exists will effectively transfer any easements that may be required to support a future laboratory. This transference will be negotiated with the owner.

1.4 ***Permits*:** Permit issues include waste rock disposal, water discharge, and underground construction.

### 2 Insurance and General Conditions

2.1 ***Insurance and trust fund***: The current South Dakota-Barrick Gold negotiations, if concluded successfully, will require the state to shoulder certain insurance obligations to relieve Barrick Gold of potential liabilities associated with the donation of its lands to public use. The amount of that insurance is not yet fixed. We enter no cost, instead electing to discuss this issue in Section G.

2.2 ***General Conditions:*** Often within a WBS General Conditions include typical construction requirements such as scheduling and supervision; performance and payment bonds; and temporary site controls and utilities. Elsewhere in this WBS we have included a contractor's markup of 15% as well as EDIA, costs for engineering, development, inspection, and administration.

Homestake will present special risks to the General Contractor because, beginning June 2003, water has been allowed to infiltrate the mine in an uncontrolled manner and ventilation has been or will be lost to lower portions of the mine. The General Contractor will be responsible for inspecting and certifying the existing infrastructure prior to reuse. The risk involved in the inspections will increase the cost of contractor's risk insurance.

Our philosophy in doing this WBS, however, has been to sequester issues connected with the consequences of flooding in Section G. The reason is that the duration and thus the severity of the effects are difficult to assess at this time: we do not know when the state will achieve a property agreement with Barrick, and whether that agreement will allow immediate intervention.
Thus we have not included General Conditions costs in this WBS, but do discuss the financial consequences (qualitatively) in Section G.

### 3 Underground development

3.1 ***Shaft and hoist upgrades***: The current proposal assumes that the South Dakota-Barrick Gold agreement will allow the laboratory to use much of the existing underground infrastructure: the hoists, if upgraded and properly maintained, will serve the laboratory well for many decades. A detail plan of upgrades for scientific access through the Ross and No. 6 Winze are presented in the appendices. A similar plan for modernizing and maintaining the Yates and No. 4 Winze as mining access is also detailed in the appendices. Other aspects of the hoist/shaft modernization for the Ross/No. 6 scientific access are discussed below.

**3.1.1** ***No. 6 Winze:*** The existing normal power systems for the #6 Winze hoist room and development areas and the shaft stations will be reused.

For redundancy, a cross tie to a new and remote normal power tiebreaker will be added per the redundancy diagram.

A utility borehole will be created parallel to the #6 shaft. The section of the Yates 12KV system twin cables that are routed down the #6 Winze will be relocated to the #6 Winze utility borehole during the #6 shaft reconfiguration. The balance of power cabling down the #6 shaft will be repositioned and re-supported within the #6 shaft during the shaft reconfiguration.

Emergency power will be provided to the #6 hoist room and area/station including the new man hoist via the Ross and the Yates head frames emergency generators and the crosstie switchgear that will be located in the 4850 development. These redundant emergency power systems will be routed down the # 6 shaft and the #6 utility borehole to supply redundant and independently routed emergency power to the 7400 laboratory facilities and elsewhere as required.

Existing normal lighting will remain in place for the stations and adjacent facilities. Life safety lighting consisting of egress fixtures will be provided throughout the reconditioned stations and development areas. The life safety lighting will be powered from normal and emergency circuits.

The existing voice, data, and leaky feeder radio communications system from the surface to the existing #6 Winze hoist room will be utilized. These systems are also currently serving the existing #6 shaft stations and the #6 shop near the 7400 level.

The existing 24 pair fiber-optic cable is routed from the Yates shaft through the #6 Winze hoist room area to the existing 7700 #6 shop. This fiber cable will remain in service.

The #6 Winze borehole will be utilized to install a redundant 24 pair fiber optic cable from the #6 Winze hoist room area to the 7400 level laboratory chambers.

The existing 100 pair copper cable routed down the #6 Winze will remain in service. The Yates 4850 machine shop 200 pair voice cable will be extended across the 4850 drift to the #6 Winze and down the borehole to the 7400 levels laboratory chambers. This extension will make voice communications to 7400 redundant.

The Leaky Feeder radio system extends throughout the #6 shaft, shops, hoist area, and 4850 development areas.

Existing communications cabling in the #6 shaft will be repositioned and re-supported within the shaft during the shaft reconfiguration.

***3.1.2 Ross shaft****:* The existing normal power systems for the Ross head frame and the shaft stations will be reused. Loop redundancy will be accomplished with the tiebreaker that is being added per the No. 6 Winze scope.

A utility borehole will be created parallel to the Ross Shaft. The section of the Ross 12KV system twin cables that are routed down the Ross shaft will be relocated to the utility borehole during shaft reconfiguration. The balance of power cabling down the Ross shaft will be repositioned and re-supported within the shaft during the shaft reconfiguration.

Emergency power will be provided to the Ross headframe and area/station including the new man hoist via the Ross and the Yates surface emergency generators and the subsurface crosstie/switchgear that will be located in the 4850 development facilities.

These on-site and redundant emergency power systems will respectively be routed down the Yates shaft and the Ross utility borehole to supply redundant and independently routed emergency power to the 4850 development facilities. The systems will also be looped through the 4850 tiebreaker switchgear to backfeed each other per the power redundancy diagram.

Existing normal lighting will remain in place for the stations and adjacent facilities. Life safety lighting consisting of egress fixtures will be provided throughout the reconditioned stations and development areas. The life safety lighting will be powered from normal and emergency circuits.

The existing voice and leaky feeder radio communications system from the Yates headend equipment to the existing Ross head frame will be utilized.

A new 24 pair fiber-optic cable for data communications will be routed from the through the Ross utility borehole to the existing 4850 #6 Winze headframe area. In addition, a new 24 pair fiber-optic cable for data communications will be routed from the Yates surface facility to the Ross headframe area.

The existing 100 pair copper cable routed down the Ross shaft will remain in service.

The existing Leaky Feeder radio system extends throughout the Ross shaft, shops, and hoist area.

Existing communications cabling in the Ross shaft will be repositioned and re-supported within the shaft during the shaft reconfiguration.

3.2 ***7400 Facilities****:* The detailed plan, costs, and schedule of the 7400-level excavations described in the Facilities Development section are given in the appendices.

3.3 ***8000 Facilities***: The detailed plan, costs, and schedule of the 8000-level excavations for geomicrobiology (the microbiology room and drilling area) are given in the appendices.

3.4 ***4850 Facilities***: The detailed plan, costs, and schedule of the 4850-level excavations described in the Facilities Development section are given in the appendices.

3.5 ***Other excavations***: The other excavations, described in detail in the appendices, will improve the Ross/No. 6 connection (the transfer station for scientific loads going to the deep laboratory levels) and the drift connecting the No. 4 and No. 6 Winzes on the 7400-ft level. Other work focuses of the new primary ventilation circuits and excavation of boreholes for the pump column and electrical work

***3.5.3, 3.5.6, 3.5.7 Ventilation***: The Oro Hondo fan is currently located on top of the Oro Hondo shaft. Although this shaft should be adequate for the foreseeable future, it is not the best location for the fan long-term. (There is excessive sloughing of rock, which must be mucked out of the bottom of the shaft.) The Oro Hondo fan will be relocated to the concrete-lined No. 5 shaft, which should serve the subsurface needs for the expected life of the laboratory.

Two separate and distinct ventilation circuits will be provided. The Ross Ventilation Circuit will be utilized to ventilate and cool the laboratory facilities at the 7400 Level. Likewise, the Yates Ventilation Circuit will be utilized to provide ventilation and cooling to the 4850 Level facilities. Both circuits will merge at the #31 exhaust shaft and continue to the relocated Oro Hondo fan at the top of No. 5 shaft.

Air control doors will be installed in strategic locations to allow the airflow throughout the subsurface areas to be properly directed and controlled. Provisions have been made for the installation of approximately 200 air control doors.

Drifts not utilized for the subsurface facilities will have concrete bulkheads installed to minimize ventilation requirements. The bulkheads will be designed with "trapped" drain lines to minimize hydrostatic pressure buildup behind the bulkheads.

Approximately 240,000 cfm of ventilation air will be brought down the Ross shaft and No. 6 Winze, of which 20,000 cfm will be used to ventilate the pumping facilities at the 6950 level and 40,000 cfm will be used to ventilate the pumping facilities at the 8000 level. The Laboratory facilities at the 7400 level will receive 180,000 cfm, of which 30,000 cfm will be diverted for mining purposes. The balance of air will be utilized for ventilating and cooling of the laboratory facilities. From the 7400 level facility, the air will travel up No. 7 shaft, through drifts on the 6950 and 7100 levels to #31 exhaust where it will merge with the Yates Ventilation Circuit prior to being exhausted to the Oro Hondo fan at the top of No. 5 shaft.

Approximately 130,000 cfm of ventilation air will be brought down the Yates shaft for use in cooling and ventilating the 4850 level facilities. The air will continue from these facilities across the 4850 level, down No. 7 shaft and across the 7400 level to #31 exhaust. There it will merge with the Ross Ventilation Circuit prior to being exhausted to the Oro Hondo fan, at the top of No. 5 shaft.

3.6 ***Waste rock handling***: The appendices include a schedule of work and costs for installation of a conveyor system off the Yates shaft, so that waste rock going up the Yates can be transferred to the conveyor and moved underground to the Open Cut. This system, while costly, is essential to the construction of the megadetector. Building this system at the start of construction will also lead to lower rock disposal costs during construction of the 7400- and 4850-ft levels.

3.7 ***Core drilling****:* The core drilling of the 4850- and 7400-ft levels, coupled with laboratory testing, will provide the data the geotechnical modelers will need to verify the room designs, optimize their stability, etc.

3.8 ***Other****:* The engineering, mobilization, and demobilization costs have been calculated for the underground construction detailed in the appendices.

3.9 ***Lower campus systems***: This is a major cost with important uncertainties. The overall costs have been based on a scaling of the systems costs for SNO and similar recent projects, taking into account differences between those projects and Homestake. Lower campus systems includes the costs of bringing all utilities, communications, and safety systems to the laboratory areas. It also includes all of the finishing costs of the rooms not already accounted for in 3.2, 3.3, and 3.4. (As detailed in the appendices, the 3.2-4 costs include excavation, shotcreting, mineguard coating, concrete floors, track, and cabling.)

Thus lower campus systems includes all of the steps required to transform rough rooms into scientific laboratories, with the required ventilation and clean room engineering, with all utilities and process engineering, with safety systems, etc. Among similar projects, the one most similar to the Homestake 7400-ft level development is SNO and SNOLab: the collaboration provided its finishing costs. These included ventilation, chillers, overhead crane, emergency generators, radon removal systems, the liquid nitrogen plant, electrical and lighting systems, process piping, ultrapure water, fiber optics, certain chemistry capabilities, and civil engineering connected with painting, establishing clean conditions, and relocating and reconnecting facilities. The net cost for SNO is \$5.6M for 17,200 $m^3$ of space, or \$325/$m^3$ (US dollars), the figure we have used for the 7400-ft level.

A reasonable estimate for the cost savings on the 4850-ft level, where the needs are somewhat less severe and the access simpler, is \$250/$m^3$. In contrast, we use \$600/$m^3$ for the 8000-ft level because the limit volume being developed will necessarily increase the cost per cubic meter. Also, unfortunately, the electrical and other systems on the 8000-ft level are certain to be damaged by flooding (see Section G). (South Dakota is hopeful that flooding can be reversed before waters reach the 7400-ft level.)

We want to stress that our plans for lower campus systems are relatively mature, as can be seen from the description below. Unfortunately we have concluded that site uncertainties (particularly the possibility that the state agreement with Barrick might rule out our use of serviceable existing systems) make it unreasonable for us to cost these systems at this time. We stress that we are prepared to provide such costing in a very short time, once the conditions of the site transfer are known and scientific access restored.

***3.9.0 General Systems Issues***: Below we address some of the planned approaches to low-campus systems for Homestake:

*3.9.0.1 Power delivery to laboratory areas*

*3.9.0.1.1 General power*: Primary site power is supplied from the 69KV East (Yates) Substation, the 69KV Ross Substation and the 69KV Oro Hondo Substation. These primary-power substations are currently owned and operated by Barrick and are fed from Black Hills Power & Light's (BHP&L) Kirk switchyard located near the old Kirk power plant.

The Yates shaft and surrounding surface facilities are currently supplied from the East substation. The Ross shaft and surrounding surface facilities are currently supplied from the Ross Substation and the Oro Hondo Substation. The Oro Hondo Substation also supplies power to the mine ventilation Oro Hondo fans and surrounding surface facilities.

The Oro Hondo Substation will be removed from service because the Oro Hondo fans will be relocated to the No. 5 shaft mouth. To facilitate this, site modifications will include a new overland 69KV line from the BHP&L Kirk switchyard to the East substation. In addition, the Oro Hondo feeders to the Ross shaft 12 KV system will be removed and new feeders will be installed from the Ross Substation to the 300 level in the Ross Shaft. Some Oro Hondo transformers and gear will be relocated to the No. 5 shaft and some will be relocated to the Ross Substation. Very little 69KV Substation gear will be required because future loads will be very similar to the existing. The removal of this substation will reduce maintenance costs and provide additional on-site power reliability.

The majority of the existing site electrical infrastructure equipment that will remain in service for Laboratory operations has been upgraded within the last 10 years with state-of-the-art equipment.

*3.9.0.1.2 Normal power*: Significant subsurface electrical infrastructure will remain in place and be utilized for facility operations such as seepage pumping and miscellaneous services. The majority of the existing site electrical equipment that will remain in service for laboratory operations has been upgraded within the past ten years with state-of-the-art distribution and protection equipment. Most of the distribution system is at 12.4 kV with some existing 2.4 kV for miscellaneous pumping and infrastructure support services. Power cabling routings will be revised as outlined in the shaft reconfiguration sections.

The 4850 and 7400 development facilities will be provided normal power from two cross-tied switchgear lineups. An existing normal power switchgear lineup will be reconfigured to serve as one lineup and a new lineup with a tiebreaker will be installed at the other location. These lineups will be operating at 12KV and will be between 600-800 amperes in capacity.

Normal power for the laboratory chambers will be 480 Volt. The chambers will receive normal power from both of the normal power switchgear lineups. An automatic transfer switch located at the chamber main panel will be used to select the available source. The chamber main normal panel will distribute the normal power to non-essential loads within the chamber.

Infrastructure and utility loads located outside the laboratory chambers like fans and HVAC loads will be sourced from an 800 amp, 480V distribution panelboard located adjacent to each 12KV switchgear lineup.

The laboratories developed areas (mainly 4850 and 7400 levels) normal power will be fully redundant from either the Yates 12KV system or the Ross 12 KV system. The subsurface infrastructure will have more than two times the anticipated initial subsurface laboratory's normal power requirements facilitating relatively low-cost expansion capabilities.

*3.9.0.1.3 Emergency power*: The 4850 and 7400 development facilities will be provided emergency power from two cross-tied switchgear lineups. One lineup will be equipped with a tiebreaker. These lineups will be operating at 12KV and will be 200 amperes in capacity.

Emergency power for the laboratory chambers will be 480 Volt. The chambers will receive emergency power from both of the emergency power switchgear lineups. An automatic transfer switch located at the chamber main panel will be used to select the available source. The chamber main emergency panel will distribute the emergency power to critical and life safety loads within the chamber.

Infrastructure and utility loads located outside the laboratory chambers that are consider life safety (access lighting) will be sourced from a 400 amp, 480V distribution panelboard located adjacent to each 12KV switchgear lineup.

The laboratories developed areas (mainly 4850 and 7400 levels) emergency power will be fully redundant from either the Yates headframe emergency generator or the Ross headframe emergency generator.

Power reliability to the No. 6 Winze hoists will be increased with the addition of new emergency power feeds and transfer switches.

*3.9.0.2 Access lighting*

*3.9.0.2.1 Normal lighting*: Normal lighting in each access way will consist of a wet location wall-mounted fluorescent indirect lighting that will be installed intermittently along the drift as required to provide an minimum of 5 foot-candelas at the floor. Higher traffic areas will be equipped with additional lighting as appropriate for the task and/or traffic patterns. In addition, Low-bay metal halide fixtures will be utilized in developed areas like machine shops and utilities.

*3.9.0.2.2 Emergency lighting*: Emergency lighting each access way will consist of both egress and exit lighting. The exit lights will consist of battery-powered Lightpanel Technologies ¼ watt LightPanel fixtures (or equal) charged from the space lighting circuit. The egress lighting will be powered from the emergency generator life safety power system. The access emergency lighting shall be accomplished utilizing the normal lighting fixtures and equipping them with emergency ballast. The emergency fixtures shall be designed to provide sufficient lighting to satisfy egress lighting code minimums. Approximately 1/5 of the normal access way fixtures will be powered from the life safety system.

*3.9.0.3 Data and communications*

*3.9.0.3.1 General issues*: Homestake currently has state-of-the-art communications and data transmission systems throughout the Yates and Ross shafts.

The current systems consist of a copper cable system for voice communications, fiber optics for data transmission, and a leaky-feeder radio system. The leaky-feeder radio system will be utilized for operations and backup communications. Each communication media will be extended to 7,400 and looped through the 4,850 and 7400 drifts for redundancy.

*3.9.0.3.2 Voice*: The existing infrastructure provides a minimum of 100 pair of copper conductors throughout the proposed footprint for landline voice communications. As each phase of the project is performed, voice communications will be extended.

*3.9.0.3.3 Voice/Radio*: The Leaky Feeder radio system is just a few years old and is one of the most flexible and reliable mine-ready radio communication systems on the market. As each phase of the project is completed, the leaky-feeder radio system will be extended to each area of underground occupancy. The system includes a telephone interface, which provides for communications with any telephone on the surface.

*3.9.0.3.4 Data*: Data communication will consist of two infrastructures, internal and external. Both infrastructures currently have existing equipment and connectivity that will be expanded to help provide more redundancy and speed to the remote locations of the complex. As each internal phase of the project is completed, the fiber optics and copper communications will be extended. At the completion of the developments, complete redundant data communications will be achieved throughout the internal facilities hubs

The internal infrastructure will consist of data communication internal to the complex. Due to the nature of the facility, with distance and quality of environment being primary issues, a fiber backbone will be constructed to each of the participating data locations (chambers, hub shops, etc).

Strategically placing enterprise modular switches in several locations will be the key to handle the redundancy and constraints of length. The current fiber backbone consists of 24 pair 62.5 micron multimode run to existing primary locations. New installation of fiber per the communications diagram will also be required to extend the existing internal infrastructure to the new laboratories.

Primary hub and data exchange locations will be connected to the fiber backbone. These areas will include underground operations, new surface lab facilities, 7,400 level machine shops, detector chambers, the No. 6 Winze/Shop, and support shops as required.

Each location will contain a gigabit switch attached to the fiber backbone. These switches are designed for the enterprise network that will be required. The gigabit switches contain the highest bandwidth available on the market today. Features like ‘quality of service’ and ‘spanning-tree algorithm’ will automatically adjust routes due to congestion or failure, VLANs will help segment the network traffic and load balancing will allow for the most redundancy available.

The proposed external infrastructure teamed with the proposed internal infrastructure will ascertain the highest speeds and redundancy available. These features matched with possible future expansion available at the complex will allow the opportunity to obtain the highest quality of service.

*3.9.0.3.5 Public address system*: Subsurface PA systems shall provide paging coverage on the developed levels in the areas normally occupied including the lab chambers. As development is completed, the system will be extended. Two separate and localized systems will be installed; at the 4850 development and at the 7400 development.

*3.9.0.4 Monitoring and controls*

*3.9.0.4.1 Infrastructure controls*: The existing site infrastructure FIXDMACS control and monitoring systems include a tried and tested PLC based control and monitoring system. This system has four nodes and includes operator interfaces in the existing mine office and an underground operator interface.

FIXDMACS is used to monitor the entire mine, control pumps, fans, etc., including a carbon-monoxide sensor system. The system power is backed up with inverters for uninterruptible power and increased reliability. The system has approximately 50 process control screens and additional control interface templates and equipment status screens. The system provides vital process information to the facility operators for optimized control. It demonstrates high quality graphics of the name-brand operator interface software - Intellution.

As the process systems are extended, the controls will be reprogrammed and extended to monitor and control the subsurface equipment.

*3.9.0.4.2 Subsurface fire detection*: The lab structures will be equipped with a single new addressable fire detection and alarm system that is fiber linked to the main fire alarm panel at the surface. This fiber communication will eventually be accomplished through redundant paths. The main fire alarm panel at the surface will serve as the fire alarm command center for the entire facility.

The subsurface fire detection system will also include a remote fire alarm annunciator, door holder/closers, manual pull stations, smoke detection and audio/visual notification devices throughout the corridors, smoke detection and audio/visual notification in the labs and shops. As each Phase of the project adds facilities, the system will be extended to monitor the new areas.

*3.9.0.4.3 Subsurface gas detection*: The existing mining infrastructure monitors CO with several detectors located throughout the mine. These detectors will be relocated to the new ventilation paths as required and continue to report to the extended control system.

In addition, a multi-gas multi-channel gas monitoring system will be provided for the detector chambers areas. This system will report to the control system and the fire detection system command center.

*3.9.0.4.4 Subsurface security/access system*: The subsurface security/access system shall consist of a card reader system with card readers installed at all laboratory entrances and exit doors. In addition, access to the facility will be monitored by video cameras at each primary level hoisting station.

***3.91 4850 level chambers systems***

*3.9.1.1 Electrical*
*3.9.1.1.1 Normal power*: Normal power will be redundantly supplied to each chambers main distribution panel from the utility facilities tiebreaker gear.

The main distribution panel will be located at the entrance to the chamber. A typical choice is a 480/277-volt, three-phase, 200-ampere panel equipped with in-surge suppression. A larger main panel or additional distribution panels (with or without surge suppression) can be installed as required depending on the power requirements and the size/shape of the specific chamber. The chamber main panel will typically be utilized to power all chamber 480/277-volt loads. This will not include any substantial HVAC or facilities infrastructure loads.

The main panel voltage will be stepped down with localized transformers to supply 120 volts power panels (with surge suppression as may be required) to serve computer loads, ancillary 120 loads and convenience receptacles. These transformers will be located as dictated by the experiments and loads that they serve.

*3.9.1.1.2 Emergency power*: Emergency power will be redundantly supplied to each chambers emergency power panel from the utility facilities tiebreaker gear. Emergency power will be available within 10 seconds of a normal power failure. The generators integral fuel storage will be sized for minimum of 8 hours generation. Where necessary by load type, localized uninterruptible power supplies (UPS) will be utilized to provide continuous power to non-interruptible loads.

The emergency panel will be located adjacent to the main distribution panel at the entrance to the chamber. A typical choice is a 480/277-volt, three-phase, 100-ampere panel equipped with in-surge suppression. A larger emergency panel can be installed as required depending on the power requirements of the specific chamber.

Life safety and critical power circuit separation will originate at the chamber emergency panel. The *life safety* circuits will supply loads (typically fire detection and annunciation and lighting for safe entrance and egress) that are required to assure life safety. This will not typically include any HVAC or facilities infrastructure loads unless they are deemed necessary for preservation of life. The *critical* emergency circuits will supply loads (typically equipment and communications) that are not required to assure life safety, but are desired to maintain equipment, experiments, or information systems. This will not include any substantial HVAC or facilities infrastructure loads. The critical emergency loads will be selectively shed in a case of insufficient emergency power for life safety.

The emergency panel voltage will be stepped down with localized transformers to supply 120 volts power panels (with surge suppression as may be required) to serve critical or life safety loads. These transformers will be located as dictated by the experiments and loads that they serve.

*3.9.1.2 4850 level chambers lighting*
*3.9.1.2.1 Normal lighting*: Normal lighting in each chamber will consist of a combination of lighting solutions. Wet location wall-mounted fluorescent indirect lighting will be installed around the perimeter of the chamber, open-type suspended industrial high-bay metal halide will be utilized over the main chamber areas, and fluorescent task lighting over walkways, and desk spaces and in other specified areas. An average of 60 foot-candelas will be provided at 30 inches above the finished floor.

*3.9.1.2.2 Emergency lighting*: Emergency lighting in each chamber will consist of both egress and exit lighting. The exit lights will consist of battery-powered Lightpanel Technologies ¼ watt LightPanel fixtures (or equal) charged from the space lighting circuit. The egress lighting will be powered from the emergency generator life-safety power system. These fixtures shall be instant-on 40-watt biax (qty of six lamps/fixture) low-bay fixtures or Induction Lamp (Sylvania Icetron or equal) low-bay fixtures. The emergency fixtures shall be designed to provide sufficient lighting to satisfy egress lighting code minimums.

*3.9.1.3 4850 level chambers communications*
*3.9.1.3.1 General:* The existing communication systems infrastructure consist of a copper cable system for voice communications, fiber optics for data transmission, and a leaky-feeder radio system. The leaky-feeder radio system will be utilized for operations and backup communications. For redundancy purposes, provisions have been made to extend and loop each of these systems at the 4850 level and the 7400 level.

*3.9.1.3.2 Voice*: Each chamber will be equipped with copper-connectivity landline voice communications. As each development is created, voice communications will be extended.

*3.9.1.3.3 Voice/radio*: Each chamber will have the leaky-feeder system extended to it to be primarily used for facility operations radio communications.

*3.9.1.3.4 Data*: A high-speed fiber switch will be located in each chamber that requires high-speed the service. These high-speed facility switches will attached to the fiber backbone at Hubs located on the developed area level.

*3.9.1.3.5 Public address*: Subsurface PA systems shall provide paging coverage on the developed levels in the areas normally occupied including the lab chambers.

*3.9.1.4 4850 level chambers space monitoring*
*3.9.1.4.1 Gas monitoring*: Each chamber will be equipped with a multi-gas monitor connected to the mine-controls system. These signals will allow for active gas purging control. They will also be monitored in the fire command center located on the surface. Monitoring will facilitate the broadcast of appropriate alarms and warnings.

*3.9.1.4.2 Fire detection*: Each chamber will be equipped with full area coverage fire detection and occupant notification to satisfy the building codes. The system components for each chamber will consist of addressable photoelectric smoke detectors, pull stations, horn/strobe annunciation devices, heat detectors, and sprinkler monitoring flow and tamper switches. These signals will be communicated to the fire command center located on the surface. Central command will initiate the broadcast of appropriate alarms and warnings as may be required for voice evacuation. Certain signals will also be communicated to the central controls system to allow for active gas purging and/or smoke control.

*3.9.1.4.3 Security and access*: The subsurface security/access system shall consist of a card reader system with card readers installed at all laboratory entrances and exit doors.

*3.9.1.5 4850 level chambers HVAC*
*3.9.1.5.1 General*: Personnel access to the 4850 Level facility will be achieved through a pressurized air lock near the Ross Shaft to help maintain cleanliness of the facility. Materials and equipment access will be achieved through an air lock at the east end of the main corridor and will be large enough to accommodate detector crates as well as equipment needed to transport the container.

Ventilation air for the facility will be provided from an intake structure located near the equipment air lock at the eastern end of the main corridor. The intake structure will be ducted to a Primary AHU, which will filter and precondition the ventilation air prior to delivery to individual AHU's serving the accelerator chamber and ancillary spaces. Adequate ventilation will be provided to dilute radon generated by the excavation as well as meet ASHRAE

Standard 62-2001 (*Ventilation for Acceptable Indoor Air Quality*) guidelines for specified size and occupancy of the facility. Excess outdoor air from the Yates shaft will bypass the facility by means of a bypass drift.

*3.9.1.5.2 Chilled water system*: A new chiller plant will be installed on the 4850 Level off the Ross shaft sized to handle the internal heat gain from the laboratory facility as well as heat gain from the surrounding rock structure and autocompression of the air. The existing dewatering system will be utilized to pump water to an existing 200,000-gallon sump located near the Ross shaft on the 5000 Level. A weir will be installed in the sump to divide it into a "cold well" side and a "hot well" side, allowing overflow from the "cold well" side to the "hot well" side. A portion of the water from the "cold well" side will be pumped to the chiller plant for use as condensing water. Condenser water from the chillers will be pumped down to the "hot well" side of sump and then pumped to the surface utilizing the existing pumping infrastructure. Any makeup water required for this system will be obtained from an existing 6" industrial water line from surface.

The new chiller plant will be utilized to cool a new screw air compressor, cool a new air dryer, provide cooling water for coils in the air handling units and chill domestic water to 13° C through a shell and tube heat exchanger.

*3.9.1.5.3 Primary AHU*: A preliminary estimate of the initial ventilation requirement for the 4850 Level facilities is 6,500 cfm, with a total future requirement of 12,000 cfm. Therefore the Primary AHU will be sized for 12,000 cfm and fitted with a VFD to allow initial operation at 6,500 cfm. The AHU will be fitted with HEPA filtration and no heating coil will be required.

*3.9.1.5.4 Accelerator chamber HVAC:* The accelerator chamber will incorporate a dedicated AHU system and will be outfitted with HEPA filtration to maintain the chamber to "white room" standards. Air in the chamber will be recirculated to maintain ambient conditions to a maximum of 18° C, 60% RH. Air from the Primary AHU will provide ventilation air as well as positive pressurization in the chamber to maintain cleanliness from corridors.

The accelerator chamber will have a method of egress at each end. One exit will connect to the main corridor and the other to the ventilation air bypass drift. Pressurization/purge fans for this chamber will be ducted to a point in the bypass drift that is "downwind" from the detector chamber exit. The pressurization/purge fan will be equipped with a VFD. Normal operation of the fan will provide enough exhaust to maintain positive pressurization in the accelerator chamber. In the event evacuation needs to be made to the bypass drift, the personnel will walk "upstream" in the bypass drift to the next chamber, where they can again enter the conditioned environment.

*3.9.1.5.5 Ancillary laboratory spaces*: Ancillary laboratory spaces will incorporate dedicated AHU systems. Portions of this space will be outfitted with HEPA filtration to maintain the laboratory spaces to "White Room" standards. Air in the ancillary laboratory spaces will be recirculated to maintain ambient conditions to a maximum of 18° C, 60% RH. Air from the Primary AHU will provide ventilation air and provide positive pressurization in the laboratory spaces to maintain cleanliness separation from corridors.

Refuge areas will be incorporated into the Ancillary support facilities near the Ross Shaft.

*3.9.1.5.6 Ductwork and devices*: An allowance has been made for fabrication and installation of 50,000 pounds of aluminum ductwork and air control devices.

*3.9.1.5.7 Corridors:* Corridors will incorporate dedicated AHU systems and will be outfitted with high efficiency filtration to maintain the corridors to "White Room" standards. Air in the corridors will be recirculated to maintain ambient conditions to a maximum of 18° C, 60% RH. Air from the Primary AHU will provide ventilation air.

*3.9.1.5.8 Smoke control:* Each AHU's operation will be controlled to pressurize all corridors and refuge areas associated with a facility evacuation to contain smoke and move it toward exhaust fans. Smoke will be diverted as much as possible to the exhaust drift, relying on the bypass ventilation system to remove products of combustion, while maintaining clear evacuation paths.

3.9.1.5.9 Emergency purge ventilation: In the event of a hazardous spill or toxic gas release in the accelerator chamber, a motorized damper will open. The pressurization/purge fan in the chamber will switch to purge mode (VFD to 100%) and draw air from the Primary AHU system exhausting the contaminated air into the bypass drift. The contaminated air will be diluted and carried off through the exhaust drift to the #31 exhaust shaft and vented through the Oro Hondo exhaust fan. The purge system will allow the accelerator chamber to be cleared in 15 minutes.

*3.9.1.6 4850 level chambers plumbing*

*3.9.1.6.1 Detector chambers*: Each chamber will be provided with toilet facilities, mop service basins, water coolers, eyewash and emergency shower stations. Domestic hot and cold water will be provided at each toilet room location. Each fixture group will discharge into sewage ejector, which will pump sewage to a holding tank near the Ross shaft.

*3.9.1.6.2 Ancillary support facilities:* Plumbing systems will be designed as necessary to meet building function as well as all federal, state and local codes and regulations. Special plumbing systems such as acid waste, compressed gas delivery systems, centralized water treatment, etc. will be provided within laboratory and assembly areas that support the needs of research activities. Each fixture group will discharge into sewage ejector, which will pump sewage via overhead lines to a common collection facility.

*3.9.1.6.3 Common waste collection facility*: Waste from each ejector sump will be pumped to a holding tank at a central collection point located on the 4850 level near the Ross shaft. The sewage holding tank will be emptied to a portable tank on an as-needed basis.

*3.9.1.6.4 Domestic water systems:* A new domestic water storage tank will be installed on the 4550 Level of sufficient size and capacity to provide for the domestic water needs of the facility. Domestic water will flow from the tank via gravity feed to a mechanical room located on the 4550 level.

A shell and tube heat exchanger located in the mechanical room will generate domestic cold water, chilled to 13° C by the chilled water loop. Electric storage water heaters located in the mechanical room will generate domestic hot water. An aquastat-controlled recirculating pump will control flow through domestic hot water recirculating piping. This will maintain the required service temperature in the domestic hot water distribution piping. Domestic water piping will be type L hard copper with lead-free solder and will run overhead.

*3.9.1.7 4850 level chambers fire protection*

*3.9.1.7.1 General:* Fire protection to the 4850 level of the underground laboratory will consist primarily of a sprinkler system served by the industrial water source within the facility. All areas containing combustible materials will be protected by either wet, dry, or gaseous fire-protection systems to satisfy the requirements of use. The requirements of the National Fire Protection Association (NFPA) Standards and Guidelines will be met as they apply to this facility.

Refuge areas will be used within the facility for staging areas to prepare for evacuation in the event of a fire or other emergency. These areas will contain surface communications equipment and self-contained breathing apparatus for use during evacuation. Refuge areas will employ clean agent fire protection systems that will allow occupants to function during emergency events even if the area requires fire suppression.

*3.9.1.7.2 Water storage tank*: A 100,000-gallon fire protection reservoir located on the 4550 Level and fed by Industrial Water from the surface will serve the underground Laboratory Facility. A gravity feed system will pressurize the sprinkler piping and deliver flow to covered areas from the reservoir. The system will be able to deliver flow for at least 90 minutes, if not indefinitely.

*3.9.1.7.3 Distribution and sprinklers*: Areas of low hazard such as corridors and support areas will be covered by wet sprinkler systems designed for light hazard occupancy in accordance with NFPA 13. Areas of higher combustible material loading such as storage rooms and laboratories will be covered by wet sprinkler systems designed for ordinary hazard occupancy in accordance with NFPA 13.  The water source will be designed to accommodate wet sprinkler systems serving extra hazard occupancies as may arise throughout the life of the facility.

For equipment requiring higher levels of protection from water damage, double interlocked preaction fire suppression systems will be installed. This type of system employs automatic sprinklers attached to a piping system charged with air.  A detection system will allow the release of water when two independent detection devices have confirmed fire conditions.

*3.9.1.7.4 FM 200*: Where highly sensitive or irreplaceable electronic equipment is housed, a gaseous fire suppression system designed for computer rooms will be employed as a first line of defense against fire, supplemented by a double interlocked fire preaction system.

*3.9.1.8 4850 level process piping*
*3.9.1.8.1 General*: During the Baseline Definition process further discussions between engineers and physicists (especially those designing the accelerator) will be needed to clarify experimental requirements for process piping.

*3.9.1.8.2 Bulk fluid storage*: It is assumed that cryogenic liquids used in detector labs will be brought into the labs from the surface in portable dewars and either transferred to experimental equipment or used directly from dewars. It is also assumed that all large volume toxic/flammable liquids or suffocating gasses will be transported to and maintained in double containment containers when in detector lab.

*3.9.1.8.3 Bulk fluid drainage and spill containment*: Large volume liquid spills could be directed down the bypass tunnel to be contained at a lower level drift until such time that they can be recovered.

***3.9.2 7400 level chambers systems***
*3.9.2.1 Electrical*
*3.9.2.1.1 Normal power*: See 3.9.1.1.1.

*3.9.2.1.2 Emergency power*: See 3.9.1.1.2.

*3.9.2.2 7400 level chambers lighting*
*3.9.2.2.1 Normal lighting*: See 3.9.1.2.1.

*3.9.2.2.2 Emergency lighting*: See 3.9.1.2.2.

*3.9.2.3 7400 level chambers communications*
*3.9.2.3.1 General*: See 3.9.1.3.1.

*3.9.2.3.2 Voice*: See 3.9.1.3.2.

*3.9.2.3.3 Voice/radio*: See 3.9.1.3.3.

*3.9.2.3.4 Data*: See 3.9.1.3.4.

*3.9.2.3.5 Public address*: See 3.9.1.3.5.

*3.9.2.4 7400 level chambers space monitoring*
*3.9.2.4.1 Gas monitoring*: See 3.9.1.4.1.

*3.9.2.4.2 Fire detection*: See 3.9.1.4.2.

*3.9.2.4.3 Security and access*: See 3.9.1.4.3.

*3.9.2.5 7400 level chambers HVAC*

*3.9.2.5.1 General*: Entrance to the 7400 level facility will be achieved through a pressurized air lock near the No. 6 Winze shaft to help maintain cleanliness of the facility. The air lock will be large enough to accommodate detector crates as well as equipment needed to transport the container.

Ventilation air for the facility will be provided from an intake structure located near the No. 6 Winze. The intake structure will be ducted to a primary Air Handling Unit (AHU), which will filter and precondition the ventilation air prior to delivery to individual AHU's serving the detector chambers and ancillary spaces. Adequate ventilation will be provided to dilute radon generated by the excavation as well as meet ASHRAE Standard 62-2001 (*Ventilation for Acceptable Indoor Air Quality*) guidelines for specified size and occupancy of the facility. Excess outdoor air from the No. 6 Winze shaft will bypass the facility by means of a bypass drift and will provide cooling for two air-cooled chillers located near No. 7 Shaft.

*3.9.2.5.2 Chilled water system*: A new chilled water plant consisting of two air-cooled 300 ton chillers will be installed on the 7400 level within the new facility. It is sized to handle the internal heat gain from the laboratory facility as well as heat gain from the surrounding rock structure and autocompression of the ventilation air. The air-cooled chillers will be located near the No. 7 shaft rejecting heat to the bypass and exhaust air from the 7400 level facilities.

The new chiller plant will be utilized to chill coils in the air handling units and screw air compressor/dryer system, and will be circulated through a shell and tube heat exchanger to cool the domestic water to 13° C.

*3.9.2.5.3 Primary AHU*: The preliminary estimate of the initial ventilation requirement for the 7400 level facilities is 31,700 cfm, with a total future requirement of 52,500 cfm. Therefore the Primary AHU will be sized for 52,500 cfm and fitted with a Variable Frequency Drive (VFD) to allow initial operation at 31,700 cfm. The AHU will be fitted with High Efficiency Particulate Air (HEPA) filtration and no heating coil will be required.

*3.9.2.5.4 7400 level chambers HVAC*: Each detector chamber will incorporate a dedicated AHU system and will be outfitted with HEPA filtration to maintain the chamber to "white room" standards. Air in the chamber will be recirculated to maintain ambient conditions to a maximum of 18° C, 60% RH. Air from the Primary AHU will provide ventilation air as well as positive pressurization in the chamber to maintain cleanliness from corridors.

Each detector laboratory will have a method of egress at each end. One exit will connect to the main corridor and the other to the ventilation air bypass drift. Pressurization/purge fans for each detector chamber will be ducted to a point in the bypass drift that is "downwind" from the detector chamber exit. Each pressurization/purge fan will be equipped with a VFD. Normal operation of the fans will provide enough exhaust to maintain positive pressurization in each detector chamber. In the event evacuation needs to be made to the bypass drift, the personnel will walk "upstream" in the bypass drift to the next chamber, where they can again enter the conditioned environment.

In the event there is a problem associated with the No. 6 Winze shaft, egress can be made to either No. 4 Winze or a series of existing ramps to the 4850 Level. Once at the 4850 Level, personnel can utilize either the Ross or Yates shafts for egress to the surface.

*3.9.2.5.5 Ancillary laboratory spaces*: Ancillary laboratory spaces will incorporate dedicated AHU systems. Portions of this space will be outfitted with HEPA filtration to maintain the laboratory spaces to "White Room" standards. Air in the ancillary laboratory spaces will be recirculated to maintain ambient conditions to a maximum of 18° C, 60% RH. Air from the Primary AHU will provide ventilation air and provide positive pressurization in the laboratory spaces to maintain cleanliness separation from corridors.

Refuge areas will be incorporated into the Ancillary support facilities near the No. 6 Winze egress shaft.

*3.9.2.5.6 Corridors*: See 3.9.1.5.7.

*3.9.2.5.7 Ductwork and devices:* An allowance has been made for fabrication and installation of 140,000 pounds of aluminum ductwork and air control devices.

*3.9.2.5.8 Cleanrooms*: The underground receiving area, within the ancillary laboratory space, will open to a staging Class 100,000 area for unpacking and repacking of containers. Located near the staging area will be individual Class 10,000 parts cleaning rooms with clean benches for rebag of assemblies. A Class 10,000 gowning room with airlock will be located off the staging clean room with access to a assembly Class 100 clean room for individual detector final assembly and repack for trip to or from the detector bays. Cleanrooms will be prefabricated unless high ceilings or special cranes are required. In such cases cleanrooms will be built in place.

*3.9.2.5.9 Smoke control*: See 3.9.1.5.8.

*3.9.2.5.10 Emergency purge ventilation*: In the event of a hazardous spill or toxic gas release in any of the detector laboratories, motorized dampers will open, the pressurization/purge fan in each laboratory chamber (other than the LLCF and General Purpose Hall, which will utilize dedicated purge fans). The switch to purge mode (VFD to 100%) will draw air from the Primary AHU system exhausting the contaminated air into the bypass drift. The contaminated air will be diluted and carried off through the exhaust drift, up through the No. 7 shaft and vented through the Oro Hondo exhaust fan. This purge system will allow all of the detector chambers to be cleared in 15 minutes, except the General Purpose Hall, which will take approximately 22 minutes to clear.

*3.9.2.6 7400 level chambers plumbing*
*3.9.2.6.1 Detector chamber laboratories*: Each detector chamber laboratory will be provided with toilet facilities, mop service basins, water coolers, eyewash and emergency shower stations. Domestic hot and cold water will be provided at each toilet room location. Each fixture group will discharge into sewage ejector, which will pump sewage to a holding tank near the No. 6 shaft.

*3.9.2.6.2 Ancillary support facilities*: See 3.9.1.6.2.

*3.9.2.6.3 Common waste collection facility*: Waste from each ejector sump will be pumped to a holding tank at a central collection point located on the 7400 level near the No. 6 Winze shaft. The sewage holding tank will be emptied to a portable tank on an as-needed basis for transport to the surface.

*3.9.2.6.4 Domestic water systems*: A new domestic water storage tank will be installed on the 7100 level of sufficient size and capacity to provide for the domestic water needs of the facility. Domestic water will flow from the tank via gravity feed to a mechanical room located on the 7400 level.

A shell and tube heat exchanger located in the mechanical room will generate domestic cold water, chilled to 13° C by the chilled water loop. Electric storage water heaters located in the mechanical room will generate domestic hot water. An aquastat controlled recirculating pump will control flow through domestic hot water recirculating piping to maintain required service temperature in the domestic hot water distribution piping. Domestic water piping will be type L hard copper with lead-free solder and will run overhead.

*3.9.2.7 7400 chambers fire protection*
*3.9.2.7.1 General:* See 3.9.1.7.1.

*3.9.2.7.2 Water storage tank*: A 100,000-gallon fire protection reservoir located on the 7100 level and fed by Industrial Water from the surface will serve the underground Laboratory Facility. A gravity feed system will pressurize the sprinkler piping and deliver flow to covered areas from the reservoir. The system will be able to deliver flow for at least 90 minutes, if not indefinitely.

*3.9.2.7.3 Distribution and sprinklers:* See 3.9.1.7.3.

*3.9.2.7.4 FM 200*: See 3.9.1.7.4.

*3.9.2.8 7200 level process piping*

3.9.2.8.1 *General:* This category requires extensive iteration with the planned users of the laboratory chambers, a process that will take place during the Baseline Definition. Topics that will be added to this WBS include a reverse osmosis system, compressed gasses, air compressors, and compressed gas distribution. Coordination with the Low Level Counting Facility, which is being designed with certain excess capacities, will be important.

*3.9.2.8.2 Bulk fluid storage*: See 3.9.1.8.2.

*3.9.2.8.3 Bulk fluid drainage and spill containment:* See 3.9.1.8.3.

*3.9.2.9 7400 level internal structures*
*3.9.2.9.1 General:* The physics community addressed some of the physical support facilities, such as bridge cranes, in the Lead workshops. (The results were included in our Science Book.) However, as with process piping, further iterations – engineers to physicists and the reverse -- during the Baseline Definition will be essential in defining individual chamber needs.

*3.9.2.9.2 Prefabricated cleanrooms*: Experimental clean rooms within each detector chamber will be assumed to consist of prefabricated, site-assembled rooms with dedicated air conditioning, filtration and fan systems to meet the individual needs of specific projects. The clean rooms will be positively pressurized with respect to the detector chamber to maintain cleanliness separation from the detector chamber. These lab specific cleanrooms could be designed to standards as low as Class 100,000 or as high as Class 1/10 depending upon the requirements of the investigators. Electrical power, chilled water and outside air would be provided in the detector chamber to service the experimental clean rooms. The experimental clean rooms will be provided as needed and capitalized against the particular experimental installation.

### *3.9.3 8000 level chambers systems*

The earth scientists have provided their room and drilling space needs, but have not yet determined the systems requirements for their geomicrobiology space on the 8000-ft level. As noted earlier, we believe costs will be dominated by the work to bring systems to the 8000-ft level, not by chamber work.

## 4 Surface Development for Science

***4.1 Demolishing existing structures:*** A total of $5.25M is provided for the demolishing existing structures. The Homestake site includes a number of warehouses and shop structures of little historical or practical value. This item will fund the removal of these structures as well as the landscaping and re-vegetation of the site as a scientific laboratory and as a site of major public interest.

***4.2 Roads and parking:*** A total of $1.625M has been provided for construction of roads and parking. We discussed in the Facilities Development section the specific road and parking improvements that will be need for access to the Ross shaft (where parking is currently quite limited) and to the Visitor Experience Center (which we are placing near the existing city visitor center). We noted earlier the opportunities for providing expanded parking for this facility.

NUSEL will generate two to three times as much traffic as recent operations at the Homestake Mine. Furthermore, visitor traffic will consist primarily of drivers unfamiliar with the area, arriving by car, bus, motorhome, trailers, etc. Significant improvements are required in the roadway layout and in the number and size of parking spaces. While we have suggested sites for the main laboratory building and the Visitor Experience Center in this Reference Design, the Baseline Definition decision on sites cannot be made until after a property agreement, as a survey by experienced civil engineers and architects is essential to building placement. But conceptually, we envision a roadway scheme that will separate staff and visiting scientists, public visitors, and freight traffic, likely leading them to their destinations via routes at different elevation levels along the hillside that separates Homestake from Lead. We hope to provide on-site parking for about 300 vehicles, which should be sufficient for 9 months of the year. NUSEL and Lead will discuss summer parking options. Possibilities include expanded parking on Mill Street, across from the Open Cut Lead visitor center, and various off-site parking arrangements with Lead and Deadwood coupled to shuttle buses.

The six-acre Mill Street site is shown on site Figure #2 in the Facilities Development section. The site would require the acquisition and demolition of some structures as well as site grading.

***4.3 Science and Administration building:*** The space requirements for NUSEL are based on comparisons made with Gran Sasso, as discussed in the Facilities Development section. We have also taken guidance from Sandro Bettini, who just finished his term as Gran Sasso director. Bettini pointed out that Gran Sasso is about to erect a new three-story office building because space is too limited, given the demands of future long-baseline and other neutrino experiments. Bettini also stressed that Gran Sasso's assembly areas are too limiting. One would anticipate that assembly needs of a vertical access facility would exceed those of a horizontal facility. As there are significant uncertainties in the costing discussed below, we have used a 40% for the building.

Building site possibilities have been discussed previously. Approximately two acres of property at the top of Mill Street near the Ross head frame could be made available by relocating or acquiring and demolishing some residential structures (see the site diagram labeled Figure #1 in the Facilities Development section). Grading would require cutting the top of the hill to create a level building site. Rock and soil from this excavation will be utilized to fill a nearby depression, creating a level building site for additional facilities or parking.

***4.3.1 Office space in the Science Building***: The office space portion of the Science building, 68.4K gross square feet (gsf), has been costed at $150/sf, which is at the upper end of the range for Rapid City office construction. However this building must be designed to last 50 years, and should have a ``signature'' appearance consistent with the outreach and historical preservation goals of NUSEL.

***4.3.2 Laboratory space***: 36.0K gsf of laboratory space has been costed at $225/gsf, which is at the low end of a range that can extend, for highly technical laboratory space, to above $1000/gsf. However, this space includes substantial areas for glass, machine, and electronics shops, where a lower number is not inappropriate. Furthermore we include substantial funding for upper campus systems in a separate item, which will be used to address utilities and clean room needs of technical areas within the science building.

***4.3.3 Assembly space****:* The 56.2K gsf has been costed at $120/gsf. We are hopeful that several of the Barrick building identified in the Facilities Development section can be renovated to provided the needed assembly and warehouse space. However, given that civil engineering inspections of these buildings have not been allowed to date, we have prudently costed this space as new.

Very likely some a good portion of the assembly space will not be located in the main Science Building, but in an area picked for convenience to the underground. The Ross Pipe Shop could be demolished and replaced with a new Receiving and Assembly facility. This new facility would incorporate clean assembly space and be adjacent to the Ross head frame, allowing easy access to the subsurface facilities.

Most of the underground civil excavation will utilize skid-mounted compressors and a dedicated air compressor for the science activities will be installed at the subsurface facility. Moving the air compressors underground will allow the Ross Compressor facility to be utilized for other purposes. As the facility has an overhead crane, it is another candidate for conversion to a warehouse.

***4.4 Upper campus systems***

**4.4.1** ***Offsite electrical infrastructure****:* All site power is distributed from the utility-owned Kirk switchyard. The Kirk switchyard is fed from High Voltage Lines from the Yellow Creek Substation and the Lookout Substation that are currently configured to supply up to 32 Megawatts to the mine site. The available capacity is many times that which is expected for site development and laboratory operations. The Homestake Mine is a "primary customer" which means that it owns, operates and controls the existing on-site power distribution required for all mine facilities.

The Utility (Black Hill Power and Light) substations are very reliable because they are relatively new and are "redundant loop" powered substations. No system changes are anticipated.

**4.4.2** ***Communications***: The external infrastructure will consist of data communication external to the complex. Black Hills FiberCom supports the current connectivity to the complex with a direct fiber connection. The current speed of the connection is an OC-1 (51.84 Mbps) supported by a SONET (Synchronous Optical Network) Hierarchy.

Redundancy comes from several paths should the fiber be severed in any location. Future expansion of the current infrastructure can obtain OC-3 (155Mbps) speed if necessary. This expansion would allow remote users to obtain speeds as if they were connected inside the complex infrastructure. This is not planned at this time.

**4.4.3** ***Security and access***: Site perimeter fencing and site lighting shall serve as the first intruder deterrent systems. The surface security/access system shall be a site-wide system that will consist of a card reader systems installed at all new-structure exterior doors, movement and proximity detection, and video monitoring.

Access to the facility shall be limited to dedicated entrance gates. Each site entrance gate shall be equipped with a card reader system and will be monitored and recorded by a digital video system.

**4.4.4** ***Site lighting***: Site lighting will be added at selective site perimeter fencing, on new and existing surface facilities, and at new parking locations.

The perimeter fencing lighting shall be accomplished with dusk to dawn utility fixtures on 20 ft high round steel fencing poles. These fixtures will have integral photocells.

The new Parking and Roadway lighting shall be commercial grade metal halide fixtures on 35 ft high square tapered steel poles. These fixtures will be controlled with remote controls including time clocks and photocells.

**4.4 5** ***Clean rooms***: Clean rooms will be provided with high-efficiency particulate air (HEPA) or ultralow penetration air (ULPA) filtration, humidity control, temperature control and pressurization control to maintain interior conditions. Clean rooms associated with the laboratory and assembly spaces will be staged with regard to level of cleanliness. For example, from a common Class 10,000 corridor, material and staff will enter the Class 100 to Class 1/10 clean rooms.

The Class 1 to 100 clean rooms will incorporate laminar airflow with low wall returns where possible to allow for heavy floor loads. These rooms will include Class 10,000 rooms for gowning, staging, and cleaning of exterior envelopes of clean assemblies. A white room for welding fabrication and controlled assembly will connect to the clean area as well. Clean benches in the staging areas will be provided to pre-clean assemblies prior to entry into clean spaces for final assembly. Pressurized airlocks will be incorporated between rooms of differing classification and will be large enough to contain the largest assembly.

High bay clean rooms, over 20-foot clear height, will be provided with overhead cranes and monorails to assist in assembly of large detectors. Special clean room protocols will be followed to ensure that the overhead fixtures and equipment do not contaminate the critical items during assembly.

Low ceiling height clean rooms will be of prefabricated type permitting flexibility of assembly and revision to meet the needs of the researcher. High bay clean rooms will be site fabricated and designed to provide a maximum flexibility of use.

Deionized water and clean compressed air will be provided in the clean lab areas for parts cleaning and laboratory use. Laboratory clean rooms will have pass through openings for parts transfer and utility service for cryogenic gasses or compressed gasses. Lab configurations will strive to keep the cylinders and dewars outside of the lab spaces where possible. Spill containment and special exhaust will be provided in selected clean rooms to ensure safety from hazardous and suffocating materials.

Clean detectors and supplies will be bagged and placed in a transfer container in the surface facility assembly cleanrooms for their trip to the subsurface detector labs. These containers will then be placed on transfer carts and towed to the Yates lift staging room. The containers will be moved to the air-conditioned lift car and the cart brakes set for the trip down to the detector level of the facility.

***4.4.6 Central energy plant***: To accommodate the large and varying load throughout the surface campus, a central energy plant will be developed to provide chilled water, steam and heating hot water to the complex. Electric water chillers and gas-fired boilers will be the primary equipment housed in the energy plant. Another readily available source of heat is the water pumped from the underground facilities (steady flow rate of 350 – 500 gpm after drift development ceases). However a feasibility study will be required to determine if energy recovered from the Mine Water through the use of heat pumps is an economical solution to heating some of the facilities at the Ross complex.
.

***4.4.7 Normal power/transformers***: Most of the existing surface support facilities will not require any new electrical service transformers or other electrical equipment. A standard oil-filled utility grade service transformer will be installed to serve each new on-site laboratory facility. Each new facility will have a 480-volt main power panel sized per the 2002 NEC and powered from the Ross or East Substations. The local utility will provide a standard oil filled service transformer to serve each new off-site laboratory facility. Each new facility will have a 480 volt main power panel sized per the 2002 NEC and powered from the local utility grid.

***4.4.8 Emergency power***: A generator will be supplied for the surface facilities that will require substantial emergency power for Life Safety or Critical Equipment loads.

Surface facility generators will be located outdoors in a weatherproof enclosure with an integral fuel tank. The fuel supply shall be sized for 2 hours minimum full load operation. Each generator shall be sized for communication loads, critical experiment loads, elevator loads, fire pump loads, and life safety loads including egress and exit lighting.

***4.4.9 Lighting***: The new facilities interiors will be equipped with normal energy efficient fluorescent lighting as well as emergency lighting consisting of combination entrance and exit fixtures. Exterior wall fixtures and pole fixtures will be commercial grade 100% cutoff metal halide fixtures.

***4.4.10 Security***: Facility security will be tied into the site-wide security network. Each new facility will have a card reader access system and some level video monitoring. Each exiting critical “maintenance” facility will be equipped with video monitoring and recording.

***4.4.11 Fire detection***: The new facilities will be equipped with addressable Fire detection systems.

***4.4.12 External communications***: Each of the new surface facilities will be connected to the Black Hills FiberCom Network through the existing direct fiber connection currently located in the machine shop.

***4.4.13 Internal communications***: The new surface facilities will be wired to EIA Category 6 (cat 6) standards unless faster connectivity (fiber) is required to certain jack locations. Dual cat 6 data jacks will be located throughout the facility rooms and will be wired to an intermediate communications patch panel (IDF) on each level. Each IDF will be connected by fiber back to a main distribution patch panel (MDF). The MDF will serve as the connection point to the Black Hill FiberCom Network via the fiber backbone.

Facility paging will be accomplished through the Building Phone system.

***4.4.14 HVAC***: Heating, ventilation and air conditioning will be accomplished throughout the surface facilities with individual AHUs and terminal devices with hydronic heating and cooling coils (or heat pumps if Mine Water is used as a heating source). These AHUs will be utilized to condition and distribute the appropriate amounts of outside air to ventilate the space. Air within each space will be filtered to meet the operating standards of cleanliness dictated by facility function.

Clean rooms within the space will be served by dedicated clean room air handling systems. Air will be filtered by high efficiency filtration systems and distributed at an air change rate appropriate for the room class served.

Special ventilation systems will be provided to serve scavenger and hood exhaust systems. Exhaust systems and hoods will be designed to contain and remove contaminants from the space and discharge with dilution and/or filtration in a safe location. Shielding of hoods and special rooms to filter nuclear or electromagnetic radiation will also be available on an as-needed basis.

The complex will utilize state of the art direct digital control systems to control all building functions associated with heating, air conditioning and ventilation. This will also allow building function to be monitored from remote locations to increase maintenance efficiency and trouble shooting ability. This system will allow a high level of comfort and safety control as well as increased energy efficiency. Controls for the subsurface mechanical and pumping systems could also be tied into this system.

***4.4.15 Plumbing***: Plumbing systems will be designed as necessary to meet building function as well as all federal, State and local codes and regulations. Special plumbing systems such as acid waste, compressed gas delivery systems, centralized water treatment, etc. will be provided within laboratory and assembly areas that support the needs of research activities.

***4.4.16 Fire protection***: Fire protection will be provided to meet the requirements of the National Fire Protection Association as well as local code. It is anticipated that gaseous fire protection agents will be used to protect critical equipment and procedures as well as pre-action and wet sprinkler coverage of the entire building.

**5 Site Operations and Maintenance**

***5.1 Capital equipment*****:** Funding is provided to replace major equipment needed for operations and maintenance of the site.

***5.2 Operations***: The costs shown are based on known operations of Homestake. This is the cost of keeping the mine open in a mothballed fashion. Thus all additional costs included in section 3 of this WBS are measured with respect to this baseline. This separation of operations and mining costs is presented in detail in the appendices, broken down for both the 7400-ft and 4850-ft levels.

***5.3 Operations personnel***: These costs are also very well known, based on actual operations personnel costs for Homestake. The detailed spreadsheets are presented in the appendices. Note that operator costs (5.3.8) are almost halved by the hoist modernization proposed in WBS section 3.1.

***5.4 Maintenance personnel***: These costs are also detailed in the appendices. Note that the year-five operations and maintenance costs (5.2+5.3+5.4) are $8.2M: this is the level of support needed after the improvements of WBS section 3.

**6 Science operations**

***6.1 Capital equipment:*** The capital equipment shown is the equipping of the low-level counting facility, which was described in some detail in the Facilities Development section. In the appendices a spreadsheet is provided detailing the costs.

***6.2 Science operations personnel:*** The WBS takes the personnel levels described and justified in the Facilities Development section and proposes a gradual increase of the effort, until the desired level is reached in year five. This process is described in detail in a spreadsheet included in the appendices. The year-five cost is $6.5M for the scientists and science support staff.

***6.3 Other science operations:*** Representative costs for travel, supplies, publications, consultants (for advisory committees), and computer services are given. The budget is not a generous one: domestic travel, for example, is budgeted at $2,000/y for 25 scientists at or above the postdoc level, plus $10K for visitors.

**7 Detector Operations**

***7.1 Detector operations personnel:*** The WBS assumes that major activities in detector installation are underway by the beginning of year three: key personnel on brought on board at the beginning of year two. The personnel level corresponds to that presented in the Facilities Development plan, and the assumed salaries are shown in the WBS. A more detailed spreadsheet is included in the appendices.

***7.2 Other detector operations:*** Reasonable computer services and supplies are budgeted, and a modest amount of travel.

**8 Director's Operations**

***8.1 Site development group:*** Within the Director's Office a modest team of four will handle site development issues. This team will work with the city of Lead and the state on traffic and other logistical problems associated with the creation of NUSEL and its outreach program. It will address major site issues associated with historical preservation and with integrating NUSEL's new buildings into the site. It will handled shared maintenance issues – Barrick will continue to hold much of its site, and many road, utility, and ecological issues will need to be tackled jointly.

***8.2 Administrative office:*** This office, within the Director's Office, will handle all of the standard personnel, financial, visitor liaison, and public relations functions of the national laboratory. By prevailing standards (e.g., Gran Sasso), the 9 FTEs is very lean. This office is also the home of the security force. A security guard/watchman is needed 24/7, and we consider two officers to be a minimum staff during normal operating hours.

***8.3 Contract services:*** Among the services that NUSEL will likely contract out (in addition to the construction activities described elsewhere) are cafeteria, custodial, and MSHA training services.

***8.4 Other operations:*** Funds for Director's Office operations – travels, supplies, publications, and consultants – are provided. As this office handles public relations, substantial funding is provided for publications.

| WBS-1: LABORATORY SCIENCE CONSTRUCTION AND OPERATIONS | | | | | | | | |
|---|---|---|---|---|---|---|---|---|
| | | | | | | | | |
| | **Year 1** | **Year 2** | **Year 3** | **Year 4** | **Year 5** | **Years 1-5** | | **Salary** |
| | | | | | | | | **basis** |
| **1 Property** | | | | | | | | **($/FTE/y)** |
| | | | | | | | | |
| 1.1 Property acquisition | $500,000 | $0 | $0 | $0 | $0 | $500,000 | | |
| 1.2 Environmental | $200,000 | $0 | $0 | $0 | $0 | $200,000 | | |
| 1.3 Easements | $200,000 | $0 | $0 | $0 | $0 | $200,000 | | |
| 1.4 Permits | $300,000 | $0 | $0 | $0 | $0 | $300,000 | | |
| Subtotals | $1,200,000 | $0 | $0 | $0 | $0 | $1,200,000 | | |
| Contingency (50%) | $600,000 | $0 | $0 | $0 | $0 | $600,000 | | |
| | | | | | | | | |
| **Totals Property** | $1,800,000 | $0 | $0 | $0 | $0 | $1,800,000 | | |
| | | | | | | | | |
| **2 Insurance/Gen. Conditions** | | | | | | | | |
| 2.1 Insurance/trust fund | $0 | $0 | $0 | $0 | $0 | $0 | | |
| 2.2 General conditions | $0 | $0 | $0 | $0 | $0 | $0 | | |
| Contingency (50%) | $0 | $0 | $0 | $0 | $0 | $0 | | |
| | | | | | | | | |
| **Totals Insurance/Gen. Cond.** | $0 | $0 | $0 | $0 | $0 | $0 | | |
| | | | | | | | | |
| **3 Underground development** | | | | | | | | |
| | | | | | | | | |
| 3.1 Shaft/hoist upgrades | | | | | | | | |
| 3.1.1 Yates shaft upgrade | $3,586,000 | $0 | $0 | $0 | $0 | $3,586,000 | | |
| 3.1.2 #4 Winze upgrade | $1,652,000 | $300,000 | $0 | $0 | $0 | $1,952,000 | | |
| 3.1.3 Ross shaft reconfiguration | $3,849,000 | $4,888,000 | $2,987,000 | $0 | $0 | $11,724,000 | | |
| 3.1.4 #6 Winze reconfiguration | $91,000 | $811,000 | $5,306,000 | $2,579,000 | $0 | $8,787,000 | | |
| Subtotals access | $9,178,000 | $5,999,000 | $8,293,000 | $2,579,000 | $0 | $26,049,000 | | |
| Contractor markup (15%) | $1,376,700 | $899,850 | $1,243,950 | $386,850 | $0 | $3,907,350 | | |
| Excise tax (2%) | $211,094 | $137,977 | $190,739 | $59,317 | $0 | $599,127 | | |
| Contingency (40%) | $3,671,200 | $2,399,600 | $3,317,200 | $1,031,600 | $0 | $10,419,600 | | |
| Totals access | $14,436,994 | $9,436,427 | $13,044,889 | $4,056,767 | $0 | $40,975,077 | | |
| | | | | | | | | |
| 3.2 7400 facilities | | | | | | | | |
| 3.2.2 Central hall/entrance | $598,000 | $1,195,000 | $598,000 | $0 | $0 | $2,391,000 | | |
| 3.2.3 Exhaust/utility drift | $473,000 | $946,000 | $473,000 | $0 | $0 | $1,892,000 | | |
| 3.2.4 Main exhaust drift | $466,000 | $466,000 | $0 | $0 | $0 | $932,000 | | |
| 3.2.5 Car wash/change | $154,000 | $0 | $0 | $0 | $0 | $154,000 | | |
| 3.2.6 Utilities room | $544,000 | $0 | $0 | $0 | $0 | $544,000 | | |
| 3.2.7 GeoLab | $285,000 | $0 | $0 | $0 | $0 | $285,000 | | |
| 3.2.8 LLC Hall | $0 | $915,000 | $915,000 | $0 | $0 | $1,830,000 | | |
| 3.2.9 Hall B (large hall) | $0 | $2,026,000 | $1,959,000 | $0 | $0 | $3,985,000 | | |
| 3.2.10 Dark matter #1 | $298,000 | $99,000 | $0 | $0 | $0 | $397,000 | | |
| 3.2.11 Double beta decay | $271,000 | $271,000 | $0 | $0 | $0 | $542,000 | | |
| 3.2.12 Dark matter #2 | $0 | $398,000 | $0 | $0 | $0 | $398,000 | | |
| 3.2.13 Solar neutrinos | $0 | $173,000 | $518,000 | $0 | $0 | $691,000 | | |
| 3.2.14 Offices/seminar room | $0 | $300,000 | $0 | $0 | $0 | $300,000 | | |
| 3.2.15 Interior machine shop | $0 | $170,000 | $0 | $0 | $0 | $170,000 | | |
| 3.2.16 Exterior machine shop | $0 | $0 | $0 | $0 | $0 | $0 | | |
| 3.2.17 Refrig/fan/coil room | $340,000 | $0 | $0 | $0 | $0 | $340,000 | | |
| 3.2.18 Lunch/refuge | $202,000 | $0 | $0 | $0 | $0 | $202,000 | | |
| Subtotals 7400 facilities | $3,631,000 | $6,959,000 | $4,463,000 | $0 | $0 | $15,053,000 | | |
| Contractor markup (15%) | $544,650 | $1,043,850 | $669,450 | $0 | $0 | $2,257,950 | | |
| Excise tax (2%) | $83,513 | $160,057 | $102,649 | $0 | $0 | $346,219 | | |
| Contingency (25%) | $907,750 | $1,739,750 | $1,115,750 | $0 | $0 | $3,763,250 | | |
| Totals 7400 facilities | $5,166,913 | $9,902,657 | $6,350,849 | $0 | $0 | $21,420,419 | | |
| | | | | | | | | |
| 3.3 8000 facilities | | | | | | | | |
| 3.3.1 Anaerobic glove room | $0 | $58,000 | $0 | $0 | $0 | $58,000 | | |
| 3.3.2 Drilling platform | $0 | $114,000 | $0 | $0 | $0 | $114,000 | | |
| Subtotals 8000 facilities | $0 | $172,000 | $0 | $0 | $0 | $172,000 | | |
| Contractor markup (15%) | $0 | $25,800 | $0 | $0 | $0 | $25,800 | | |
| Excise tax (2%) | $0 | $3,956 | $0 | $0 | $0 | $3,956 | | |
| Contingency (25%) | $0 | $43,000 | $0 | $0 | $0 | $43,000 | | |
| Totals 8000 facilities | $0 | $244,756 | $0 | $0 | $0 | $244,756 | | |
| | | | | | | | | |
| 3.4 4850 facilities | | | | | | | | |
| 3.4.1 4850 lab access | $277,000 | $555,000 | $1,110,000 | $277,000 | $0 | $2,219,000 | | |
| 3.4.2 4850 materials storage | $92,000 | $0 | $0 | $0 | $0 | $92,000 | | |
| 3.4.3 Lunch/refuge | $34,000 | $0 | $0 | $0 | $0 | $34,000 | | |
| 3.4.4 Accelerator room | $147,000 | $587,000 | $245,000 | $0 | $0 | $979,000 | | |
| 3.4.5 Clean room | $0 | $74,000 | $0 | $0 | $0 | $74,000 | | |
| 3.4.6 Lab vent/utility | $184,000 | $184,000 | $367,000 | $0 | $0 | $735,000 | | |
| Subtotals 4850 facilities | $734,000 | $1,400,000 | $1,722,000 | $277,000 | $0 | $4,133,000 | | |
| Contractor markup (15%) | $110,100 | $210,000 | $258,300 | $41,550 | $0 | $619,950 | | |
| Excise tax (2%) | $16,882 | $32,200 | $39,606 | $6,371 | $0 | $95,059 | | |
| Contingency (25%) | $183,500 | $350,000 | $430,500 | $69,250 | $0 | $1,033,250 | | |
| Totals 4850 facilities | $1,044,482 | $1,992,200 | $2,450,406 | $394,171 | $0 | $5,881,259 | | |
| | | | | | | | | |
| 3.5 Other excavation | | | | | | | | |
| 3.5.1 4850 Yates station | $0 | $107,000 | $0 | $0 | $0 | $107,000 | | |
| 3.5.2 4850 Ross/#6 connection | $815,000 | $815,000 | $0 | $0 | $0 | $1,630,000 | | |
| 3.5.3 4850 Vent drift to #5 shaft | $0 | $0 | $94,000 | $94,000 | $0 | $188,000 | | |
| 3.5.4 #4-#6 Winze connection | $334,000 | $0 | $0 | $0 | $0 | $334,000 | | |
| 3.5.5 Pump/Electrical Boreholes | $1,908,000 | $327,000 | $0 | $0 | $0 | $2,235,000 | | |
| 3.5.6 #5 Shaft vent. Conversion | $0 | $0 | $0 | $374,000 | $0 | $374,000 | | |
| 3.5.7 Mine doors and walls | $84,000 | $168,000 | $0 | $0 | $0 | $252,000 | | |
| Subtotals other excavation | $3,141,000 | $1,417,000 | $94,000 | $468,000 | $0 | $5,120,000 | | |
| Contractor markup (15%) | $471,150 | $212,550 | $14,100 | $70,200 | $0 | $768,000 | | |
| Excise tax (2%) | $72,243 | $32,591 | $2,162 | $10,764 | $0 | $117,760 | | |
| Contingency (25%) | $785,250 | $354,250 | $23,500 | $117,000 | $0 | $1,280,000 | | |
| Totals other excavation | $4,469,643 | $2,016,391 | $133,762 | $665,964 | $0 | $7,285,760 | | |
| | | | | | | | | |
| 3.6 Waste rock handling | | | | | | | | |
| 3.6.1 Yates waste conveyor | $2,547,000 | $2,260,000 | $0 | $0 | $0 | $4,807,000 | | |
| Subtotals waste rock handling | $2,547,000 | $2,260,000 | $0 | $0 | $0 | $4,807,000 | | |
| Contractor markup (15%) | $382,050 | $339,000 | $0 | $0 | $0 | $721,050 | | |
| Excise tax (2%) | $58,581 | $51,980 | $0 | $0 | $0 | $110,561 | | |
| Contingency (25%) | $636,750 | $565,000 | $0 | $0 | $0 | $1,201,750 | | |
| Totals waste rock handling | $3,624,381 | $3,215,980 | $0 | $0 | $0 | $6,840,361 | | |
| | | | | | | | | |
| 3.7 Core drilling | | | | | | | | |
| 3.7.1 4850 coring | $72,000 | $0 | $0 | $0 | $0 | $72,000 | | |
| 3.7.2 7400 coring | $216,000 | $0 | $0 | $0 | $0 | $216,000 | | |
| Subtotals core drilling | $288,000 | $0 | $0 | $0 | $0 | $288,000 | | |
| Contractor markup (15%) | $43,200 | $0 | $0 | $0 | $0 | $43,200 | | |
| Excise tax (2%) | $6,624 | $0 | $0 | $0 | $0 | $6,624 | | |
| Contingency (25%) | $72,000 | $0 | $0 | $0 | $0 | $72,000 | | |
| Totals core drilling | $409,824 | $0 | $0 | $0 | $0 | $409,824 | | |
| 3.8 Other | | | | | | | | |
| 3.8.1 Mobilization | $556,000 | $0 | $0 | $0 | $0 | $556,000 | | |
| 3.8.2 Demobilization | $0 | $0 | $0 | $0 | $111,000 | $111,000 | | |
| 3.8.3 Engineering | $834,000 | $0 | $0 | $0 | $0 | $834,000 | | |
| Subtotals other | $1,390,000 | $0 | $0 | $0 | $111,000 | $1,501,000 | | |
| Contractor markup (15%) | $208,500 | $0 | $0 | $0 | $16,650 | $225,150 | | |
| Excise tax (2%) | $31,970 | $0 | $0 | $0 | $2,553 | $34,523 | | |
| Contingency (25%) | $347,500 | $0 | $0 | $0 | $27,750 | $375,250 | | |
| Totals other | $1,977,970 | $0 | $0 | $0 | $157,953 | $2,135,923 | | |
| | | | | | | | | |
| 3.9 Lower campus systems | | | | | | | | |
| 3.9.1 4850: 9700m^3@$250/m^3 | $0 | $0 | $970,000 | $1,455,000 | $0 | $2,425,000 | | |
| 3.9.2 7400:74765m^3@$325/m^3 | $0 | $1,869,000 | $11,215,000 | $9,346,000 | $1,869,000 | $24,299,000 | | |
| 3.9.3 8000: 350m^3@$600/m^3 | $0 | $210,000 | $0 | $0 | $0 | $210,000 | | |
| Subtotals low campus systems | $0 | $2,079,000 | $12,185,000 | $10,801,000 | $1,869,000 | $26,934,000 | | |
| Contractor markup (15%) | $0 | $311,850 | $1,827,750 | $1,620,150 | $280,350 | $4,040,100 | | |
| Excise tax (2%) | $0 | $47,817 | $280,255 | $248,423 | $42,987 | $619,482 | | |
| Contingency (40%) | $0 | $831,600 | $4,874,000 | $4,320,400 | $747,600 | $10,773,600 | | |
| Totals low campus systems | $0 | $3,270,267 | $19,167,005 | $16,989,973 | $2,939,937 | $42,367,182 | | |
| | | | | | | | | |
| 3.10 Sealing unused areas | | | | | | | | |
| 3.10.1 Sealing | $400,000 | $160,000 | $165,000 | $85,000 | $0 | $810,000 | | |
| Subtotals sealing | $400,000 | $160,000 | $165,000 | $85,000 | $0 | $810,000 | | |
| Contractor markup (15%) | $60,000 | $24,000 | $24,750 | $12,750 | $0 | $121,500 | | |
| Excise tax (2%) | $9,200 | $3,680 | $3,795 | $1,955 | $0 | $18,630 | | |
| Contingency (50%) | $200,000 | $80,000 | $82,500 | $42,500 | $0 | $405,000 | | |
| Totals sealing | $669,200 | $267,680 | $276,045 | $142,205 | $0 | $1,355,130 | | |
| | | | | | | | | |
| Totals 1:10 | $31,799,407 | $30,346,358 | $41,422,956 | $22,249,080 | $3,097,890 | $128,915,691 | | |
| EDIA (8%) | $2,543,953 | $2,427,709 | $3,313,836 | $1,779,926 | $247,831 | $10,313,255 | | |
| | | | | | | | | |
| **Totals Undergrd Development** | $34,343,360 | $32,774,067 | $44,736,792 | $24,029,006 | $3,345,721 | $139,228,946 | | |
| | | | | | | | | |
| **4 Surface Develop. Science** | | | | | | | | |
| | | | | | | | | |
| 4.1 Demolishing Ext. Structures | $1,000,000 | $1,030,000 | $1,591,000 | $1,639,000 | $0 | $5,260,000 | | |
| 4.2 Roads, parking | $0 | $309,000 | $318,000 | $492,000 | $506,000 | $1,625,000 | | |
| 4.3 Science/Admin building | | | | | | | | |
| 4.3.1 68.4K gsf office @ $150 | $0 | $2,052,000 | $3,078,000 | $3,078,000 | $2,052,000 | $10,260,000 | | |
| 4.3.2 36.0K gsf lab @ $250 | $0 | $1,800,000 | $2,700,000 | $2,700,000 | $1,800,000 | $9,000,000 | | |
| 4.3.3 56.25K gsf assem @ $120 | $0 | $1,350,000 | $2,025,000 | $2,025,000 | $1,350,000 | $6,750,000 | | |
| 4.4 Upper campus systems | $300,000 | $309,000 | $955,000 | $983,000 | $675,000 | $3,222,000 | | |
| Subtotal surface sci. develop. | $1,300,000 | $6,850,000 | $10,667,000 | $10,917,000 | $6,383,000 | $36,117,000 | | |
| Excise tax (2%) | $26,000 | $137,000 | $213,340 | $218,340 | $127,660 | $722,340 | | |
| Contingency (40%) | $520,000 | $2,740,000 | $4,266,800 | $4,366,800 | $2,553,200 | $14,446,800 | | |
| Totals | $1,846,000 | $9,727,000 | $15,147,140 | $15,502,140 | $9,063,860 | $51,286,140 | | |
| EDIA (8%) | $147,680 | $778,160 | $1,211,771 | $1,240,171 | $725,109 | $4,102,891 | | |
| | | | | | | | | |
| **Totals Surf. Develop. Science** | $1,993,680 | $10,505,160 | $16,358,911 | $16,742,311 | $9,788,969 | $55,389,031 | | |
| | | | | | | | | |
| **5 Site Operations/Maintenance** | | | | | | | | |
| | | | | | | | | |
| 5.1 Capital equipment | $500,000 | $500,000 | $0 | $0 | $0 | $1,000,000 | | |
| Contingency (40%) | $200,000 | $200,000 | $0 | $0 | $0 | $400,000 | | |
| Subtotals | $700,000 | $700,000 | $0 | $0 | $0 | $1,400,000 | | |
| | | | | | | | | |
| 5.2 Operations | | | | | | | | |
| 5.2.1 Haulage | $150,000 | $150,000 | $150,000 | $150,000 | $150,000 | $750,000 | | |
| 5.2.2 Other mine operating | $147,000 | $147,000 | $147,000 | $147,000 | $147,000 | $735,000 | | |
| 5.2.3 Mine general | $279,000 | $279,000 | $279,000 | $279,000 | $279,000 | $1,395,000 | | |
| 5.2.4 Ventilation/cooling | $691,000 | $691,000 | $691,000 | $691,000 | $691,000 | $3,455,000 | | |
| 5.2.5 Hoists/shafts | $507,000 | $507,000 | $507,000 | $507,000 | $507,000 | $2,535,000 | | |
| 5.2.6 Waste water | $88,000 | $88,000 | $88,000 | $88,000 | $88,000 | $440,000 | | |
| 5.2.7 Administration/general | $617,000 | $617,000 | $617,000 | $617,000 | $617,000 | $3,085,000 | | |
| Subtotals | $2,479,000 | $2,479,000 | $2,479,000 | $2,479,000 | $2,479,000 | $12,395,000 | | |
| Contingency (25%) | $619,750 | $619,750 | $619,750 | $619,750 | $619,750 | $3,098,750 | | |
| Subtotals | $3,098,750 | $3,098,750 | $3,098,750 | $3,098,750 | $3,098,750 | $15,493,750 | | |
| | | | | | | | | |
| 5.3 Operations personnel | | | | | | | | |
| 5.3.1 Operations manager | $100,000 | $100,000 | $100,000 | $100,000 | $100,000 | $500,000 | | $100.00 |
| 5.3.2 Other managers (2) | $150,000 | $150,000 | $150,000 | $150,000 | $150,000 | $750,000 | | $75,000 |
| 5.3.3 Engineers (2.25-3) | $146,250 | $195,000 | $195,000 | $195,000 | $195,000 | $926,250 | | $65,000 |
| 5.3.4 Technicians (1.75-3) | $78,750 | $135,000 | $135,000 | $135,000 | $135,000 | $618,750 | | $45,000 |
| 5.3.5 Supervisors (1.75-2) | $96,250 | $110,000 | $110,000 | $110,000 | $110,000 | $536,250 | | $55,000 |
| 5.3.6 Secretarial/clerical (1) | $25,000 | $25,000 | $25,000 | $25,000 | $25,000 | $125,000 | | $25,000 |
| 5.3.7 Accounting (1.5-2) | $67,500 | $90,000 | $90,000 | $90,000 | $90,000 | $427,500 | | $45,000 |
| 5.3.8 Operators (42.75-22) | $1,496,250 | $1,443,750 | $1,190,000 | $822,500 | $770,000 | $5,722,500 | | $35,000 |
| Subtotals | $2,160,000 | $2,248,750 | $1,995,000 | $1,627,500 | $1,575,000 | $9,606,250 | | |
| Inflation (+3%) | $2,160,000 | $2,316,213 | $2,116,496 | $1,778,413 | $1,772,676 | $10,143,798 | | |
| Subtotals (+ 30% benefits) | $2,808,000 | $3,011,076 | $2,751,444 | $2,311,937 | $2,304,479 | $13,186,937 | | |
| Contingency (10%) | $280,800 | $301,108 | $275,144 | $231,194 | $230,448 | $1,318,694 | | |
| Subtotals | $3,088,800 | $3,312,184 | $3,026,589 | $2,543,131 | $2,534,927 | $14,505,631 | | |
| | | | | | | | | |
| 5.4 Maintenance personnel | | | | | | | | |
| 5.4.1 Mech. Plant Engineer | $65,000 | $65,000 | $65,000 | $65,000 | $65,000 | $325,000 | | $65,000 |
| 5.4.2 Supervisors (2-3) | $137,500 | $165,000 | $165,000 | $165,000 | $165,000 | $797,500 | | $55,000 |
| 5.4.3 Maintenance planner | $45,000 | $45,000 | $45,000 | $45,000 | $45,000 | $225,000 | | $45,000 |
| 5.4.4 Electricians/Techs(14.5-16) | $580,000 | $640,000 | $640,000 | $640,000 | $640,000 | $3,140,000 | | $40,000 |
| Subtotals | $827,500 | $915,000 | $915,000 | $915,000 | $915,000 | $4,487,500 | | |
| Inflation (+3%) | $827,500 | $942,450 | $970,724 | $999,845 | $1,029,841 | $4,770,359 | | |
| Subtotals (+ 30% benefits) | $1,075,750 | $1,225,185 | $1,261,941 | $1,299,799 | $1,338,793 | $6,201,467 | | |
| Contingency (10%) | $107,575 | $122,519 | $126,194 | $129,980 | $133,879 | $620,147 | | |
| Subtotals | $1,183,325 | $1,347,704 | $1,388,135 | $1,429,779 | $1,472,672 | $6,821,614 | | |
| | | | | | | | | |
| Total direct costs | $8,070,875 | $8,458,637 | $7,513,473 | $7,071,660 | $7,106,349 | $38,220,994 | | |
| Indirect costs (15%) | $1,210,631 | $1,268,796 | $1,127,021 | $1,060,749 | $1,065,952 | $5,733,149 | | |
| | | | | | | | | |
| **Total, direct+indirect** | $9,281,506 | $9,727,433 | $8,640,494 | $8,132,408 | $8,172,302 | $43,954,143 | | |
| | | | | | | | | |
| **6 Science Operations** | | | | | | | | |
| | | | | | | | | |
| 6.1 Capital equipment | | | | | | | | |
| 6.1.1 LLCF equipment | $100,000 | $680,000 | $2,950,000 | $3,040,000 | $3,045,000 | $9,815,000 | | |
| Contingency (40%) | $40,000 | $272,000 | $1,180,000 | $1,216,000 | $1,218,000 | $3,926,000 | | |
| Subtotals | $140,000 | $952,000 | $4,130,000 | $4,256,000 | $4,263,000 | $13,741,000 | | |
| | | | | | | | | |
| 6.2 Science Op. Personnel | | | | | | | | |
| 6.2.1 Director (0.7) | $126,000 | $126,000 | $126,000 | $126,000 | $126,000 | $630,000 | | $180,000 |
| 6.2.1 Assoc. Directors (1-2) | $150,000 | $300,000 | $300,000 | $300,000 | $300,000 | $1,350,000 | | $150,000 |
| 6.2.2 Senior scientists (0-4) | $0 | $0 | $250,000 | $500,000 | $500,000 | $1,250,000 | | $125,000 |
| 6.2.3 Associate scientists (0-4) | $0 | $0 | $200,000 | $400,000 | $400,000 | $1,000,000 | | $100,000 |
| 6.2.4 Assistant scientists (0-4) | $0 | $0 | $150,000 | $300,000 | $300,000 | $750,000 | | $75,000 |
| 6.2.5 Postdocs (0-10) | $0 | $0 | $180,000 | $360,000 | $450,000 | $990,000 | | $45,000 |
| 6.2.6 Graduate students (0-10) | $0 | $0 | $80,000 | $160,000 | $200,000 | $440,000 | | $20,000 |
| 6.2.7 Undergraduates (0-5) | $0 | $0 | $30,000 | $60,000 | $75,000 | $165,000 | | $15,000 |
| 6.2.8 Librarians (0-2) | $0 | $40,000 | $80,000 | $80,000 | $80,000 | $280,000 | | $40,000 |
| 6.2.9 Machinists (0-4) | $0 | $40,000 | $80,000 | $160,000 | $160,000 | $440,000 | | $40,000 |
| 6.2.10 Draftsman (0-1) | $0 | $35,000 | $35,000 | $35,000 | $35,000 | $140,000 | | $35,000 |
| 6.2.11 Electronics (0-3) | $0 | $40,000 | $80,000 | $120,000 | $120,000 | $360,000 | | $40,000 |
| 6.2.12 Glass (0-1) | $0 | $40,000 | $40,000 | $40,000 | $40,000 | $160,000 | | $40,000 |
| 6.2.13 LLC (0-2) | $0 | $60,000 | $120,000 | $120,000 | $120,000 | $420,000 | | $60,000 |
| 6.2.14 Chemistry (0-5) | $0 | $50,000 | $100,000 | $150,000 | $250,000 | $550,000 | | $50,000 |
| 6.2.15 Computing (0-5) | $0 | $80,000 | $200,000 | $200,000 | $200,000 | $680,000 | | $40,000 |
| 6.2.16 Receiving (0-2) | $25,000 | $50,000 | $50,000 | $50,000 | $50,000 | $225,000 | | $25,000 |
| 6.2.17 Secretarial (0-4) | $25,000 | $50,000 | $100,000 | $100,000 | $100,000 | $375,000 | | $25,000 |
| Subtotals | $326,000 | $911,000 | $2,201,000 | $3,261,000 | $3,506,000 | $10,205,000 | | |
| Inflation (+3%) | $326,000 | $938,330 | $2,335,041 | $3,563,383 | $3,946,034 | $11,108,788 | | |
| Subtotals (+ 30% benefits) | $423,800 | $1,219,829 | $3,035,553 | $4,632,398 | $5,129,844 | $14,441,424 | | |
| Contingency (10%) | $42,380 | $121,983 | $303,555 | $463,240 | $512,984 | $1,444,142 | | |
| Subtotals | $466,180 | $1,341,812 | $3,339,108 | $5,095,637 | $5,642,828 | $15,885,566 | | |
| | | | | | | | | |
| 6.3 Other Science Operations | | | | | | | | |
| 6.3.1 Domestic travel | $14,000 | $16,000 | $36,000 | $56,000 | $60,000 | $182,000 | | |
| 6.3.2 Foreign travel | $4,000 | $6,000 | $18,000 | $30,000 | $30,000 | $88,000 | | |
| 6.3.3 Supplies | $10,000 | $15,000 | $20,000 | $25,000 | $30,000 | $100,000 | | |
| 6.3.5 Publications | $10,000 | $15,000 | $20,000 | $25,000 | $30,000 | $100,000 | | |
| 6.3.6 Consultants (PACs) | $10,000 | $20,000 | $20,000 | $20,000 | $20,000 | $90,000 | | |
| 6.3.5 Computer services | $10,000 | $30,000 | $30,000 | $30,000 | $30,000 | $130,000 | | |
| Subtotals | $58,000 | $102,000 | $144,000 | $186,000 | $200,000 | $690,000 | | |
| Contingency (40%) | $23,200 | $40,800 | $57,600 | $74,400 | $80,000 | $276,000 | | |
| Subtotals | $81,200 | $142,800 | $201,600 | $260,400 | $280,000 | $966,000 | | |
| | | | | | | | | |
| Total direct costs | $687,380 | $2,436,612 | $7,670,708 | $9,612,037 | $10,185,828 | $30,592,566 | | |
| Indirect costs (15%) | $103,107 | $365,492 | $1,150,606 | $1,441,806 | $1,527,874 | $4,588,885 | | |
| | | | | | | | | |
| **Total, direct+indirect** | $790,487 | $2,802,104 | $8,821,315 | $11,053,843 | $11,713,703 | $35,181,451 | | |
| | | | | | | | | |
| **7 Detector Operations** | | | | | | | | |
| | | | | | | | | |
| 7.1 Detector opers. personnel | | | | | | | | |
| 7.1.1 Assistant manager (0-1) | $0 | $0 | $75,000 | $75,000 | $75,000 | $225,000 | | $75,000 |
| 7.1.2 Mech/process eng. (0-2) | $0 | $65,000 | $130,000 | $130,000 | $130,000 | $455,000 | | $65,000 |
| 7.1.3 Maintenance mech. (0-2) | $0 | $0 | $80,000 | $80,000 | $80,000 | $240,000 | | $40,000 |
| 7.1.4 Profess. operators (0-3) | $0 | $40,000 | $40,000 | $120,000 | $120,000 | $320,000 | | $40,000 |
| 7.1.5 Undergr. assembly (0-3) | $0 | $0 | $40,000 | $120,000 | $120,000 | $280,000 | | $40,000 |
| 7.1.6 Large assembly (0-3) | $0 | $0 | $40,000 | $120,000 | $120,000 | $280,000 | | $40,000 |
| 7.1.7 Transportation (0-1) | $0 | $0 | $35,000 | $35,000 | $35,000 | $105,000 | | $35,000 |
| 7.1.8 Secretary (0-1) | $0 | $25,000 | $25,000 | $25,000 | $25,000 | $100,000 | | $25,000 |
| Subtotals | $0 | $130,000 | $465,000 | $705,000 | $705,000 | $2,005,000 | | |
| Inflation (+3%) | $0 | $133,900 | $493,319 | $770,373 | $793,484 | $2,191,075 | | |
| Subtotals (+ 30% benefits) | $0 | $174,070 | $641,314 | $1,001,484 | $1,031,529 | $2,848,397 | | |
| Contingency (10%) | $0 | $17,407 | $64,131 | $100,148 | $103,153 | $284,840 | | |
| Subtotals | $0 | $191,477 | $705,445 | $1,101,633 | $1,134,682 | $3,133,237 | | |
| | | | | | | | | |
| 7.2 Other detector operations | | | | | | | | |
| 7.2.1 Computer services | $0 | $30,000 | $35,000 | $40,000 | $45,000 | $150,000 | | |
| 7.2.2 Domestic travel | $0 | $0 | $4,000 | $6,000 | $6,000 | $16,000 | | |
| 7.2.3 Foreign travel | $0 | $0 | $2,000 | $3,000 | $3,000 | $8,000 | | |
| 7.2.4 Materials and supplies | $0 | $25,000 | $50,000 | $75,000 | $100,000 | $250,000 | | |
| Subtotals | $0 | $55,000 | $91,000 | $124,000 | $154,000 | $424,000 | | |
| Contingency (40%) | $0 | $22,000 | $36,400 | $49,600 | $61,600 | $169,600 | | |
| Subtotals | $0 | $77,000 | $127,400 | $173,600 | $215,600 | $593,600 | | |
| | | | | | | | | |
| Total direct costs | $0 | $268,477 | $832,845 | $1,275,233 | $1,350,282 | $3,726,837 | | |
| Indirect costs (15%) | $0 | $40,272 | $124,927 | $191,285 | $202,542 | $559,026 | | |
| | | | | | | | | |
| **Total, direct+indirect** | $0 | $308,749 | $957,772 | $1,466,518 | $1,552,824 | $4,285,862 | | |
| | | | | | | | | |
| **8 Director's Operations** | | | | | | | | |
| | | | | | | | | |
| 8.1 Site development group | | | | | | | | |
| 8.1.1 Manager (0.8) | $48,000 | $48,000 | $48,000 | $48,000 | $48,000 | $240,000 | | $60,000 |
| 8.1.2 Assistants (1.6) | $56,000 | $56,000 | $56,000 | $56,000 | $56,000 | $280,000 | | $35,000 |
| 8.1.3 Secretarial/clerical(0.8) | $20,000 | $20,000 | $20,000 | $20,000 | $20,000 | $100,000 | | $25,000 |
| Subtotals | $124,000 | $124,000 | $124,000 | $124,000 | $124,000 | $620,000 | | |
| Inflation (+3%) | $124,000 | $127,720 | $131,552 | $135,498 | $139,563 | $658,333 | | |
| Subtotals (+30% benefits) | $161,200 | $166,036 | $171,017 | $176,148 | $181,432 | $855,833 | | |
| Contingency (10%) | $16,120 | $16,604 | $17,102 | $17,615 | $18,143 | $85,583 | | |
| Subtotals | $177,320 | $182,640 | $188,119 | $193,762 | $199,575 | $941,416 | | |
| | | | | | | | | |
| 8.2 Administrative office | | | | | | | | |
| 8.2.1 Admin. Office director (1) | $70,000 | $70,000 | $70,000 | $70,000 | $70,000 | $350,000 | | $70,000 |
| 8.2.2 Personnel officer (1) | $60,000 | $60,000 | $60,000 | $60,000 | $60,000 | $300,000 | | $60,000 |
| 8.2.3 Financial officer (1) | $60,000 | $60,000 | $60,000 | $60,000 | $60,000 | $300,000 | | $60,000 |
| 8.2.4 Visitor liaison officer (1) | $0 | $60,000 | $60,000 | $60,000 | $60,000 | $240,000 | | $60,000 |
| 8.2.5 Public inform. officer (1) | $0 | $60,000 | $60,000 | $60,000 | $60,000 | $240,000 | | $60,000 |
| 8.2.6 Office assistants (4) | $60,000 | $120,000 | $120,000 | $120,000 | $120,000 | $540,000 | | $30,000 |
| 8.2.7 Security guards (8) | $120,000 | $240,000 | $240,000 | $240,000 | $240,000 | $1,080,000 | | $30,000 |
| Subtotals | $370,000 | $670,000 | $670,000 | $670,000 | $670,000 | $3,110,000 | | |
| Inflation (+3%) | $370,000 | $690,100 | $710,803 | $732,127 | $754,091 | $3,257,121 | | |
| Subtotals (+30% benefits) | $481,000 | $897,130 | $924,044 | $951,765 | $980,318 | $4,234,257 | | |
| Contingency (25%) | $120,250 | $224,283 | $231,011 | $237,941 | $245,080 | $1,058,564 | | |
| Subtotals | $601,250 | $1,121,413 | $1,155,055 | $1,189,707 | $1,225,398 | $5,292,822 | | |
| | | | | | | | | |
| 8.3 Contract services | | | | | | | | |
| 8.3.1 Cafeteria | $100,000 | $200,000 | $200,000 | $200,000 | $200,000 | $900,000 | | |
| 8.3.2 Custodial | $50,000 | $50,000 | $100,000 | $150,000 | $150,000 | $500,000 | | |
| 8.3.3 MSHA training | $35,000 | $35,000 | $35,000 | $35,000 | $35,000 | $175,000 | | |
| Subtotals | $185,000 | $285,000 | $335,000 | $385,000 | $385,000 | $1,575,000 | | |
| Inflation (+3%) | $185,000 | $293,550 | $355,401.50 | $420,700 | $433,321 | $1,687,972 | | |
| Subtotals (+30% benefits) | $240,500 | $381,615 | $462,021.95 | $546,910 | $563,317 | $2,194,364 | | |
| Contractor's markup (15%) | $36,075 | $57,242 | $69,303 | $82,036 | $84,498 | $329,155 | | |
| Subtotals | $276,575 | $438,857 | $531,325 | $628,946 | $647,815 | $2,523,519 | | |
| Contingency (40%) | $110,630 | $175,543 | $212,530 | $251,579 | $259,126 | $1,009,407 | | |
| Subtotals | $387,205 | $614,400 | $743,855 | $880,525 | $906,941 | $3,532,926 | | |
| | | | | | | | | |
| 8.4 Other operations | | | | | | | | |
| 8.4.1 Domestic travel | $22,000 | $26,000 | $26,000 | $26,000 | $26,000 | $126,000 | | |
| 8.4.2 Foreign travel | $12,000 | $16,000 | $16,000 | $16,000 | $16,000 | $76,000 | | |
| 8.4.3 Supplies | $15,000 | $25,000 | $25,000 | $25,000 | $25,000 | $115,000 | | |
| 8.4.4 Publications | $40,000 | $50,000 | $50,000 | $50,000 | $50,000 | $240,000 | | |
| 8.4.5 Consultants | $20,000 | $20,000 | $20,000 | $20,000 | $20,000 | $100,000 | | |
| 8.4.6 Computer services | $7,000 | $10,000 | $10,000 | $10,000 | $10,000 | $47,000 | | |
| Subtotals | $116,000 | $147,000 | $147,000 | $147,000 | $147,000 | $704,000 | | |
| Contingency (40%) | $46,400 | $58,800 | $58,800 | $58,800 | $58,800 | $281,600 | | |
| Subtotals | $162,400 | $205,800 | $205,800 | $205,800 | $205,800 | $985,600 | | |
| | | | | | | | | |
| Total direct costs | $1,328,175 | $2,124,252 | $2,292,829 | $2,469,794 | $2,537,714 | $10,752,764 | | |
| Indirect (15% - no contract) | $141,146 | $226,478 | $232,346 | $238,390 | $244,616 | $1,082,976 | | |
| | | | | | | | | |
| **Total, direct+indirect** | $1,469,321 | $2,350,730 | $2,525,175 | $2,708,184 | $2,782,330 | $11,835,739 | | |
| | | | | | | | | |
| | | | | | | | | |
| **LABORATORY COST SUMMARY: SCIENCE CONSTRUCTION AND OPERATIONS** | | | | | | | | |
| | | | | | | | | |
| **Property** | $1,800,000 | $0 | $0 | $0 | $0 | $1,800,000 | | |
| **Insurance/Gen. Conditions** | $0 | $0 | $0 | $0 | $0 | $0 | | |
| **Underground Development** | $34,343,360 | $32,774,067 | $44,736,792 | $24,029,006 | $3,345,721 | $139,228,946 | | |
| **Surface Develop. Science** | $1,993,680 | $10,505,160 | $16,358,911 | $16,742,311 | $9,788,969 | $55,389,031 | | |
| **Site Operations/Maintenance** | $9,281,506 | $9,727,433 | $8,640,494 | $8,132,408 | $8,172,302 | $43,954,143 | | |
| **Science Operations** | $790,487 | $2,802,104 | $8,821,315 | $11,053,843 | $11,713,703 | $35,181,451 | | |
| **Detector Operations** | $0 | $308,749 | $957,772 | $1,466,518 | $1,552,824 | $4,285,862 | | |
| **Director's Operations** | $1,469,321 | $2,350,730 | $2,525,175 | $2,708,184 | $2,782,330 | $11,835,739 | | |
| | | | | | | | | |
| **TOTALS** | **$49,678,353** | **$58,468,242** | **$82,040,460** | **$64,132,271** | **$37,355,848** | **$291,675,174** | | |

**WBS-1: BUDGET SUMMARY**

| | FY06 | FY07 | FY08 | FY09 | FY10 | FY06-10 |
|---|---|---|---|---|---|---|
| Total senior faculty | $303,600 | $482,658 | $1,197,332 | $1,954,452 | $2,013,085 | $5,951,126 |
| Postdoctoral associates | $0 | $0 | $210,058 | $432,720 | $557,127 | $1,199,905 |
| Graduate students | $0 | $0 | $93,359 | $192,320 | $247,612 | $533,291 |
| Undergraduates | $0 | $0 | $35,010 | $72,120 | $92,854 | $199,984 |
| Senior professional staff | $565,300 | $736,759 | $846,386 | $871,778 | $897,931 | $3,918,154 |
| Technical staff | $625,625 | $1,319,945 | $1,843,844 | $2,103,499 | $2,290,410 | $8,183,324 |
| Secret./receiv,/adm. asst. | $315,350 | $512,528 | $586,253 | $603,841 | $621,956 | $2,639,928 |
| Operators | $2,283,875 | $2,406,209 | $2,409,834 | $2,328,874 | $2,333,743 | $11,762,535 |
| | | | | | | |
| Total salaries | $4,093,750 | $5,458,099 | $7,222,077 | $8,559,604 | $9,054,718 | $34,388,248 |
| | | | | | | |
| Fringe benefits | $1,228,125 | $1,637,430 | $2,166,623 | $2,567,881 | $2,716,416 | $10,316,474 |
| | | | | | | |
| **Total salaries+benefits** | $5,321,875 | $7,095,528 | $9,388,700 | $11,127,485 | $11,771,134 | $44,704,722 |
| | | | | | | |
| Equipment | $840,000 | $1,652,000 | $4,130,000 | $4,256,000 | $4,263,000 | $15,141,000 |
| Domestic travel | $50,400 | $58,800 | $92,400 | $123,200 | $128,800 | $453,600 |
| Foreign travel | $22,400 | $30,800 | $50,400 | $68,600 | $68,600 | $240,800 |
| Materials and supplies | $35,000 | $91,000 | $133,000 | $175,000 | $217,000 | $651,000 |
| Publications | $70,000 | $91,000 | $98,000 | $105,000 | $112,000 | $476,000 |
| Consulting | $42,000 | $56,000 | $56,000 | $56,000 | $56,000 | $266,000 |
| Computer services | $23,800 | $98,000 | $105,000 | $112,000 | $119,000 | $457,800 |
| Security | $195,000 | $401,700 | $413,751 | $426,164 | $438,948 | $1,875,563 |
| Contract services | $387,205 | $614,400 | $743,855 | $880,525 | $906,941 | $3,532,926 |
| Utilities/operations costs | $3,098,750 | $3,098,750 | $3,098,750 | $3,098,750 | $3,098,750 | $15,493,750 |
| | | | | | | |
| Total other direct costs | $4,764,555 | $6,192,450 | $8,921,156 | $9,301,238 | $9,409,039 | $38,588,439 |
| | | | | | | |
| **Total direct costs** | $10,086,430 | $13,287,979 | $18,309,856 | $20,428,723 | $21,180,173 | $83,293,161 |
| Noncont. ind. costs (15%) | $1,454,884 | $1,901,037 | $2,634,900 | $2,932,230 | $3,040,985 | $11,964,035 |
| **Total direct+indir costs** | $11,541,314 | $15,189,015 | $20,944,756 | $23,360,953 | $24,221,158 | $95,257,196 |
| | | | | | | |
| Property/insurance costs | $1,800,000 | $0 | $0 | $0 | $0 | $1,800,000 |
| Construction contracts | $36,337,040 | $43,279,227 | $61,095,703 | $40,771,317 | $13,134,690 | $194,617,977 |
| | | | | | | |
| **Total costs** | $49,678,354 | $58,468,242 | $82,040,459 | $64,132,270 | $37,355,848 | $291,675,173 |

## WBS-2: LABORATORY EDUCATION/OUTREACH CONSTRUCTION AND OPERATIONS

**1 Property:** These costs are discussed in WBS-1. Our assumption is that the environmental assessment will be done initially for the entire site, and that the education and outreach program will not need to budget for this activity. However, the Visitor Experience Center and the parking it will require could involve substantial property acquisition, easement, and permitting costs.

**2 Insurance:** We anticipate that the lab will have to acquire additional liability insurance because of increased visitor traffic due to outreach and education activities. However we enter no cost because this will depend on the details of the property arrangement between NUSEL and the state of South Dakota and/or Barrick Gold.

**3 Underground (near-surface) Facility:** The conceptual proposal described a rather ambitious underground experience for visitors. This has been scaled back in the current proposal, in favor of an improved surface Visitor Experience Center that we feel presents fewer liability problems. We believe, however, that a good experience can be provided that helps to illustrate the region's geology and mining history, as well as Homestake's potential for science. Our concept is an underground entrance from the visitor center into a near-surface drift constructed to illustrate the geology of the site and mining techniques, as well as the use of underground space for science experiments.

Because of the proposed Mill Street site for the Visitor Experience Center, a relatively short drift could lead visitors from the Visitor Experience Center, underground, to an opening on the Open Cut, a rather spectacular open pit that marks the point where the Homestake gold deposit intersected the surface. A viewing area could be constructed at that point. This would make a rather spectacular exit for the Visitor Experience Center.

The proposed entrance to the underground experience will use a conventional elevator. A second elevator would return the visitors to the surface, near Lead's visitor center and its Open Cut viewing area. Care would be taken to provide all the necessary safety facilities and conveniences for the general public, including younger school children. The underground tours will be guided.

For this project $300,000 is provided for vertical access, the construction of a borehole from the visitor center to the 300-ft level and the equipping of that access with an elevator. Our estimate of the cost of constructing a drift from the base of this elevator to the existing 300-ft drift – a total distance of about 700 ft – is $350,000. We also provide $500,000 to tie back to the Ross shaft via the Powder Tunnel, so that access to the science shaft is also available. This last development is an option, and could be removed. Finally, we provide $350,000 for safety and display engineering along this underground walk.

### 4 Surface Development for Outreach

**4.1 *Demolishing existing structures*:** The budget provides $500,000 for demolishing existing structures on the Mill Street site and on areas designated for visitor parking.

**4.2 *Roads and parking*:** As noted in the WBS-1 discussion, an effort will be made to route traffic to the NUSEL site efficiently, with separate routes for staff, visitors, and deliveries. A significant fraction of the costs with be associated with the Visitor Experience Center because that will, by far, constitute the greatest source of traffic. As noted before, the road and parking engineering will include off-site parking and shuttle bus options that will help minimize the impact on both NUSEL and Lead. A total of $1.5M is budgeted for anticipated costs.

**4.3 *Visitor Experience Center*:** As discussed in the Facilities Development section, we estimate that the Visitor Experience Center will require about 80,000 gsf to accommodate an expected 200,000-400,000 visitors/year. This construction is budgeted at $150/sf, which we believe is adequate to finish the building with excellent display areas and a versatile theater. These parameters are not unlike those from the Lawrence Hall of Science, which accommodates about 260,000 visitors per year and places great emphasis on visiting school groups.

## 5 Outreach and Education Operations

**5.1** ***Outreach and Education personnel*****:** The personnel required to staff the Outreach and Education Office was discussed in the Facilities Development section. We anticipate that the Visitor Experience Center program will begin at the start of year five. The staff is increased slowly, according to the timeliness of various activities, with the Office Director, the manager of the Visitor Center, and the manager for computing and networking being brought onboard earlier, so that they can be involved in the physical planning of the Visitor Center. The major staff categories are the display design staff (3), the K-12 education staff (3), the visitor center staff (6), and the web/interactive display staff (4). These align with the major goals of the NUSEL education and outreach program. The year-five personnel costs are $4.6M.

**5.2** ***Other operations*****:** Modest travel and supplies and substantial publication costs are covered by this budget. The documentation/publication costs will cover brochures and other public information materials the Center produces to introduce itself to the region and nation.

## 6 Director's Operations

**6.1** ***Site development group:*** This group, also positioned in the Director's Office, is described in WBS-1 in section 8.1. The Visitor Experience Center, and more generally outreach and education, is expected to play a major role in site planning and in NUSEL-Lead and NUSEL-regional interactions. Thus 20% of this group's effort is considered WBS-2 effort.

**6.2** ***Contract services:*** The Visitor Experience Center will have a major impact on the cafeteria and on NUSEL's custodial needs. These contract costs are discussed in WBS-1, section 8.3. We have made an estimate of the Visitor Experience Center's share in these costs, beginning in year five.

**WBS-2: LABORATORY EDUCATION/OUTREACH CONSTRUCTION AND OPERATIONS**

| | Year 1 | Year 2 | Year 3 | Year 4 | Year 5 | Years 1-5 | | Salary basis ($/FTE/y) |
|---|---|---|---|---|---|---|---|---|
| **1 Property** | | | | | | | | |
| 1.1 Property acquisition | $200,000 | $0 | $0 | $0 | $0 | $200,000 | | |
| 1.2 Environmental | $0 | $0 | $0 | $0 | $0 | $0 | | |
| 1.3 Easements | $50,000 | $0 | $0 | $0 | $0 | $50,000 | | |
| 1.4 Permits | $100,000 | $0 | $0 | $0 | $0 | $100,000 | | |
| Subtotals | $350,000 | $0 | $0 | $0 | $0 | $350,000 | | |
| Contingency (50%) | $175,000 | $0 | $0 | $0 | $0 | $175,000 | | |
| | | | | | | | | |
| **Totals Property** | $525,000 | $0 | $0 | $0 | $0 | $525,000 | | |
| | | | | | | | | |
| **2 Insurance** | | | | | | | | |
| 2.1 Liability insurance | $0 | $0 | $0 | $0 | $0 | $0 | | |
| Contingency (50%) | $0 | $0 | $0 | $0 | $0 | $0 | | |
| | | | | | | | | |
| **Totals Insurance** | $0 | $0 | $0 | $0 | $0 | $0 | | |
| | | | | | | | | |
| **3 Underground (near surface) Facility** | | | | | | | | |
| | | | | | | | | |
| 3.1 Near-surface visitor facility | | | | | | | | |
| 3.1.1 Ramp/drift excavation | $0 | $0 | $450,000 | $900,000 | $0 | $1,350,000 | | |
| 3.1.2 Safety/conveniences | $0 | $0 | $0 | $350,000 | $0 | $350,000 | | |
| Subtotals | $0 | $0 | $450,000 | $1,250,000 | $0 | $1,700,000 | | |
| Contractor markup (15%) | $0 | $0 | $67,500 | $187,500 | $0 | $255,000 | | |
| Excise tax (2%) | $0 | $0 | $10,350 | $28,750 | $0 | $39,100 | | |
| Contingency (40%) | $0 | $0 | $180,000 | $500,000 | $0 | $680,000 | | |
| Totals near-surface facility | $0 | $0 | $707,850 | $1,966,250 | $0 | $2,674,100 | | |
| EDIA (8%) | $0 | $0 | $56,628 | $157,300 | $0 | $213,928 | | |
| | | | | | | | | |
| **Totals Near-surface Facility** | $0 | $0 | $764,478 | $2,123,550 | $0 | $2,888,028 | | |
| | | | | | | | | |
| **4 Surface Devel. Outreach** | | | | | | | | |
| | | | | | | | | |
| 4.1 Demolishing Ext. Structures | $0 | $500,000 | $0 | $0 | $0 | $500,000 | | |
| 4.2 Roads, parking | $0 | $500,000 | $500,000 | $500,000 | $0 | $1,500,000 | | |
| 4.3 Visitor Experience Center | | | | | | | | |
| 4.3.1 80.0K gsf office @ $150 | $0 | $2,400,000 | $3,600,000 | $3,600,000 | $2,400,000 | $12,000,000 | | |
| Subtotal surface sci. develop. | $0 | $3,400,000 | $4,100,000 | $4,100,000 | $2,400,000 | $14,000,000 | | |
| Excise tax (2%) | $0 | $68,000 | $82,000 | $82,000 | $48,000 | $280,000 | | |
| Contingency (40%) | $0 | $1,360,000 | $1,640,000 | $1,640,000 | $960,000 | $5,600,000 | | |
| Totals | $0 | $4,828,000 | $5,822,000 | $5,822,000 | $3,408,000 | $19,880,000 | | |
| EDIA (8%) | $0 | $386,240 | $465,760 | $465,760 | $272,640 | $1,590,400 | | |
| | | | | | | | | |
| **Totals Surf. Devel. Outreach** | $0 | $5,214,240 | $6,287,760 | $6,287,760 | $3,680,640 | $21,470,400 | | |
| | | | | | | | | |
| **5. Outreach/Education Oper.** | | | | | | | | |
| | | | | | | | | |
| 5.1 Outreach/Educ. Personnel | | | | | | | | |
| 5.1.1 Director (0.2) | $36,000 | $36,000 | $36,000 | $36,000 | $36,000 | $180,000 | | $180,000 |
| 5.1.2 Director, Ed/Out Office | $80,000 | $80,000 | $80,000 | $80,000 | $80,000 | $400,000 | | $80,000 |
| 5.1.3 Manager, Visitor Center | $0 | $70,000 | $70,000 | $70,000 | $70,000 | $280,000 | | $70,000 |
| 5.1.4 Manager, computing/net | $0 | $70,000 | $70,000 | $70,000 | $70,000 | $280,000 | | $70,000 |
| 5.1.5 Manager, K-12 education | $0 | $0 | $70,000 | $70,000 | $70,000 | $210,000 | | $70,000 |
| 5.1.6 Display design staff (0-3) | $0 | $0 | $0 | $105,000 | $105,000 | $210,000 | | $35,000 |
| 5.1.7 K-12 education staff (0-3) | $0 | $0 | $0 | $52,500 | $105,000 | $157,500 | | $35,000 |
| 5.1.8 Visitor Center staff (0-6) | $0 | $0 | $0 | $90,000 | $180,000 | $270,000 | | $30,000 |
| 5.1.9 Web/int. display staff (0-4) | $0 | $0 | $90,000 | $90,000 | $180,000 | $360,000 | | $45,000 |
| 5.1.10 Secretarial/clerical (1-2) | $25,000 | $25,000 | $50,000 | $50,000 | $50,000 | $200,000 | | $25,000 |
| Subtotals | $141,000 | $281,000 | $466,000 | $713,500 | $946,000 | $2,547,500 | | |
| Inflation (+3%) | $141,000 | $289,430 | $494,379 | $779,661 | $1,064,731 | $2,769,201 | | |
| Subtotals (+ 30% benefits) | $183,300 | $376,259 | $642,693 | $1,013,559 | $1,384,151 | $3,599,962 | | |
| Contingency (10%) | $18,330 | $37,626 | $64,269 | $101,356 | $138,415 | $359,996 | | |
| Subtotals | $201,630 | $413,885 | $706,963 | $1,114,915 | $1,522,566 | $3,959,958 | | |
| | | | | | | | | |
| 5.2 Other Out/Ed Operations | | | | | | | | |
| 5.2.1 Domestic travel | $2,000 | $6,000 | $8,000 | $8,000 | $8,000 | $32,000 | | |
| 5.2.2 Supplies | $2,000 | $6,000 | $8,000 | $16,000 | $18,000 | $50,000 | | |
| 5.2.3 Documentation/Publication | $2,000 | $6,000 | $8,000 | $16,000 | $18,000 | $50,000 | | |
| 5.2.4 Consultants | $10,000 | $12,000 | $14,000 | $14,000 | $14,000 | $64,000 | | |
| 5.2.5 Computer services | $2,000 | $6,000 | $8,000 | $18,000 | $18,000 | $52,000 | | |
| Subtotals | $18,000 | $36,000 | $46,000 | $72,000 | $76,000 | $248,000 | | |
| Contingency (40%) | $7,200 | $14,400 | $18,400 | $28,800 | $30,400 | $99,200 | | |
| Subtotals | $25,200 | $50,400 | $64,400 | $100,800 | $106,400 | $347,200 | | |
| | | | | | | | | |
| Total direct costs | $226,830 | $464,285 | $771,363 | $1,215,715 | $1,628,966 | $4,307,158 | | |
| Indirect costs (15%) | $34,025 | $69,643 | $115,704 | $182,357 | $244,345 | $646,074 | | |
| | | | | | | | | |
| **Total, direct+indirect** | $260,855 | $533,928 | $887,067 | $1,398,072 | $1,873,311 | $4,953,232 | | |
| | | | | | | | | |
| **6 Director's Operations** | | | | | | | | |
| | | | | | | | | |
| 6.1 Site development group | | | | | | | | |
| 6.1.1 Manager (0.2) | $12,000 | $12,000 | $12,000 | $12,000 | $12,000 | $60,000 | | $60,000 |
| 6.1.2 Assistants (0.4) | $14,000 | $14,000 | $14,000 | $14,000 | $14,000 | $70,000 | | $35,000 |
| 6.1.3 Secretarial/clerical(0.2) | $5,000 | $5,000 | $5,000 | $5,000 | $5,000 | $25,000 | | $25,000 |
| Subtotals | $31,000 | $31,000 | $31,000 | $31,000 | $31,000 | $155,000 | | |
| Inflation (+3%) | $31,000 | $31,930 | $32,888 | $33,875 | $34,891 | $164,583 | | |
| Subtotals (+30% benefits) | $40,300 | $41,509 | $42,754 | $44,037 | $45,358 | $213,958 | | |
| Contingency (10%) | $4,030 | $4,151 | $4,275 | $4,404 | $4,536 | $21,396 | | |
| Subtotals | $44,330 | $45,660 | $47,030 | $48,441 | $49,894 | $235,354 | | |
| | | | | | | | | |
| 6.2 Contract services | | | | | | | | |
| 6.2.1 Cafeteria | $0 | $0 | $0 | $0 | $100,000 | $100,000 | | |
| 6.2.2 Custodial | $0 | $0 | $0 | $0 | $75,000 | $75,000 | | |
| Subtotals | $0 | $0 | $0 | $0 | $175,000 | $175,000 | | |
| Inflation (+3%) | $0 | $0 | $0 | $0 | $196,964 | $196,964 | | |
| Subtotals (+30% benefits) | $0 | $0 | $0 | $0 | $256,053 | $256,053 | | |
| Contractor's markup (15%) | $0 | $0 | $0 | $0 | $38,408 | $38,408 | | |
| Subtotals | $0 | $0 | $0 | $0 | $294,461 | $294,461 | | |
| Contingency (40%) | $0 | $0 | $0 | $0 | $117,784 | $117,784 | | |
| Subtotals | $0 | $0 | $0 | $0 | $412,246 | $412,246 | | |
| | | | | | | | | |
| Total direct costs | $44,330 | $45,660 | $47,030 | $48,441 | $462,140 | $647,600 | | |
| Indirect (15% - no contract) | $6,650 | $6,849 | $7,054 | $7,266 | $7,484 | $35,303 | | |
| | | | | | | | | |
| **Total, direct+indirect** | $50,980 | $52,509 | $54,084 | $55,707 | $469,624 | $682,903 | | |
| | | | | | | | | |
| **LABORATORY COST SUMMARY: EDUCATION AND OUTREACH CONSTRUCTION AND OPERATIONS** | | | | | | | | |
| | | | | | | | | |
| **Property** | $525,000 | $0 | $0 | $0 | $0 | $525,000 | | |
| **Insurance** | $0 | $0 | $0 | $0 | $0 | $0 | | |
| **Underground (near surface)** | $0 | $0 | $764,478 | $2,123,550 | $0 | $2,888,028 | | |
| **Surface Devel. Ed/Outreach** | $0 | $5,214,240 | $6,287,760 | $6,287,760 | $3,680,640 | $21,470,400 | | |
| **Ed/Outreach Operations** | $260,855 | $533,928 | $887,067 | $1,398,072 | $1,873,311 | $4,953,232 | | |
| **Director's Operations** | $50,980 | $52,509 | $54,084 | $55,707 | $469,624 | $682,903 | | |
| | | | | | | | | |
| **TOTALS** | **$836,834** | **$5,800,677** | **$7,993,389** | **$9,865,089** | **$6,023,574** | **$30,519,563** | | |

**WBS-2: BUDGET SUMMARY**

| | FY0 | FY0 | FY0 | FY0 | FY1 | FY06-10 |
|---|---|---|---|---|---|---|
| Total senior faculty | $0 | $0 | $0 | $0 | $0 | $0 |
| Postdoctoral associates | $0 | $0 | $0 | $0 | $0 | $0 |
| Graduate students | $0 | $0 | $0 | $0 | $0 | $0 |
| Undergraduates | $0 | $0 | $0 | $0 | $0 | $0 |
| Senior professional staff | $140,800 | $303,644 | $394,443 | $406,276 | $418,464 | $1,663,627 |
| Technical staff | $0 | $0 | $105,029 | $405,675 | $705,694 | $1,216,398 |
| Secret./receiv,/adm. asst. | $48,400 | $49,852 | $80,522 | $82,938 | $85,426 | $347,138 |
| Operators | $0 | $0 | $0 | $0 | $0 | $0 |
| Total salaries | $189,200 | $353,496 | $579,994 | $894,889 | $1,209,584 | $3,227,163 |
| Fringe benefits | $56,760 | $106,049 | $173,998 | $268,467 | $362,875 | $968,149 |
| **Total salaries+benefits** | $245,960 | $459,545 | $753,992 | $1,163,355 | $1,572,460 | $4,195,312 |
| Equipment | $0 | $0 | $0 | $0 | $0 | $0 |
| Domestic travel | $2,800 | $8,400 | $11,200 | $11,200 | $11,200 | $44,800 |
| Foreign travel | $0 | $0 | $0 | $0 | $0 | $0 |
| Materials and supplies | $2,800 | $8,400 | $11,200 | $22,400 | $25,200 | $70,000 |
| Publications | $2,800 | $8,400 | $11,200 | $22,400 | $25,200 | $70,000 |
| Consulting | $14,000 | $16,800 | $19,600 | $19,600 | $19,600 | $89,600 |
| Computer services | $2,800 | $8,400 | $11,200 | $25,200 | $25,200 | $72,800 |
| Security | $0 | $0 | $0 | $0 | $0 | $0 |
| Contract services | $0 | $0 | $0 | $0 | $412,246 | $412,246 |
| Utilities/operations costs | $0 | $0 | $0 | $0 | $0 | $0 |
| Total other direct costs | $25,200 | $50,400 | $64,400 | $100,800 | $518,646 | $759,446 |
| **Total direct costs** | $271,160 | $509,945 | $818,392 | $1,264,155 | $2,091,105 | $4,954,758 |
| Noncontract ind. costs | $40,674 | $76,492 | $122,759 | $189,623 | $251,829 | $681,377 |
| **Total direct+indirect costs** | $311,834 | $586,437 | $941,151 | $1,453,779 | $2,342,934 | $5,636,135 |
| Property/insurance costs | $525,000 | $0 | $0 | $0 | $0 | $525,000 |
| Construction contracts | $0 | $5,214,240 | $7,052,238 | $8,411,310 | $3,680,640 | $24,358,428 |
| **Total costs** | $836,834 | $5,800,677 | $7,993,389 | $9,865,089 | $6,023,574 | $30,519,563 |

**WBS-3: LABORATORY MANAGEMENT**

**1 Management Site Office**

**1.1** ***Personnel*****:** If URA or a similar management entity holds the cooperative agreement for NUSEL, it will be essential for that entity to have an onsite office. This will allow the manager to work with its two principal partners, the NUSEL Director and the State of South Dakota Site Office. WBS-3 provides a Management Site Office director, an administrative assistant, and some personnel support for the DC office.

**1.2** ***Other costs:*** Funds for travel, supplies, documentation/publication, consultants, insurance, and audit services are provided. Substantial travel and consultant funding will be needed for Board of Governors meetings and for periodic reviews on NUSEL construct and operations. One-sixth of URA corporate General and Administrative costs is attributed to NUSEL. This assumes that NUSEL will be about 20% the size of Fermilab. The DOE requires that URA corporate G&A costs be shared proportionately among all contracts. Details in applying this policy to NUSEL will depend on what, if any, involvement or arrangements DOE might have with NSF and URA on NUSEL. No URA management fee is entered. This fee will be negotiated with NSF.

**2 South Dakota "Landlord's" Office**

**1.1.** ***Personnel:*** We envision that, in the memorandum of understanding between management and the site office, adequate funding will be provided to the South Dakota so that it can maintain a site office. This will allow the state to monitor NUSEL activities to make sure they conform to the site agreement and to other state laws.

***1.2 Other costs:*** Funding for standard activities is provided. Note that state participation in the Board of Governors and its relevant subcommittees, as described in the management plan, will be funded by management.

**WBS-3: LABORATORY MANAGEMENT**

| | Year 1 | Year 2 | Year 3 | Year 4 | Year 5 | Years 1-5 | | Salary basis ($/FTE/y) |
|---|---|---|---|---|---|---|---|---|
| **1 Management office** | | | | | | | | |
| | | | | | | | | |
| 1.1 Personnel | | | | | | | | |
| 1.1.1 Laboratory director (0.1) | $18,000 | $18,000 | $18,000 | $18,000 | $18,000 | $90,000 | | $180,000 |
| 1.1.2 Site office director (1.0) | $120,000 | $120,000 | $120,000 | $120,000 | $120,000 | $600,000 | | $120,000 |
| 1.1.3 Site office adm. asst. (1.0) | $30,000 | $30,000 | $30,000 | $30,000 | $30,000 | $150,000 | | $30,000 |
| 1.1.4 Other office personnel | $50,000 | $50,000 | $50,000 | $50,000 | $50,000 | $250,000 | | |
| Subtotals | $218,000 | $218,000 | $218,000 | $218,000 | $218,000 | $1,090,000 | | |
| Inflation (+3%) | $218,000 | $224,540 | $231,276 | $238,214 | $245,361 | $1,157,392 | | |
| Subtotals (+ 30% benefits) | $283,400 | $291,902 | $300,659 | $309,679 | $318,969 | $1,504,609 | | |
| Contingency (25%) | $70,850 | $72,976 | $75,165 | $77,420 | $79,742 | $376,152 | | |
| Subtotals | $354,250 | $364,878 | $375,824 | $387,099 | $398,711 | $1,880,761 | | |
| | | | | | | | | |
| 1.2 Other costs | | | | | | | | |
| 1.2.1 BOO meetings | $75,000 | $75,000 | $75,000 | $75,000 | $75,000 | $375,000 | | |
| 1.2.2 Site office supplies | $20,000 | $20,000 | $20,000 | $20,000 | $20,000 | $100,000 | | |
| 1.2.3 Documentation/Publication | $20,000 | $20,000 | $20,000 | $20,000 | $20,000 | $100,000 | | |
| 1.2.4 Consultants | $10,000 | $10,000 | $10,000 | $10,000 | $10,000 | $50,000 | | |
| 1.2.5 User Exec. Com. Travel | $15,000 | $15,000 | $15,000 | $15,000 | $15,000 | $75,000 | | |
| 1.2.6 Other travel to/from DC | $10,000 | $10,000 | $10,000 | $10,000 | $10,000 | $50,000 | | |
| 1.2.7 Insurance | $30,000 | $30,000 | $30,000 | $30,000 | $30,000 | $150,000 | | |
| 1.2.8 Internal audit function | $60,000 | $60,000 | $60,000 | $60,000 | $60,000 | $300,000 | | |
| 1.2.9 Additional DC costs | $10,000 | $10,000 | $10,000 | $10,000 | $10,000 | $50,000 | | |
| 1.2.10 URA corp. G&A costs | $300,000 | $300,000 | $300,000 | $300,000 | $300,000 | $1,500,000 | | |
| 1.2.11 Total fee (to be determ.) | $0 | $0 | $0 | $0 | $0 | $0 | | |
| 1.2.12 Audit | $50,000 | $50,000 | $50,000 | $50,000 | $50,000 | $250,000 | | |
| Subtotals | $600,000 | $600,000 | $600,000 | $600,000 | $600,000 | $3,000,000 | | |
| Contingency (25%) | $150,000 | $150,000 | $150,000 | $150,000 | $150,000 | $750,000 | | |
| Subtotals | $750,000 | $750,000 | $750,000 | $750,000 | $750,000 | $3,750,000 | | |
| | | | | | | | | |
| Total direct costs | $1,104,250 | $1,114,878 | $1,125,824 | $1,137,099 | $1,148,711 | $5,630,761 | | |
| Indirect costs (15%) | $165,638 | $167,232 | $168,874 | $170,565 | $172,307 | $844,614 | | |
| **Total, direct+indirect** | $1,269,888 | $1,282,109 | $1,294,697 | $1,307,663 | $1,321,018 | $6,475,376 | | |
| | | | | | | | | |
| **2 Landlord (SD) site office** | | | | | | | | |
| | | | | | | | | |
| 2.1 Personnel | | | | | | | | |
| 2.1.1 Site office director (1.0) | $120,000 | $120,000 | $120,000 | $120,000 | $120,000 | $600,000 | | $120,000 |
| 2.1.2 Site office adm. Asst. (1.0) | $30,000 | $30,000 | $30,000 | $30,000 | $30,000 | $150,000 | | $30,000 |
| 2.1.3 Other personnel | $50,000 | $50,000 | $50,000 | $50,000 | $50,000 | $250,000 | | |
| Subtotals | $200,000 | $200,000 | $200,000 | $200,000 | $200,000 | $1,000,000 | | |
| Inflation (+3%) | $200,000 | $206,000 | $212,180 | $218,545 | $225,102 | $1,061,827 | | |
| Subtotals (+30% benefits) | $260,000 | $267,800 | $275,834 | $284,109 | $292,632 | $1,380,375 | | |
| Contingency (25%) | $65,000 | $66,950 | $68,959 | $71,027 | $73,158 | $345,093.83 | | |
| Subtotals | $325,000 | $334,750 | $344,793 | $355,136 | $365,790 | $1,725,469 | | |
| | | | | | | | | |
| 2.2 Other costs | | | | | | | | |
| 1.2.1 Domestic travel | $20,000 | $20,000 | $20,000 | $20,000 | $20,000 | $100,000 | | |
| 1.2.2 Supplies | $20,000 | $20,000 | $20,000 | $20,000 | $20,000 | $100,000 | | |
| 1.2.3 Documentation | $10,000 | $10,000 | $10,000 | $10,000 | $10,000 | $50,000 | | |
| 1.2.4 Computer services | $10,000 | $10,000 | $10,000 | $10,000 | $10,000 | $50,000 | | |
| Subtotals | $60,000 | $60,000 | $60,000 | $60,000 | $60,000 | $300,000 | | |
| Contingency (40%) | $24,000 | $24,000 | $24,000 | $24,000 | $24,000 | $120,000 | | |
| Subtotals | $84,000 | $84,000 | $84,000 | $84,000 | $84,000 | $420,000 | | |
| | | | | | | | | |
| Total direct costs | $409,000 | $418,750 | $428,793 | $439,136 | $449,790 | $2,145,469 | | |
| Indirect costs (15%) | $61,350 | $62,813 | $64,319 | $65,870 | $67,469 | $321,820 | | |
| | | | | | | | | |
| **Total, direct+indirect** | $470,350 | $481,563 | $493,111 | $505,007 | $517,259 | $2,467,290 | | |
| | | | | | | | | |
| | | | | | | | | |
| **LABORATORY COST SUMMARY: MANAGEMENT** | | | | | | | | |
| | | | | | | | | |
| **Management site office** | $1,269,888 | $1,282,109 | $1,294,697 | $1,307,663 | $1,321,018 | $6,475,376 | | |
| **Landlord (SD) site office** | $470,350 | $481,563 | $493,111 | $505,007 | $517,259 | $2,467,290 | | |
| | | | | | | | | |
| **TOTALS** | **$1,740,238** | **$1,763,672** | **$1,787,809** | **$1,812,670** | **$1,838,277** | **$8,942,665** | | |

**WBS-3: BUDGET SUMMARY**

| | FY06 | FY07 | FY08 | FY09 | FY10 | FY06-10 |
|---|---|---|---|---|---|---|
| Total senior faculty | $22,500 | $23,175 | $23,870 | $24,586 | $25,324 | $119,456 |
| Postdoctoral associates | $0 | $0 | $0 | $0 | $0 | $0 |
| Graduate students | $0 | $0 | $0 | $0 | $0 | $0 |
| Undergraduates | $0 | $0 | $0 | $0 | $0 | $0 |
| Senior professional staff | $425,000 | $437,750 | $450,883 | $464,409 | $478,341 | $2,256,383 |
| Technical staff | $0 | $0 | $0 | $0 | $0 | $0 |
| Secret./receiv,/adm. asst. | $75,000 | $77,250 | $79,568 | $81,955 | $84,413 | $398,185 |
| Operators | $0 | $0 | $0 | $0 | $0 | $0 |
| | | | | | | |
| Total salaries | $522,500 | $538,175 | $554,320 | $570,950 | $588,078 | $2,774,023 |
| | | | | | | |
| Fringe benefits | $156,750 | $161,453 | $166,296 | $171,285 | $176,424 | $832,207 |
| | | | | | | |
| **Total salaries+benefits** | $679,250 | $699,628 | $720,616 | $742,235 | $764,502 | $3,606,230 |
| | | | | | | |
| Equipment | $0 | $0 | $0 | $0 | $0 | $0 |
| Domestic travel | $153,000 | $153,000 | $153,000 | $153,000 | $153,000 | $765,000 |
| Foreign travel | $0 | $0 | $0 | $0 | $0 | $0 |
| Materials and supplies | $65,500 | $65,500 | $65,500 | $65,500 | $65,500 | $327,500 |
| Publications | $39,000 | $39,000 | $39,000 | $39,000 | $39,000 | $195,000 |
| Consulting | $12,500 | $12,500 | $12,500 | $12,500 | $12,500 | $62,500 |
| Computer services | $14,000 | $14,000 | $14,000 | $14,000 | $14,000 | $70,000 |
| Security | $0 | $0 | $0 | $0 | $0 | $0 |
| Contract services | $512,500 | $512,500 | $512,500 | $512,500 | $512,500 | $2,562,500 |
| Utilities/operations costs | $0 | $0 | $0 | $0 | $0 | $0 |
| | | | | | | |
| Total other direct costs | $796,500 | $796,500 | $796,500 | $796,500 | $796,500 | $3,982,500 |
| | | | | | | |
| **Total direct costs** | $1,475,750 | $1,496,128 | $1,517,116 | $1,538,735 | $1,561,002 | $7,588,730 |
| Noncontract ind. costs | $221,363 | $224,419 | $227,567 | $230,810 | $234,150 | $1,138,310 |
| **Total direct+indirect costs** | $1,697,113 | $1,720,547 | $1,744,684 | $1,769,545 | $1,795,152 | $8,727,040 |
| | | | | | | |
| Property/insurance costs | $43,125 | $43,125 | $43,125 | $43,125 | $43,125 | $215,625 |
| Construction contracts | $0 | $0 | $0 | $0 | $0 | $0 |
| | | | | | | |
| **Total costs** | $1,740,238 | $1,763,672 | $1,787,809 | $1,812,670 | $1,838,277 | $8,942,665 |

## G. MINE STATUS: DEWATERING AND SITE TRANSFER ISSUES

Here we discuss briefly the status of the site and of property availability, issues that have important implications for this proposal.

In early June 2003 the Homestake Mine owner, Barrick Gold of Toronto, turned off the pumps that dewater the lower portions of the mine. This decision was opposed by our collaboration because we felt that continued maintenance of the site would be a less expensive alternative while also protecting the mine's science potential. However Barrick expressed concerns about the unknown timescale and outcome of the NUSEL approval process within the NSF. The resulting flooding is a slow process: roughly 25 years will be required to completely fill the mine. We estimate that in 18 months water will reach the important 7400-ft level. (The uncertainty on this estimate is at least a factor of two.)

Within our collaboration the earth scientists expect, with the buildup of water pressure at the 8000-ft level, that water will be forced into fissures in the rock. This recharging could threaten significant portions of the geomicrobiology program. (The water originates from higher levels in the mine, and thus disturbs the microbial conditions of the 8000-ft level.) The collaboration's hydrologists are also concerned. Prior to flooding the mine was in a steady-state condition that made it amenable to hydrological modeling. Calculations by Brian McPherson, an earth scientist who has advised our collaboration, now indicate that accurate modeling will be much more difficult. McPherson estimates that a year of measurements and theory will be needed before the feasibility of quantitative modeling, under these new conditions, can be assessed.

Physics concerns involve damage to the mine's infrastructure due to the flooding and due to the loss of ventilation in the lower portions of the mine. As ambient temperatures at depth are well above 120 F, the mine will be exposed to hot and humid conditions for a prolonged period – most likely at least two years. The task of dewatering is not trivial: severely flooded deep mines in South Africa have been dewatered, but the cost and environmental engineering requirements were substantial. After dewatering, there will be concerns about ground support conditions as well as the potential of trapped water. Thus a careful program of inspections and, quite possibly, repairs will have to be conducted. The contractor who does the underground work – both mine rehabilitation and NUSEL construction – will face additional hazards. This is why we included (but did not cost) General Conditions in the WBS-1: we should expect to pay some premium to the contractor because of adverse mine conditions. But we cannot, at this point, estimate what this additional cost for reconstruction/construction will be.

Clearly a key issue is early dewatering. If South Dakota can intervene before water reaches the 7400-ft level (the base of the #4 Winze), it will be very helpful to NUSEL. While the state has resources for dewatering ($10M provided in FY02), this is unlikely to be sufficient to complete the job. Given the alternative – additional delays and additional reconstruction – we hope NSF will consider aiding South Dakota.

The other great uncertainty affecting NUSEL-Homestake is the absence of a site agreement between South Dakota and Barrick Gold. South Dakota has stated its willingness to accept the needed portions of the Homestake site. Barrick Gold has stated its willingness to donate the property to the state. The difficulty, unresolved now for 2.5 years, is the liability protection Barrick Gold has requested. We included an item in WBS-1 for insurance, but did not estimate a cost. The costs the state will shoulder will be passed on to NUSEL, as NUSEL is the only reason the site is being transferred. Thus the required annual insurance premium will be an addition operations cost. It is also likely, over NUSEL's lifetime, that additional contributions will be required in order to build an Environmental Trust Fund for the site. This Trust would fund insurance and maintenance of the site after NUSEL has finished its life span.

Recently there have been public discussions indicating that Barrick Gold may ask that some of the existing underground infrastructure not be used by NUSEL. This may be an effort to limit company liability. The low cost of underground access in this proposal can be attributed to efficient use of Homestake infrastructure. Some of the underground facilities, as we have noted, are just a few years old. If replacing underground infrastructure proves to be a necessary condition for transfer of the site, this Reference Design will have to be modified. This will make NUSEL more expensive, but it may also allow us to improve designs (which will now be science specific).

While the flooding is a setback and possible loss of infrastructure a concern, we emphasize that Homestake remains a strong NUSEL site, regardless. A recent NSF geotechnical committee noted that the known integrity of Homestake rock is of utmost importance. It is one of the few locations in the US where large cavities have been excavated and studied.

However, clearly it is better if the dewatering commences soon and the useful underground infrastructure of Homestake is preserved. This will result in lower construction costs and shorter time to first science. NSF will be helping the science community if it can be active in such issues.